\documentclass[12pt]{article}
\pdfoutput=1

\usepackage[height=8.85in,width=6.45in]{geometry}

\usepackage[T1]{fontenc}
\usepackage{times}
\usepackage{microtype}

\usepackage{amsmath,amsfonts,amssymb}
\usepackage{bbm}
\usepackage{mathrsfs}
\usepackage{array}
\usepackage{booktabs}
\usepackage{placeins}
\usepackage{graphicx}
\usepackage[font=small,labelfont=bf]{caption}

\usepackage{tocloft}
\makeatletter
\let\qf@startsection\@startsection
\renewcommand{\@startsection}[6]{%
  \qf@startsection{#1}{#2}{#3}{#4}{#5}{#6\boldmath}}
\makeatother

\usepackage[dvipsnames,svgnames]{xcolor}
\definecolor{PaperLink}{RGB}{155,70,70}
\usepackage{tikz}
\usetikzlibrary{decorations,decorations.pathreplacing,decorations.markings,arrows.meta}
\usepackage[numbers,sort&compress]{natbib}
\usepackage[colorlinks,allcolors=PaperLink,linktocpage]{hyperref}
\newcommand{\figref}[2]{\hyperref[#1]{\ref*{#1}#2}}

\hypersetup{
  pdfstartview={FitH},
  pdftitle={Chern-Simons Interfaces and Minimal Models from Quantum Groups},
  pdfauthor={Suriyah Rajalingam Kannagi and Sergio Sanjurjo}
}

\numberwithin{equation}{section}
\newcommand{\fsu}{\mathfrak{su}}
\newcommand{\fu}{\mathfrak{u}}
\newcommand{\fgl}{\mathfrak{gl}}
\newcommand{\fsl}{\mathfrak{sl}}

\newcommand{\bZ}{\mathbb{Z}}
\newcommand{\bR}{\mathbb{R}}
\newcommand{\bC}{\mathbb{C}}
\newcommand{\bP}{\mathbb{P}}

\newcommand{\cA}{\mathcal{A}}
\newcommand{\cO}{\mathcal{O}}
\newcommand{\cC}{\mathcal{C}}
\newcommand{\cH}{\mathcal{H}}
\newcommand{\cL}{\mathcal{L}}

\newcommand{\qpsi}{\Psi}
\newcommand{\bqpsi}{\bar{\qpsi}}
\newcommand{\bpsi}{\bar{\psi}}
\newcommand{\bchi}{\bar{\chi}}

\newcommand{\NumericsOperatorSize}{12}

\begin{document}
\bibliographystyle{JHEP}
\hypersetup{pageanchor=false}

\begin{titlepage}
\vspace*{8mm}
\begin{center}

\vspace{6mm}

{\LARGE\bfseries\boldmath Chern--Simons Interfaces and Minimal Models \\ from Quantum Groups\par}

\vspace{8mm}

{\large Suriyah Rajalingam Kannagi\textsuperscript{1,2}\qquad Sergio Sanjurjo\textsuperscript{1,2}\par}

\vspace{8mm}

{\small\itshape\textsuperscript{1}Perimeter Institute for Theoretical Physics,\\[1mm] Waterloo, Ontario, N2L 2Y5, Canada\par}

\vspace{4mm}

{\small\itshape\textsuperscript{2}Department of Physics and Astronomy, University of Waterloo,\\[1mm] Waterloo, Ontario, N2L 3G1, Canada\par}

\vspace{5mm}

\end{center}

\vspace{3mm}

\begin{abstract}
\noindent
Coupling two-dimensional fundamental Dirac fermions to $SU(N)_k$ Chern--Simons theory produces non-chiral interfaces that in the infrared flow to a compact boson and a $W_N$ minimal model. We propose a description of these interfaces as continuum limits of $q$-deformed fermion chains. The $q$-deformation is achieved by coupling a chain of one-dimensional fundamental complex fermions to $SU(N)_k$ Chern--Simons theory, keeping the underlying topological gauge field continuous. The operator algebra is controlled by quantum groups, which provide an algebraic perspective on the usual categorical formulation of anyon chains. As an application, we derive a Lieb--Schultz--Mattis constraint involving baryon-number and lattice translation, obstructing a unique symmetric gapped ground state. At unit filling, we gap out the compact boson sector with a Mott term and obtain standard anyon chain Hamiltonians previously studied in the literature. This provides an anomaly argument for the existence of $N$ copies of minimal model sectors permuted by lattice translation in the infrared.
\end{abstract}

\vfill
\noindent\rule{0.3\textwidth}{0.4pt}\\[1.5mm]
{\footnotesize\hypersetup{urlcolor=black}%
 \href{mailto:srajalingamkannagi@perimeterinstitute.ca}{srajalingamkannagi@perimeterinstitute.ca}\\[0.5mm]
 \href{mailto:ssanjurjo@perimeterinstitute.ca}{ssanjurjo@perimeterinstitute.ca}\par}
\end{titlepage}
\hypersetup{pageanchor=true}

\setcounter{tocdepth}{2}
\pagenumbering{roman}
{\small\pagestyle{empty}\tocloftpagestyle{empty}\tableofcontents\clearpage}
\pagenumbering{arabic}

\section{Introduction}
\label{sec:introduction}

Gauge theories are ubiquitous in physics, and at low energies they are often strongly coupled. A particularly simple example is discrete gauge theory, which has no propagating bulk degrees of freedom and is therefore topological. When coupled to matter, it only affects the global structure of the matter and leaves its local dynamics, such as its stress tensor, unchanged. In contrast, gauge theories such as $SU(N)$ Yang--Mills theory are not topological and affect the local degrees of freedom of the matter they are coupled to. Three-dimensional $SU(N)_k$ Chern--Simons theory~\cite{Witten1989:QuantumFieldTheory} has no propagating bulk degrees of freedom and is therefore topological, but, unlike a discrete gauge theory, it changes the stress tensor of the matter it is coupled to, thereby affecting quantities such as scaling dimensions and spins. It is thus a natural intermediate between discrete gauge theories and non-topological gauge theories. The topological nature of the gauge field allows us to ask whether critical matter coupled to Chern--Simons theory can be described by discretizing only the matter, while keeping the gauge field continuous. We pursue this question for one-dimensional chains, whose proposed low-energy limits describe two-dimensional non-chiral interfaces in three-dimensional Chern--Simons theory.

We implement this idea by coupling a lattice of fundamental complex fermions, governed by the familiar nearest-neighbor hopping Hamiltonian, to $SU(N)_k$ Chern--Simons theory. This deforms the operator algebra of the free fermions into that of a quantum group with deformation parameter $q=\exp[i\pi/(k+N)]$. Based on anomaly arguments and numerical evidence, we conjecture that the low-energy spectrum of these $q$-deformed fermion chains is that of a compact boson and a $W_N$ minimal model glued together in a specific way. For $q$ a root of unity, corresponding to unitary Chern--Simons theory, we find null states in the $q$-fermion Fock space. Such null states are expected, as in the BRST formalism of gauge theory, and here they are closely related to the representation theory of quantum groups at roots of unity.

We emphasize that most of our analysis is another perspective on the well-established field of anyon chains \cite{Feiguin2007:InteractingAnyons,Trebst2008:FibonacciAnyonModels,Gils2013:AnyonicQuantumSpin}. These works use the data of the underlying modular tensor category to determine the Hamiltonian. In contrast, we provide an algebraic understanding of these Hamiltonians through the action of quantum-group generators on Hilbert spaces that are representations of the quantum group. There is also a long history of connections between minimal models and quantum groups at roots of unity \cite{PasquierSaleur1990:CommonStructures,JimboKunibaMiwaOkado1988:AnFaceModels,BazhanovReshetikhin1990:RSOSCFT}, which are closely related to some of the ideas presented in this paper. Itinerant non-Abelian anyons, that is, anyon chains with a direct sum of anyons at every site, have also been found to have critical spectra with total central charge $1+c_{\rm neutral}$ \cite{Poilblanc2012:FractionalizationItinerantAnyons,Poilblanc2013:ItinerantAnyons}.

\paragraph{Summary.} We give here a high-level summary of the main points of the paper, for a quick grasp of its contents.
\begin{itemize}
  \item \textit{Hilbert space.} We consider a chain of $m$ sites, each carrying a complex fermion in the fundamental representation of $SU(N)$. When coupled to Chern--Simons theory, the local Fock space of each fermion becomes a sum of the antisymmetric representations $\omega_r$, $r=0,\ldots,N$, of the quantum group $U^{\mathrm{L}}_q(\fsl_N)$ at the root of unity $q=\exp[i\pi/(k+N)]$, and the Fock space of the chain is the tensor product of $m$ such local spaces,
  \[
  \begin{aligned}
    \Lambda_q^\bullet\bigl(\bC_q^N\bigr)&=\bigoplus_{r=0}^{N}\Lambda_q^r\bigl(\bC_q^N\bigr)\cong\bigoplus_{r=0}^{N}\mathcal{L}_{\omega_r},\\
    \Lambda_q^\bullet\bigl(\bC_q^N\otimes\bC_q^m\bigr)&\cong\Lambda_q^\bullet\bigl(\bC_q^N\bigr)^{\otimes m}.
  \end{aligned}
  \]
  For large $m$, the $q$-fermion Fock space contains null states. To identify these null states we construct a map $\mu(q)$ from $SU(N)_k$ fusion trees to $q$-fermion states and identify the null states with those trees whose labels are non-integrable, i.e., those that leave the level-$k$ alcove. We also conjecture a positive-semidefinite inner product whose zero-norm states are exactly the null states.

  \item \textit{Hamiltonian.} The gauge-invariant operators acting on this Hilbert space include a second quantum group, $U^{\mathrm{L}}_q(\fgl_m)$, whose elements we interpret as mesons built from fundamental Wilson lines stretched between fermion worldlines. Since we consider $SU(N)_k$ Chern--Simons theory, we also have baryon operators. We explicitly construct these operators and derive their algebra with the mesons. In analogy with the free fermion Hamiltonian, we consider a nearest-neighbor hopping Hamiltonian whose terms are the mesons corresponding to the simple-root generators $E_i$ and $F_i$ of $U^{\mathrm{L}}_q(\fgl_m)$ \cite{Chun2017:JunctionsSurfaceOperators,Cautis2014:WebsQuantumSkew,Gaiotto2025:DbranesPlanar}, which satisfy $E_i^\dagger=F_i$ in the inner product above. Explicitly, the Hamiltonian is
  \begin{equation}
    H=-t\sum_i(E_i+F_i)-\mu F,
    \label{eq:intro-hamiltonian}
  \end{equation}
  where $F = \sum^{m}_{i=1} n_i$ is the total fermion number. For the periodic chain the meson part of the operator algebra is promoted to the quantum loop group $U^{\mathrm{L}}_q(L\fsl_m)$, and we decompose the Hilbert space on the circle into irreducible representations of this algebra.

  \item \textit{Symmetries.} Since the chain lives naturally in a three-dimensional bulk, it has non-invertible symmetries, which on the periodic chain are closed Wilson lines running parallel to the chain and form the Verlinde algebra of $\text{Rep}(SU(N)_k)$. In addition, the periodic Hamiltonian has translation symmetry $\bZ^{\text{Trans}}$ and a baryon-number symmetry $U(1)_B=U(1)_F/\bZ_N$. We derive a mixed Lieb--Schultz--Mattis (LSM) anomaly between translation and baryon number \cite{Lieb1961:TwoSolubleModels,Yamanaka1997:Luttinger1D,Oshikawa2000:CommensurabilityExcitationGap,ChengSeiberg2023:LSMLuttingerTHooft,Seifnashri2024:LSMAnomaliesGauging} from the operator that inserts a $2\pi$ baryon flux,
  \[
    U^B_{2\pi}=\mathbf{J}_{\frac{1}{2}}^{\dagger}\exp\!\left(\frac{2\pi i}{mN}\sum_{j=1}^{m}j n_j\right),
  \]
  where $\mathbf{J}_{\frac{1}{2}}$ is the simple-current Wilson line of $SU(N)_k$ wrapped around the bond between sites $m$ and $1$, which acts as a local Verlinde operator \cite{Verlinde1988:FusionRulesModular,Gaiotto2014:OpenVerlindeLines}. This anomaly excludes a unique symmetric gapped ground state at every filling $0<\nu<N$, where $\nu$ is the number of fermions per site.

  \item \textit{Continuum limit.} Let $M^+_{N,k-1}$ be the non-chiral interface   obtained by coupling a two-dimensional fundamental Dirac fermion to $SU(N)_k$ Chern--Simons theory \cite{GaiottoZeng2026:InterfaceMinimalModel}. The spectrum of this interface is that of the diagonal rational CFT of $U(1)^{\mathrm{lat}}_N\times\mathcal{W}_{N,k-1}$, a compact boson glued to the $W_N$ minimal model whose chiral algebra is the coset \cite{Bais1988:CosetConstructionExtendedVirasoro,BouwknegtSchoutens1993:WSymmetry}
  \[
  \mathcal{W}_{N,k-1}=\frac{SU(N)_{k-1}\times SU(N)_1}{SU(N)_k}.
  \]
  This infrared fixed point can be understood by regularizing the non-chiral Dirac fermion into two chiral halves, each of which becomes a chiral interface between two different Chern--Simons theories, $SU(N)_{k-1}$ and $SU(N)_k$, across which the chiral fermions raise the level by one \cite{GaiottoRapcak2019:VertexAlgebrasCorner,GaiottoZeng2026:InterfaceMinimalModel}, as sketched in Figure~\ref{fig:cs-slab}. Each half carries the central charge
  \[
        c_{N,k-1}=1+c_{\mathcal{W}_{N,k-1}}=1+(N-1)\left(1-\frac{N(N+1)}{(N+k-1)(N+k)}\right).
  \]
  Table~\ref{tab:intro-examples} lists examples, with $c_{\rm total}=c_{N,k-1}$ and $c_{\rm neutral}=c_{\mathcal{W}_{N,k-1}}$.
\begin{figure}[!htbp]
  \centering
  \includegraphics{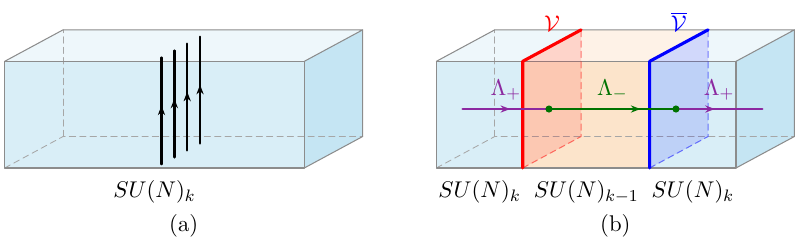}
\caption{(a) The fermions of the chain, drawn as black worldlines, in $SU(N)_k$ Chern--Simons theory. (b) In the infrared they flow to the non-chiral interface $M^+_{N,k-1}$, which can be regularized to chiral halves $\mathcal{V}$ and $\overline{\mathcal{V}}$ enclosing a region of $SU(N)_{k-1}$, with local operators attached to Wilson line $\Lambda_+$ in the exterior theory and $\Lambda_-$ in the emergent interior theory}
  \label{fig:cs-slab}
\end{figure}

  By the operator-state correspondence, the states of this interface are identified with operators that may live at the ends of bulk Wilson lines, see Figure~\figref{fig:cs-slab}{b}. This interface construction should be contrasted with the usual sandwich construction of a coset CFT $G/H$ \cite{FuchsRunkelSchweigert2002:TFTConstructionI,KapustinSaulina2011:SurfaceOperators3dTFT,Fuchs2013:BicategoriesBoundarySurfaceDefects}, in which the chiral and anti-chiral algebras are boundaries, the region between them is the modular tensor category $\mathrm{MTC}(G/H)$ that hosts the Verlinde lines of the CFT, while the outside is the vacuum. In our case the chiral and anti-chiral algebras live on genuine interfaces rather than boundaries. Interface operators are labeled by the triples $(\Lambda_-, \omega_r ; \Lambda_+)$ satisfying the constraint
  \[
  |\Lambda_+|-|\Lambda_-|\equiv r\pmod N .
  \]
  We conjecture based on anomaly arguments and numerical evidence that in the continuum our lattice $q$-fermion chain flows to $M^+_{N,k-1}$ only up to exactly marginal deformations, which are generated along the RG flow and replace $U(1)^{\mathrm{lat}}_N$ by $U(1)_{\tilde R^2}$. We also describe the case where one dimensional lattice fermions are coupled to $U(1)_k$ Chern--Simons theory ($N=1$) in Appendix~\ref{app:u1-chain}. In this case, we find an exact realization of the $q$-fermion Hamiltonian as a free-fermion chain with a gauged $\bZ_k$. Using the usual map to the integrable XXZ chain \cite{Pereira2007:XXZStructureFactor}, we give an explicit Hamiltonian that flows to any radius in the range $0<\tilde R^2<2k^2$, including those predicted by the chiral interface construction above, which removes for the case of $U(1)_k$ the caveat that the lattice and the interface agree only up to exactly marginal deformations.

  \item \textit{Anomaly matching.} In the continuum, lattice translation becomes an emanant internal symmetry, and the exact Wilson lines of the external $SU(N)_k$ Chern Simons theory become topological defects of the interface \cite{BuicanGromov2017:AnyonicChainsDefects,Belletete2020:TopologicalDefectsPeriodicRSOS}. The mixed anomaly between $U(1)_B$ and translation is matched by a mixed anomaly between $U(1)_B$ and the emanant axial rotation, which we find $M^+_{N,k-1}$ saturates \cite{ChengSeiberg2023:LSMLuttingerTHooft}. We further show that $U(1)_B$, $\bZ^{\text{Trans}}$ and the Wilson lines $\text{Rep}(SU(N)_k)$ forbid every relevant scalar deformation of $M^+_{N,k-1}$ that the RG flow could generate, at least for fillings close to half filling, so the interface is a stable fixed point protected by these symmetries \cite{Feiguin2007:InteractingAnyons,Pfeifer2012:TranslationTopologyCriticality,Gils2013:AnyonicQuantumSpin}. This argument does not forbid the exactly marginal current-current coupling $J_LJ_R$ \cite{ForsteRoggenkamp2003:CurrentCurrent}, which is generated and moves the radius $\tilde R$ away from its free-fermion value $\tilde{R}_0=\sqrt N$. Close to half filling, we estimate the change in $\tilde R$ perturbatively in $\theta$, where $q=e^{i\theta}$. We also point out the special case of $N=2$ at $\nu=1$, which possesses an exact lattice $\eta$-pairing symmetry that enlarges $U(1)_B$ to $SO(3)$ and pins the radius to the free-fermion self-dual value $\tilde{R}_0=\sqrt2$.

  \item \textit{Mott limit.} At unit filling, we add to the Hamiltonian the Mott term
  \[
    H=-t\sum_i(E_i+F_i)+U\sum_{i=1}^{m}(n_i-1)^2,
  \]
  and as $U$ passes a critical value $U_c$ the chain undergoes a Berezinskii--Kosterlitz--Thouless transition \cite{Berezinskii1971:DestructionLongRangeOrderI,Berezinskii1972:DestructionLongRangeOrderII,KosterlitzThouless1973:OrderingMetastability,Assaraf1999:MetalInsulatorSUN} that gaps out the charge sector. For large $U$, second-order perturbation theory gives the effective Hamiltonian
  \[
    H_{\rm eff}=-\frac{t^2}{U}\sum_i\bP^{(\omega_1)}E_iF_i\bP^{(\omega_1)}=-\frac{[2]_q\,t^2}{U}\sum_i\Pi_i^{(\omega_2)},
  \]
  where $\bP^{(\omega_1)}$ projects onto one fermion per site and $\Pi_i^{(\omega_2)}$ projects two neighboring fundamentals onto their $q$-antisymmetric channel. This is exactly the anyon chain of $SU(N)_k$ fundamentals, in which the generators $[2]_q\Pi_i^{(\omega_2)}$ obey the Temperley--Lieb algebra for $N=2$ \cite{TemperleyLieb1971:PercolationColouring,Feiguin2007:InteractingAnyons,Trebst2008:FibonacciAnyonModels} and, more generally, the Hecke algebra \cite{Jimbo1986:qAnalogueHeckeYangBaxter,JimboKunibaMiwaOkado1988:AnFaceModels}. With no exactly marginal coupling left, $H_{\rm eff}$ flows to a direct sum of $N$ copies of $\mathcal{W}_{N,k-1}$. The LSM anomaly of the gapless charged phase is transmuted into a mixed anomaly between translation and the simple-current loop $\mathbf{J}$ around a bond, which now commutes with $H_{\rm eff}$ and becomes a topological point operator generating a $\bZ_N$ one-form symmetry. Translation emanates to a discrete symmetry $\tau$ that permutes the $N$ copies in the infrared, and together $\tau$ and $\mathbf{J}$ obey the relation,
  \[
    \tau\,\mathbf{J}\,\tau^{-1}=\omega\,\mathbf{J},\qquad \omega=e^{2\pi i/N}.
  \]
  We refer the reader to Section~\ref{sec:projector-continuum} for a more detailed discussion. Figure~\ref{fig:qf-anomaly-rectangle} summarizes the anomaly matching at unit filling.
\end{itemize}

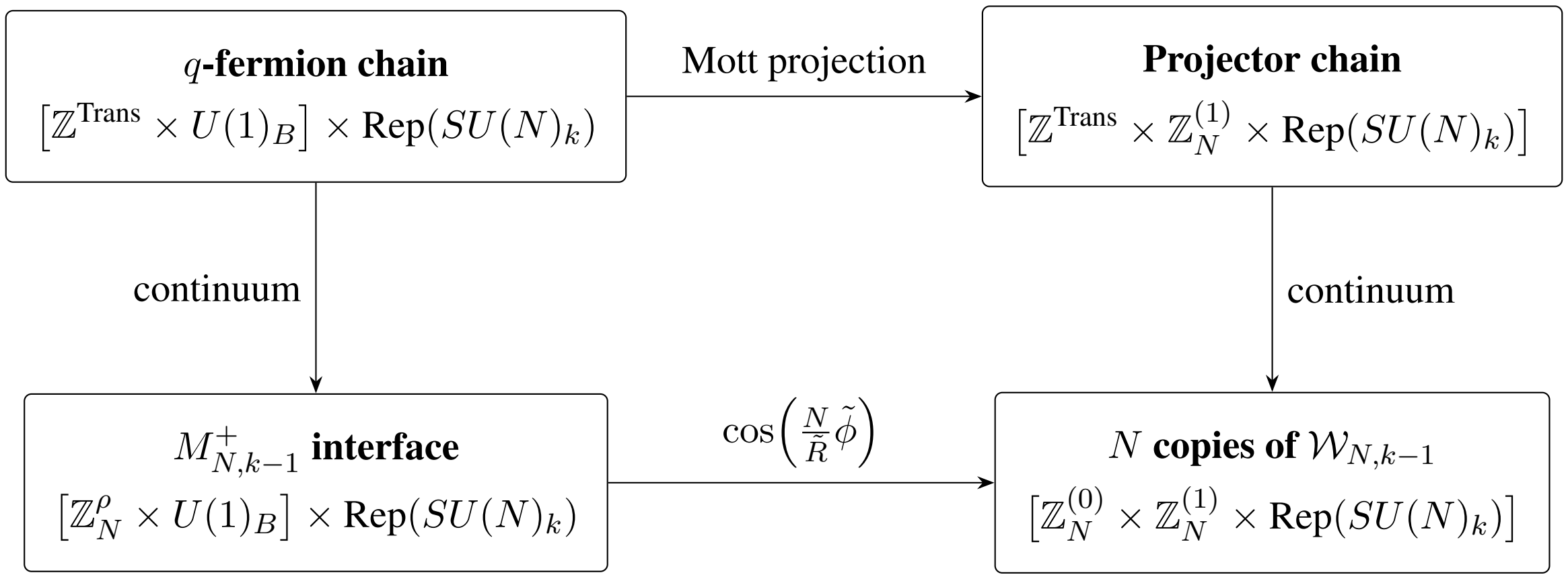
\begin{figure}[t]
  \centering
  \begin{tikzpicture}[
    >=Stealth,
    box/.style={draw, rounded corners=2pt, inner sep=8pt,
                minimum width=46mm, minimum height=15mm,
                align=center, font=\small},
    lbl/.style={font=\small, align=center}]
    \node[box] (ul) at (0,0) {$\begin{gathered}
      q\textbf{-fermion chain}\\
      \bigl[\mathbb{Z}^{\text{Trans}}\times U(1)_B\bigr]\times\text{Rep}(SU(N)_k)
    \end{gathered}$};
    \node[box] (ur) at (9.4,0) {$\begin{gathered}
      \textbf{Projector chain}\\
      \bigl[\mathbb{Z}^{\text{Trans}}\times\mathbb{Z}^{(1)}_N\times\text{Rep}(SU(N)_k)\bigr]
    \end{gathered}$};
    \node[box] (ll) at (0,-3.8) {$\begin{gathered}
      M^+_{N,k-1}\textbf{ interface}\\
      \bigl[\mathbb{Z}^{\rho}_N\times U(1)_B\bigr]\times\text{Rep}(SU(N)_k)
    \end{gathered}$};
    \node[box] (lr) at (9.4,-3.8) {$\begin{gathered}
      N\textbf{ copies of }\mathcal{W}_{N,k-1}\\
      \bigl[\mathbb{Z}^{(0)}_N\times\mathbb{Z}^{(1)}_N\times\text{Rep}(SU(N)_k)\bigr]
    \end{gathered}$};
    \draw[->] (ul) -- node[lbl,above]{Mott projection} (ur);
    \draw[->] (ll) -- node[lbl,above]
      {$\cos\!\left(\tfrac{N}{\tilde R}\tilde\phi\right)$} (lr);
    \draw[->] (ul) -- node[lbl,left]{continuum} (ll);
    \draw[->] (ur) -- node[lbl,right]{continuum} (lr);
  \end{tikzpicture}
\caption{Anomaly matching for the $q$-fermion chain at $\nu=1$, where the projector chain is $H_{\rm eff}$. Square brackets group the symmetries with mixed anomalies. Here $\mathbb{Z}^{\rho}_N$ is generated by $\rho=e^{-i\pi Q_B}\tau$ in the charged theory, while $\mathbb{Z}^{(0)}_N$ denotes the neutral emanant translation; $\mathbb{Z}^{(1)}_N$ is the one-form symmetry generated by $\mathbf{J}$, and $\tilde\phi$ is the dual boson of radius $\tilde R$.}
  \label{fig:qf-anomaly-rectangle}
\end{figure}

\begin{table}[h]
  \centering
  \caption{Examples of the proposed charged continuum theory and its unit-filling Mott limit.}
  \begin{tabular}{@{}cccc@{}}
    \toprule
    Bulk theory & Neutral CFT & $c_{\rm total}$ & $c_{\rm neutral}$ \\
    \midrule
    $SU(2)_2$ & Ising & $3/2$ & $1/2$ \\
    $SU(2)_3$ & Tricritical Ising & $17/10$ & $7/10$ \\
    $SU(3)_2$ & Three-state Potts & $9/5$ & $4/5$ \\
    \bottomrule
  \end{tabular}
  \label{tab:intro-examples}
\end{table}

\paragraph{Outlook.}
We conclude by outlining several directions for future work. This paper treats the Mott transition at unit filling, but other integer fillings $\nu=p$ are also interesting. At these fillings a large Mott term projects every site onto $\omega_p$ and gives the projector Hamiltonians given in Appendix~\ref{sec:other-fillings}. Their infrared theories could be identified as for unit filling, by finding the first symmetry-allowed operators of $M^+_{N,k-1}$ to become relevant, and compared with the continuum limit of these projector chains.

Another related direction is to consider $q$-fermions that transform not in the fundamental representation but in a general representation $R$ of $U^{\mathrm{L}}_q(\fsl_N)$. Following the expectations of this paper, the continuum limit should then be a compact boson together with a generalized diagonal Goddard--Kent--Olive (GKO) coset \cite{GoddardKentOlive1986:UnitaryVirasoro},
\begin{equation}
 \frac{SU(N)_{k-s}\times SU(N)_s}{SU(N)_k},
 \label{eq:outlook-diagonal-coset}
\end{equation}
with the level $s$ set by $R$. For the fundamental representation, $s=1$ and the coset reduces to $\mathcal{W}_{N,k-1}$.

Another interesting direction is to consider $N_F$ flavors of $q$-fermions, whose free counterparts have an $SU(N_F)$ flavor symmetry, and to understand which theory describes their low-energy limit. Since the currents of the free fermions contain $U(1)\times SU(N)_{N_F}\times SU(N_F)_N$ as a conformal subalgebra \cite{DiFrancescoMathieuSenechal1997:CFT}, a natural candidate is the coset of Equation~\eqref{eq:outlook-diagonal-coset} with $s=N_F$, together with a compact boson and $SU(N_F)_N$ flavor currents \cite{GaiottoZeng2026:InterfaceMinimalModel}.

From the algebraic point of view, taking $q$ at a root of unity is a choice, which we made to reach the $W_N$ minimal models, and one can equally take $q$ generic and study the low-energy limit. The Hamiltonian must then be compatible with the corresponding $*$-structure of the quantum group. For real $q>0$, the compact $*$-structure compatible with the coproduct of Equation~\eqref{eq:q-fermion-coproduct} is
\[
  E_i^\dagger=F_iK_i,\qquad F_i^\dagger=K_i^{-1}E_i,\qquad K_i^\dagger=K_i,
\]
where $K_i=q^{n_i-n_{i+1}}$. A simple Hamiltonian could be,
\[
  H=-t\sum_i\bigl(E_i+F_iK_i\bigr)-\mu F ,
\]
which is Hermitian and has a real spectrum. In the continuum, the chain at generic $q$ should flow to an interface in analytically continued Chern--Simons theory \cite{Witten2011:AnalyticContinuationCS}, which itself lives at the boundary of the Kapustin--Witten twist of $\mathcal{N}=4$ super Yang--Mills theory \cite{KapustinWitten2007:ElectricMagneticLanglands,Witten2012:FivebranesKnots,GaiottoRapcak2019:VertexAlgebrasCorner}, and it would be interesting to identify the chiral algebra that controls it. Numerically, this set up is simpler than at the root of unity, since for real $q$ there are no null states. The periodic Hilbert space, however, becomes infinite-dimensional, since the label $\alpha$ in Equation~\eqref{eq:periodic-Hilbert-space} runs over all dominant weights of $SU(N)$ rather than the level-$k$ integrable weights.

Another important direction is to generalize our construction to two-dimensional lattice systems, with Wilson-line hoppings between neighboring sites in both directions. On the ring the hoppings generate the quantum loop algebra $U^{\mathrm{L}}_q(L\fsl_m)$ of Equation~\eqref{eq:periodic-quantum-loop}, and it would be interesting to identify the algebra they generate on a two-dimensional lattice, possibly related to the quantum toroidal algebra \cite{Hernandez2009:QuantumToroidal}. If behavior similar to what we found here persists, such lattices could be a window into Chern--Simons theories coupled to matter in $2+1$ dimensions and the infrared fixed points.

\begin{figure}[t]
\centering
\begin{tikzpicture}[line cap=round, line join=round, font=\small, >=stealth,
  site/.style={circle, fill=black, inner sep=1.55pt},
  osite/.style={circle, draw=black, fill=white, line width=0.6pt, inner sep=1.4pt},
  leg/.style={line width=1pt, legc},
  zig/.style={line width=1pt, zigc}]
  \definecolor{legc}{RGB}{0,114,178}
  \definecolor{zigc}{RGB}{213,94,0}
  \def\a{0.9}
  \def\h{0.9}
  \def\s{0}
  \def\D{1.1}
  \pgfmathsetmacro{\w}{4*\a+\s}
  \draw[zig] (0,\h/2) \foreach \i in {0,...,3}{ -- (\i*\a+\s,-\h/2) -- ({(\i+1)*\a},\h/2) } -- (4*\a+\s,-\h/2);
  \draw[leg] (0,\h/2) -- (4*\a,\h/2);
  \draw[leg] (\s,-\h/2) -- (4*\a+\s,-\h/2);
  \foreach \i in {0,...,4}{
    \node[site] at (\i*\a,\h/2) {};
    \node[osite] at (\i*\a+\s,-\h/2) {};
  }
  \node[legc, anchor=south] at (0.5*\a,\h/2+0.06) {$t$};
  \node[zigc, anchor=east] at (\s/2-0.08,0) {$t'$};
  \coordinate (O) at (\w+0.85,0);
  \draw[->, thick] (O) -- ++(1.15,\D-0.1) node[pos=0.55, anchor=south east] {$t'=0$};
  \draw[->, thick] (O) -- ++(1.15,-\D+0.1) node[pos=0.55, anchor=north east] {$t=0$};
  \pgfmathsetmacro{\X}{\w+0.85+1.15+0.35}
  \begin{scope}[shift={(\X,\D)}]
    \draw[leg] (0,\h/2) -- (4*\a,\h/2);
    \draw[leg] (\s,-\h/2) -- (4*\a+\s,-\h/2);
    \foreach \i in {0,...,4}{
      \node[site] at (\i*\a,\h/2) {};
      \node[osite] at (\i*\a+\s,-\h/2) {};
    }
    \draw[decorate, decoration={brace, amplitude=5pt}]
      (\w+0.27,\h/2+0.12) -- (\w+0.27,-\h/2-0.12);
    \node[anchor=west, align=left] at (\w+0.47,0) {$\mathcal{W}_{N,k-1}^{\otimes 2}$\\two copies};
  \end{scope}
  \begin{scope}[shift={(\X,-\D)}]
    \draw[zig] (0,0) -- (\w,0);
    \foreach \j in {0,2,...,8}{\node[site] at (\j*\w/9,0) {};}
    \foreach \j in {1,3,...,9}{\node[osite] at (\j*\w/9,0) {};}
    \node[anchor=west, align=left] at (\w+0.47,0) {$\mathcal{W}_{N,k-1}$\\one copy};
  \end{scope}
\end{tikzpicture}
\caption{A ladder of two $q$-fermion chains, with hopping $t$ along the legs and zigzag hopping $t'$ between them. Switching off $t'$ leaves two decoupled chains with two copies of the neutral sector $\mathcal{W}_{N,k-1}$, while switching off $t$ leaves one zigzag chain of twice the length with a single copy.}
\label{fig:outlook-ladder}
\end{figure}
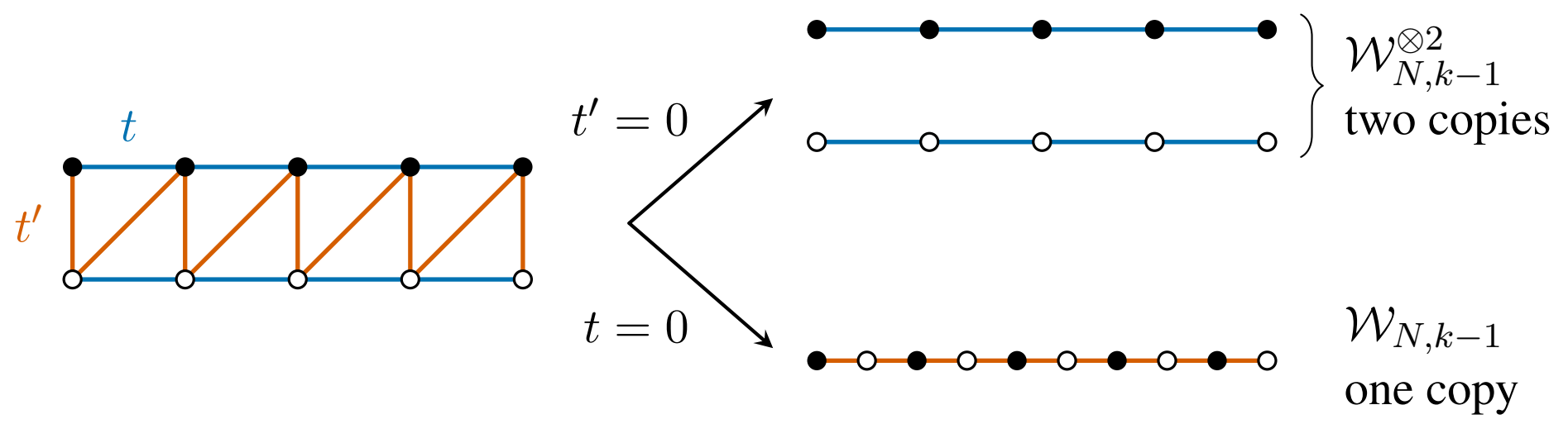

As an intermediate step toward such lattices, one can consider stacking several chains and turning on interactions between them, as studied for ladders of non-Abelian anyons \cite{Ludwig2011:TwoDimensionalQuantumLiquids,Soni2016:DopedAnyonLadders,Ayeni2018:LadderBraidedAnyons,Kirchner2026:FibonacciAnyonsLadder}. The simplest example is the two-leg ladder of Figure~\ref{fig:outlook-ladder}. With only the hopping $t$ along the legs, it consists of two decoupled chains with two copies of the neutral sector, $\mathcal{W}_{N,k-1}^{\otimes 2}$, while with only the zigzag hopping $t'$ it is a single chain of twice the length which in the continuum limit should flow to a single copy of  $\mathcal{W}_{N,k-1}$. It would be interesting to understand the RG flow between these two limits, and more generally the flows $\mathcal{W}_{N,k-1}^{\otimes n_1}\to\mathcal{W}_{N,k-1}^{\otimes n_2}$ with $n_2<n_1$ that stacks several layers of chains and in spirit could engineer RG flows between multi-factor interfaces in \cite{GaiottoZeng2026:InterfaceMinimalModel}.

On the holographic side, the interface duality of \cite{GaiottoZeng2026:InterfaceMinimalModel} embeds these minimal models \cite{Gaberdiel2011:AdS3Dual} in topological string holography and proposes the dual to be a coisotropic and an anti-coisotropic D-brane of the A-model topological string. In our construction the interfaces are instead engineered from one-dimensional fermions, whose Wilson lines are expected to be dual to three-dimensional D-branes of the A-model \cite{Gaiotto2025:DbranesPlanar}. It would be interesting to understand how the dual of the interfaces arises from these 3d branes.

For $N\geq2$ we believe that the $q$-fermion chains are not integrable; however, their Mott limit at unit filling is integrable \cite{PasquierSaleur1990:CommonStructures,JimboKunibaMiwaOkado1988:AnFaceModels,BlakeneyCorcoranDeLeeuw2026:ConstrainedIntegrability}. It would be interesting to understand our construction and the emergence of integrability in the Mott limit, possibly by connecting them to four-dimensional Chern--Simons theory \cite{CostelloWittenYamazaki2018:GaugeIntegrabilityII}.

\paragraph{Outline of the paper.}
The plan of the rest of the paper is as follows. In \textbf{Section~\ref{sec:free-fermion-review}} we review the free fermion chain, its continuum limit, its anomalies and its Mott transition. This review serves as a direct precedent for the rest of the paper, and readers familiar with this material can skip it. In \textbf{Section~\ref{sec:q-fermion-chain}} we construct the $q$-fermion chain. We introduce the $q$-fermion Fock space and the map relating it to the fusion-tree basis, and obtain the physical Hilbert space. We consider both the open and the periodic chain, and work out the symmetries and anomalies of the $q$-fermion chain. In \textbf{Section~\ref{sec:continuum-limit}} we construct the non-chiral interfaces, closely following \cite{GaiottoZeng2026:InterfaceMinimalModel}, identify the spectrum of the continuum limit in terms of $U(1)\times\mathcal{W}_{N,k-1}$, and match the symmetries and anomalies of the lattice. We also discuss the charge radius, which is not fixed by symmetry in general. In \textbf{Section~\ref{sec:mott-projection}} we take the large-$U$ limit at unit filling to obtain the familiar anyon-chain Hamiltonians found in literature, and discuss their continuum limit, symmetries and anomalies. The \textbf{Appendices} contain material that complements the main text. Appendix~\ref{app:uqglm-relations} reviews the finite and affine quantum groups and the quantum loop algebras used in the paper, together with their representation theory at roots of unity. Appendix~\ref{app:u1-chain} treats the exactly solvable case of the $U(1)_k$ chain. Appendix~\ref{app:su2k} works out the case of $SU(2)_k$ in detail. Appendix~\ref{app:baryon-algebra} contains the baryon operators and their algebra with the mesons. Appendix~\ref{app:derivations} collects the essential derivations for the $q$-fermion chain used in the main text. Finally, in Appendix~\ref{app:numerical-methods} we present numerical evidence for the claims made in the paper.

\section{Review of the free fermion chain}
\label{sec:free-fermion-review}
Consider a chain of $m$ sites, with $N$ one-dimensional complex fermions at each site. The site and flavor labels are $i,j=1,\ldots,m$ and $a,b=1,\ldots,N$, respectively. The creation operators $c_i^{a\dagger}$ transform in the fundamental representation of $SU(N)$ and obey the canonical anticommutation relations
\begin{equation}
  \{c_i^a,c_j^b\}=0,
  \qquad
  \{c_i^{a\dagger},c_j^{b\dagger}\}=0,
  \qquad
  \{c_i^a,c_j^{b\dagger}\}=\delta_{ij}\delta^{ab} .
  \label{eq:canonical-anticommutation}
\end{equation}
The algebra generated by these creation and annihilation operators is the complex Clifford algebra $\text{Cl}_{2Nm}(\bC)$. With the Hermitian structure $\bigl(c_i^a\bigr)^*=c_i^{a\dagger}$, it has a unique irreducible representation up to unitary equivalence \cite{Berezin1966:MethodSecondQuantization}. This is the fermionic Fock space obtained by choosing a vacuum $\lvert0\rangle$ with $c_i^a\lvert0\rangle=0$ for all $i$ and $a$,
\begin{equation}
  \mathcal{H}
  =
  \Lambda^\bullet (\bC^N\otimes\bC^m)
  \cong
  \bigotimes_{i=1}^m\Lambda^\bullet\bC^N .
  \label{eq:ff-fock-space}
\end{equation}
The tensor product is graded, the grading supplying the signs between different sites that make $c_i^a$ and $c_j^b$ anticommute for $i\neq j$. The Fock representation realizes the Clifford algebra \textit{faithfully},\footnote{An algebra $\cA$ is realized faithfully by a representation $\rho:\cA\rightarrow\text{End}_{\bC}(\mathcal{H})$ if $\ker\rho=\{0\}$.}
\begin{equation}
  \text{Cl}_{2Nm}(\bC)
  \cong
  \text{End}_{\bC}\bigl(\mathcal{H}\bigr) \cong \text{Mat}_{2^{Nm}}(\bC).
  \label{eq:ff-clifford-endomorphisms}
\end{equation}
The creation operators alone generate the exterior algebra $\Lambda^\bullet (\bC^N\otimes\;\bC^m)$, which is isomorphic to $\mathcal{H}$ as a vector space. We now single out the symmetry subalgebras of $\text{Cl}_{2Nm}(\bC)$ that are used throughout the paper. Consider the fermion bilinears constructed by summing over flavor indices,
\begin{equation}
    \widetilde{G}_{ij}
    =
    \sum_{a=1}^N c_i^{a\dagger}c_j^a,
    \qquad
    [\widetilde{G}_{ij},\widetilde{G}_{k\ell}]
    =
    \delta_{jk}\widetilde{G}_{i\ell}
    -
    \delta_{\ell i}\widetilde{G}_{kj}.
  \label{eq:ff-site-bilinears}
\end{equation}
These $\widetilde{G}_{ij}$ generate the Lie algebra $\fgl_m(\bC)$, which we call the \textit{site} algebra. Its traceless part
\begin{equation}
    G_{ij}
    =
    \widetilde{G}_{ij}-\frac{\delta_{ij}}{m} \sum_{k=1}^m \widetilde{G}_{kk},
    \qquad
    \sum_i G_{ii}=0,
  \label{eq:ff-traceless-site-generators}
\end{equation}
generates $\fsl_m(\bC)$. In particular, the simple-root and Cartan generators are
\begin{equation}
    E_i=G_{i,i+1},
    \qquad
    F_i=G_{i+1,i},
    \qquad
    n_i
    =
    \widetilde{G}_{ii}
    \qquad
    h_i
    =
    n_i-n_{i+1} ,
    \label{eq:ff-EF}
\end{equation}
Similarly, summing over sites rather than over flavors, we have the operators $\widetilde{Q}^{ab}=\sum_{i=1}^m c_i^{a\dagger}c_i^b$, which span the flavor algebra $\fgl_N(\bC)$, and their traceless parts $\fsl_N(\bC)$ are spanned by $Q^{ab}$. There is also the total fermion number operator,
\begin{equation}
    F
    =
    \sum_{i=1}^m\sum_{a=1}^N
    c_i^{a\dagger}c_i^a.
    \label{eq:ff-number}
\end{equation}
It spans the common center of both algebras, $\fgl_m(\bC)=\fsl_m(\bC)\oplus\bC F$ and $\fgl_N(\bC)=\fsl_N(\bC)\oplus\bC F$, since $F=\sum_i\widetilde{G}_{ii}=\sum_a\widetilde{Q}^{aa}$. All three of these algebras commute with each other,
\begin{equation}
    [G_{ij},Q^{ab}]=0, \qquad [G_{ij},F]=0, \qquad [Q^{ab},F]=0.
  \label{eq:ff-commuting-symmetry-algebras}
\end{equation}

Hermitian conjugation acts on these bilinears by $\widetilde{G}_{ij}^{\dagger}=\widetilde{G}_{ji}$ and $(\widetilde{Q}^{ab})^{\dagger}=\widetilde{Q}^{ba}$, so their anti-Hermitian elements span the compact real forms $\fu(m)$ and $\fu(N)$, which integrate to unitary actions of $U(m)$ and $U(N)$ on $\mathcal{H}$ with both $U(1)$ factors generated by $F$. The three algebras therefore act simultaneously on the Fock space as
\begin{equation}
  \fsl_N\oplus\fsl_m\oplus \fgl_{1}
  \label{eq:ff-symmetry-lie-algebra}
\end{equation}
or multiplicatively by their associated universal enveloping algebra
\begin{equation}
U\bigl(\fsl_N\bigr)\otimes U\bigl(\fsl_m\bigr)\otimes U\bigl(\fgl_{1}\bigr) .
  \label{eq:ff-symmetry-enveloping-algebra}
\end{equation}
This commuting action allows a multiplicity-free decomposition of the Fock space into irreducible representations of the three algebras. This is the statement of skew Howe duality \cite{Howe1989:RemarksClassicalInvariant,ChengWang2012:DualitiesRepresentations}:
\begin{equation}
\Lambda^\bullet(\bC^N\otimes\bC^m)
\cong
\bigoplus_{\lambda\subseteq(m^N)}
\cL^{\fsl_N}_{\overline\lambda}
\otimes
\cL^{\fsl_m}_{\overline{\lambda^{\mathrm T}}}
\otimes
\cL^{\fgl_1}_{|\lambda|}.
\label{eq:skew_howe_sl}
\end{equation}
Here $\cL^{\mathfrak{g}}_{\mu}$ denotes the irreducible $\mathfrak{g}$-module of highest weight $\mu$, so that $\cL^{\fgl_1}_{|\lambda|}$ is the one-dimensional module on which $F$ acts by $|\lambda|$. The label $\lambda\subseteq(m^N)$ is a Young diagram with at most $N$ rows and $m$ columns, $\lambda^{\mathrm T}$ is its transpose, and $|\lambda|$ is the number of boxes. The label $\overline\lambda$ is obtained by deleting full columns of height $N$, while $\overline{\lambda^{\mathrm T}}$ is obtained independently from the transposed Young diagram $\lambda^{\mathrm T}$ by deleting full columns of height $m$. The three labels together determine $\lambda$, since $|\lambda|$ fixes the number of deleted columns, so the decomposition is multiplicity-free. Later we will be interested in the gauge-invariant subalgebra that survives once the $SU(N)$ symmetry is gauged. As a precursor we ask the same question for the free fermion chain, keeping only the operators invariant under the global $SU(N)$ flavor rotations\footnote{This can be obtained by coupling the one-dimensional fermions to gauge fields on the links and taking the $g=0$ limit~\cite{KogutSusskind1975:HamiltonianLatticeGauge,Banuls2017:SU2LatticeGaugeBasis}.}
\begin{equation}
\text{Cl}_{2Nm}(\bC)^{SU(N)}
  \label{eq:ff-flavor-invariant-clifford-algebra}
\end{equation}
The gauge-invariant operators clearly include the meson operators $\widetilde{G}_{ij}$, which span the whole of $\fgl_m(\bC)$; and because the invariance imposed is $SU(N)$ and not $U(N)$, the determinant representation is trivial as well, so we also have baryons \cite{Weyl1946:ClassicalGroups,KraftProcesi1996:ClassicalInvariantPrimer,Fliss2026:FortuityComplexityQuarkModel},
\begin{equation}
\begin{aligned}
  B^{\dagger}_{i_1,i_2,\ldots,i_N} &= \epsilon_{a_1, a_2, \ldots, a_N} c_{i_1}^{a_1\dagger} c_{i_2}^{a_2\dagger} \cdots c_{i_N}^{a_N\dagger}
\end{aligned}
  \label{eq:ff-baryon-creation-operator}
\end{equation}
Their algebra with the mesons is described in Appendix~\ref{app:free-baryons}.

\subsection{Hopping chain and continuum limit}
\label{sec:free-fermion-dynamics}
Now consider a nearest-neighbor hopping Hamiltonian with a chemical potential on this fermionic Fock space,
\begin{equation}
  H_{0}
  =
  -t\sum_{i}
  \left(E_i+F_i\right) - \mu F,
  \label{eq:ff-hopping}
\end{equation}
with periodic boundary conditions $c_{i+m}^a=c_i^a$ and even $m$, so that $E_m:=\widetilde{G}_{m,1}$ and $F_m:=\widetilde{G}_{1,m}$.\footnote{In this free-fermion review we work with $m\in2\bZ$ and periodic boundary conditions. The dependence on the parity of $m$ and other boundary conditions can be found in the literature~\cite{ChengSeiberg2023:LSMLuttingerTHooft,SeibergShao2024:MajoranaChainIsing,Chatterjee2025:QuantizedAxialCharge}.} For generic $\mu$, the Hamiltonian is invariant under the ultraviolet symmetry
\begin{equation}
 G_{\text{UV}} = (U(N) \rtimes \bZ_2^{\mathsf{T}}) \times D_m, \qquad D_m = \bZ_m^{\text{Trans}} \rtimes \bZ_2^{\mathsf{P}}
 \label{eq:ff-onsite-symmetry}
\end{equation}
where in $U(N) = (SU(N) \times U(1)_{F})/\bZ_N$ the centers of the flavor and charge groups are identified and $\bZ_2^{\mathsf{T}}$ is time reversal and $\bZ_2^{\mathsf{P}}$ is lattice reflection. At half filling, $\mu=0$, separating each complex fermion into two Majorana fermions $c^a_j = \frac{i^{-j}}{\sqrt{2}}(a^a_j + i b^{a}_j)$, the symmetry enlarges to
\begin{equation}
  G^{\mu=0}_{\text{UV}}=O(2N)\rtimes(D_m\times\bZ_2^{\mathsf{T}})
  \label{eq:ff-half-filling-uv-symmetry}
\end{equation}
with the particle-hole symmetry $\mathsf{C}c^{a}_{j}\mathsf{C}^{-1}=(-1)^{j}c^{a\dagger}_{j}$ in $O(2N)$. There is also a translation of a single Majorana fermion $\mathsf{T}_{b} b^a_{j} \mathsf{T}^{-1}_{b} = b^a_{j+1}$, with $b^a_{j+m}=(-1)^{m/2}b^a_j$ (and likewise for $a^a_j$) and the other Majoranas fixed, which preserves the Hamiltonian at half filling \cite{SeibergShao2024:MajoranaChainIsing,Chatterjee2025:QuantizedAxialCharge}. In momentum space,
\begin{equation}
  H_{0}=\sum_{a=1}^N\sum_{n=-m/2}^{m/2-1}\bigl(-2t\cos k-\mu\bigr)c_k^{a\dagger}c_k^a ,\qquad k=\frac{2\pi n}{m} ,
  \label{eq:ff-momentum-space-hamiltonian}
\end{equation}
and for $|\mu|<2t$ the low-energy modes lie near the Fermi points $k=\pm k_F$, with $2t\cos k_F=-\mu$. In the continuum limit, the one-dimensional complex fermions can be decomposed into slowly varying left- and right-moving continuum fields,
\begin{equation}
  \frac{c_j^a}{\sqrt{a_0}}
  \sim
  e^{-i k_F j}\psi_L^a(x_j)
  +e^{i k_F j}\psi_R^a(x_j).
  \label{eq:lattice-continuum-fields}
\end{equation}
Here $a_0$ is the lattice spacing and $x_j=ja_0$.\footnote{Both chiralities are required by the Nielsen--Ninomiya theorem \cite{Nielsen1981:AbsenceNeutrinosI,Nielsen1981:AbsenceNeutrinosII}.} The Hamiltonian in Equation~\eqref{eq:ff-hopping} becomes
\begin{equation}
  H_{0}
  =v_F\sum_{a=1}^N\int dx\,
  \left[
    \psi_L^{a\dagger}(i\partial_x)\psi_L^a
    +\psi_R^{a\dagger}(-i\partial_x)\psi_R^a
  \right],
  \qquad
  v_F=2t a_0\sin \left(k_F\right).
  \label{eq:ff-continuum-hamiltonian}
\end{equation}
Therefore the low-energy theory consists of $N$ massless complex Dirac fermions. The local operator algebra of the theory is described by the product
\begin{equation}
\mathrm{Ff}^{N}_{\bC} \otimes \overline{\mathrm{Ff}}^{N}_{\bC}
  \label{eq:ff-continuum-chiral-algebras}
\end{equation}
of the vertex operator superalgebra (VOSA) $\mathrm{Ff}^{N}_{\bC}$ of $N$ chiral complex fermions and its anti-chiral copy $\overline{\mathrm{Ff}}^{N}_{\bC}$.\footnote{Here we work in the Neveu--Schwarz sector. $(\mathrm{Ff}^{N}_{\bC})_{\text{NS}}\otimes\overline{(\mathrm{Ff}^{N}_{\bC})_{\text{NS}}}$ is the non-chiral spin CFT built on these chiral algebras. The Ramond sector is its $(-1)^F$-twisted module.}
The singular operator product expansions are
\begin{equation}
  \psi_L^{a\dagger}(z)\,\psi_L^{b}(w)
  \;\sim\;
  \psi_{L}^{b}(z)\,\psi_{L}^{a\dagger}(w)
  \sim
  \frac{\delta^{ab}}{z-w}.
  \label{eq:free-fermion-ope}
\end{equation}

From \cite{Witten1984:NonAbelianBosonization,Affleck1986:ExactCriticalExponents,Affleck1988:CriticalBehaviourSUN}, $N$ free Dirac fermions can be bosonized to a compact boson $U(1)_N$ and an $SU(N)_1$ WZW model, also called the $U(N)_1$ WZW model, with the identification
\begin{equation}
  \psi_L^{a}(z) = :e^{\frac{i}{\sqrt{N}}\phi_L(z)}:g^{a}(z) ,
  \label{eq:free-fermion-bosonization}
\end{equation}
where $g^{a}(z)$ transforms in the fundamental of $SU(N)_1$ and $\phi_L(z)$ can be thought of as the chiral component of the compact boson $\phi = \phi_L + \phi_R$ with radius $R=2/\tilde R$, with the dual compact boson $\tilde\phi = \phi_L - \phi_R$ of radius $\tilde R$. For free fermions, $\tilde R_0=\sqrt N$ and $R_0=2/\sqrt N$.

Let $J=i\sqrt{N}\,\partial\phi_L$ be the $U(1)$ current, normalized as
\begin{equation}
    J(z)J(w)\sim \frac{N}{(z-w)^2} ,
  \label{eq:ff-charge-current-ope}
\end{equation}
so that $\psi^a_L$ has charge one, and let $U(1)^{\text{lat}}_N$ be the lattice vertex algebra extension of $U(1)_N$ Kac--Moody algebra by the baryon operators $:e^{\frac{i}{\sqrt{N}}Q\phi_L(z)}:$ of charge $Q\in N\bZ$.\footnote{The generating baryons have conformal weight $N/2$, so the extension is a vertex algebra for even $N$ and a vertex superalgebra for odd $N$ \cite{Kac1998:VertexAlgebrasBeginners}.} The free fermion chiral algebra then decomposes into modules of the chiral subalgebra $U(1)^{\text{lat}}_N\times SU(N)_1$, with the $U(1)$ charge fixing the $SU(N)_1$ label \cite{Frenkel1981:BosonFermionCorrespondence}:
\begin{equation}
  \mathrm{Ff}_{\bC}^{N} = \bigoplus_{r=0}^{N-1} \mathcal{U}_r \otimes \mathcal{L}_{\omega_r} ,
  \label{eq:free-fermion-decomposition}
\end{equation}
where $\mathcal{U}_r$ are the modules of $U(1)^{\text{lat}}_N$ and $\mathcal{L}_{\omega_r}$ are the modules of $SU(N)_1$. The combined modules are mutually (super)local. The two-dimensional Dirac fermion theory is diagonal with respect to $\mathrm{Ff}_{\bC}^{N}$ but not with respect to $U(1)^{\text{lat}}_N\times SU(N)_1$, since every pair of left and right labels appears,
\begin{equation}
  \mathrm{Ff}^{N}_{\bC}\otimes\overline{\mathrm{Ff}}^{N}_{\bC}
  =
  \bigoplus_{r,s=0}^{N-1}
  \bigl(\mathcal{U}_r\otimes\mathcal{L}_{\omega_r}\bigr)\otimes
  \bigl(\overline{\mathcal{U}}_s\otimes\overline{\mathcal{L}}_{\omega_s}\bigr) .
  \label{eq:free-fermion-decomposition-2d}
\end{equation}
The continuum theory has a larger infrared symmetry than the lattice symmetry of Equation~\eqref{eq:ff-onsite-symmetry}. Since each chirality can be rotated and charge conjugated independently,\footnote{The full emergent symmetry is in fact larger. Writing the $N$ Dirac fermions as $2N$ Majorana fermions gives $O(2N)_L\times O(2N)_R$. The bilinear currents $:\psi^a\psi^b:$ and $:\psi^{a\dagger}\psi^{b\dagger}:$ supplement the $U(N)$ currents to generate $\widehat{\mathfrak{so}}(2N)_1$ in each chirality.}
\begin{equation}
  G_{\mathrm{IR}} = \Bigl[\bigl(U(N)_L\rtimes\bZ_2^{\mathsf{C}_L}\bigr)\times\bigl(U(N)_R\rtimes\bZ_2^{\mathsf{C}_R}\bigr)\Bigr]\rtimes\bigl(\bZ_2^{\mathsf{P}}\times\bZ_2^{\mathsf{T}}\bigr) ,
  \label{eq:ff-continuum-symmetry}
\end{equation}
where we have included only the internal symmetries and the discrete spacetime symmetries. The lattice symmetry of Equation~\eqref{eq:ff-onsite-symmetry} maps $U(N)$ to the diagonal subgroup $U(N)_V$ and $\mathsf{P}$, $\mathsf{T}$ to their continuum counterparts, while lattice translation flows to a continuum translation $P$ combined with an \textit{emanant} symmetry, which will be discussed in Section~\ref{sec:free-fermion-anomalies}. The charge transformations form the subgroup,
\begin{equation}
  G^{\text{charge}}_{\mathrm{IR}} = U(1)_L \times U(1)_R \subset G_{\mathrm{IR}}
  \label{eq:ff-continuum-charge-symmetry}
\end{equation}

\subsection{Anomalies}
\label{sec:free-fermion-anomalies}
The gaplessness of the free fermion chain in the infrared is consistent with the anomaly of the lattice model in Equation~\eqref{eq:ff-hopping}. This is the Lieb--Schultz--Mattis (LSM) anomaly \cite{Lieb1961:TwoSolubleModels,Yamanaka1997:Luttinger1D,Oshikawa2000:CommensurabilityExcitationGap}, which can be understood as a mixed 't~Hooft anomaly between lattice translation and an internal symmetry \cite{ChengSeiberg2023:LSMLuttingerTHooft,Seifnashri2024:LSMAnomaliesGauging}. Away from special fillings, it forbids a unique gapped ground state that preserves both symmetries. Let the filling, defined as the number of fermions per site, be\footnote{We will treat the chemical potential $\mu$ as an external tuning parameter, which will be used to fix the desired filling $\nu$ of the ground state.}
\begin{equation}
\nu
=
\frac{\langle F \rangle}{m}
=
\frac{N k_F}{\pi}.
  \label{eq:ff-filling-fermi-momentum}
\end{equation}
With \(T\,c_j^a\,T^{-1}=c_{j+1}^a \) the LSM anomaly for this $U(1)_F$-symmetric system can be expressed as the following projective commutation relation:
\begin{equation}
  T\,U_{2\pi}\,T^{-1}
  =
  e^{-2\pi i \nu}\,U_{2\pi} ,
  \qquad
  U_{2\pi}
  =
  \exp\!\left(\frac{2\pi i}{m}\sum_{j=1}^{m} j n_j\right) ,
  \label{eq:lsm-anomaly-u1}
\end{equation}
where $U_{2\pi}$ is the twist operator that implements a large gauge transformation of $U(1)_F$. Although written as an exponential, $U_{2\pi}$ lies inside $\mathrm{Cl}_{2Nm}(\mathbb{C})$, as $n_j=\sum_a c_j^{a\dagger}c_j^a$ has a finite spectrum $\{0,\ldots,N\}$, so the exponential can be represented by a polynomial in the $n_j$. It can be shown \cite{Lieb1961:TwoSolubleModels,Yamanaka1997:Luttinger1D,Oshikawa2000:CommensurabilityExcitationGap} that $U_{2 \pi}| \text{GS} \rangle$ is a state with energy $O(m^{-1})$ above the ground-state energy and has a lattice momentum that is shifted by $2\pi\nu$ from $|\text{GS} \rangle$. Therefore, for $\nu\notin\bZ$ the twisted state is orthogonal to the ground state, so the system must either have degenerate ground states or be gapless. A stronger statement \cite{Yamanaka1997:Luttinger1D} can be phrased for the $N$-flavor fermion chain. By making use of the full $U(N)$ symmetry,\footnote{In fact, this argument only requires the subgroup $U(1)^N\rtimes\bZ_N \subset U(N)$.} by twisting one flavor at a time with the operator
\begin{equation}
U^a_{2\pi}
  =
  \exp\!\left(\frac{2\pi i}{m}\sum_{j=1}^{m} j\, c_j^{a\dagger}c_j^a\right)
  \qquad
  T\,U^a_{2\pi}\,T^{-1}
  =
  e^{-2\pi i \nu^a}\,U^a_{2\pi} ,
  \qquad
  \nu^a=\frac{\langle \widetilde{Q}^{aa}\rangle}{m} .
\label{eq:lsm-anomaly-suN}
\end{equation}
If the $U(N)$ symmetry is not spontaneously broken by the ground state, then all the flavors are equally filled, so that $\nu^a = \frac{\nu}{N}$. Therefore, Equation~\eqref{eq:lsm-anomaly-suN} implies for $0<\nu<N$ that either translation is spontaneously broken or the system is gapless.

In the infrared, the $N$-flavor Dirac fermion theory saturates this anomaly. Lattice translation $T$ for filling $\nu = \tfrac{a}{b}$ with $\gcd(a,b)=1$ emanates~\cite{ChengSeiberg2023:LSMLuttingerTHooft,SeibergShao2024:MajoranaChainIsing} to a discrete subgroup of the internal symmetry $G_{\text{IR}}$,\footnote{Since at low energies the eigenvalue of $e^{ia_0P}$ tends to one for momenta much smaller than $1/a_0$, conservation of $T$ implies conservation of $\tau$.}
\begin{equation}
T = e^{i a_0 P} \tau, \qquad \tau = e^{-i k_F (Q_L - Q_R)}, \qquad \langle \tau \rangle = \mathbb{Z}^\tau_{n_\tau}, \quad n_\tau = \frac{2Nb}{\gcd(2N,a)}.
  \label{eq:ff-rational-filling-translation}
\end{equation}
Since $\tau^{Nb/\gcd(N,a)} = (-1)^{aF/\gcd(N,a)} \in U(N)_V$, it is more convenient to consider instead the subgroup,
\begin{equation}
\rho = e^{-i k_F F}\tau = e^{-2 i k_F Q_L}, \qquad \langle \rho \rangle = \mathbb{Z}^\rho_{n_\rho} \qquad n_\rho = \frac{Nb}{\gcd(N,a)}.
  \label{eq:ff-rational-filling-chiral-translation}
\end{equation}
Then the anomaly is realized as a mixed anomaly between $U(1)_V$ and $\mathbb{Z}_{n_\rho}^\rho$ for Equation~\eqref{eq:lsm-anomaly-u1} and mixed anomaly between $U(N)_V$ and $\mathbb{Z}_{n_\rho}^\rho $ for Equation~\eqref{eq:lsm-anomaly-suN}. At $\mu=0$, the exact symmetry of translating all $b^a_j$ with the $a^a_j$ fixed emanates to right charge conjugation, $\mathsf{T}_{\text{Maj}} = \mathsf{C}_R e^{i a_0 P_b}$ ($P_b$ is the momentum of the $b$ sector), which makes $\mathsf{C}_{R}$ and $\mathsf{C}_{L}$ emanant symmetries as particle-hole $\mathsf{C} = \mathsf{C}_L \mathsf{C}_R$ is exact at half filling.

It is instructive to ask whether any relevant perturbations preserving $G_{\text{UV}} \subset G_{\text{IR}}$ could destabilize the free fermion phase. Any perturbation can be written as \cite{Assaraf1999:MetalInsulatorSUN}
\begin{equation}
  e^{-i k_F (p - q) j } : \exp \left( \frac{i}{\sqrt{N}} ( p \phi_L + q \phi_R ) \right) : \Phi_{\omega_p} \bar{\Phi}_{\omega_q} + \text{descendant terms} + \text{h.c.} ,
  \label{eq:ff-mott-general-term}
\end{equation}
where $\Phi_{\omega_p}(\bar{\Phi}_{\omega_q})$ are primaries of $SU(N)_1 (\overline{SU(N)}_1)$, with labels understood modulo $N$. The crystal-momentum prefactor comes directly from the crystal momenta $\mp k_F$ of $\psi_L$ and $\psi_R$ in Equation~\eqref{eq:lattice-continuum-fields}, and the $U(1)_N$ and $SU(N)_1$ primaries are correlated as in Equation~\eqref{eq:free-fermion-decomposition}. Conservation of $U(N)_V$ demands $p = - q$ and invariance under $\tau$ requires
\begin{equation}
p \in \frac{N b}{\gcd(N, a)}\mathbb{Z}
  \label{eq:ff-allowed-umklapp-charge}
\end{equation}
For $p \neq 0$, the scaling dimension of the primary in Equation~\eqref{eq:ff-mott-general-term} is
\begin{equation}
\Delta_{p} = \underbrace{\frac{p^2}{N}}_{U(1)_N} + \underbrace{\frac{r(N-r)}{N}}_{SU(N)_1}
\geq p
  \label{eq:ff-umklapp-dimension-bound}
\end{equation}
where $r = p \; \text{mod} \; N$. The minimal nontrivial value of $p$ is
\begin{equation}
p_{\ast} = \frac{N b}{\gcd(N, a)} \implies \begin{cases} p_{\ast} \geq 2 & \text{if } \nu = \frac{N}{2}, \\ p_{\ast} \geq 3 & \text{else} \end{cases}
  \label{eq:ff-minimal-umklapp-charge}
\end{equation}
Therefore, there are no relevant operators for $p \neq 0$ and only a marginal operator exists for $\nu = N/2$,
\begin{equation}
g :\exp{ \left( \frac{2i}{\sqrt{N}} ( \phi_L - \phi_R ) \right) }: \Phi_{\omega_2} \bar{\Phi}_{\omega_{N-2}} + \text{h.c.}
  \label{eq:ff-half-filling-charge-two-perturbation}
\end{equation}
which is \textit{marginally relevant} for all $|g| \neq 0$. However, this operator is not invariant under emanant $\mathsf{C}_R$, hence not generated.\footnote{Equivalently, it is charged under the lattice axial charge $Q_A=\mathsf{T}_{\text{Maj}}\bigl(F-\tfrac{Nm}{2}\bigr)\mathsf{T}_{\text{Maj}}^{-1}$, which commutes with $H_0$ at $\mu=0$ and emanates to the continuum axial charge $Q_L-Q_R$, as studied recently in \cite{Chatterjee2025:QuantizedAxialCharge}.}

For $p=0$, there are several operators of interest that are built from the currents. The current $J_L+J_R$ is \textit{relevant} and shifts the chemical potential since we tune it externally, so we ignore it. The combinations $J_L-J_R$, ${:}J_L^2{:}-{:}J_R^2{:}$ and ${:}J^A_LJ^A_L{:}-{:}J^A_RJ^A_R{:}$ are odd under parity or time reversal and are not generated. What remains are the four parity-even singlets of dimension two, and, with $J_{L,R}=\sum_a\psi^{a\dagger}_{L,R}\psi^a_{L,R}$ and $J^A_{L,R}=\psi^\dagger_{L,R}T^A\psi_{L,R}$ the real-space densities, the most general such Hamiltonian is
\begin{equation}
\begin{gathered}
  H_c=\int dx\,\Bigl[\frac{\pi u_c}{N}\bigl({:}J_L^2{:}+{:}J_R^2{:}\bigr)+g_c\,J_LJ_R\Bigr],
  \\
  H_s=\int dx\,\Bigl[\frac{2\pi u_s}{N+1}\bigl({:}J^A_LJ^A_L{:}+{:}J^A_RJ^A_R{:}\bigr)+g_s\,J^A_LJ^A_R\Bigr],
\end{gathered}
  \label{eq:ff-marginal-hamiltonian}
\end{equation}
with $u_c=u_s=v_F$ and $g_c=g_s=0$ at the free point. The charge sector is a compact boson with velocity and radius
\begin{equation}
  v_c=\sqrt{u_c^2-\left(\frac{Ng_c}{2\pi}\right)^2},
  \qquad
  \tilde R=\frac{2}{R}=\sqrt N\left(\frac{2\pi u_c+Ng_c}{2\pi u_c-Ng_c}\right)^{1/4} .
  \label{eq:ff-charge-radius}
\end{equation}
The radius depends only on the ratio $\gamma_c=g_c/u_c$. For $\gamma_c>0$ the radius $\tilde R=2/R$ of the dual boson $\tilde\phi$ increases, and $\gamma_c<0$ does the reverse. Since $g_c=0$ at the free fermion point and changing the filling through $\mu F$ affects only $u_c$, the radius stays at $\tilde R=\sqrt N$ for every filling. In the neutral sector, $u_s$ is the velocity of the $SU(N)_1$ modes, while $g_sJ^A_LJ^A_R$ runs, and it is \textit{marginally irrelevant} for $g_s<0$ and \textit{marginally relevant} for $g_s>0$ \cite{Assaraf1999:MetalInsulatorSUN}. To rule out the generation of these terms, we appeal to the trivial fact that the lattice Hamiltonian of Equation~\eqref{eq:ff-hopping} is \text{free}. Every mode occupation number is conserved, $[n_k, H_0] = 0$, and their local combinations form an infinite tower of conserved charges, which in the infrared flow to the $W_{1+\infty}$.\footnote{These charges are $Q_r=\sum_{j,a}\bigl(c^{a\dagger}_jc^a_{j+r}+\text{h.c.}\bigr)=\sum_k2\cos(rk)\,n_k$ and $\widetilde Q_r=-i\sum_{j,a}\bigl(c^{a\dagger}_jc^a_{j+r}-\text{h.c.}\bigr)=\sum_k2\sin(rk)\,n_k$ \cite{GrabowskiMathieu1995:ConservationLaws,EsslerFagotti2016:QuenchDynamics}. Expanded around $\pm k_F$ they become the zero modes of the bilinears ${:}\psi^{a\dagger}\partial^{n}\psi^{a}{:}$ of spin $n+1$, which generate the $W_{1+\infty}$ algebra of free fermions \cite{AwataFukumaMatsuoOdake1995:W1plusInfinity,FrenkelKacRadulWang1995:W1plusInfinityGlN}.} which neither $J_LJ_R$ nor $J^A_LJ^A_R$ commutes with. These charges exist only because $H_0$ is quadratic and generic interactions break them. We now look at such an example.
\enlargethispage{3pt}
\subsection{The Mott transition}
\label{sec:ff-mott}
When $\nu \in \mathbb{Z}$ there is no anomaly that obstructs a gapped ground state preserving $U(1)_F$ and $T$, that is, there is no obstruction to gapping out the compact boson part of the free fermion without introducing any degeneracy. As a concrete example, consider a chemical potential $\mu$ such that the ground state is at unit filling $\nu = 1$ and add the following repulsive Mott term \cite{Assaraf1999:MetalInsulatorSUN}, or charging energy, to the Hamiltonian in Equation~\eqref{eq:ff-hopping},
\begin{equation}
  H
  =
  H_{0} + U \sum_{j=1}^m (n_j-1)^{2} ,
  \label{eq:ff-mott-hamiltonian}
\end{equation}
This interaction respects the symmetries of Equation~\eqref{eq:ff-onsite-symmetry}.\footnote{The case of $N=2$ is also integrable and has $U_c=0$ at unit filling \cite{LiebWu1968:Hubbard}.} For $N>2$ and sufficiently small $U > 0$ this Hamiltonian flows to a CFT made from the modules of $U(1)_{\tilde R^2} \times SU(N)_1$, with a shifted level from that of the free fermions, since there is no symmetry that forbids $J_LJ_R$ from being generated. For an arbitrary radius $\tilde{R} \in \mathbb{R}$, the decomposition in Equation~\eqref{eq:free-fermion-decomposition} has to be modified and the theory is no longer rational. Since $J_LJ_R$ does not break the $U(1)_V \times U(1)_A$ symmetry, the operators can still be labeled by $Q_V$ and $Q_A$ and the gluing with $SU(N)_1$ is still meaningful, with only the conformal dimensions of operators changing. Concretely, let $R$ be the modified radius of the compact boson $\phi=\phi_L+\phi_R$, whose shift symmetry is $U(1)_V$, and $\tilde R=2/R$ be the radius of the dual boson $\tilde\phi=\phi_L-\phi_R$, with $R_0=2/\sqrt N$ at the free fermion point. Let $U(1)_{\tilde R^2}$ be the Kac--Moody algebra with current $J=i\tilde R\,\partial\phi_L$, and
\begin{equation}
J(z)J(w)\sim\tilde R^2/(z-w)^2
  \label{eq:ff-deformed-current-ope}
\end{equation}
The deformation of the radius keeps the charges $Q_V=p+q$ and $Q_A=p-q$ of the operator in Equation~\eqref{eq:ff-mott-general-term} and its gluing with $SU(N)_1$ labels fixed while changing the vertex operator,
\begin{equation}
  {:}e^{\frac{i}{\sqrt N}(p\phi_L+q\phi_R)}{:}\,\Phi_{\omega_p}\bar{\Phi}_{\omega_q}
  \;\longrightarrow\;
  {:}\exp\Bigl[\tfrac{i}{NR}\,Q_V\,\phi+\tfrac{iR}{4}\,Q_A\,\tilde\phi\Bigr]{:}\,\Phi_{\omega_p}\bar{\Phi}_{\omega_q},
  \label{eq:ff-deformed-vertex}
\end{equation}
whose scaling dimension is
\begin{equation}
  \Delta_{p,q}(R)=\frac{Q_V^2}{N^2R^2}+\frac{R^2Q_A^2}{16}+\frac{r(N-r)+s(N-s)}{2N}
  \label{eq:ff-deformed-dimension}
\end{equation}
where $r, s = p , q \mod N$. Its spin
\begin{equation}
h-\bar h=\frac{Q_VQ_A}{2N}+\frac{r(N-r)-s(N-s)}{2N}
  \label{eq:ff-deformed-conformal-spin}
\end{equation}
is independent of $R$. The local operators of this irrational theory organize into the infinite sum,
\begin{equation}
  \bigoplus_{p,q\in\bZ}\bigl(\mathcal F_{\lambda_L}\otimes\mathcal L_{\omega_r}\bigr)\otimes\bigl(\overline{\mathcal F}_{\lambda_R}\otimes\overline{\mathcal L}_{\omega_s}\bigr),
  \qquad
  \lambda_{L,R}=\frac{\tilde R^2Q_V}{2N}\pm\frac{Q_A}{2},
  \label{eq:ff-deformed-decomposition}
\end{equation}
where $\mathcal F_\lambda$ is the $U(1)_{\tilde R^2}$ module of weight $\lambda^2/2\tilde R^2$, and $r=p$, $s=q$ modulo $N$. At the free fermion point, $\lambda_{L, R} = p, q$, and the modules $\mathcal F_p$ with $p\equiv r\bmod N$ can be grouped into the modules $\mathcal U_r$ of $U(1)^{\text{lat}}_{N}$ and recover Equation~\eqref{eq:free-fermion-decomposition-2d}. The Mott term generates all the neutral descendants of Equation~\eqref{eq:ff-marginal-hamiltonian}, with $\gamma_c\propto U$, so for $U>0$ the radius $\tilde R=2/R$ of the dual boson increases, $\tilde R>\sqrt N$.\footnote{At tree level, with the lattice spacing set to one, $U\sum_j(n_j-1)^2$ gives $u_c=v_F+(N-1)U/\pi$, $u_s=v_F-U/\pi$, $g_c=2U(N-1)/N$ and $g_s=-4U$. Therefore $\tilde R=\sqrt N\bigl(1+\tfrac{2(N-1)U}{\pi v_F}\bigr)^{1/4}$.} There is also the running term $J^A_LJ^A_R$ which is generated, but comes with $g_s\propto -U$, so for $U>0$ it is marginally irrelevant and does not affect the low-energy theory. For sufficiently large $U > U_{\text{c}}$ the system undergoes a Mott transition of the Berezinskii--Kosterlitz--Thouless type that gaps out the charge mode. We now describe this transition from two perspectives, first from the lattice by expanding in perturbations of $t/U$ and from the continuum by identifying the relevant operators that are generated. We will also follow how the anomalies of the free fermion are inherited along the way.

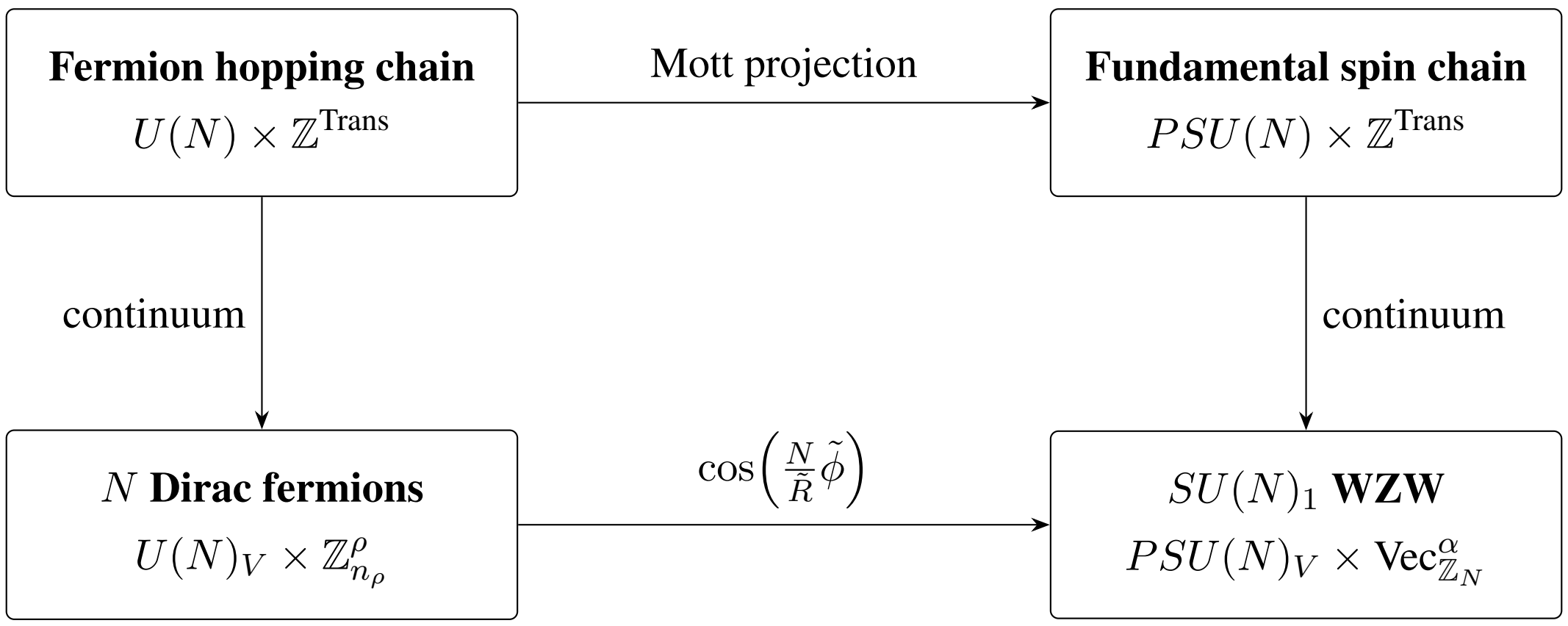
\begin{figure}[!htbp]
  \centering
  \begin{tikzpicture}[
    >=Stealth,
    box/.style={draw, rounded corners=2pt, inner sep=8pt,
                minimum width=46mm, minimum height=15mm,
                align=center, font=\small},
    lbl/.style={font=\small, align=center}]
    \node[box] (ul) at (0,0) {$\begin{gathered}
      \textbf{Fermion hopping chain}\\
      U(N) \times \mathbb{Z}^{\text{Trans}}
    \end{gathered}$};
    \node[box] (ur) at (9.4,0) {$\begin{gathered}
      \textbf{Fundamental spin chain}\\
      PSU(N)\times\mathbb{Z}^{\text{Trans}}
    \end{gathered}$};
    \node[box] (ll) at (0,-3.8) {$\begin{gathered}
      N\textbf{ Dirac fermions}\\
      U(N)_V \times \mathbb{Z}^{\rho}_{n_\rho}
    \end{gathered}$};
    \node[box] (lr) at (9.4,-3.8) {$\begin{gathered}
      SU(N)_1\textbf{ WZW}\\
      PSU(N)_V \times \text{Vec}^{\alpha}_{\mathbb{Z}_N}
    \end{gathered}$};
    \draw[->] (ul) -- node[lbl,above]{Mott projection} (ur);
    \draw[->] (ll) -- node[lbl,above]
      {$\cos\!\left(\tfrac{N}{\tilde R}\tilde\phi\right)$} (lr);
    \draw[->] (ul) -- node[lbl,left]{continuum} (ll);
    \draw[->] (ur) -- node[lbl,right]{continuum} (lr);
  \end{tikzpicture}
\caption{Anomaly matching for the $N$-flavor free fermion chain at $\nu=1$. Here $\alpha=0$ for odd $N$ and $\alpha=N/2$ for even $N$.}
  \label{fig:ff-anomaly-rectangle}
\end{figure}

In the large-$U$ limit of Equation~\eqref{eq:ff-mott-hamiltonian}, the low-energy effective Hamiltonian is obtained by projecting onto the subspace where every site carries the fundamental $\bC^N$ of $SU(N)$ and performing perturbation theory. Let $\mathbb P^{(\omega_1)}$ be the operator that projects onto the singly occupied subspace $\mathcal{H}_{\mathrm{Mott}}\cong(\bC^N)^{\otimes m} \subset \Lambda^{\bullet} \bC^{Nm}$. On this subspace, introduce the spin operators
\begin{equation}
  S_j^A=\sum_{a,b=1}^N (T^A)_{ab}\widetilde Q_j^{ab},
  \qquad
  \text{tr}(T^AT^B)=\frac12\delta^{AB},
  \qquad A=1,\ldots,N^2-1,
  \label{eq:su-n-generators}
\end{equation}
where $T^A$ are Hermitian generators of $\fsu(N)$ in the fundamental representation. The effective Hamiltonian on this subspace is obtained by treating $H_0$ as the perturbation and performing second-order degenerate perturbation theory. We have
\begin{equation}
\begin{aligned}
  E_jF_j=F_jE_j
  &=\frac{N-1}{N}-2\sum_A S_j^AS_{j+1}^A
   =2\Pi^{(\omega_2)}_{j,j+1},
\end{aligned}
  \label{eq:ff-mott-exchange-projector}
\end{equation}
where $\Pi^{(\omega_2)}$ projects onto the subspace of the antisymmetric representation $\Lambda^2\bC^N$ in $(\bC^N)^{\otimes 2}$. Discarding the overall constants, the effective Hamiltonian is then
\begin{equation}
\begin{aligned}
  H_{\mathrm{eff}}^{(2)}
  &=-\frac{1}{2U}\,\mathbb P^{(\omega_1)}H_0\bigl(1-\mathbb P^{(\omega_1)}\bigr)H_0\mathbb P^{(\omega_1)}\\
  &=-J\sum_{j=1}^m\Pi^{(\omega_2)}_{j,j+1}
   =J\sum_{j=1}^m\sum_A S_j^AS_{j+1}^A,
  \qquad J=\frac{2t^2}{U}>0 .
\end{aligned}
\label{eq:ff-mott-effective}
\end{equation}
This is the antiferromagnetic $SU(N)$ Heisenberg chain.\footnote{This theory is Bethe-ansatz solvable \cite{Sutherland1975:ModelMulticomponentQuantum}, even though the Hamiltonian~\eqref{eq:ff-mott-hamiltonian} is not integrable for $N>2$. We will see similar behavior in the $q$-deformed case.} In the continuum limit this flows to the $SU(N)_1$ WZW CFT \cite{Affleck1986:ExactCriticalExponents,Affleck1988:CriticalBehaviourSUN}, where as expected the compact boson mode $U(1)_N$ is gapped out. The operator algebra of the projected chain is generated by the spin operators $S_j^A$. The generators of $U(1)_F$ and the center of $SU(N)$ act trivially on this algebra. So the \textit{faithfully} acting internal symmetry of the Hamiltonian~\eqref{eq:ff-mott-effective} is the subgroup $PSU(N)=SU(N)/\bZ_N$. However, the states themselves at unit filling transform under the fundamental representation of $SU(N)$, and therefore carry a projective representation of $PSU(N)$,
\begin{equation}
x | \Omega \rangle = e^{2\pi i/N} | \Omega \rangle, \qquad x = e^{2\pi i/N}\mathbbm{1}_N \in Z( SU(N) ) \cong \bZ_N
  \label{eq:ff-mott-site-center-action}
\end{equation}
The anomaly of Equation~\eqref{eq:lsm-anomaly-suN} is now captured by the mixed anomaly between $PSU(N)$ and lattice translation $\bZ^{\text{Trans}}$ \cite{Yao2019:AnomalyMatchingSUN}.\footnote{The mixed anomaly lies in the summand $H^2(BPSU(N),U(1))\cong\bZ_N$ of $H^3(B(PSU(N)\times\bZ^{\text{Trans}}),U(1))$. This group classifies the projective representations of $PSU(N)$} In the infrared, lattice translation acts as
\begin{equation}
T = e^{i a_0 P}\tau,
\qquad
\tau:g\mapsto\omega g,
\qquad
\omega=e^{2\pi i/N},
  \label{eq:ff-mott-translation-action}
\end{equation}
where $g$ is the WZW field transforming in the fundamental representation of $SU(N)_1$ and $\tau$ is the internal symmetry operator which can be identified with the generating invertible Verlinde line $\mathcal{D}$ of $SU(N)_1$. Translation thus emanates to the fusion category $\mathrm{Vec}^{\alpha}_{\bZ_N}$, with an \textit{emergent} anomaly specified by the cohomology class $\alpha \in H^3(\bZ_N,U(1))$.\footnote{For $SU(N)_1$, $\alpha=0$ for odd $N$ and $\alpha=N/2\in\bZ_N$ for even $N$ \cite{ChengSeiberg2023:LSMLuttingerTHooft}. This is an ``emergent anomaly'' in the terminology of \cite{Yao2019:AnomalyMatchingSUN}: its pullback along $\bZ^{\text{Trans}}\to\bZ_N^\tau$ is trivial, since $H^3(\bZ,U(1))=0$. This should be contrasted with the LSM anomaly which is encoded by the mixed anomaly between $PSU(N)$ and lattice translation $\bZ^{\text{Trans}}$.} In parallel with Equation~\eqref{eq:ff-continuum-symmetry}, the $SU(N)_1$ WZW theory has the global symmetry group
\begin{equation}
  G^{\mathrm{WZW}}_{\mathrm{IR}} = \Bigl[\bigl(SU(N)_L\times SU(N)_R\bigr)/\bZ_N^{\mathrm{diag}}\rtimes\bZ_2^{\mathsf{C}}\Bigr]\rtimes\bigl(\bZ_2^{\mathsf{P}}\times\bZ_2^{\mathsf{T}}\bigr),
  \qquad
  (U_L,U_R):g\mapsto U_LgU_R^\dagger ,
  \label{eq:wzw-continuum-symmetry}
\end{equation}
Relative to the free fermion, $U(1)_V$ acts trivially since the surviving local operators have $Q_V=0$, and $U(1)_A$ is broken to the center $\bZ_N^{\tau}$ generated by $\tau$. The lattice $U(N)$ of Equation~\eqref{eq:ff-onsite-symmetry} therefore maps to $PSU(N)_V$ and the LSM anomaly is captured as a mixed anomaly between the $PSU(N)_V$ and $\bZ_{N}^\tau$, where $U \in PSU(N)_V$ acts as $(U, U)$ and $\tau$ is represented by $(\omega\mathbbm{1},\mathbbm{1})$~\cite{Yao2019:AnomalyMatchingSUN,ChengSeiberg2023:LSMLuttingerTHooft,Seifnashri2024:LSMAnomaliesGauging}. The Mott transition can also be explained from a purely continuum point of view. At $\nu = 1$, Equation~\eqref{eq:ff-mott-general-term} becomes
\begin{equation}
  e^{-i \frac{2\pi p}{N} j } : \exp \left( \frac{i p}{\tilde R} \tilde\phi \right) : \Phi_{\omega_p} \bar{\Phi}_{\omega_{N-p}} + \text{descendant terms} + \text{h.c.} ,
  \label{eq:ff-mott-neutral-term}
\end{equation}
Invariance under $\bZ^{\tau}_{N}$ imposes the constraint that $p \in N \bZ$. For the first nontrivial value $p = N$, we have the operator
\begin{equation}
:\cos\left( \frac{N}{\tilde R} \tilde\phi \right): + \; \text{descendant terms}
\label{eq:ff-mott-umklapp}
\end{equation}
which has scaling dimension $\Delta=N^2/\tilde R^2=N^2R^2/4$. As $U$ increases so does $\tilde R$, and once $\tilde R$ exceeds $N/\sqrt{2}$, at $U=U_c$, this operator becomes relevant and triggers a BKT transition that gaps out the charge mode. Figure~\ref{fig:ff-anomaly-rectangle} summarizes the anomaly matching discussed in this section. We note here that even though the term in Equation~\eqref{eq:ff-mott-umklapp} has $N$ minima, because of the gluing of modules in Equation~\eqref{eq:free-fermion-decomposition}, the $N$ minima are identified and we have a single copy of the $SU(N)_1$ WZW theory in the infrared. The physics at other fillings can also be discussed; we refer the reader to the literature for more details \cite{Yamanaka1997:Luttinger1D, Assaraf1999:MetalInsulatorSUN}.

\section{The \texorpdfstring{$q$}{q}-fermion chain}
\label{sec:q-fermion-chain}

\begin{figure}[!htbp]
  \centering
  \begin{tikzpicture}[>=stealth,
    worldline/.style={very thick},
    up/.style={postaction={decorate, decoration={markings,
      mark=at position 0.55 with {\arrow{>}}}}},
    rung/.style={thick,
      postaction={decorate, decoration={markings,
        mark=at position 0.58 with {\arrow{>}}}}},
    junction/.style={circle, fill, inner sep=1.4pt},
    num/.style={font=\small, inner sep=5pt}
  ]
    \begin{scope}
      \draw[worldline,up] (0,-1.3) -- (0,0);
      \draw[worldline,up] (0,0) -- (0,1.3);
      \draw[worldline,up] (2,-1.3) -- (2,0);
      \draw[worldline,up] (2,0) -- (2,1.3);
      \draw[rung] (0,0) -- (2,0);
      \node[junction] at (0,0) {};
      \node[junction] at (2,0) {};
      \node[above] at (1,0.12) {$E_i$};
      \node[num, left] at (0,-0.7) {$s_i$};
      \node[num, left] at (0, 0.7) {$s_i-1$};
      \node[num, right] at (2,-0.7) {$s_{i+1}$};
      \node[num, right] at (2, 0.7) {$s_{i+1}+1$};
      \node[below] at (0,-1.3) {$i$};
      \node[below] at (2,-1.3) {$i+1$};
    \end{scope}
    \begin{scope}[xshift=6cm]
      \draw[worldline,up] (0,-1.3) -- (0,0);
      \draw[worldline,up] (0,0) -- (0,1.3);
      \draw[worldline,up] (2,-1.3) -- (2,0);
      \draw[worldline,up] (2,0) -- (2,1.3);
      \draw[rung] (2,0) -- (0,0);
      \node[junction] at (0,0) {};
      \node[junction] at (2,0) {};
      \node[above] at (1,0.12) {$F_i$};
      \node[num, left] at (0,-0.7) {$s_i$};
      \node[num, left] at (0, 0.7) {$s_i+1$};
      \node[num, right] at (2,-0.7) {$s_{i+1}$};
      \node[num, right] at (2, 0.7) {$s_{i+1}-1$};
      \node[below] at (0,-1.3) {$i$};
      \node[below] at (2,-1.3) {$i+1$};
    \end{scope}
  \end{tikzpicture}
\caption{The fundamental Wilson line (rungs) stretched between the sites $i$ and $i+1$. Time moves upward. The rung carrying a fermion from site $i+1$ to site $i$ corresponds to the operator $E_i$. Although $E_i$ appears oriented from left to right in this diagram, we will see from its action on states that it moves a fermion from site $i+1$ to site $i$. Here the dual labels are $s_i=N-n_i$. Similarly, the rung carrying a fermion from site $i$ to site $i+1$ corresponds to the operator $F_i$ in $U^{\mathrm{L}}_q(\fgl_m)$.}
  \label{fig:EF-Wilson-rungs}
\end{figure}
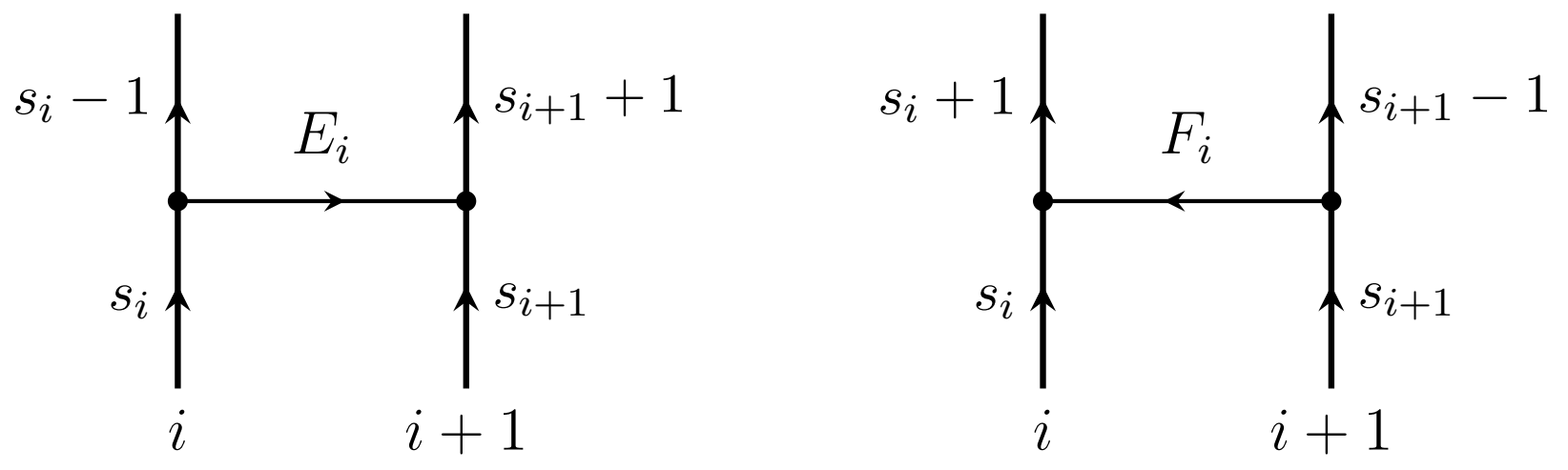

The free complex fermions of Section~\ref{sec:free-fermion-review} can be viewed as $m$ particle worldlines in the vacuum, coupled by the Hamiltonian of Equation~\eqref{eq:ff-hopping}. We now define $q$-fermions as these worldlines coupled to Chern--Simons theory, so that each of them becomes a Wilson line \cite{Witten1989:QuantumFieldTheory}. Let $\gamma_i \subset M_3$ be the worldlines of fermions $\psi_i(\tau)$ in a three-manifold $M_3$. We have the action
\begin{equation}
  S[A,\{\psi_i\}]
  =
  S_{\mathrm{CS}}[A]
  +\sum_{j=1}^m\int_{\gamma_j}
  \psi^{a \dagger}_{j}\,iD^{ab}\,\psi_{j}^b,
  \qquad
  D^{ab}
  =\delta^{ab}\,d
   -iA^{ab} ,
  \label{eq:multiworldline-action}
\end{equation}
where $\psi_i^{a \dagger}$ transforms under the fundamental representation of $SU(N)$. Consider the system in $S^2\times\bR$, with the worldlines parallel to the time direction $\bR$ and located equidistantly along the $x$-axis in the patch containing the origin. Since the Hilbert space of Chern--Simons theory on $S^2$ is trivial \cite{Witten1989:QuantumFieldTheory}, the only nontrivial states come from the Fock space of the fermions. Quantizing the fermions on site $i$ gives the local Fock space,
\begin{equation}
\mathcal{H}_{\gamma_i}=\Lambda^\bullet\bC^N=\bigoplus_{r_i=0}^N\Lambda^{r_i}\bC^N,
  \label{eq:cs-worldline-hilbert-space}
\end{equation}
and the total Hilbert space is naively the tensor product,
\begin{equation}
  \mathcal{H}
  =\bigotimes_{i=1}^m\mathcal{H}_{\gamma_i}
  =\bigotimes_{i=1}^m\Lambda^\bullet\bC^N .
  \label{eq:cs-worldline-tensor-product}
\end{equation}
On $S^2$ the constraint $F_{xy}=0$ requires that we retain only the $SU(N)$-invariant part of this Hilbert space. However, when we work in a definite gauge color sector $a$, we simply place an additional anyon $a^\vee$ of compensating charge at infinity, so that the total representation is $SU(N)$ trivial. In the quantum theory the gauge group is replaced by a quantum group, and we emphasize that we regard this Hilbert space not as a representation of the original Lie group $SU(N)$, but as a representation of Lusztig's quantum group $U^{\mathrm{L}}_q(\fsl_N)$ \cite{Lusztig1990:QuantumGroupsRootsOf1}, which we call the \textit{color} quantum group. More concretely, Chern--Simons theory is the topological quantum field theory described by the modular tensor category obtained as the semisimplification of the category of tilting modules of the quantum group $U^{\mathrm{L}}_q(\fsl_N)$ at $q$ a root of unity \cite{Witten1989:QuantumFieldTheory,ReshetikhinTuraev1991:Invariants3Manifolds,Andersen1992:TensorProductsTilting,AndersenParadowski1995:FusionCategories}. 
We define
\begin{equation}
\mathcal{C}_{N, k} := \operatorname{Tilt}\!\left(U^{\mathrm{L}}_q(\mathfrak{sl}_N)\right)/\text{Negl} .
  \label{eq:cs-semisimplified-tilting-category}
\end{equation}
Here \(\operatorname{Tilt}(U^{\mathrm{L}}_q(\mathfrak{sl}_N))\) denotes the category of finite-dimensional type-\(1\) tilting modules, and \(\operatorname{Negl}\) is the tensor ideal of negligible morphisms. We refer the reader to Appendix~\ref{app:quantum-groups-at-root-of-unity} for details.  The local Hilbert space at a site $i$, viewed as a representation of $U^{\mathrm{L}}_q(\fsl_N)$, is written as
\begin{equation}
  \Lambda_q^\bullet(\bC_q^N) = \bigoplus_{r=0}^N \Lambda_q^r(\bC_q^N),
   \qquad q=\exp\left(\frac{\pi i}{k+N}\right),
  \label{eq:local-qfock}
\end{equation}
whose summands $\Lambda_q^r(\bC_q^N)$ are irreducible representations. With the Hilbert space in hand, it is natural to ask which gauge-invariant operators act on it, and hence which algebra it is a representation of. Chern--Simons theory has no gauge-invariant local operators, and the fermions $\psi_i$ are not gauge invariant either, as they transform under the gauge group. The natural gauge-invariant operators supported on a thin time slice $S^2\times(-\epsilon, \epsilon)$ are mesons constructed out of fermion bilinears,
\begin{equation}
  M_{ij}
  :=
  \psi^{ a \dagger}_{j}
  \left[
\text{P}\exp\!\left(i\int_{\ell_{ij}}A\right)
  \right]^{ab}
  \psi_{i}^b,
  \qquad
  \ell_{ij}:i\longrightarrow j ,
  \label{eq:meson-bilinear}
\end{equation}
where $\ell_{ij}$ is a path from site $i$ to site $j$, so that $M_{ij}$ moves a fermion from site $i$ to site $j$. These mesons are fundamental Wilson lines attached to the one-dimensional fermion worldlines at sites $i$ and $j$ by trivalent topological junction operators which belong to the spaces
\begin{equation}
\begin{gathered}
\text{Hom}_{\mathcal{C}_{N,k}}(\Lambda_q^\ell(\bC_q^N) \otimes \bC^N_q, \Lambda_q^{\ell+1}(\bC_q^N)),\\
\text{Hom}_{\mathcal{C}_{N,k}}(\Lambda_q^\ell(\bC_q^N) \otimes (\bC^N_q)^*, \Lambda_q^{\ell-1}(\bC_q^N)).
\end{gathered}
  \label{eq:cs-fermion-junction-spaces}
\end{equation}
Therefore a fundamental Wilson line stretched between adjacent sites $i$ and $i+1$ transfers a fermion from one worldline to the other, in the direction set by its orientation, as in Figure~\ref{fig:EF-Wilson-rungs}. The algebra of these Wilson lines ending on punctures taken modulo isotopy and the local skein relations forms the $U^{\mathrm{L}}_q(\fsl_N)$ skein algebra of the punctured sphere, where the product of two operators is obtained by stacking one slice on top of the other.\footnote{For closed Wilson lines, which do not end on the worldlines, the algebra in the $q\to1$ limit reduces to the commutative algebra of gauge-invariant functions of the flat connection on the punctured sphere, while the open Wilson lines ending on the worldlines reduce to the fermion bilinears of Section~\ref{sec:free-fermion-review}.}

It was shown in \cite{Chun2017:JunctionsSurfaceOperators,Cautis2014:WebsQuantumSkew,Gaiotto2025:DbranesPlanar} that the skein algebra of the punctured sphere is a representation of Lusztig's quantum group $U^{\mathrm{L}}_q(\fgl_m)$, with nearest-neighbor fundamental Wilson lines obeying precisely the relations of the simple-root generators $E_i$ and $F_i$ of $U^{\mathrm{L}}_q(\fgl_m)$, collected in Appendix~\ref{app:uqglm-relations}, with $E_i\propto M_{i+1,i}$ and $F_i\propto M_{i,i+1}$. The site operators $n_i$ read off the fermion number at site $i$. Their exponentials $L_i=q^{n_i}$ serve as the Cartan generators.\footnote{This exponential is a well-defined element of the operator algebra. Since $n_i$ has the finite spectrum $\{0,1,\ldots,N\}$, $L_i$ is a polynomial in $n_i$.} We call $U^{\mathrm{L}}_q(\fgl_m)$ the \textit{site} quantum group.

Just like in the free fermion chain, the color and site quantum groups have a commuting action,
\begin{equation}
  U^{\mathrm{L}}_q(\fsl_N)
  \curvearrowright
   (\Lambda_q^\bullet(\bC_q^N))^{\otimes m} \cong \Lambda_q^\bullet(\bC_q^N\otimes\bC_q^m)
  \curvearrowleft
  U^{\mathrm{L}}_q(\fgl_m).
  \label{eq:qfockspace}
\end{equation}
Following \cite{Cautis2014:WebsQuantumSkew,Queffelec2015:SkeinModulesAffineExtensions}, an analogous statement of \textit{quantum} skew Howe duality can be made for generic $q$,\footnote{We emphasize that this skew Howe duality statement is for generic $q$ \cite{Cautis2014:WebsQuantumSkew}. In unitary Chern--Simons theory, however, $q$ is a root of unity, which complicates this relation, as we will discuss in Section~\ref{sec:fusion-tree-basis}. Also note that $U^{\mathrm{L}}_q(\fgl_m)\cong U_q(\fgl_m)$ for $q$ not a root of unity.}
\begin{equation}
  \Lambda_q^\bullet(\bC_q^N\otimes\bC_q^m)
  \cong
  \bigoplus_{\lambda\subseteq(m^N)}
  \mathcal{L}^{U^{\mathrm{L}}_q(\fsl_N)}_{\overline\lambda}
  \otimes
  \mathcal{L}^{U^{\mathrm{L}}_q(\fsl_m)}_{\overline{\lambda^\top}}
  \otimes
  \mathcal{L}^{\fgl_1}_{|\lambda|},
  \label{eq:q-skew-howe-sl}
\end{equation}

where we restrict to the subalgebra $U^{\mathrm{L}}_q(\fsl_m) \subset U^{\mathrm{L}}_q(\fgl_m)$ and retain the remaining information about the representation through the $F$-grading, with central element $L_1L_2\cdots L_m=q^{F}$. The bar $\bar{\cdot}$ follows the same convention as in Equation~\eqref{eq:skew_howe_sl}. Since we are working with $SU(N)_k$, besides the mesons described by $U^{\mathrm{L}}_q(\fgl_m)$ we also have baryon operators. These are $SU(N)$-invariant operators that move between the different $U^{\mathrm{L}}_q(\fgl_m)$ representations in Equation~\eqref{eq:q-skew-howe-sl} with the same $U^{\mathrm{L}}_q(\fsl_N)$ representation $\overline{\lambda}$ but a different fermion number $|\lambda|$. The algebra of baryons and mesons is described in Appendix~\ref{app:baryon-algebra}.

We now write down a gauge-invariant Hamiltonian for the $q$-fermion chain, that is, one built purely out of operators from $U^{\mathrm{L}}_q(\fgl_m)$ and baryons. Inspired by the free fermion chain, we write down the $q$-analog of Equation~\eqref{eq:ff-hopping},
\begin{equation}
  H^{q}_{0}= -t\sum_{i=1}^{m-1}(E_i+F_i) - \mu F,
  \label{eq:path-qfermion-hamiltonian}
\end{equation}
where $E_i$ and $F_i$ are the simple-root generators of $U^{\mathrm{L}}_q(\fsl_m)$ and $F=\sum_{j=1}^{m}n_j$ is the fermion number operator. The number operators $n_j$ can be constructed directly from the generators of $U^{\mathrm{L}}_q(\fgl_m)$. We work with open boundary conditions for now. To understand the action of these operators on the Hilbert space we will introduce explicitly the elements of the $q$-exterior algebra.

\subsection{The \texorpdfstring{$q$}{q}-fermion Fock space}
\label{sec:q-fock-space}
Let $V$ be the tensor product of the fundamental representations of $U^{\mathrm{L}}_q\!\left(\fsl_N\right)$ and $U^{\mathrm{L}}_q\!\left(\fsl_m\right)$,
\begin{equation}
    V
    =
    \bC_q^N\otimes\bC_q^m .
  \label{eq:q-fermion-color-site-module}
\end{equation}
We choose the basis of one-particle states to be
\begin{equation}
    \bqpsi_i^a
    :=
    \bqpsi^a\otimes\bqpsi_i
    \in V .
  \label{eq:q-fermion-tensor-product-basis}
\end{equation}
To construct multiparticle states, we first define the tensor algebra of $V$,
\begin{equation}
    T(V)
    :=
    \bigoplus_{r\geq 0}V^{\otimes r},
  \label{eq:q-fermion-tensor-algebra}
\end{equation}
and then quotient by the two-sided ideal generated by the $q$-symmetric quadratic subspace $S_q^2(V)$ \cite{BerensteinZwicknagl2008:BraidedSymmetricExterior}:
\begin{equation}
    \Lambda_q^\bullet(V)
    :=
    T(V)\Big/\bigl\langle S_q^2(V)\bigr\rangle .
    \label{eq:q-fermion-exterior-algebra}
\end{equation}
The tensor products are taken over the field $\bC(q)$. The quadratic subspace $S_q^2(V)$ determines the $q$-fermion anticommutation relations. Let $s_{ij}:= \text{sgn}(i-j)$, the defining relations in the quotient algebra $\Lambda_q^\bullet(V)$ are
\begin{equation}
\begin{aligned}
\bqpsi_i^a \otimes \bqpsi_j^b
    ={}&
    -
    \bigl(1-\delta_{ij}\delta_{ab}\bigr)
    q^{-s_{ij}\delta_{ab}-s_{ab}\delta_{ij}}
    \bqpsi_j^b \otimes \bqpsi_i^a
    \\
    &+
    \frac{q-q^{-1}}{2}
    \Bigl[
        s_{ij}\bigl(1-\delta_{ab}\bigr)
        -
        s_{ab}\bigl(1-\delta_{ij}\bigr)
    \Bigr]
    \bqpsi_j^a \otimes \bqpsi_i^b .
\end{aligned}
\label{eq:q-fermion-braided-antisymmetry}
\end{equation}
The derivation of Equation~\eqref{eq:q-fermion-braided-antisymmetry} is given in Appendix~\ref{app:q-fermion-braided-antisymmetry}. Note that, unlike the free fermion relations, the $q$-fermion anticommutation relations contain a term that mixes the color and site indices. Since the Fock space carries an algebra structure, $\Lambda_q^\bullet(V)$ acts on itself by left multiplication and we get a sense of operator--state correspondence, with operators and states identified by $x\leftrightarrow x|0\rangle$, and $\bqpsi_i^a$ can be regarded as a $q$-fermion creation operator.\footnote{This $\bqpsi_i^a$ should be distinguished from the microscopic fermion creation operator $\psi_i^{a\dagger}$ coupled to the Chern--Simons gauge field in Equation~\eqref{eq:multiworldline-action}. We keep using $\otimes$ for the product of $\Lambda_q^\bullet(V)$.} Because of the nontrivial nature of $q$-fermion anticommutation relations in Equation~\eqref{eq:q-fermion-braided-antisymmetry}, we choose a Poincar\'e--Birkhoff--Witt (PBW) basis for the $q$-fermion Fock space,
\begin{equation}
    \bqpsi_{i_1}^{a_1} \otimes
    \bqpsi_{i_2}^{a_2} \otimes
    \cdots \otimes
    \bqpsi_{i_r}^{a_r}|0\rangle,
    \qquad
    (i_1,a_1)
    <
    (i_2,a_2)
    <
    \cdots
    <
    (i_r,a_r),
    \label{eq:q-fermion-pbw-basis}
\end{equation}
where pairs are ordered lexicographically, first by site and then by color; the relations in Equation~\eqref{eq:q-fermion-braided-antisymmetry} bring any product to this order, since the mixed term exchanges sites while keeping the colors in place.\footnote{The order is dictated by the relations: the mixed term exchanges only sites, so sorting site-first terminates; color-first would not.} We now describe the action of the color and site quantum groups on $V$. The generators of the site quantum group $U^{\mathrm{L}}_q(\fgl_m)$ act on the basis of $V$ as
\begin{equation}
\begin{split}
    E_i(\bqpsi^{a}_{j})&=\delta_{i+1,j}\bqpsi^{a}_{i},\\
  F_i(\bqpsi^{a}_{j})&=\delta_{ij}\bqpsi^{a}_{i+1},\\
  L_i(\bqpsi^{a}_{j})&=q^{\delta_{ij}}\bqpsi^{a}_{j},\\
  K_i(\bqpsi^{a}_{j})&=q^{\delta_{ij}-\delta_{i+1,j}}\bqpsi^{a}_{j},
\end{split}
\label{eq:q-fermion-site-cartan-action}
\end{equation}
where $K_i:=L_iL_{i+1}^{-1}$. Their action on $T(V)$ is given by the coproduct. On an $r$-fold product the iterated coproduct gives\footnote{We have used the shorthand $\Delta^{(r-1)}$ for the $(r-1)$-fold iterated coproduct.}
\begin{equation}
\begin{split}
  \Delta^{(r-1)}(E_i)
  &=\sum_{p=1}^{r}
    1^{\otimes(p-1)}\otimes E_i\otimes K_i^{\otimes(r-p)},
  \\
  \Delta^{(r-1)}(F_i)
  &=\sum_{p=1}^{r}
    (K_i^{-1})^{\otimes(p-1)}\otimes F_i\otimes 1^{\otimes(r-p)},
  \\
  \Delta^{(r-1)}(L_j)
  &=L_j^{\otimes r}, \qquad
  \Delta^{(r-1)}(K_i)=K_i^{\otimes r} .
\end{split}
\label{eq:q-fermion-coproduct}
\end{equation}
To distinguish the color generators from the site generators, we denote the Chevalley generators of the color quantum group $U^{\mathrm{L}}_q(\fsl_N)$ by $X^a_{\mathrm{col}}$, with $X=E,F,K$ and $a=1,\ldots,N-1$. Their action and coproducts take the same form as Equations~\eqref{eq:q-fermion-site-cartan-action} and~\eqref{eq:q-fermion-coproduct}, with the roles of site and color indices interchanged, and since we are in $SU(N)_k$ Chern--Simons theory, there is no color analog of the $L_i$. The two actions commute as they act on different tensor factors of $V$. Both quantum group actions descend to $\Lambda_q^\bullet(V)$ as they preserve the quadratic ideal $\langle S_q^2(V)\rangle$. It also carries a natural fermion-number grading, $F(\xi)=r\,\xi$ for $\xi\in\Lambda_q^r(V)$. The $q$-fermion Fock space has dimension
\begin{equation}
\dim_{\bC(q)} \Lambda_q^\bullet\bigl(\bC_q^N\otimes\bC_q^m\bigr) = 2^{Nm},
  \label{eq:q-fermion-fock-dimension}
\end{equation}
the same as in the free fermion case \cite{Hayashi1990:qAnaloguesCliffordWeyl,Cautis2014:WebsQuantumSkew}. Therefore the $q$-deformation does not change the vector-space dimension, only the anticommutation relations and the action of the generators.

At the root of unity $q=\exp(\pi i/(k+N))$, however, the $q$-fermion Fock space is not the physical Hilbert space of Chern--Simons theory, as it suffers from the appearance of null states, as expected in gauge theories, which we will describe in the next section.

\subsection{The fusion-tree basis}
\label{sec:fusion-tree-basis}

To motivate the null states, we highlight the following observation: the color and site quantum groups have exactly the same parameter $q$. If we write the $q$ of $U^{\mathrm{L}}_q(\fsl_m)$ as $q=\exp\left(\tfrac{\pi i}{k'+m}\right)$, then the equality of the two exponentials implies
\begin{equation}
  k'=\frac{k+N}{1-2r(k+N)}-m,
  \qquad r\in\bZ.
  \label{eq:q-fermion-site-level-ambiguity}
\end{equation}
The only solution with integer $k'$ is $r=0$, namely
\begin{equation}
  k'=k+N-m .
  \label{eq:q-fermion-site-level-canonical}
\end{equation}
Therefore $k'>0$ if and only if $m<k+N$. The Kazhdan--Lusztig--Finkelberg equivalence \cite{KazhdanLusztig1993:TensorStructuresI,KazhdanLusztig1993:TensorStructuresII,KazhdanLusztig1994:TensorStructuresIII,KazhdanLusztig1994:TensorStructuresIV,Finkelberg1996:EquivalenceFusionCategories,Andersen1992:TensorProductsTilting,AndersenParadowski1995:FusionCategories} gives
\begin{equation}
  \mathcal{C}_{N,k}=\operatorname{Tilt}\!\left(U^{\mathrm{L}}_q(\fsl_N)\right)\big/\operatorname{Negl}
  \simeq
  \operatorname{Rep}(SU(N)_k)
  \label{eq:klf-equivalence}
\end{equation}
as modular tensor categories, where $\operatorname{Rep}(SU(N)_k)$ is the category of integrable representations of the affine Lie algebra $\widehat{\fsl}_N$ at level $k$. Its simple objects, the Wilson lines, are labeled by the level-$k$ integrable highest weights
\begin{equation}
P_k^+
=
\left\{
\lambda=\sum_{i=1}^{N-1}\lambda_i\omega_i
\;\middle|\;
\lambda_i\in\bZ_{\geq0},
\quad
\sum_{i=1}^{N-1}\lambda_i\leq k
\right\},
  \label{eq:q-fermion-integrable-alcove}
\end{equation}
where $\omega_i$ are the fundamental weights of $\fsl_N$, and the corresponding simple object is denoted $\mathcal{L}_\lambda$ which possesses no null vectors. Each such $\lambda$ is a Young diagram with at most $N-1$ rows and at most $k$ columns. Fusion is given by the Kac--Walton formula \cite{Walton1990:FusionRulesWZW}: a Young diagram in the Littlewood--Richardson product that is wider than $k$ is reflected back into the level-$k$ alcove by affine Weyl reflections and contributes with their overall sign, see Appendix~\ref{app:quantum-groups-at-root-of-unity}.

Applied to the site quantum group, the equivalence of \ref{eq:klf-equivalence} requires $k'>0$, while in the large-$m$ limit of interest, $m>k+N$, the level $k'$ is negative and the positive-level equivalence does not apply.  A feature of the representation theory of $U^{\mathrm{L}}_q(\fsl_m)$ at a root of unity $q=\exp(\pi i/\ell)$ with $\ell\leq m$ is the appearance of null states / proper submodules in the representations $\mathcal{L}^{U^{\mathrm{L}}_q(\fgl_m)}_{\lambda^\top}$ of the  quantum skew Howe duality of Equation~\eqref{eq:q-skew-howe-sl},
\[
  \mathcal{L}^{U^{\mathrm{L}}_q(\fsl_N)}_{\overline\lambda}
  \otimes
  \mathcal{L}^{U^{\mathrm{L}}_q(\fgl_m)}_{\lambda^\top} \subset \Lambda_q^\bullet(\bC_q^N\otimes\bC_q^m)
\]
specializing it to a Weyl module $\Delta_q(\lambda^\top)$ from a simple module which has no proper submodules, even when $\overline\lambda$ is a level-$k$ integrable highest weight of $U^{\mathrm{L}}_q(\fsl_N)$. A more thorough and precise explanation of the null states, and of how quantum skew Howe duality is modified at the root of unity before taking the quotient, is given in Appendix~\ref{app:quantum-groups-at-root-of-unity}, with examples.\footnote{The modding out of null vectors in these $U^{\mathrm{L}}_q(\fsl_m)$ representations appears to be the lattice analog of the familiar removal of null vectors in the minimal models of $W_N$ algebras \cite{FateevLukyanov1988:ZnSymmetryModels,Bais1988:CosetConstructionExtendedVirasoro}.}
In our setup, however, a stronger condition sets in before the inequality $m<k+N$ is violated. The full $q$-fermion Fock space in Equation~\eqref{eq:qfockspace} is free of null vectors only when $m\leq k$. To identify these null states, we must choose a Hermitian form compatible with the $*$-structure of the quantum group.\footnote{A $*$-structure is a conjugate-linear involutive anti-automorphism of the quantum group; the compatible Hermitian form is fixed in Section~\ref{sec:orthonormalization-qfermion-basis}.} As we will show explicitly in Appendix~\ref{sec:non-existence-of-annihilation-operator}, for $N\geq2$ and $m\geq k+1$ the $q$-fermion creation operators do not possess Hermitian adjoints on the physical quotient, so the norms of the states cannot be computed in the usual free fermion way. To circumvent this problem, we use the equivalent description of the system in terms of anyons of $SU(N)_k$ Chern--Simons theory. Canonical quantization \cite{Witten1989:QuantumFieldTheory,ElitzurEtAl1989:CanonicalQuantization} identifies the physical Hilbert space of Chern--Simons theory on a manifold with punctures $\Sigma_{g, n}$ with the space of conformal blocks of the $SU(N)_k$ WZW model, equivalently the fusion space of the corresponding Wilson lines \cite{MooreSeiberg1989:ClassicalQuantumCFT}. A basis of this space is given by the fusion trees \cite{Feiguin2007:InteractingAnyons,Trebst2008:FibonacciAnyonModels}, and in this realization the distinction between null and physical states is immediate, as we now describe.

\begin{figure}[!htbp]
  \centering
  \begin{tikzpicture}[>=stealth,
    trunk/.style={thick},
    leg/.style={thick},
    mid/.style={postaction={decorate, decoration={markings,
      mark=at position 0.58 with {\arrow{>}}}}},
    vtx/.style={circle, fill, inner sep=1.5pt},
    lab/.style={font=\small}
  ]
    \draw[trunk,mid] (-1.0,0) -- (0.6,0);
    \draw[trunk,mid] ( 0.6,0) -- (2.3,0);
    \draw[trunk,mid] ( 2.3,0) -- (4.0,0);
    \draw[trunk,mid] ( 5.2,0) -- (6.6,0);
    \draw[trunk,mid] ( 6.6,0) -- (8.0,0);
    \node at (4.6,0) {$\cdots$};
    \foreach \x in {0.6,2.3,6.6}{\node[vtx] at (\x,0) {};}
    \draw[leg,mid] (0.6,1.35) -- (0.6,0);
    \draw[leg,mid] (2.3,1.35) -- (2.3,0);
    \draw[leg,mid] (6.6,1.35) -- (6.6,0);
    \node[lab, above] at (0.6,1.35) {$\omega_{r_1}$};
    \node[lab, above] at (2.3,1.35) {$\omega_{r_2}$};
    \node[lab, above] at (6.6,1.35) {$\omega_{r_m}$};
    \node[lab, below] at (-0.2,-0.08) {$\lambda^{(0)}$};
    \node[lab, below] at ( 1.45,-0.08) {$\lambda^{(1)}$};
    \node[lab, below] at ( 3.15,-0.08) {$\lambda^{(2)}$};
    \node[lab, below] at ( 5.9,-0.08) {$\lambda^{(m-1)}$};
    \node[lab, below] at ( 7.3,-0.08) {$\lambda^{(m)}$};
  \end{tikzpicture}
\caption{The left-associated fusion tree is used throughout. Horizontal edges carry the intermediate labels $\lambda^{(i)}\in P_k^+$ and the vertical legs carry the local objects $\mathcal{L}_{\omega_{r_i}}$, $r_i\in\{0,\ldots,N\}$. Each trivalent vertex carries a topological junction operator in $\text{Hom}_{\mathcal{C}_{N,k}}(\mathcal{L}_{\lambda^{(i-1)}}\otimes \mathcal{L}_{\omega_{r_i}}, \mathcal{L}_{\lambda^{(i)}})$, whose space is one-dimensional since fusion with the antisymmetric powers is multiplicity-free \cite{Walton1990:FusionRulesWZW,GoodmanWenzl1990:LRHeckeRootsUnity}. Note that the vertical lines point downward. An upward-pointing fusion tree can be obtained by dualizing the representations from $r_i$ to $N-r_i$.}
  \label{fig:fusion-tree-topology}
\end{figure}
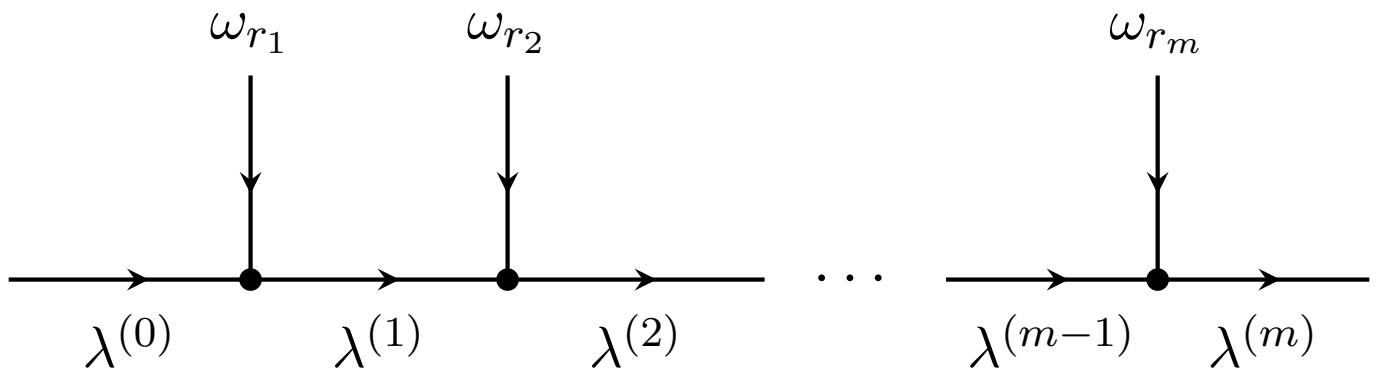

At every lattice site $i$, we have the direct sum of antisymmetric power representations of $SU(N)_k$ along with a grading corresponding to the fermion number:
\begin{equation}
\mathcal{H}_{i}:= \bigoplus_{j=0}^{N}\mathcal{L}_{\omega_j},
\label{eq:local-hilbert-space}
\end{equation}
where $\mathcal{L}_{\omega_0} \cong \mathcal{L}_{\omega_N} \cong \mathbbm{1}$ as simple objects of $SU(N)_k$. These have dimension\footnote{Here we mean the vector space dimension, not to be confused with the quantum dimension $\dim_q$.}
\begin{equation}
  \dim_{\bC} \mathcal{L}_{\omega_i}=\binom{N}{i},
  \label{eq:q-fermion-exterior-module-dimension}
\end{equation}
so the full Hilbert space for one site has dimension $2^{N}$. However, for $m$ sites the naive $2^{Nm}$ is not the dimension of the physical Hilbert space, since the local spaces are tensored not in $\text{Vec}$ but in $\mathcal{C}_{N,k}$.\footnote{$\text{Vec}$ is the symmetric tensor category of finite-dimensional vector spaces. $\text{Rep}(SU(N))$, the category of representations of $SU(N)$, has a forgetful functor to $\text{Vec}$ compatible with the tensor product, whereas $\text{Rep}(SU(N)_k)$ has no forgetful functor compatible with its braiding.} The simple objects also carry the $U(1)_F$ charges,
\begin{equation}
F(\mathcal{L}_{\omega_i})=i .
  \label{eq:q-fermion-exterior-module-charge}
\end{equation}
A general fusion tree begins at a charge $\lambda^{(0)}\in P_k^+$, fuses successively with the local objects $\mathcal{L}_{\omega_{r_i}}$, where $r_i\in\{0,\ldots,N\}$ is the fermion number at site $i$, and terminates at $\lambda^{(m)}\in P_k^+$, as denoted by the diagram
\begin{equation}
\mathcal{L}_{\lambda^{(0)}}
\xrightarrow{\;\mathcal{L}_{\omega_{r_1}}\;}
\mathcal{L}_{\lambda^{(1)}}
\xrightarrow{\;\mathcal{L}_{\omega_{r_2}}\;}
\cdots
\xrightarrow{\;\mathcal{L}_{\omega_{r_m}}\;}
\mathcal{L}_{\lambda^{(m)}},
\label{eq:fusion-arrow-diagram}
\end{equation}
where
\begin{equation}
  \mathcal{L}_{\lambda^{(i+1)}} \subset \mathcal{L}_{\lambda^{(i)}} \otimes \mathcal{L}_{\omega_{r_{i+1}}} .
  \label{eq:q-fermion-path-admissibility}
\end{equation}
Since the fusion with the antisymmetric powers $\mathcal{L}_{\omega_j}$ is multiplicity-free, the fusion diagram in Equation~\eqref{eq:fusion-arrow-diagram} uniquely determines a state in the space
\begin{equation}
  \mathcal{H}_{(\lambda^{(0)},\lambda^{(m)})}
  =
  \text{Hom}_{\mathcal{C}_{N,k}}\!\Bigl(
    \bigl(\cdots\bigl(\bigl(\mathcal{L}_{\lambda^{(0)}}\otimes\mathcal{H}_1\bigr)
      \otimes\mathcal{H}_2\bigr)\otimes\cdots\bigr)\otimes\mathcal{H}_m
    ,\;
    \mathcal{L}_{\lambda^{(m)}}
  \Bigr),
  \label{eq:fusion-tree-state-space}
\end{equation}
up to a normalization factor, which will be fixed in Section~\ref{sec:orthonormalization-qfermion-basis}. The topology of the fusion tree, that is, the order of fusion specified by the brackets in Equation~\eqref{eq:fusion-tree-state-space}, is shown in Figure~\ref{fig:fusion-tree-topology}.\footnote{This is a common choice; a different bracketing gives a different basis of the same Hilbert space, related to this one by the $F$-symbols.}

We will denote such a state corresponding to a fusion tree by
\begin{equation}
|\lambda^{(0)}, r_1 ; \lambda^{(1)}, r_2 ; \cdots; \lambda^{(m-1)}, r_m ; \lambda^{(m)}\rangle .
\label{eq:fusion-tree-state}
\end{equation}
We can associate to the state in Equation~\eqref{eq:fusion-tree-state} a whole $\dim_{\bC} \mathcal{L}_{\lambda^{(m)}}$ multiplet of states transforming in $\mathcal{L}_{\lambda^{(m)}}$ of $U_q^{\mathrm{L}}(\fsl_N)$, which we denote by $\widehat{\mathcal{H}}_{(\lambda^{(0)},\lambda^{(m)})}:=\mathcal{L}_{\lambda^{(m)}}\otimes\mathcal{H}_{(\lambda^{(0)},\lambda^{(m)})}$; contracting with the compensating anyon ${\lambda^{(m)}}^{\vee}$ at infinity singles out one gauge-invariant state from this multiplet, which is the gauge-invariant state of the Chern--Simons theory. The total $U(1)_F$ charge of the state is given by the sum of the $U(1)_F$ charges of the local objects $\mathcal{L}_{\omega_{r_i}}$ along the fusion tree,
\begin{equation}
  Q = \sum_{i=1}^{m} r_i .
  \label{eq:q-fermion-path-total-charge}
\end{equation}

We can therefore classify the Hilbert space further into sectors labeled by the charges $\lambda^{(0)}$ and $\lambda^{(m)}$ of the two external lines and by the total charge $Q\in\{0,\ldots,Nm\}$,
\begin{equation}
  \mathcal{H}_{(\lambda^{(0)},\lambda^{(m)})} (Q) .
  \label{eq:q-fermion-fixed-charge-hilbert-space}
\end{equation}
These sector labels are not all independent, because of the identification of the $\bZ_N$ grading of the $U^{\mathrm{L}}_q(\fsl_N)$ representation by $N$-ality/simple current with a $\bZ_{N}$ subgroup of $U(1)_F$.\footnote{This is an analogous statement of $ \frac{SU(N) \times U(1)_F}{\bZ_N}$ for free fermions.} Consequently, a sector can be nonzero only if
\begin{equation}
    \mathcal{H}_{(\lambda^{(0)},\lambda^{(m)})}(Q)\neq 0
    \quad\Longrightarrow\quad
    Q\equiv
    |\lambda^{(m)}|-|\lambda^{(0)}|
    \pmod N ,
    \label{eq:fermion-nality-selection}
\end{equation}
where $|\lambda|$ is the number of boxes in the $\fsl_N$ Young diagram.

To identify the null states of the $q$-fermions it is useful to consider the space of unrestricted fusion trees of $\mathrm{Rep}\!\left(SU(N)\right) \cong \cC_{N, \infty}$, with simple objects corresponding to any irreducible representation of $SU(N)$. Their intermediate labels $\lambda^{(i)}$ are arbitrary $SU(N)$ highest weights, represented by Young diagrams with at most $N-1$ rows and with no restriction on their column width. We denote the corresponding unrestricted Hilbert space as $\widehat{\mathcal{H}}^{\infty}$. In a sector with fixed initial and final charges, we have
\begin{equation}
\mathcal{H}^{\infty}_
{\left(\lambda^{(0)},\lambda^{(m)}\right)}
=
\text{Hom}_{\cC_{N, \infty}}\!
\!\left(
\bigl(\cdots\bigl(
\mathcal{L}_{\lambda^{(0)}}\otimes\mathcal{H}_1
\bigr)\otimes\cdots\bigr)\otimes\mathcal{H}_m,
\mathcal{L}_{\lambda^{(m)}}
\right).
\label{eq:prequotient-fusion-path-space}
\end{equation}
Since the fusion rules of $\mathcal{C}_{N,k}$ are a truncation of those of $\mathcal{C}_{N,\infty}$, every admissible finite level-$k$ fusion tree is in particular an unrestricted one,
\begin{equation}
  \mathcal{H}_{(\lambda^{(0)},\lambda^{(m)})} \subseteq \mathcal{H}^{\infty}_{(\lambda^{(0)},\lambda^{(m)})} .
  \label{eq:q-fermion-physical-path-inclusion}
\end{equation}
The null states are spanned by the unrestricted trees that are not admissible, so that, as vector spaces,
\begin{equation}
  \mathcal{H}^{\infty}_{(\lambda^{(0)},\lambda^{(m)})}=\mathcal{H}_{(\lambda^{(0)},\lambda^{(m)})}\oplus\mathcal{R}_{(\lambda^{(0)},\lambda^{(m)})},
  \label{eq:sector-null-decomposition}
\end{equation}
with $\mathcal{R}_{(\lambda^{(0)},\lambda^{(m)})}$ the span of the non-admissible trees of the sector. In the original setup of the $q$-fermion Fock space in Equation~\eqref{eq:q-fermion-exterior-algebra}, when the initial charge is taken to be trivial, $\lambda^{(0)}=\mathbbm{1}$, we have the identification
\begin{equation}
\Lambda_q^\bullet\bigl(\bC_q^N \otimes \bC_q^m\bigr)
\cong
\bigoplus_{\lambda^{(m)} \in P^+}
\mathcal{L}_{\lambda^{(m)}}\otimes\mathcal{H}^{\infty}_{(\mathbbm{1},\,\lambda^{(m)})} =: \widehat{\mathcal{H}}^{\infty} .
\label{eq:unrestricted-space-identification}
\end{equation}
We note that this is an isomorphism  as vector spaces, not of $U^{\mathrm{L}}_q(\fsl_N)$-modules. \footnote{This isomorphism will be specified in \ref{sec:path-to-qfermion}} As we describe in Appendix~\ref{app:quantum-groups-at-root-of-unity}, the $U^{\mathrm{L}}_q(\fsl_N)$-module structure of $\Lambda_q^\bullet\bigl(\bC_q^N \otimes \bC_q^m\bigr)$ is a direct sum of indecomposable tilting modules $T_q(\lambda)$ of $U^{\mathrm{L}}_q(\fsl_N)$ tensored with multiplicity spaces $M_\lambda$,
\begin{equation}
\Lambda_q^\bullet\bigl(\bC_q^N \otimes \bC_q^m\bigr)
\cong
\bigoplus_{\lambda}
T_q(\lambda)\otimes M_\lambda ,
\label{eq:fock-tilting-decomposition}
\end{equation}
where $\lambda$ runs over the highest weights of the tilting summands in this tensor product.
The physical Hilbert space given by fusion trees of $\mathcal{C}_{N,k}$ is then identified with
\begin{equation}
  \widehat{\mathcal{H}}^{\mathrm{phys}}
  \cong
 \bigoplus_{\lambda^{(m)} \in P_k^+} \mathcal{L}_{\lambda^{(m)}}\otimes\mathcal{H}_{(\mathbbm{1},\,\lambda^{(m)})} \cong \frac{\Lambda_q^\bullet\bigl(\bC_q^N \otimes \bC_q^m\bigr)}{ \mathcal{R}_k}
 \label{eq:physical-Hilbert-space-qfermion}
 \end{equation}
where $\mathcal{R}_k\cong\bigoplus_{\lambda^{(m)}\in P^+}\mathcal{L}_{\lambda^{(m)}}\otimes\mathcal{R}_{(\mathbbm{1},\lambda^{(m)})}$ is the space of all null states in the $q$-fermion Fock space.\footnote{Note that we are working with $\widehat{\mathcal{H}}$ here and have not isolated the gauge-invariant state from the multiplet. It is also important to note that gauge invariance and zero norm are independent conditions.} We can define the inner product,
\begin{equation}
  \langle \gamma^{(0)}, s_1 ; \cdots ; \gamma^{(m)}
  | \lambda^{(0)}, r_1 ; \cdots ; \lambda^{(m)} \rangle
  = g \prod_{j = 1}^{m} \delta_{(s_j, r_j)} \prod_{i=0}^{m} \delta_{(\gamma^{(i)}, \lambda^{(i)})}
  \label{eq:path-qfermion-inner-product-orthogonality}
\end{equation}
since fusion with antisymmetric powers is multiplicity-free, and $g$ should vanish whenever a highest-weight label $\lambda^{(i)}$ lies outside the level-$k$ alcove. We will provide a more detailed construction of this Hermitian form in Section~\ref{sec:orthonormalization-qfermion-basis}. It can be seen directly from the fusion tree that no null states occur when $m\leq k$, provided that the tree begins at the vacuum, $\lambda^{(0)}=\mathbbm{1}$. Define
\begin{equation}
\ell(\lambda)
:=
\left\langle\lambda,\theta^\vee\right\rangle
=
\sum_{a=1}^{N-1}\lambda_a,
  \label{eq:q-fermion-weight-level}
\end{equation}
for a highest weight $\lambda$, where $\theta$ is the highest root of $A_{N-1}$. $\ell(\lambda)$ measures the width of the corresponding Young diagram. The Pieri rule used to describe the fusion of Young diagrams states that fusion with an antisymmetric representation $\mathcal{L}_{\omega_r}$ adds a vertical strip of $r$ boxes. This operation can increase the column width of the Young diagram by at most one. Consequently, every step of the fusion tree satisfies
\begin{equation}
\ell\!\left(\lambda^{(j+1)}\right)
\leq
\ell\!\left(\lambda^{(j)}\right)+1.
  \label{eq:q-fermion-path-level-step-bound}
\end{equation}
Starting from $\lambda^{(0)}=\mathbbm{1}$,
\begin{equation}
\ell\!\left(\lambda^{(j)}\right)\leq j.
  \label{eq:q-fermion-path-level-vacuum-bound}
\end{equation}
If $m\leq k$, then every intermediate weight of a fusion tree satisfies
\begin{equation}
\ell\!\left(\lambda^{(j)}\right)
\leq j
\leq m
\leq k,
  \label{eq:q-fermion-vacuum-untruncated-range}
\end{equation}
and hence $\lambda^{(j)}\in P_k^+$ for all $ j=0,\ldots,m$. Therefore the unrestricted Hilbert space $\mathcal{H}^{\infty}_{(\mathbbm{1},\lambda^{(m)})}$ contains no null states and coincides with the physical Hilbert space $\mathcal{H}_{(\mathbbm{1},\,\lambda^{(m)})}$. 
In this case $m\leq k$ also satisfies $m<k+N$, so the site level $k'$ of Equation~\eqref{eq:q-fermion-site-level-canonical} is positive, and the quantum skew Howe duality of Equation~\eqref{eq:q-skew-howe-sl} holds,
\begin{equation}
 \Lambda_q^\bullet(\bC_q^N\otimes\bC_q^m)
 \cong\bigoplus_{\lambda\subseteq(m^N)}
 \mathcal{L}^{U^{\mathrm{L}}_q(\fsl_N)}_{\overline\lambda}\otimes\mathcal{L}^{U^{\mathrm{L}}_q(\fgl_m)}_{\lambda^\top},
 \label{eq:q-skew-howe-root-of-unity}
\end{equation}
with $\mathcal{L}^{U^{\mathrm{L}}_q(\fgl_m)}_{\lambda^\top}$ being the integrable representation of $\text{Rep}(SU(m)_{k'})$. More generally, for a fusion tree beginning at a nontrivial initial charge $\lambda^{(0)}$, the corresponding sufficient condition is
\begin{equation}
\ell\!\left(\lambda^{(0)}\right)+m\leq k.
  \label{eq:q-fermion-endpoint-untruncated-range}
\end{equation}

However, for large $m$ the Fock space does contain null states. The skew Howe duality statement of Equation~\eqref{eq:q-skew-howe-root-of-unity} can be made after quotienting them out. Using Equation~\eqref{eq:physical-Hilbert-space-qfermion}, we have
\begin{equation}
 \bigl(\Lambda_q^\bullet(\bC_q^N\otimes\bC_q^m)\bigr)_{\mathrm{phys}}
 =\frac{\Lambda_q^\bullet(\bC_q^N\otimes\bC_q^m)}{\mathcal{R}_k}
 \cong\bigoplus_{\substack{\lambda\subseteq(m^N)\\\overline\lambda\in P_k^+}}
 \mathcal{L}^{U^{\mathrm{L}}_q(\fsl_N)}_{\overline\lambda}\otimes \cH_{(1, \overline{\lambda})}(|\lambda|),
 \label{eq:q-skew-howe-physical}
\end{equation}
where only the color labels in the alcove survive. The multiplicity space $\mathcal{H}_{(\mathbbm{1},\overline\lambda)}(|\lambda|)$ can be shown to be the $U^{\mathrm{L}}_q(\fgl_m)$-representation of highest weight $\lambda^\top$ with its null vectors modded out, that is, the simple quotient of the Weyl module $\Delta_q(\lambda^\top)$, see Appendix~\ref{app:quantum-groups-at-root-of-unity},
\begin{equation}
\mathcal{H}_{(\mathbbm{1},\lambda^{(m)})}(Q) \cong \cL^{U^{\mathrm{L}}_q(\fgl_m)}_{\lambda^\top} ,
  \label{eq:q-fermion-vacuum-endpoint-skew-howe}
\end{equation}

For a general initial charge $\lambda^{(0)}$, tensoring Equation~\eqref{eq:q-skew-howe-physical} with $\mathcal{L}_{\lambda^{(0)}}$ in $\mathcal{C}_{N,k}$ gives
\begin{equation}
\begin{aligned}
&\left(\mathcal{L}_{\lambda^{(0)}} \otimes
\Lambda_q^\bullet(\bC_q^N \otimes \bC_q^m)\right)_{\mathrm{phys}}
\cong
\bigoplus_{\lambda^{(m)} \in P_k^+}
\mathcal{L}^{U^{\mathrm{L}}_q(\fsl_N)}_{\lambda^{(m)}}
\otimes
\left(
\bigoplus_{\substack{\lambda\subseteq(m^N)\\\overline\lambda\in P_k^+}}
\mathcal{N}^{\lambda^{(m)}}_{\lambda^{(0)}\overline{\lambda}}\;
\mathcal{L}^{U^{\mathrm{L}}_q(\fgl_m)}_{\lambda^\top}
\right),
\end{aligned}
  \label{eq:q-fermion-endpoint-tensor-decomposition}
\end{equation}
where the subscript again denotes the physical quotient and $\mathcal{N}^{\lambda^{(m)}}_{\lambda^{(0)}\overline\lambda}$ are the fusion coefficients of $\mathcal{C}_{N,k}$. Each sector is therefore a semisimple $U^{\mathrm{L}}_q(\fgl_m)$-module,
\begin{equation}
\mathcal{H}_{(\lambda^{(0)},\lambda^{(m)})}(Q) \cong \bigoplus_{\substack{\lambda\subseteq(m^N),\ |\lambda|=Q\\\overline\lambda\in P_k^+}} \mathcal{N}^{\lambda^{(m)}}_{\lambda^{(0)}\overline\lambda}\; \mathcal{L}^{U^{\mathrm{L}}_q(\fgl_m)}_{\lambda^\top} ,
  \label{eq:q-fermion-general-endpoint-skew-howe}
\end{equation}
which reduces to Equation~\eqref{eq:q-fermion-vacuum-endpoint-skew-howe} for $\lambda^{(0)}=\mathbbm{1}$. The simplest case, $SU(2)_k$, is worked out in Appendix~\ref{app:su2k-fusion-tree-example}.

The number of states in the physical Hilbert space of a given sector can be counted by the standard transfer-matrix method of anyon chains \cite{Feiguin2007:InteractingAnyons,Trebst2008:FibonacciAnyonModels}, with the local object replaced by the direct sum of antisymmetric powers. Let $\mathcal{N}_{ab}^c$ be the fusion numbers satisfying
\begin{equation}
a \otimes b = \bigoplus_{c} \mathcal{N}_{ab}^c \; c \qquad \forall a,b,c \in P_k^+ .
  \label{eq:q-fermion-fusion-coefficients}
\end{equation}
Define the fusion matrix
\begin{equation}
\left(N_a\right)^c_{b} = \mathcal{N}_{ab}^{c} .
  \label{eq:q-fermion-fusion-matrices}
\end{equation}
Define the generating matrix $M(x)$ as
\begin{equation}
M(x)
=
\sum_{r=0}^{N}x^r N_{\omega_r}
=
(1+x^N)I
+
\sum_{r=1}^{N-1}x^r N_{\omega_r}.
  \label{eq:q-fermion-occupation-transfer-matrix}
\end{equation}
Then the number of states from $\alpha$ to $\beta$ with $m$ sites and total $U(1)_F$ charge $Q$ is
\begin{equation}
\dim\widehat{\mathcal{H}}_{(\alpha,\beta)}(Q)
=
\dim(\mathcal{L}_{\beta})[x^Q]
\left(M(x)^m\right)_{\beta,\alpha}.
  \label{eq:q-fermion-fixed-charge-path-dimension}
\end{equation}
\begin{equation}
  \dim\widehat{\mathcal{H}}_{(\alpha,\beta)} = \sum_{Q} \dim\widehat{\mathcal{H}}_{(\alpha,\beta)}(Q) = \dim(\mathcal{L}_{\beta}) (M(1)^m)_{\beta,\alpha} .
  \label{eq:q-fermion-total-path-dimension}
\end{equation}

\subsection[The \texorpdfstring{$q$}{q}-fermion basis and the site quantum-group action]{The \texorpdfstring{$q$}{q}-fermion basis and the action of \texorpdfstring{$U^{\mathrm{L}}_q(\fsl_m)$}{Uq(fslm)} on fusion trees}
\label{sec:path-to-qfermion}

From Section~\ref{sec:fusion-tree-basis}, it is clear that the fusion-tree basis directly specifies the states of the physical Hilbert space. However, the action of the $U^{\mathrm{L}}_q(\fgl_m)$ quantum group generators is naturally defined on $\Lambda_q^{\bullet}(\bC_q^N \otimes \bC_q^m)$. To understand the action of $U^{\mathrm{L}}_q(\fgl_m)$ on the physical Hilbert space, we require a map relating the fusion-tree basis and the $q$-fermion basis. We begin by defining a map, invertible for generic $q$,\footnote{At a root of unity, tensor products of $U^{\mathrm{L}}_q(\fsl_N)$ modules need not be semisimple, so the map $\mu$ need not remain invertible.}
\begin{equation}
  \mu(q) : \widehat{\mathcal{H}}^{\infty} \longrightarrow \Lambda_q^{\bullet}(\bC_q^N \otimes \bC_q^m) ,
  \label{eq:q-fermion-path-multiplication-map}
\end{equation}
where $\widehat{\mathcal{H}}^{\infty}$ includes the color multiplets, as in Equation~\eqref{eq:unrestricted-space-identification}, and the fusion trees start at the vacuum. We construct $\mu(q)$ iteratively, that is, site by site, sending each fusion tree to a PBW-ordered state in the $q$-fermion Fock space. Consider a fusion tree and suppose the state has been specified up to site $j<m$. At each position $j$ there is a $\dim_{\bC}{(\mathcal{L}_{\lambda^{(j)}})}$-fold multiplicity of states, but these are easy to obtain once the highest weight of the $U^{\mathrm{L}}_q(\fsl_N)$ representation is given by iteratively applying the lowering operators $F^a_{\mathrm{col}}$, see\ Equation~\eqref{eq:q-fermion-site-cartan-action}. Let $\Omega^{\text{hw}}_{j}$ be the highest-weight vector representing the tree in Equation~\eqref{eq:path-qfermion-fusion-tree} until site $j$,
\begin{equation}
\underbrace{
\mathcal{L}_{\lambda^{(0)}}
\xrightarrow{\ \mathcal{L}_{\omega_{r_1}}\ }
\cdots
\xrightarrow{\ \mathcal{L}_{\omega_{r_j}}\ }
\mathcal{L}_{\lambda^{(j)}}
}_{\textstyle \Omega^{\text{hw}}_{j} \in \mathcal{H}_{(\lambda^{(0)},\lambda^{(j)})}}
\xrightarrow{\ \mathcal{L}_{\omega_{r_{j+1}}}\ }
\mathcal{L}_{\lambda^{(j+1)}}
\cdots
\xrightarrow{\ \mathcal{L}_{\omega_{r_m}}\ }
\mathcal{L}_{\lambda^{(m)}} ,
\label{eq:path-qfermion-fusion-tree}
\end{equation}
satisfying the highest-weight condition,
\begin{equation}
\begin{aligned}
E^a_{\mathrm{col}}(\Omega^{\text{hw}}_{j})&=0, \\
K^a_{\mathrm{col}}(\Omega^{\text{hw}}_{j})&=q^{\langle\lambda^{(j)},\alpha_a^\vee\rangle}\,\Omega^{\text{hw}}_{j},
\end{aligned}
  \label{eq:q-fermion-path-highest-weight-action}
\end{equation}
for $a=1,\ldots,N-1$. Let $r_j = p$. Then the states of the representation $\mathcal{L}_{\omega_p} \cong \Lambda_q^{p} \bC_q^N$ at site $j$ are given by\footnote{Here we have suppressed the tensor product $\otimes$ for brevity.}
\begin{equation}
\bchi^{S}_{j}
:=
\bqpsi^{s_1}_{j}\bqpsi^{s_2}_{j}\cdots\bqpsi^{s_p}_{j} , \quad \text{where} \quad S = \{s_1<\cdots<s_p\}\subset\{1,\ldots,N\} .
  \label{eq:q-fermion-site-exterior-monomial}
\end{equation}
The highest-weight vector of this representation is
\begin{equation}
  \bchi^{\text{hw}}_{j}:=\bqpsi^{1}_{j}\bqpsi^{2}_{j}\cdots\bqpsi^{p}_{j},
  \label{eq:q-fermion-site-highest-weight-monomial}
\end{equation}
satisfying
\begin{equation}
\begin{aligned}
E^a_{\mathrm{col}}(\bchi^{\text{hw}}_{j})&=0,
 \\
K^a_{\mathrm{col}}(\bchi^{\text{hw}}_{j})&=q^{\langle\omega_p,\alpha_a^\vee\rangle}\,\bchi^{\text{hw}}_{j}
=q^{\delta_{ap}}\,\bchi^{\text{hw}}_{j},
\end{aligned}
  \label{eq:q-fermion-site-highest-weight-action}
\end{equation}
for $a=1,\ldots,N-1$. The remaining states of the representation are obtained by applying the lowering operators $F^a_{\mathrm{col}}$ to the highest-weight state. Since $F^a_{\mathrm{col}}$ moves one color index from $a$ to $a+1$, reaching $S=\{s_1<\cdots<s_p\}$ from $\{1,\ldots,p\}$ requires raising each $l$ to $s_l$. Define
\begin{equation}
\mathcal{F}^l_{\mathrm{col}} := F^{s_l-1}_{\mathrm{col}}\cdots F^{l}_{\mathrm{col}} .
  \label{eq:q-fermion-color-lowering-string}
\end{equation}
Then the state corresponding to $S$ is given by
\begin{equation}
  \bchi^{S}_{j}
  =
  \mathcal{F}^1_{\mathrm{col}} \mathcal{F}^2_{\mathrm{col}} \cdots \mathcal{F}^p_{\mathrm{col}}\bigl(\bchi^{\text{hw}}_{j}\bigr) ,
  \label{eq:path-qfermion-lowered-states}
\end{equation}
As $\mathcal{L}_{\omega_p}$ is minuscule, every weight space is one-dimensional, so any other lowering word either vanishes or reproduces the same state. Therefore the words in Equation~\eqref{eq:path-qfermion-lowered-states} specify all the states of $\mathcal{L}_{\omega_p}$. Although $\mathcal{L}_{\omega_0} \cong \mathcal{L}_{\omega_N} \cong \mathbbm{1}$ as simple objects of $\cC_{N, k}$ at site $j$, the two correspond to tensoring by different $q$-fermion representatives,
\begin{equation}
\begin{aligned}
\mathcal{L}_{\omega_0} &\longrightarrow \mathbbm{1}_j, \\
  \mathcal{L}_{\omega_N} &\longrightarrow \bqpsi^{1}_j \bqpsi^{2}_j \cdots \bqpsi^{N}_j ,
\end{aligned}
  \label{eq:q-fermion-trivial-exterior-embeddings}
\end{equation}
The highest-weight state at site $j+1$ corresponding to $\lambda^{(j+1)}$ is built by appending the states of the representation $\mathcal{L}_{\omega_{r_{j+1}}}$ to the right of $\Omega^{\text{hw}}_{j}$ and of its descendants and taking a linear combination of those terms of weight $\lambda^{(j+1)}$, and imposing the highest-weight condition. Schematically, then, the highest-weight state at site $j+1$ is given by
\begin{equation}
\Omega^{\text{hw}}_{j+1}
=
\sum_{\substack{A \in \mathscr{A} \\ S \in \mathscr{S}}}
C^{A,S}_{j}\,
\bigl(\mathcal{F}^A_{\mathrm{col}}\Omega^{\text{hw}}_{j}\bigr)
\otimes \bchi^{S}_{j+1},
\label{eq:path-qfermion-recursion}
\end{equation}
where $S\in\mathscr{S}$ runs over the $r_{j+1}$-element subsets of $\{1,\ldots,N\}$ and $\mathcal{F}^A_{\mathrm{col}}\Omega^{\text{hw}}_{j} \in \mathcal{L}_{\lambda^{(j)}}$, $A\in\mathscr{A}$, where $\mathscr{A}$ is the set of all lowering words of $\mathcal{L}_{\lambda^{(j)}}$ constructed from $F^a_{\mathrm{col}}$. Each term in the sum has weight $\lambda^{(j+1)}$,
\begin{equation}
\mathrm{wt}(\mathcal{F}^A_{\mathrm{col}}\Omega^{\text{hw}}_{j}) + \mathrm{wt}(\bchi^{S}_{j+1}) = \lambda^{(j+1)}
  \label{eq:q-fermion-path-weight-matching}
\end{equation}
The highest-weight state of $\mathcal{L}_{\lambda^{(j+1)}}$ is obtained by imposing the highest-weight condition on $\Omega^{\text{hw}}_{j+1}$, which for generic $q$, fixes the vector up to an overall normalization.
  \begin{equation}
\begin{aligned}
E^a_{\mathrm{col}}\bigl(\Omega^{\text{hw}}_{j+1}\bigr)&=0,
\\
K^a_{\mathrm{col}}\bigl(\Omega^{\text{hw}}_{j+1}\bigr)
&=q^{\langle\lambda^{(j+1)}, \;\alpha_a^\vee\rangle}\,\Omega^{\text{hw}}_{j+1} .
\end{aligned}
\label{eq:path-qfermion-highest-weight-equations}
\end{equation}
For the overall scalar we choose the following convention. We set the coefficient of the unlowered term $\Omega^{\text{hw}}_{j}\otimes\bchi^{S}_{j+1}$, with $ \nu_S :=\text{wt}(\bchi^{S}_{j+1})=\lambda^{(j+1)}-\lambda^{(j)}$, equal to\footnote{In \cite{Elias2015:LightLadders} they pick the leading coefficient of $\Omega^{\text{hw}}_{j}\otimes\bchi^{S}_{j+1}$ to be $\pm q^{n}c_{\lambda,S}$ for some integer $n$. At $q=e^{i\pi/(k+N)}$ the overall phase $\pm q^{n}$ does not change the norm, which is of primary interest here.}
\begin{equation}
\begin{gathered}
 C_{j}^{0, S}
 :=
 \frac{\displaystyle\prod_{\substack{\alpha\in\Phi^{+}\\ \langle\nu_S,\alpha^{\vee}\rangle=-1}}[\langle\lambda^{(j)}+\rho,\alpha^{\vee}\rangle-1]_q}
      {\displaystyle\prod_{\substack{\alpha\in\Phi^{+}\\ \langle\nu_S,\alpha^{\vee}\rangle=+1}}[\langle\rho,\alpha^{\vee}\rangle]_q}
=
 \frac{\displaystyle\prod_{\substack{1\leq k<l\leq N\\ \delta_{kl}=-1}}[a_{kl}-1]_q}
      {\displaystyle\prod_{\substack{1\leq k<l\leq N\\ \delta_{kl}=+1}}[\,l-k\,]_q},
 \label{eq:path-qfermion-step-normalization}
\\
 a_{kl}=l-k+\sum_{\ell=k}^{l-1}\lambda^{(j)}_\ell,
 \qquad
 \delta_{kl}:=\langle\nu_S,\alpha_{kl}^{\vee}\rangle=\mathbf{1}_S(k)-\mathbf{1}_S(l),
\\
 \mathbf{1}_S(k)=\begin{cases}1, & k\in S,\\ 0, & k\notin S.\end{cases}
\end{gathered}
\end{equation}
where $\Phi^{+}$ is the set of positive roots of $\fsl_N$, described in the orthogonal coordinates as $\alpha_{kl}=\varepsilon_k-\varepsilon_l$, $k<l$. The vector $\rho = \frac{1}{2}\sum_{\alpha\in\Phi^{+}}\alpha$ is the Weyl vector, $\lambda^{(j)}_\ell$ are the Dynkin labels of $\lambda^{(j)}$, and $[n]_q = \tfrac{q^n - q^{-n}}{q - q^{-1}}$ is the $q$-number. Some comments are in order for this choice.
\begin{itemize}
  \item This term is always present at every junction and is unique. To prove this, suppose there were no term of the form $\Omega^{\text{hw}}_{j}\otimes\bchi^{S}_{j+1}$, and among the terms of the sum consider those whose first factor $\mathcal{F}^A_{\mathrm{col}}\Omega^{\text{hw}}_{j}$ has the largest weight, say $\lambda^{(j)}-\beta$. Since $\Delta(E^a_{\mathrm{col}})=E^a_{\mathrm{col}}\otimes K^a_{\mathrm{col}}+1\otimes E^a_{\mathrm{col}}$, the first term acts on the first factor and raises its weight to $\lambda^{(j)}-\beta+\alpha_a$, and for some $a$ the result is nonzero, since a vector annihilated by every $E^a_{\mathrm{col}}$ would be a highest-weight vector of weight $\lambda^{(j)}$, which is $\Omega^{\text{hw}}_{j}$. As there is no other term whose first factor has a larger weight by assumption, this term cannot be canceled and the highest-weight condition is violated. This issue is not present for the term $\Omega^{\text{hw}}_{j}\otimes\bchi^{S}_{j+1}$, since it is annihilated by $E^a_{\mathrm{col}} \otimes K^a_{\mathrm{col}}$. Uniqueness follows directly from the fact that the weight spaces of $\Lambda_q^{r} \bC^{N}_q$, and the weight space of $\Omega^{\text{hw}}_{j}$ in $\mathcal{L}_{\lambda^{(j)}}$ are one-dimensional.
  \item The denominator of this coefficient is nonzero for every $SU(N)_k$ with $k \geq 1$, so it is well defined even when $q$ is a root of unity. This choice also ensures that the vector $\Omega^{\text{hw}}_{j+1}$ is regular and nonzero at $q=e^{i\pi/(k+N)}$ whenever $\lambda^{(j)},\lambda^{(j+1)}\in P_k^+$.\footnote{For $\lambda^{(j)}\in P_k^+$ every Pieri summand of $\mathcal{L}_{\lambda^{(j)}}\otimes\Lambda^r_q\bC^N_q$ lies in the closed alcove, where Weyl modules are simple, so the tensor product is semisimple at the root of unity and its highest-weight vector of weight $\lambda^{(j+1)}$ is unique up to a scalar. The argument of the first item applies at the root of unity, so a primitive integral highest-weight vector has a unit coefficient on the unlowered term, and $C^{0,S}_{j}$ is itself a unit for $\lambda^{(j+1)}\in P_k^+$; rescaling one to the other keeps every PBW coefficient regular. For trees that leave the alcove we have checked this only in examples, in closed form for $SU(2)_k$ in Appendix~\ref{sec:su2k-action-example}.}
  \item This choice is not arbitrary and is in fact heavily inspired by the work of Elias \cite{Elias2015:LightLadders} and Martin and Spencer \cite{MartinSpencer2022:CellModulesWebs}. Their choice of junction normalization, which will be discussed in Section~\ref{sec:orthonormalization-qfermion-basis}, leads to setting $C^{0, S}_{j} = 1$, which will result in other $C^{A, S}_j$ developing poles at $|q| = 1$. Our choice of normalization is a modification of theirs which avoids this issue. We will see yet another advantage of this choice, which is conjectured to lead fusion-tree states with non-integrable labels to have zero norm.
\end{itemize}

So we have constructed a map $\mu(q)$ which associates to each unrestricted fusion tree a state in the $q$-fermion Fock space, and the generators of the quantum group act according to Equation~\eqref{eq:q-fermion-site-cartan-action} on this pre-quotient space $\Lambda_q^\bullet(\bC_q^N \otimes \bC_q^m)$. However, it is not guaranteed a priori that the generators of $U^{\mathrm{L}}_q(\fgl_m)$ descend to well-defined operators on the physical quotient of Equation~\eqref{eq:physical-Hilbert-space-qfermion}. For this descent to be well defined, it is necessary and sufficient that the action of $U^{\mathrm{L}}_q(\fgl_m)$ preserves the null subspace. Let $\rho : U^{\mathrm{L}}_q(\fgl_m) \to \mathrm{End}(\Lambda_q^\bullet(\bC_q^N \otimes \bC_q^m))$ be the representation. Then the condition is
\begin{equation}
\rho(X)\mathcal{R}_k
\subseteq
\mathcal{R}_k
\qquad
\text{for every }
X\in U^{\mathrm{L}}_q(\fgl_m).
  \label{eq:q-fermion-radical-invariance}
\end{equation}
Categorically, descent follows because negligible morphisms form a tensor ideal. In the PBW realization, it requires compatibility of the chosen positive-semidefinite Hermitian form with the $*$-structure of the quantum group, see Appendix~\ref{app:uqglm-relations}. At the root of unity, let
\begin{equation}
\langle \cdot, \cdot \rangle : \Lambda_q^\bullet(\bC_q^N \otimes \bC_q^m) \times \Lambda_q^\bullet(\bC_q^N \otimes \bC_q^m) \to \mathbb{C}
  \label{eq:q-fermion-generic-pairing}
\end{equation}
be a positive-semidefinite Hermitian form, and let the $*$-structure of the quantum group be defined by
\begin{equation}
E_i^\star=F_i,\qquad F_i^\star=E_i,\qquad L_i^\star=L_i^{-1},
\label{eq:path-qfermion-hermiticity-condition}
\end{equation}
and compatibility means that\footnote{An orthonormalization of the physical fusion-tree states is presented in Section~\ref{sec:orthonormalization-qfermion-basis}.}
\begin{equation}
\left\langle
v,\rho(X)u
\right\rangle
=
\left\langle
\rho\!\left(X^\star\right)v,u
\right\rangle
\qquad
\forall \;
u,v\in
\Lambda_q^\bullet
\!\left(
\bC_q^N\otimes\bC_q^m
\right).
\label{eq:path-qfermion-adjoint-compatibility}
\end{equation}
For a null state $u\in\mathcal{R}_k$ which is orthogonal to every state, both sides of Equation~\eqref{eq:path-qfermion-adjoint-compatibility} vanish for all $v$ at $q=e^{i\pi/(k+N)}$. Therefore
\begin{equation}
\rho(X)\,u\in\mathcal{R}_k \qquad \text{for } u \in \mathcal{R}_k.
  \label{eq:q-fermion-radical-preserved-action}
\end{equation}
When this compatibility holds, the quantum-group generators descend to well-defined operators on the quotient,
\begin{equation}
\bar{\rho}(X)[u]
:=
[\rho(X)u] \qquad \forall \; X \in U^{\mathrm{L}}_q(\fgl_m) .
  \label{eq:q-fermion-quotient-action}
\end{equation}
More generally, any operator that admits an adjoint with respect to the chosen positive-semidefinite Hermitian form preserves $\mathcal{R}_k$, and so descends to a well-defined operator on the quotient. And this is exactly the criterion that the $q$-fermion creation operators fail to satisfy. That is, left multiplication by $\bqpsi_i^a$ gives
 \begin{equation}
    \bqpsi_i^a(\mathcal{R}_k)\not\subseteq\mathcal{R}_k.
  \label{eq:q-fermion-creator-radical-obstruction}
\end{equation}
Explicit counterexamples are constructed in Appendix~\ref{sec:non-existence-of-annihilation-operator}.

For the categorical action, we compute it in the fusion-tree basis, where the quotient is simple: the null states are exactly the trees that carry a non-integrable label.\footnote{Note that a non-integrable label can occur not only in the interior of the fusion tree, but also at the final site $\lambda^{(m)}$, in which case the whole Hilbert space $\mathcal{H}^\infty_{(\lambda^{(0)},\lambda^{(m)})}$ lies in the null subspace.} From Equation~\eqref{eq:unrestricted-space-identification}, we have the quotient map,
\begin{equation}
\Pi_k:\widehat{\mathcal{H}}^{\infty}\rightarrow\widehat{\mathcal{H}}^{\mathrm{phys}}
  \label{eq:q-fermion-physical-projection}
\end{equation}
We may therefore compute the action of $X\in U^{\mathrm{L}}_q\!\left(\fgl_m\right)$ as follows. Starting from a physical fusion-tree state, we first use $\mu(q)$ at generic $q$ to express it in the $q$-fermion basis, act with $\rho(X)$, use $\mu(q)^{-1}$ to return to the unrestricted fusion-tree basis, project to the physical Hilbert space by discarding all inadmissible fusion-tree components, and finally specialize to $q=e^{i\pi/(k+N)}$:
\begin{equation}
\widehat{\mathcal{H}}^{\mathrm{phys}}
\xrightarrow{\;\mu(q)\;}
\Lambda_q^\bullet
\!\left(
\bC_q^N\otimes\bC_q^m
\right)
\xrightarrow{\;\rho(X)\;}
\Lambda_q^\bullet
\!\left(
\bC_q^N\otimes\bC_q^m
\right)
\xrightarrow{\;\mu(q)^{-1}\;}
\widehat{\mathcal{H}}^{\infty}
\xrightarrow{\;\Pi_k\;}
\widehat{\mathcal{H}}^{\mathrm{phys}}.
\label{eq:path-qfermion-action-on-physical-Hilbert-space}
\end{equation}
Thus, the action induced on the physical Hilbert space is
\begin{equation}
\bar{\rho}(X)
=
\lim_{q\to e^{i\pi/(k+N)}}
\Pi_k
\circ
\mu(q)^{-1}
\circ
\rho(X)
\circ
\left.\mu(q)\right|_{\widehat{\mathcal{H}}^{\mathrm{phys}}} ,
\label{eq:path-qfermion-action-on-physical-Hilbert-space-descending}
\end{equation}
whose matrix elements between admissible trees have a finite limit at the root of unity. We want to make clear that, although the action of the generators of $U^{\mathrm{L}}_q(\fgl_m)$ on the $q$-fermion basis is given by the coproduct and therefore looks nonlocal, acting on all sites, the PBW ordering of the $q$-fermion basis that we choose ensures that the action alters only a local region of the state. An element $E_{ij} \in U^{\mathrm{L}}_q(\fgl_m)$ acts on the factor $\mathcal{H}_i\otimes\cdots\otimes\mathcal{H}_j$, preserving the charges $\lambda^{(i-1)}$ and $\lambda^{(j)}$ entering and leaving this region and affecting only a local region of $|i-j|+1$ sites. Schematically,\footnote{Here we are using the notation $\mathcal{L}_{\lambda} \leftrightarrow \lambda$ in the fusion tree.}
\begin{equation}
\begin{gathered}
E_{ij}\;:\;
\Bigl[\,
\cdots \rightarrow \lambda^{(i-1)}
\xrightarrow{\ r_i\ } \lambda^{(i)}
\xrightarrow{\ r_{i+1}\ } \cdots
\xrightarrow{\ r_j\ } \lambda^{(j)}
\rightarrow \cdots
\,\Bigr]
\\[2pt]
\longmapsto\;
\sum_{\tilde\bullet}\;
\Bigl[\,
\cdots \rightarrow \lambda^{(i-1)}
\xrightarrow{\ \tilde r_i\ } \tilde\lambda^{(i)}
\xrightarrow{\ \tilde r_{i+1}\ } \cdots
\xrightarrow{\ \tilde r_j\ } \lambda^{(j)}
\rightarrow \cdots
\,\Bigr],
\end{gathered}
  \label{eq:q-fermion-meson-path-action}
\end{equation}
where $\tilde\bullet$ denotes the collection of local occupations $\tilde r_i,\ldots,\tilde r_j$ and $\tilde\lambda^{(i)},\ldots,\tilde\lambda^{(j-1)}$ for all admissible choices. More explicitly, the site action entering the Hamiltonian~\eqref{eq:path-qfermion-hamiltonian} includes the Cartan generators $K_j$, which act diagonally
\begin{equation}
\begin{gathered}
K_j\;:\;
\Bigl[\,
\lambda^{(j-1)}
\xrightarrow{\ \omega_{r}\ }
\lambda^{(j)}
\xrightarrow{\ \omega_{s}\ }
\lambda^{(j+1)}
\,\Bigr]
\\[2pt]
\longmapsto\;
q^{\,r-s}\,
\Bigl[\,
\lambda^{(j-1)}
\xrightarrow{\ \omega_{r}\ }
\lambda^{(j)}
\xrightarrow{\ \omega_{s}\ }
\lambda^{(j+1)}
\,\Bigr].
\end{gathered}
\label{eq:path-site-action-K-schematic}
\end{equation}
The simple-root generators $E_i$ and $F_i$ transport fermions between neighboring sites,
\begin{equation}
\begin{gathered}
E_j\;:\;
\Bigl[\,
\lambda^{(j-1)}
\xrightarrow{\ \omega_{r}\ }
\lambda^{(j)}
\xrightarrow{\ \omega_{s}\ }
\lambda^{(j+1)}
\,\Bigr]
\\[2pt]
\longmapsto\;
\sum_{\tilde\lambda^{(j)}}
e_{\tilde\lambda^{(j)}}\,
\Bigl[\,
\lambda^{(j-1)}
\xrightarrow{\ \omega_{r+1}\ }
\tilde\lambda^{(j)}
\xrightarrow{\ \omega_{s-1}\ }
\lambda^{(j+1)}
\,\Bigr]
\end{gathered}
\quad\equiv\quad
\vcenter{\hbox{\begin{tikzpicture}[scale=0.85, >=stealth, line width=0.7pt,
    dn/.style={postaction={decorate, decoration={markings,
      mark=at position 0.55 with {\arrow{>}}}}},
    lab/.style={font=\footnotesize}]
  \draw[dn] (0,0) -- (1.25,0);
  \draw[dn] (1.25,0) -- (3.05,0);
  \draw[dn] (3.05,0) -- (4.3,0);
  \foreach \x in {1.25,3.05}{
    \draw[dn] (\x,2.0) -- (\x,1.0);
    \draw[dn] (\x,1.0) -- (\x,0);
    \fill (\x,0) circle (1.7pt);
    \fill (\x,1.0) circle (1.7pt);}
  \draw[dn] (1.25,1.0) -- (3.05,1.0);
  \node[lab, above] at (2.15,1.02) {$E_j$};
  \node[lab, above] at (1.25,2.0) {$r+1$};
  \node[lab, above] at (3.05,2.0) {$s-1$};
  \node[lab, left] at (1.25,0.5) {$r$};
  \node[lab, right] at (3.05,0.5) {$s$};
  \node[lab, below] at (0.62,-0.02) {$\lambda^{(j-1)}$};
  \node[lab, below] at (2.15,-0.02) {$\lambda^{(j)}$};
  \node[lab, below] at (3.68,-0.02) {$\lambda^{(j+1)}$};
\end{tikzpicture}}}
\label{eq:path-site-action-E-schematic}
\end{equation}
\begin{equation}
\begin{gathered}
F_j\;:\;
\Bigl[\,
\lambda^{(j-1)}
\xrightarrow{\ \omega_{r}\ }
\lambda^{(j)}
\xrightarrow{\ \omega_{s}\ }
\lambda^{(j+1)}
\,\Bigr]
\\[2pt]
\longmapsto\;
\sum_{\tilde\lambda^{(j)}}
f_{\tilde\lambda^{(j)}}\,
\Bigl[\,
\lambda^{(j-1)}
\xrightarrow{\ \omega_{r-1}\ }
\tilde\lambda^{(j)}
\xrightarrow{\ \omega_{s+1}\ }
\lambda^{(j+1)}
\,\Bigr]
\end{gathered}
\quad\equiv\quad
\vcenter{\hbox{\begin{tikzpicture}[scale=0.85, >=stealth, line width=0.7pt,
    dn/.style={postaction={decorate, decoration={markings,
      mark=at position 0.55 with {\arrow{>}}}}},
    lab/.style={font=\footnotesize}]
  \draw[dn] (0,0) -- (1.25,0);
  \draw[dn] (1.25,0) -- (3.05,0);
  \draw[dn] (3.05,0) -- (4.3,0);
  \foreach \x in {1.25,3.05}{
    \draw[dn] (\x,2.0) -- (\x,1.0);
    \draw[dn] (\x,1.0) -- (\x,0);
    \fill (\x,0) circle (1.7pt);
    \fill (\x,1.0) circle (1.7pt);}
  \draw[dn] (3.05,1.0) -- (1.25,1.0);
  \node[lab, above] at (2.15,1.02) {$F_j$};
  \node[lab, above] at (1.25,2.0) {$r-1$};
  \node[lab, above] at (3.05,2.0) {$s+1$};
  \node[lab, left] at (1.25,0.5) {$r$};
  \node[lab, right] at (3.05,0.5) {$s$};
  \node[lab, below] at (0.62,-0.02) {$\lambda^{(j-1)}$};
  \node[lab, below] at (2.15,-0.02) {$\lambda^{(j)}$};
  \node[lab, below] at (3.68,-0.02) {$\lambda^{(j+1)}$};
\end{tikzpicture}}}
\label{eq:path-site-action-F-schematic}
\end{equation}
Here $r=r_j$ and $s=r_{j+1}$ are the fillings at the sites, and both sums run over the admissible intermediate charges $\tilde\lambda^{(j)}$.\footnote{We note that in the diagrams in Equations~\eqref{eq:path-site-action-E-schematic} and~\eqref{eq:path-site-action-F-schematic} we see that even though $E_j$ is oriented to the right it moves a fermion to the left, from site $j+1$ to site $j$ and vice versa for $F_j$.} The coefficients $e_{\tilde\lambda^{(j)}}$ and $f_{\tilde\lambda^{(j)}}$ are to be obtained from Equation~\eqref{eq:path-qfermion-action-on-physical-Hilbert-space-descending}. The case of $SU(2)_k$ is worked out explicitly in Appendix~\ref{sec:su2k-action-example}. An alternative way to compute the action of the quantum-group generators on the fusion trees is to use directly the $F$-matrices and the bubble-closing relation of Equation~\eqref{eq:standard-junction-orthogonality} of the unitary modular tensor category (UMTC) $\cC_{N,k}$, with a gauge chosen to reproduce the action of the $U^{\mathrm{L}}_q(\fgl_m)$ generators on the fusion trees. This choice is discussed in Appendix~\ref{app:f-move-action}.

\subsection{The induced inner product on the physical Hilbert space}
\label{sec:orthonormalization-qfermion-basis}
As motivated in Equation~\eqref{eq:path-qfermion-inner-product-orthogonality}, the states of the $q$-fermion Fock space are orthogonal, but will not necessarily be orthonormal. The Hermiticity condition of Equation~\eqref{eq:path-qfermion-hermiticity-condition} is therefore realized only up to a diagonal Gram matrix $G$,
\begin{equation}
G\,\rho(X^\star)\,G^{-1}=\rho(X)^{\dagger},
\label{eq:path-qfermion-hermiticity-condition-gram}
\end{equation}
see Appendix~\ref{sec:su2k-action-example} for explicit examples. Orthonormal states are obtained by normalizing states by the square root of the corresponding diagonal elements of the Gram matrix. This condition for simple roots only fixes the relative norms of the states within each block, labeled by the triple $(a,b,n=r+s)$, that is, spanned by the two-site trees
\begin{equation}
  \mathcal{L}_a \xrightarrow{\ \mathcal{L}_{\omega_r}\ } \mathcal{L}_c \xrightarrow{\ \mathcal{L}_{\omega_s}\ } \mathcal{L}_b ,
  \qquad r+s=n ,
  \label{eq:gram-block-tree}
\end{equation}
with the intermediate charge $c$ and the split $(r,s)$ varying. The complete norm can then be iteratively constructed from these blocks. We now provide the explicit normalization of the topological junction operators which fixes the norms of the fusion-tree states to agree with those induced from Equation~\eqref{eq:path-qfermion-hermiticity-condition-gram}.

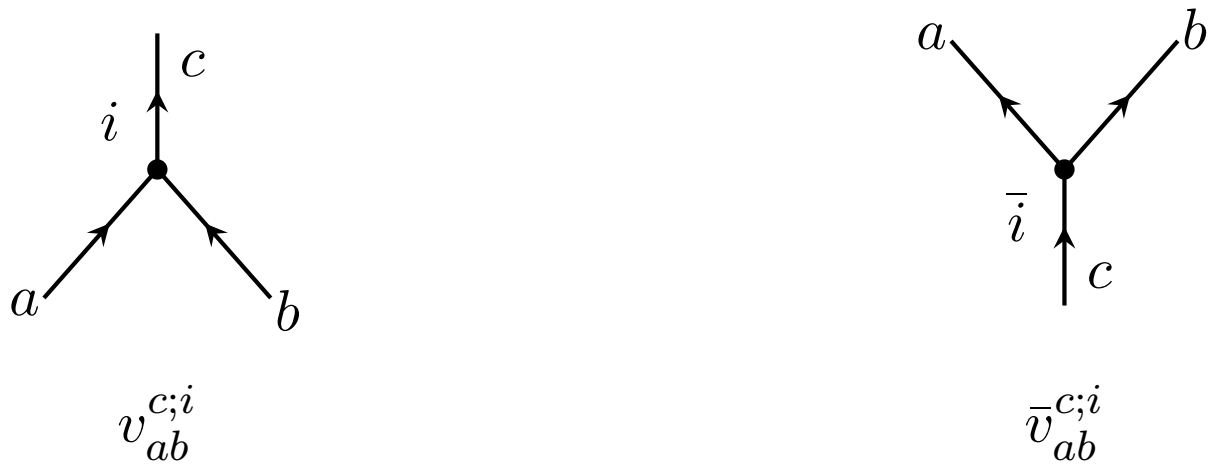
\begin{figure}[!htbp]
\centering
\begin{tikzpicture}[>=stealth,
  line/.style={thick},
  up/.style={postaction={decorate, decoration={markings,
    mark=at position 0.58 with {\arrow{>}}}}},
  vtx/.style={circle,fill,inner sep=1.35pt},
  lab/.style={font=\small}
]
  \begin{scope}
    \draw[line,up] (-0.75,-0.85)--(0,0);
    \draw[line,up] ( 0.75,-0.85)--(0,0);
    \draw[line,up] (0,0)--(0,0.9);
    \node[vtx] at (0,0) {};
    \node[lab,below left] at (-0.63,-0.65) {$a$};
    \node[lab,below right] at ( 0.63,-0.65) {$b$};
    \node[lab,right] at ( 0, 0.70) {$c$};
    \node[lab,left] at (-0.10, 0.32) {$i$};
    \node[anchor=north,font=\small] at (0,-1.28) {$v_{ab}^{c;i}$};
  \end{scope}
  \begin{scope}[xshift=6.0cm]
    \draw[line,up] (0,-0.9)--(0,0);
    \draw[line,up] (0,0)--(-0.75,0.85);
    \draw[line,up] (0,0)--( 0.75,0.85);
    \node[vtx] at (0,0) {};
    \node[lab,above left] at (-0.63, 0.65) {$a$};
    \node[lab,above right] at ( 0.63, 0.65) {$b$};
    \node[lab,right] at ( 0,-0.70) {$c$};
    \node[lab,left] at (-0.10,-0.32) {$\bar i$};
    \node[anchor=north,font=\small] at (0,-1.28) {$\bar v_{ab}^{c;i}$};
  \end{scope}
\end{tikzpicture}
\caption{The two trivalent vertices, read upward. Left: the fusion vertex $v_{ab}^{c;i}$. Right: the splitting vertex $\bar v_{ab}^{c;i}$. The index $i$ labels a basis of the multiplicity space.}
\label{fig:trivalent-vertex}
\end{figure}

\begin{figure}[!htbp]
\centering
\begin{tikzpicture}[>=stealth,
  line/.style={thick},
  up/.style={postaction={decorate, decoration={markings,
    mark=at position 0.58 with {\arrow{>}}}}},
  loop/.style={postaction={decorate, decoration={markings,
    mark=at position 0.25 with {\arrow{>}}}}},
  vtx/.style={circle,fill,inner sep=1.35pt},
  lab/.style={font=\small}
]
  \draw[line,up] (0,-1.60)--(0,-0.80);
  \draw[line,up] (0, 0.80)--(0, 1.60);
  \draw[line,up] (0,-0.80) arc[start angle=270, end angle= 90, radius=0.80];
  \draw[line,up] (0,-0.80) arc[start angle=270, end angle=450, radius=0.80];
  \node[vtx] at (0,-0.80) {};
  \node[vtx] at (0, 0.80) {};
  \node[lab,left] at (-0.92, 0) {$a$};
  \node[lab,right] at ( 0.92, 0) {$b$};
  \node[lab,right] at ( 0.10,-1.40) {$c'$};
  \node[lab,right] at ( 0.10, 1.40) {$c$};
  \node[lab,left] at (-0.11,-1.12) {$\bar j$};
  \node[lab,left] at (-0.11, 1.12) {$i$};
  \node at (1.70,0) {$=$};
  \node[anchor=west] at (2.05,0)
    {$\displaystyle\delta_{cc'}\,\delta_{ij}\,\mathcal{K}_{ab}^{c;i}$};
  \draw[line,up] (4.15,-1.60)--(4.15,1.60);
  \node[lab,right] at (4.25, 1.40) {$c$};
  \node[anchor=north,text width=5.2cm,align=center,font=\small] at (1.4,-2.20)
    {(a)};
  \draw[line,loop] (8.30,0.70) arc[start angle=90, end angle=-270, radius=0.70];
  \node[lab,left] at (7.52,0) {$a$};
  \node at (9.50,0) {$=$};
  \node[anchor=west] at (9.85,0) {$d_a$};
  \node[anchor=north,text width=4.2cm,align=center,font=\small] at (8.8,-2.20)
    {(b)};
\end{tikzpicture}
\caption{Normalization conventions for the trivalent junctions and a closed anyon loop.}
\label{fig:bubble-diagram}
\end{figure}
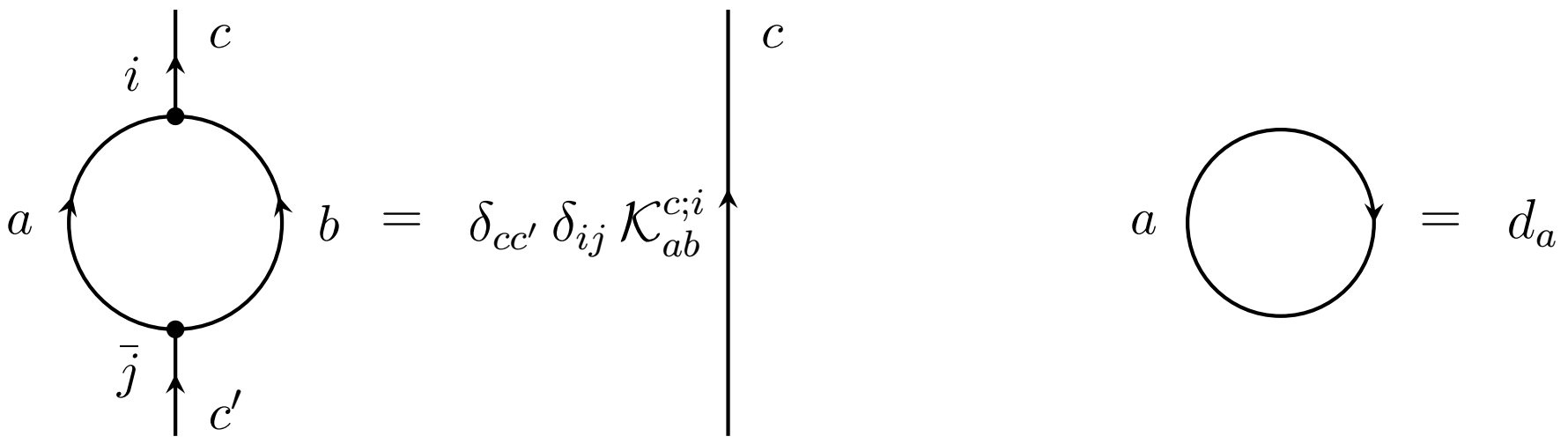

The topological junction operators living at the trivalent vertex, see Figure~\ref{fig:trivalent-vertex}, belong to the vector space,
\begin{equation}
 v_{ab}^{c;i}\in\text{Hom}_{\mathcal{C}_{N,k}}(a\otimes b,c),
 \qquad
 \bar v_{ab}^{c;i}\in\text{Hom}_{\mathcal{C}_{N,k}}(c,a\otimes b)
  \label{eq:cs-fusion-junction-bases}
\end{equation}
where $a,b,c$ are simple objects of the fusion category, the semisimplified category $\mathcal{C}_{N,k}$ of Section~\ref{sec:fusion-tree-basis} in our case, and $i$ labels a basis vector of the multiplicity space, which we may drop since the fusion spaces we work with are multiplicity-free. In a unitary fusion category the splitting vertex $\bar v$ is the adjoint of the fusion vertex $v$, so the closing of the bubble diagram shown in Figure~\ref{fig:bubble-diagram} corresponds to the inner product of these two basis vectors \cite{Kitaev2006:AnyonsExactlySolved}.
\begin{equation}
\begin{aligned}
v_{ab}^{c;i}\cdot\bar v_{ab}^{c';j}
 &=
 \delta_{cc'}\delta_{ij} \mathcal{K}_{ab}^{c;i}\,
 \text{id}_c .
\end{aligned}
\label{eq:standard-junction-orthogonality}
\end{equation}
where $\mathcal{K}_{ab}^{c;i}$ is a normalization factor of choice.\footnote{A standard choice in the literature is the isotopy-invariant normalization $\mathcal{K}^{c}_{ab}=\sqrt{d_a d_b/d_c}$ \cite{Kitaev2006:AnyonsExactlySolved}. This is not the normalization induced by the Gram matrix of Equation~\eqref{eq:path-qfermion-hermiticity-condition-gram}.}

\begin{figure}[!htbp]
\centering
\begin{tikzpicture}[>=stealth,
  line/.style={thick},
  mid/.style={postaction={decorate, decoration={markings,
    mark=at position 0.58 with {\arrow{<}}}}},
  vtx/.style={circle,fill,inner sep=1.25pt},
  lab/.style={font=\footnotesize}
]
  \draw[line,mid] (0,1.64) -- (1.3,1.64);
  \draw[line,mid] (1.3,1.64)--(2.7,1.64);
  \draw[line,mid] (2.7,1.64)--(5.5,1.64);
  \draw[line,mid] (5.5,1.64)--(6.8,1.64);
  \node[vtx] at (1.3,1.64) {};
  \node[vtx] at (2.7,1.64) {};
  \node[vtx] at (5.5,1.64) {};
  \draw[line,mid] (6.8,0)--(5.5,0);
  \draw[line,mid] (5.5,0)--(2.7,0);
  \draw[line,mid] (2.7,0)--(1.3,0);
  \draw[line,mid] (1.3,0)--(0,0);
  \node[vtx] at (1.3,0) {};
  \node[vtx] at (2.7,0) {};
  \node[vtx] at (5.5,0) {};
  \draw[line,mid] (1.3,0)--(1.3,1.64);
  \draw[line,mid] (2.7,0)--(2.7,1.64);
  \draw[line,mid] (5.5,0)--(5.5,1.64);
  \draw[line,mid] (0,0)..controls(-0.75,0.35) and (-0.75,1.29)..(0,1.64);
  \draw[line,mid] (6.8,1.64)..controls(7.55,1.29) and (7.55,0.35)..(6.8,0);
  \node[lab,right] at (1.3,0.82) {$r_1$};
  \node[lab,right] at (2.7,0.82) {$r_2$};
  \node[lab] at (4.1,0.82) {$\cdots$};
  \node[lab,right] at (5.5,0.82) {$r_m$};
  \node[lab,below] at (0.55,0) {$\lambda^{(0)}$};
  \node[lab,below] at (2.0,0) {$\lambda^{(1)}$};
  \node[lab,below] at (3.45,0) {$\lambda^{(2)}$};
  \node[lab,below] at (6.15,0) {$\lambda^{(m)}$};
  \node at (8.05,0.82) {$ = $};
  \node[anchor=west] at (8.75,0.82)
    {$\displaystyle
      d_{\lambda^{(m)}}
      \prod_{j=1}^{m}
      \mathcal{K}_{\lambda^{(j-1)},r_j}^{\lambda^{(j)}}$};
\end{tikzpicture}
\caption{The norm of a fusion tree, computed by closing the bubbles from the left.}
\label{fig:qfermion-path-norm}
\end{figure}
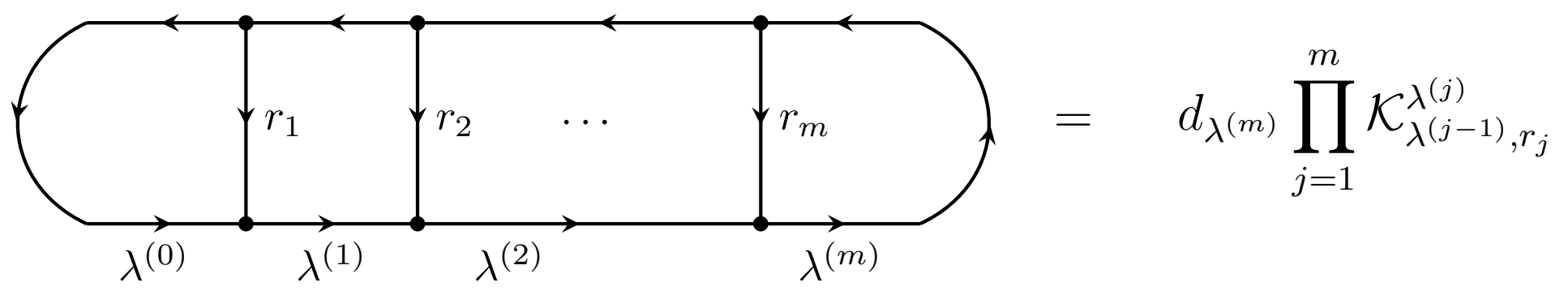

We conjecture that the junction normalization compatible with the Hermiticity condition of Equation~\eqref{eq:path-qfermion-hermiticity-condition} is (proved for $N=2$ in Appendix~\ref{sec:su2k-action-example}), for $\mu=\lambda+\nu_S$ with $\nu_S$ the weight of $\mathcal{L}_{\omega_r}$ labeled by $S\subseteq\{1,\ldots,N\}$,
\begin{equation}
\begin{aligned}
 \mathcal{K}_{\lambda,\omega_r}^{\mu}
 &:=
 \prod_{\alpha\in\Phi^{+}}
 \left(
 \frac{[\langle\lambda+\rho,\alpha^{\vee}\rangle]_q}{[\langle\rho,\alpha^{\vee}\rangle]_q}
 \right)^{\!|\langle\nu_S,\alpha^{\vee}\rangle|}
 \\
 &=
 \prod_{\substack{1\leq i<j\leq N\\ \mathbf{1}_S(i)\neq\mathbf{1}_S(j)}}
 \frac{[a_{ij}]_q}{[j-i]_q},
 \qquad
 a_{ij}
 =
 \langle\lambda+\rho,\alpha_{ij}^{\vee}\rangle
 =
 j-i+\sum_{\ell=i}^{j-1}\lambda_\ell .
\end{aligned}
 \label{eq:suN-root-product-junction-factor}
\end{equation}
The formula is stated only for fusion with the antisymmetric powers $\omega_r$, the only fusions that occur in our trees.\footnote{The resemblance of Equation~\eqref{eq:suN-root-product-junction-factor} to the quantum Weyl dimension formula is immediate,
\begin{equation}
 d_{\lambda}
 =
 \prod_{\alpha\in\Phi^{+}}
 \frac{[\langle\lambda+\rho,\alpha^\vee\rangle]_q}
      {[\langle\rho,\alpha^\vee\rangle]_q}.
 \label{eq:suN-quantum-Weyl-dimension}
\end{equation}
The quantum dimension contains every positive-root factor once, whereas the junction factor retains only those roots that have nontrivial overlap with $\nu_S$.} Fusion with the trivial object $\mathcal{L}_{\omega_0}\simeq \mathcal{L}_{\omega_N}\simeq\mathbbm{1}$ gives $\mathcal{K}=1$. Since we provide $\mathcal{K}_{\lambda, \omega_r}^{\mu;i}$ only for fusion with antisymmetric powers, we also need to choose the norm of the closed loop of an anyon $a$, which is conventionally taken to be the quantum dimension $d_a$ of the anyon, see Figure~\figref{fig:bubble-diagram}{b}.  Now consider the states on $S^2$,

\begin{equation}
 |\mathrm{P}\rangle =
 | \lambda^{(0)}, r_1 ; \cdots ; \lambda^{(m)} \rangle,
 \qquad
 |\tilde{\mathrm{P}}\rangle =
 | \tilde\lambda^{(0)}, \tilde r_1 ; \cdots ; \tilde\lambda^{(m)} \rangle .
  \label{eq:cs-fusion-path-bra-ket}
\end{equation}
From the path integral perspective, the ket $|\mathrm{P}\rangle$ lives in a ball $B^3$ whose boundary sphere is pierced by the anyons, and the bra $\langle \tilde{\mathrm{P}}|$ lives in the orientation-reversed ball. Their inner product glues the two balls along that sphere,
\begin{equation}
  B^3\cup_{S^2}\overline{B^3}\;\cong\;S^3 ,
  \label{eq:open-norm-gluing}
\end{equation}
so the overlap is a closed graph evaluated in $S^3$, with no topological subtleties.\footnote{Gluing the two balls gives the partition function $Z(S^3;\Gamma)$ of the graph. We normalize it by $Z(S^3)$, so that the state with no anyons has unit norm. Since this is a common, nonzero factor for all states, it does not affect the relative norms or matrix elements.} It is computed by successively applying Equation~\eqref{eq:standard-junction-orthogonality}, removing one bubble at a time. Starting from the left end of the chain, each step closes a bubble of the type shown in Figure~\ref{fig:bubble-diagram}, and the full sequence is drawn in Figure~\ref{fig:qfermion-path-norm}. This gives
\begin{equation}
\begin{aligned}
\langle \tilde{\mathrm{P}}|\mathrm{P}\rangle
 &=
 \prod_{j=1}^{m} \delta_{(\tilde r_j, r_j)}
 \prod_{i=0}^{m} \delta_{(\tilde\lambda^{(i)}, \lambda^{(i)})}
 \prod_{j=1}^{m}
 \mathcal{K}_{\lambda^{(j-1)},r_j}^{\lambda^{(j)}} d_{\lambda^{(m)}}.
\end{aligned}
\label{eq:norm-fusion-tree}
\end{equation}

This normalization has the following important properties.
\begin{itemize}
  \item It reproduces the elementary bubble removal relations of \cite{Cautis2014:WebsQuantumSkew} and is consistent with the reflection property of the theta graph, see Appendix~\ref{app:junction-factor-weights}.
  \item It has no poles. Its denominators are $[\langle\rho,\alpha^\vee\rangle]_q=[j-i]_q$ with $1\leq j-i\leq N-1$, which never vanish at $q=e^{i\pi/(k+N)}$ for $k \geq 1$, so $\mathcal{K}$ is finite for every label, inside or outside the alcove.
  \item It detects the null states, in the sense that every fusion tree with a label outside the alcove has zero norm. A label $\lambda$ lies in the alcove exactly when $\langle\lambda,\theta^\vee\rangle\leq k$, where $\theta$ is the highest root. Since fusion with $\mathcal{L}_{\omega_r}$ changes $\langle\lambda,\theta^\vee\rangle$ by at most one, a tree that leaves the alcove must first pass through a label $\lambda$ on its boundary,
\begin{equation}
 \langle\lambda,\theta^\vee\rangle
 =
 k+1.
 \label{eq:first-affine-wall-label}
\end{equation}
For $\fsl_N$ one has $\langle\rho,\theta^\vee\rangle=N-1$, so $a_{1N}=\langle\lambda+\rho,\theta^\vee\rangle=k+N$, and the corresponding $q$-number vanishes at the root of unity,
\begin{equation}
 [a_{1N}]_q=\frac{q^{k+N}-q^{-(k+N)}}{q-q^{-1}}=0,
 \qquad
 q=e^{i\pi/(k+N)} .
 \label{eq:wall-q-number-vanishing}
\end{equation}
This factor then enters the norm in one of two ways. If $\lambda$ is the final label, it appears in the quantum dimension $d_\lambda$ of Equation~\eqref{eq:suN-quantum-Weyl-dimension}, which therefore vanishes. If $\lambda$ is an intermediate label, junctions with $\langle\nu_S,\theta^\vee\rangle=0$ may change the label from $\lambda \rightarrow \lambda'$ but keep the pairing  $\langle\lambda'+\rho,\theta^\vee\rangle=k+N$ the same, and the first junction that changes the pairing, up or down, contributes $[a_{1N}]_q^{\,|\langle\nu_S,\theta^\vee\rangle|}=0$; if none does, the final label is still on the affine wall and therefore $d_{\lambda^{(m)}}=0$.
\end{itemize}
We also emphasize that the first-wall argument requires the initial label to lie in $P_k^+$. Therefore, when dealing with $\mathcal{L}_a\otimes\Lambda_q^\bullet(\bC_q^N\otimes\bC_q^m)$, we restrict ourselves to $a\in P^+_{k}$.

The norms of the lowered states $F^a_{\mathrm{col}}\cdots\Omega^{\text{hw}}$ follow from those of the highest-weight states, by moving each $F^a_{\mathrm{col}}$ across the inner product as $E^a_{\mathrm{col}}$ and using the defining relations of Appendix~\ref{app:uqglm-relations}. The example of $SU(2)_k$ is given explicitly in Appendix~\ref{sec:su2k-action-example}.

\subsection{The periodic chain}
\label{sec:periodic-chain}
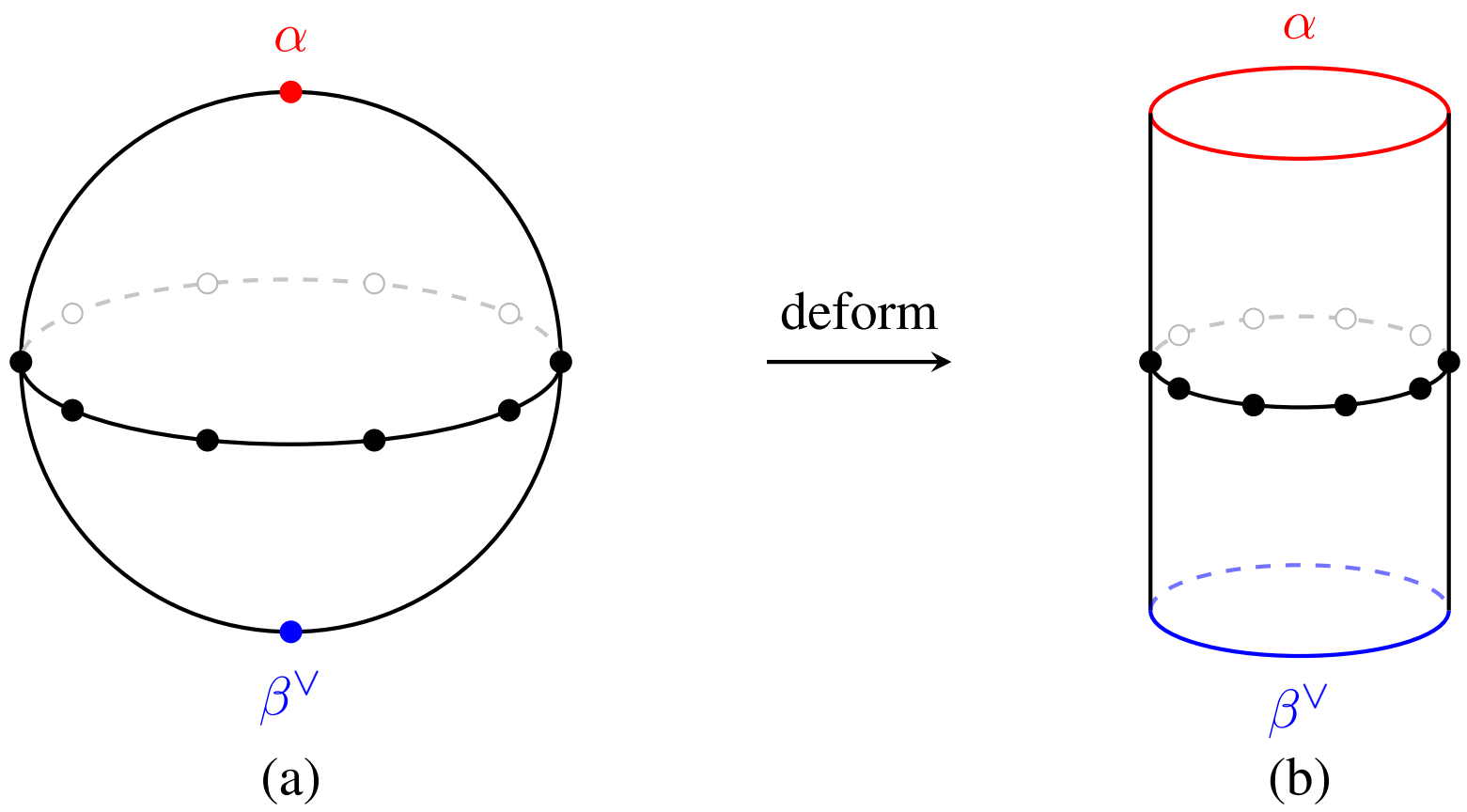
\begin{figure}[!htbp]
\centering
\begin{tikzpicture}[>=stealth,
  outline/.style={thick},
  hidden/.style={thick, gray!45, dashed},
  vtx/.style={circle, fill, inner sep=1.6pt},
  vtxb/.style={circle, draw=gray!55, fill=white, inner sep=1.4pt},
  lab/.style={font=\small}
]

  \begin{scope}[xshift=0cm]
    \def\R{1.9}
    \def\Ry{0.58}
    \draw[outline] (0,0) circle (\R);
    \draw[hidden] (\R,0) arc[start angle=0, end angle= 180, x radius=\R, y radius=\Ry];
    \draw[outline] (\R,0) arc[start angle=0, end angle=-180, x radius=\R, y radius=\Ry];
    \foreach \a in {36,72,108,144}{
      \node[vtxb] at ({\R*cos(\a)},{\Ry*sin(\a)}) {};
    }
    \foreach \a in {180,216,252,288,324,360}{
      \node[vtx] at ({\R*cos(\a)},{\Ry*sin(\a)}) {};
    }
    \node[vtx,red] at (0, \R) {};
    \node[vtx,blue] at (0,-\R) {};
    \node[lab,above,red] at (0, \R+0.14) {$\alpha$};
    \node[lab,below,blue] at (0,-\R-0.14) {$\beta^{\vee}$};
    \node[lab] at (0,-2.95) {(a)};
  \end{scope}

  \draw[->,thick] (3.35,0) -- (4.65,0);
  \node[lab,above] at (4.00,0.08) {deform};

  \begin{scope}[xshift=7.1cm]
    \def\Cx{1.05}
    \def\Cy{0.32}
    \def\H{1.75}
    \draw[outline,red] (0,\H) ellipse [x radius=\Cx, y radius=\Cy];
    \draw[outline] (-\Cx,\H) -- (-\Cx,-\H);
    \draw[outline] ( \Cx,\H) -- ( \Cx,-\H);
    \draw[thick,blue,dashed,opacity=0.55]
                   (-\Cx,-\H) arc[start angle=180, end angle= 0, x radius=\Cx, y radius=\Cy];
    \draw[outline,blue]
                   (-\Cx,-\H) arc[start angle=180, end angle=360, x radius=\Cx, y radius=\Cy];
    \draw[hidden] ( \Cx,0) arc[start angle=0, end angle= 180, x radius=\Cx, y radius=\Cy];
    \draw[outline] ( \Cx,0) arc[start angle=0, end angle=-180, x radius=\Cx, y radius=\Cy];
    \foreach \a in {36,72,108,144}{
      \node[vtxb] at ({\Cx*cos(\a)},{\Cy*sin(\a)}) {};
    }
    \foreach \a in {180,216,252,288,324,360}{
      \node[vtx] at ({\Cx*cos(\a)},{\Cy*sin(\a)}) {};
    }
    \node[lab,above,red] at (0, \H+\Cy+0.06) {$\alpha$};
    \node[lab,below,blue] at (0,-\H-\Cy-0.06) {$\beta^{\vee}$};
    \node[lab] at (0,-2.95) {(b)};
  \end{scope}

\end{tikzpicture}
\caption{The open $q$-fermion chain on the sphere. (a) The $m$ sites are equidistant punctures on the equator; each black dot carries a $\Lambda^\bullet \bC^N$. The two fixed endpoints specify the boundary conditions: $\alpha$ at the north pole (red) and $\beta^{\vee}$ at the south pole (blue). The south pole carries the dual $\beta^{\vee}$, so that the total charge on the sphere is trivial. (b) The sphere is deformed into a cylinder.}
\label{fig:open-chain-sphere}
\end{figure}

The $q$-fermion chain we have considered so far has been open. A state is a fusion tree from a fixed $\lambda^{(0)}=\alpha$ to a fixed $\lambda^{(m)}=\beta$, and the Hamiltonian in Equation~\eqref{eq:path-qfermion-hamiltonian} couples only the $m-1$ internal bonds $(i,i+1)$. The generators of $U^{\mathrm{L}}_q(\fgl_m)$ change the occupations and the internal fusion labels, while leaving $\alpha$ and $\beta$ fixed. Geometrically, these states are \textit{conformal blocks}\footnote{Here ``conformal blocks'' refers to the chiral conformal blocks of the $SU(N)_k$ WZW model, whose space is identified with the Hilbert space of Chern--Simons theory on the same punctured surface by the Chern--Simons/WZW correspondence \cite{Witten1989:QuantumFieldTheory}.} on a sphere with the $m$ site punctures along the equator and two additional punctures carrying $\alpha$ and $\beta^\vee$, as in Figure~\ref{fig:open-chain-sphere}. Removing small discs around the two distinguished punctures gives a cylinder, with the boundary labels recording the original boundary conditions.

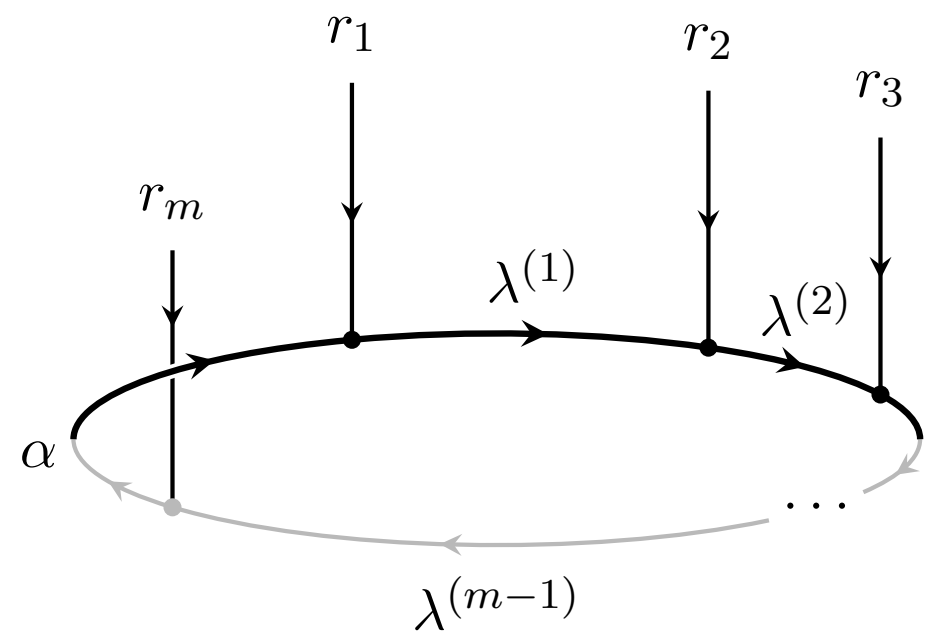
\begin{figure}[!htbp]
  \centering
  \begin{tikzpicture}[>=stealth, yscale=-1,
    front/.style={very thick},
    back/.style={thick, gray!55},
    leg/.style={thick},
    mid/.style={postaction={decorate, decoration={markings,
      mark=at position 0.55 with {\arrow{>}}}}},
    farmid/.style={postaction={decorate, decoration={markings,
      mark=at position 0.30 with {\arrow{>}}}}},
    vtx/.style={circle, fill, inner sep=1.2pt},
    vtxb/.style={circle, fill=gray!55, inner sep=1.2pt},
    lab/.style={font=\small}
  ]
    \draw[back,mid] (2.8,0)
      arc[start angle=0, end angle=30, x radius=2.8, y radius=0.7];
    \draw[back,mid] (1.800,0.536)
      arc[start angle=50, end angle=140, x radius=2.8, y radius=0.7];
    \draw[back,mid] (-2.145,0.450)
      arc[start angle=140, end angle=180, x radius=2.8, y radius=0.7];
    \node[lab] at (2.145,0.450) {$\cdots$};
    \draw[leg,farmid] (-2.145,-1.250) -- (-2.145,0.450);
    \node[vtxb] at (-2.145,0.450) {};
    \draw[white, line width=2.6pt] (-2.8,0)
      arc[start angle=180, end angle=250, x radius=2.8, y radius=0.7];
    \draw[front,mid] (-2.8,0)
      arc[start angle=180, end angle=250, x radius=2.8, y radius=0.7];
    \draw[front,mid] (-0.958,-0.658)
      arc[start angle=250, end angle=300, x radius=2.8, y radius=0.7];
    \draw[front,mid] (1.400,-0.606)
      arc[start angle=300, end angle=335, x radius=2.8, y radius=0.7];
    \draw[front] (2.538,-0.296)
      arc[start angle=335, end angle=360, x radius=2.8, y radius=0.7];
    \draw[leg,mid] (-0.958,-2.358) -- (-0.958,-0.658);
    \draw[leg,mid] ( 1.400,-2.306) -- ( 1.400,-0.606);
    \draw[leg,mid] ( 2.538,-1.996) -- ( 2.538,-0.296);
    \node[vtx] at (-0.958,-0.658) {};
    \node[vtx] at ( 1.400,-0.606) {};
    \node[vtx] at ( 2.538,-0.296) {};
    \node[lab, anchor=south] at (-0.958,-2.42) {$r_1$};
    \node[lab, anchor=south] at ( 1.400,-2.37) {$r_2$};
    \node[lab, anchor=south] at ( 2.538,-2.06) {$r_3$};
    \node[lab, anchor=south] at (-2.145,-1.31) {$r_m$};
    \node[lab, anchor=east] at (-2.76,0.10) {$\alpha$};
    \node[lab, anchor=south] at ( 0.244,-0.76) {$\lambda^{(1)}$};
    \node[lab, anchor=south] at ( 2.048,-0.54) {$\lambda^{(2)}$};
    \node[lab, anchor=north] at ( 0.000, 0.76) {$\lambda^{(m-1)}$};
  \end{tikzpicture}
\caption{Periodic fusion tree. The loop at the bottom represents the internal labels, and the prongs represent the charges at the sites.}
  \label{fig:periodic-fusion-tree}
\end{figure}

To obtain the periodic chain, we close the fusion path by identifying $\lambda^{(m)}= \lambda^{(0)} = \alpha$.\footnote{One can also choose $\lambda^{(m)} \neq \lambda^{(0)}$ and sum over all such choices of $\exists\,\delta\subset\lambda^{(m)}\otimes\lambda^{(0)\vee}$  which would correspond to twisted boundary conditions labeled by $\delta$.} The common label $\alpha$ now sits on the bond $(m,1)$, just as $\lambda^{(i)}$ sits on the bond $(i,i+1)$; the resulting periodic fusion tree is shown in Figure~\ref{fig:periodic-fusion-tree}. There is no distinguished bond on the circle, so $\alpha$ must be allowed to vary along with the other fusion labels. The periodic Hilbert space is therefore
\begin{equation}
  \mathcal{H}_{\mathrm{per}}
  =\bigoplus_{\alpha\in P_k^+}\mathcal{H}_{(\alpha,\alpha)}.
  \label{eq:periodic-Hilbert-space}
\end{equation}
The total fermion number $Q=\sum_i r_i$ is still conserved. At fixed filling, we restrict this sum to
\begin{equation}
  \mathcal{H}_{\mathrm{per}}(Q)
  =\bigoplus_{\alpha\in P_k^+}\mathcal{H}_{(\alpha,\alpha)}(Q).
  \label{eq:periodic-Hilbert-space-fixed-filling}
\end{equation}

The direct sum over $\alpha$ implements the sewing axiom of the $SU(N)_k$ modular functor~\cite{MooreSeiberg1989:ClassicalQuantumCFT,Segal2004:DefinitionCFT,BakalovKirillov2001:LecturesTensorCategories}. Cutting the punctured torus along a noncontractible circle gives a cylinder, equivalently a sphere with two additional punctures carrying dual labels $\alpha$ and $\alpha^\vee$. Sewing the oppositely oriented boundaries identifies the torus conformal-block space with the direct sum of these sphere conformal blocks $\mathcal{H}_{(\alpha,\alpha)}$.

\begin{figure}[!htbp]
\centering
\begin{tikzpicture}[
  line cap=round, line join=round, >=stealth,
  surface/.style={draw=black, thick},
  hidden/.style={draw=gray!45, thick, dashed},
  site/.style={circle, fill=black, inner sep=1.55pt},
  hsite/.style={circle, draw=gray!55, fill=white, inner sep=1.4pt},
  lab/.style={font=\small},
  acycle/.style={draw=red, thick},
  bcycle/.style={draw=blue, thick}
]
  \node[anchor=east] at (-1.38,0)
    {$\displaystyle\bigoplus_{\alpha\in P_k^+}$};

  \begin{scope}
    \def\R{0.94}
    \def\h{0.28}
    \def\H{1.45}
    \draw[surface] (-\R,-\H) -- (-\R,\H);
    \draw[surface] (\R,-\H) -- (\R,\H);
    \draw[acycle] (0,\H) ellipse[x radius=\R,y radius=\h];
    \draw[red,thick,dashed,opacity=0.55]
      (\R,-\H) arc[start angle=0,end angle=180,x radius=\R,y radius=\h];
    \draw[acycle] (-\R,-\H)
      arc[start angle=180,end angle=360,x radius=\R,y radius=\h];
    \draw[hidden] (\R,0)
      arc[start angle=0,end angle=180,x radius=\R,y radius=\h];
    \draw[surface] (-\R,0)
      arc[start angle=180,end angle=360,x radius=\R,y radius=\h];
    \foreach \u in {36,72,108,144}
      \node[hsite] at ({\R*cos(\u)},{\h*sin(\u)}) {};
    \foreach \u in {180,216,252,288,324,360}
      \node[site] at ({\R*cos(\u)},{\h*sin(\u)}) {};
    \node[lab,above=5pt,red] at (0,\H+\h) {$\alpha$};
    \node[lab,below=5pt,red] at (0,-\H-\h) {$\alpha^{\vee}$};
    \node[lab] at (0,-2.65) {(a)};
  \end{scope}

  \draw[->,black,thick] (1.42,0) -- (2.78,0);
  \node[lab,above=5pt] at (2.10,0) {sew};

  \begin{scope}[xshift=5.60cm]
    \def\R{1.70}
    \def\r{0.47}
    \def\s{0.616}
    \draw[surface]
      plot[domain=0:360,samples=121,smooth,variable=\u]
      ({\R*cos(\u)+\r*\s*cos(\u)/sqrt(\s*\s*cos(\u)^2+sin(\u)^2)},
       {\R*\s*sin(\u)+\r*sin(\u)/sqrt(\s*\s*cos(\u)^2+sin(\u)^2)}) -- cycle;
    \filldraw[fill=white,surface]
      plot[domain=0:360,samples=121,smooth,variable=\u]
      ({\R*cos(\u)-\r*\s*cos(\u)/sqrt(\s*\s*cos(\u)^2+sin(\u)^2)},
       {\R*\s*sin(\u)-\r*sin(\u)/sqrt(\s*\s*cos(\u)^2+sin(\u)^2)}) -- cycle;

    \draw[bcycle,dashed,opacity=0.55]
      ({\R-\r},0) arc[start angle=180,end angle=360,
        x radius=\r,y radius=0.370];
    \draw[bcycle]
      ({\R+\r},0) arc[start angle=0,end angle=180,
        x radius=\r,y radius=0.370];

    \draw[acycle] (0,0) ellipse[x radius=\R,y radius=1.047];
    \foreach \u in {0,36,72,108,144,180,216,252,288,324}
      \node[site] at ({\R*cos(\u)},{1.047*sin(\u)}) {};
    \node[lab,text=red,anchor=east] at (-1.85,0.70) {$a$};
    \node[lab,text=blue,anchor=west] at (2.20,0.43) {$b$};
    \node[lab] at (0,-2.65) {(b)};
  \end{scope}
\end{tikzpicture}
\caption{The periodic $q$-fermion chain on a torus. The cylinder has dual boundary labels $\alpha$ and $\alpha^{\vee}$ and sewing with a sum over the admissible labels $\alpha$ gives a torus. The site punctures lie on the red $a$-cycle, and the blue $b$-cycle winds around the cylinder.}
\label{fig:periodic-chain-torus}
\end{figure}
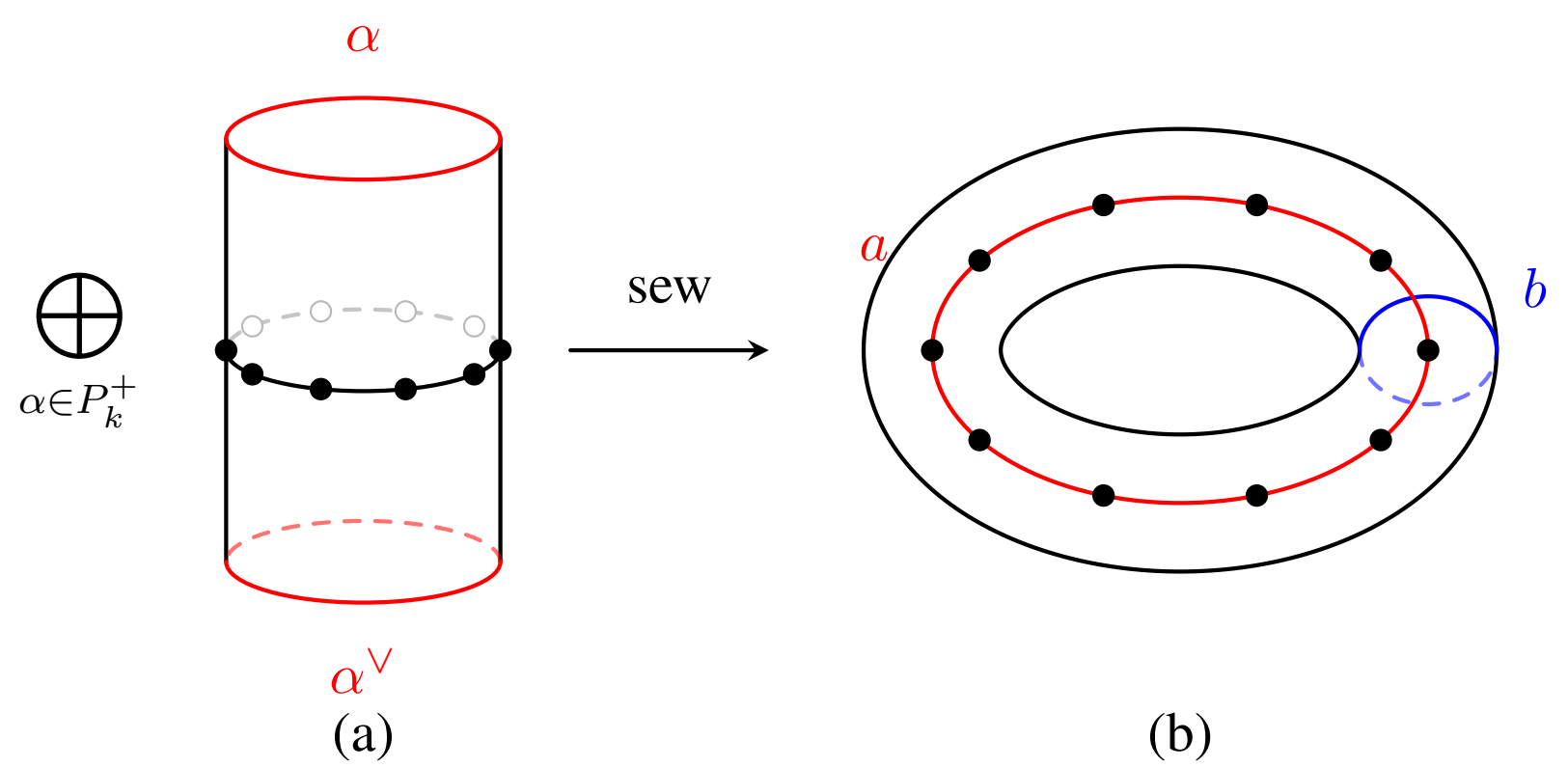

We now add the bond $(m,1)$ to the Hamiltonian. We define $E_m$, $F_m$ and $K_m=L_mL_1^{-1}$ by the same local rules as the internal bonds in Equations~\eqref{eq:path-site-action-K-schematic}--\eqref{eq:path-site-action-F-schematic}, with the site indices understood modulo $m$. In particular, $E_m$ moves one fermion from site $1$ to site $m$ and changes the fusion label between them:
\begin{equation}
\begin{aligned}
E_m\;:\quad
&\Bigl[
\lambda^{(m-1)}
\xrightarrow{\ \omega_{r_m}\ }\alpha
\xrightarrow{\ \omega_{r_1}\ }\lambda^{(1)}
\Bigr]
\\[-1pt]
&\longmapsto
\sum_{\tilde\alpha \in P^{+}_{k}}e_{\tilde\alpha}
\Bigl[
\lambda^{(m-1)}
\xrightarrow{\ \omega_{r_m+1}\ }\tilde\alpha
\xrightarrow{\ \omega_{r_1-1}\ }\lambda^{(1)}
\Bigr].
\end{aligned}
\label{eq:periodic-seam-action}
\end{equation}
Here $e_{\tilde\alpha}$ is the corresponding coefficient which is exactly the same as for any other $E_i$ and similarly for $F_i$. We also impose the extra Hermiticity condition $F_m=E_m^\dagger$, which ensures that the Hamiltonian~\eqref{eq:periodic-hamiltonian} is Hermitian. Since these extra operators can map from $\mathcal{H}_{(\alpha,\alpha)}$ to $\mathcal{H}_{(\tilde\alpha,\tilde\alpha)}$, this ensures that a sum of all individual summands in Equation~\eqref{eq:periodic-Hilbert-space} is required for the action of the periodic Hamiltonian to be closed.

We can also understand the locality of this definition directly in the $q$-fermion description. In the open chain, we choose a \textit{cut} between sites $m$ and $1$, which gave the PBW ordering of sites $1<2<\cdots<m$, in which these two sites lie at opposite ends of the ordering. We could have instead described the same state using a cut between sites $m-1$ and $m$, with ordering $m<1<2<\cdots<m-1$ and basepoint label $\lambda^{(m-1)}$. The bond $(m,1)$ is then the first bond of the Hilbert space $\cH_{(\lambda^{(m-1)}, \lambda^{(m-1)})}$, and $E_m$ and $F_m$ act by the same rules as $E_1$ and $F_1$ in the original description. Therefore, the apparent nonlocality in the original PBW ordering comes from the position of the cut. This apparent nonlocality disappears as soon as one works in the fusion-tree basis.

We have the periodic Hamiltonian $H^{q}_{\mathrm{per}}$ which runs over all $m$ bonds,
\begin{equation}
  H^{q}_{\mathrm{per}}
  =-t\sum_{i=1}^{m}(E_i+F_i)-\mu F,
  \qquad i+m\equiv i.
  \label{eq:periodic-hamiltonian}
\end{equation}
This is also analogous to the periodic anyon-chain Hamiltonians considered in the literature \cite{Feiguin2007:InteractingAnyons,Gils2013:AnyonicQuantumSpin}, which involve a local projector coupling the last anyon to the first.

At the level of the operator algebra, the closing bond extends the $U^{\mathrm{L}}_q(\fsl_m)$ part of the open-chain operator algebra of Section~\ref{sec:path-to-qfermion} to the quantum loop algebra $U^{\mathrm{L}}_q(L\fsl_m)$, which can be obtained as a quotient of the affine quantum group $U^{\mathrm{L}}_q(\widehat{\fsl}_m)$, since its central element acts trivially,
\begin{equation}
  \prod_{i=1}^{m}K_i
  =(L_1L_2^{-1})(L_2L_3^{-1})\cdots(L_mL_1^{-1})=1.
  \label{eq:periodic-level-zero}
\end{equation}
The action therefore factors through the level-zero quotient of the affine quantum group,
\begin{equation}
  U^{\mathrm{L}}_q(L\fsl_m)
  =U^{\mathrm{L}\prime}_q(\widehat{\fsl}_m)\big/\langle C-1\rangle,
  \qquad C=\prod_{i=1}^{m}K_i,
  \label{eq:periodic-quantum-loop}
\end{equation}
where the prime indicates that the degree derivation is omitted. Geometrically, the hopping generators associated with the simple roots are short Wilson lines between neighboring sites, contained in an annular neighborhood of the $a$-cycle. We can also have closed Wilson lines $Y_\mu$ running parallel to the chain, without ending on its sites. These commute with $U^{\mathrm{L}}_q(L\fsl_m)$, by virtue of the fact they can be moved into the annular neighborhood of the $a$-cycle and moved past the rungs, see Section~\ref{sec:qchain-symmetries} for details. We can also consider Wilson lines winding around the $b$-cycle that may either end on the sites or not. Together, we expect these operators to generate a finite-dimensional quotient of the quantum toroidal algebra $U^{\mathrm{L}}_q(\widehat{\widehat{\fsl}}_m)$ \cite{Hernandez2009:QuantumToroidal}. Definitions of the finite, affine and quantum loop algebras are given in Appendix~\ref{app:uqglm-relations}.

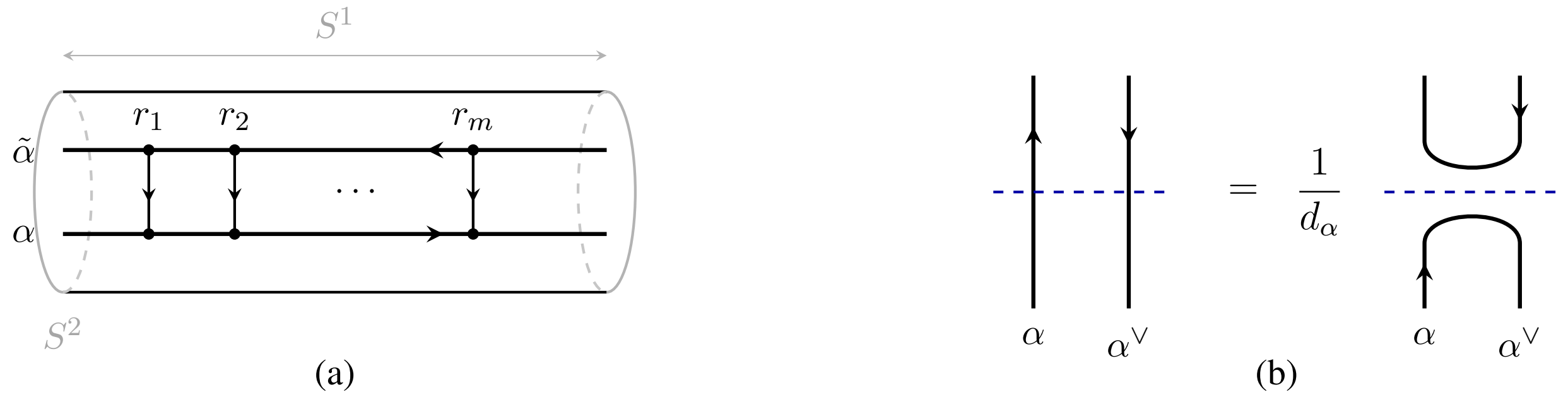
\begin{figure}[!htbp]
  \centering
  \begin{tikzpicture}[baseline=(current bounding box.south),>=stealth,
    tube/.style={thick},
    fiber/.style={thick, gray!60},
    fibreb/.style={thick, gray!45, dashed},
    strand/.style={very thick, postaction={decorate,
       decoration={markings, mark=at position 0.70 with {\arrow{>}}}}},
    strandb/.style={very thick, postaction={decorate,
       decoration={markings, mark=at position 0.70 with {\arrow{<}}}}},
    rung/.style={thick, postaction={decorate,
       decoration={markings, mark=at position 0.62 with {\arrow{>}}}}},
    vtx/.style={circle, fill, inner sep=1.2pt},
    lab/.style={font=\small}]
    \def\L{2.85}\def\Hh{1.05}\def\Ex{0.30}
    \draw[tube] (-\L,\Hh) -- (\L,\Hh);
    \draw[tube] (-\L,-\Hh) -- (\L,-\Hh);
    \draw[fibreb] (-\L,-\Hh) arc[start angle=270, end angle=450, x radius=\Ex, y radius=\Hh];
    \draw[fiber] (-\L, \Hh) arc[start angle= 90, end angle=270, x radius=\Ex, y radius=\Hh];
    \draw[fibreb] ( \L, \Hh) arc[start angle= 90, end angle=270, x radius=\Ex, y radius=\Hh];
    \draw[fiber] ( \L,-\Hh) arc[start angle=270, end angle=450, x radius=\Ex, y radius=\Hh];
    \node[lab,gray!70,anchor=north] at (-\L,-\Hh-0.12) {$S^2$};
    \draw[<->,gray!60] (-\L,\Hh+0.38) -- (\L,\Hh+0.38);
    \node[lab,gray!70,anchor=south] at (0,\Hh+0.42) {$S^1$};
    \draw[strand] (-\L,-0.44) -- (\L,-0.44);
    \draw[strandb] (-\L, 0.44) -- (\L, 0.44);
    \node[lab,anchor=east] at (-\L-0.14,-0.44) {$\alpha$};
    \node[lab,anchor=east] at (-\L-0.14, 0.44) {$\tilde\alpha$};
    \foreach \x/\nm in {-1.95/1, -1.05/2, 1.45/m}{
      \draw[rung] (\x,0.44) -- (\x,-0.44);
      \node[vtx] at (\x,-0.44) {};
      \node[vtx] at (\x, 0.44) {};
      \node[lab] at (\x,0.76) {$r_{\nm}$};}
    \node[lab] at (0.25,0) {$\cdots$};
    \node[lab] at (0,-1.92) {(a)};
  \end{tikzpicture}
  \hfill
  \begin{tikzpicture}[baseline=(current bounding box.south),>=stealth,
    strand/.style={very thick, postaction={decorate,
       decoration={markings, mark=at position 0.78 with {\arrow{>}}}}},
    strandb/.style={very thick, postaction={decorate,
       decoration={markings, mark=at position 0.78 with {\arrow{<}}}}},
    capped/.style={very thick, postaction={decorate,
       decoration={markings, mark=at position 0.18 with {\arrow{>}}}}},
    cut/.style={thick, blue!65!black, dashed},
    lab/.style={font=\small}]
    \draw[strand] (0,-1.22) -- (0,1.22);
    \draw[strandb] (1.00,-1.22) -- (1.00,1.22);
    \draw[cut] (-0.42,0) -- (1.42,0);
    \node[lab,anchor=north] at (0,-1.28) {$\alpha$};
    \node[lab,anchor=north] at (1.00,-1.28) {$\alpha^{\vee}$};
    \node at (2.20,0) {$=$};
    \node at (3.00,0) {$\dfrac{1}{d_{\alpha}}$};
    \begin{scope}[xshift=4.10cm]
      \draw[capped] (1.00,1.22) -- (1.00,0.54)
         .. controls (1.00,0.16) and (0,0.16) .. (0,0.54) -- (0,1.22);
      \draw[capped] (0,-1.22) -- (0,-0.54)
         .. controls (0,-0.16) and (1.00,-0.16) .. (1.00,-0.54) -- (1.00,-1.22);
      \draw[cut] (-0.42,0) -- (1.42,0);
      \node[lab,anchor=north] at (0,-1.28) {$\alpha$};
      \node[lab,anchor=north] at (1.00,-1.28) {$\alpha^{\vee}$};
    \end{scope}
    \node[lab] at (2.55,-1.92) {(b)};
  \end{tikzpicture}
\caption{(a) The periodic norm as a Wilson-line network in $S^2\times S^1$, drawn as $S^2\times[0,1]$ with its two end fibers identified. (b) Cutting and capping on the two-punctured sphere. Since the space in Equation~\eqref{eq:two-punctured-sphere} is one-dimensional, the two parallel strands equal $d_{\alpha}^{-1}$ times a cup and a cap, which detaches the graph from the nontrivial cycle.}
  \label{fig:periodic-norm-cut}
\end{figure}

As in the open chain of Section~\ref{sec:orthonormalization-qfermion-basis}, we can equip the states of the periodic chain with an inner product that respects the Hermitian structure, now that of $U^{\mathrm{L}}_q(L\fsl_m)$. A periodic state is naturally prepared on a solid torus $D^2\times S^1$ as in Figure~\ref{fig:periodic-chain-torus}, with the anyons piercing its boundary $T^2$. To compute the norm we reflect this solid torus and glue it to the original along that boundary,
\begin{equation}
  (D^2\times S^1)\cup_{T^2}\overline{(D^2\times S^1)}\;\cong\;S^2\times S^1 ,
  \label{eq:periodic-norm-gluing}
\end{equation}
so the inner product is the evaluation of the Wilson-line network of Figure~\figref{fig:periodic-norm-cut}{a} in the three-manifold $S^2\times S^1$.

Unlike the open chain, where the ambient $S^3$ was topologically trivial, the graph of Figure~\figref{fig:periodic-norm-cut}{a} cannot be shrunk away: its base loop carrying the fusion channels $\lambda^{(i)}$ wraps the $S^1$. We evaluate this network by cutting the manifold $S^2\times S^1$ open along a sphere transverse to the $S^1$, on the bond between sites $m$ and $1$.\footnote{The norm is independent of where we choose to cut.} Two anyons pierce this transverse $S^2$, carrying $\alpha$ and $\tilde\alpha^{\vee}$, so the cut carries the Hilbert space
\begin{equation}
  \cH\bigl(S^2;\alpha,\tilde\alpha^{\vee}\bigr)
  \;\cong\;\text{Hom}_{\cC_{N, k}}(\tilde\alpha,\alpha),
  \qquad
  \dim\cH\bigl(S^2;\alpha,\tilde\alpha^{\vee}\bigr)
  =\delta_{\alpha,\tilde\alpha} .
  \label{eq:two-punctured-sphere}
\end{equation}
Gauge invariance forces the two basepoint labels to agree, and the space is then one dimensional, spanned by the cup state $|0;\alpha,\alpha^\vee\rangle$ of norm $\langle 0;\alpha,\alpha^\vee|0;\alpha,\alpha^\vee\rangle=d_\alpha Z(S^3)$. Cutting along this sphere turns $\Gamma$, embedded as in Figure~\figref{fig:periodic-norm-cut}{a}, into an operator $\Theta_\Gamma$ on this space. The partition function is then the trace of $\Theta_\Gamma$ over the orthonormal states, which in this case is the single state $|0;\alpha,\alpha^\vee\rangle/\sqrt{Z(S^3)d_\alpha}$. This matrix element can then be thought of as the graph obtained after capping the strands on both sides of the cut, Figure~\figref{fig:periodic-norm-cut}{b},
\begin{equation}
\begin{aligned}
  Z\bigl(S^2\times S^1;\Gamma\bigr)
  &=\mathrm{Tr}_{\cH(S^2;\alpha,\alpha^\vee)}\,\Theta_\Gamma
  =\frac{1}{Z(S^3)\,d_{\alpha}}\,\langle 0;\alpha,\alpha^\vee|\,\Theta_\Gamma\,|0;\alpha,\alpha^\vee\rangle\\
  &=\frac{Z\bigl(S^3;\Gamma_{\mathrm{capped}}\bigr)}{Z(S^3)\,d_{\alpha}}
  =\frac{1}{d_{\alpha}}\,
   \bigl\langle\Gamma_{\mathrm{capped}}\bigr\rangle_{S^3},
\end{aligned}
  \label{eq:s2xs1-capping}
\end{equation}
which is embedded in a topologically trivial way,
where $\langle\,\cdot\,\rangle_{S^3}$ denotes division by $Z(S^3)$, so that $\langle\bigcirc_{\alpha}\rangle_{S^3}=d_{\alpha}$.

The capped graph $|\Gamma_{\mathrm{capped}}|$ is precisely the open-chain graph of Figure~\ref{fig:qfermion-path-norm} with $\lambda^{(m)}=\lambda^{(0)}=\alpha$. By Equation~\eqref{eq:norm-fusion-tree} it evaluates to $d_{\alpha}\prod_{j}\mathcal{K}$, and that loop factor cancels against the $1/d_{\alpha}$ of Equation~\eqref{eq:s2xs1-capping},
\begin{equation}
  \bigl\|\mathrm{P}\bigr\|^{2}_{\mathrm{per}}
  =\frac{1}{d_{\alpha}}
   \biggl[\,d_{\alpha}\prod_{j=1}^{m}
     \mathcal{K}_{\lambda^{(j-1)},r_j}^{\lambda^{(j)}}\biggr]
  =\prod_{j=1}^{m}
     \mathcal{K}_{\lambda^{(j-1)},r_j}^{\lambda^{(j)}} .
  \label{eq:periodic-path-norm-value}
\end{equation}
The periodic norm is thus the product of junction factors alone and is manifestly translation invariant and, with the junction normalization of Section~\ref{sec:orthonormalization-qfermion-basis}, satisfies the condition $E_i^\star=F_i$ for all $m$ bonds.

\subsection{Symmetries and anomalies}
\label{sec:qchain-symmetries}

We begin by describing the symmetries of the periodic chain. As alluded to earlier in the normalization discussion, the periodic chain has a natural translation symmetry which acts on the fusion tree by shifting the labels one unit to the right,
\begin{equation}
T\;:\;
\Bigl[\,
\alpha
\xrightarrow{\ \omega_{r_1}\ }
\lambda^{(1)}
\xrightarrow{\ \omega_{r_2}\ }
\cdots
\xrightarrow{\ \omega_{r_m}\ }
\alpha
\,\Bigr]
\longmapsto
\Bigl[\,
\lambda^{(m-1)}
\xrightarrow{\ \omega_{r_m}\ }
\alpha
\xrightarrow{\ \omega_{r_1}\ }
\cdots
\xrightarrow{\ \omega_{r_{m-1}}\ }
\lambda^{(m-1)}
\,\Bigr],
\label{eq:periodic-translation}
\end{equation}
that is, $r_i\mapsto r_{i-1}$ and $\lambda^{(i)}\mapsto\lambda^{(i-1)}$ with indices modulo $m$. We see that $T$ also moves the basepoint from $\alpha$ to $\lambda^{(m-1)}$, so it does not preserve the individual summands of Equation~\eqref{eq:periodic-Hilbert-space}; it is well-defined only on the full sum in Equation~\eqref{eq:periodic-Hilbert-space}, which is the torus Hilbert space used in this construction. On a punctured sphere, cyclic translation can instead be implemented using braiding and recoupling \cite{Pfeifer2012:TranslationTopologyCriticality}. The translation symmetry should be identified with the extended Dynkin diagram automorphism of $U^{\mathrm{L}}_q(L\fsl_m)$. We have
\begin{equation}
  \left[H^{q}_{\mathrm{per}},T\right]=0 ,\qquad
  T\,X_i\,T^{-1}=X_{i+1} ,\qquad
  T^{m}=1 ,
  \label{eq:periodic-translation-relations}
\end{equation}
where $X_i\in\{E_i,F_i,K_i\}$ with $i$ modulo $m$. So, just as in the free fermion case of Section~\ref{sec:free-fermion-dynamics}, the eigenstates and the corresponding local operators in the continuum can be labeled by their lattice momentum. The lattice momenta take the values
\begin{equation}
  k= \frac{2 \pi n}{m}\quad \text{for} \quad 0\leq n\leq m-1.
  \label{eq:q-fermion-periodic-momentum}
\end{equation}

\begin{figure}[!htbp]
  \centering
  \begin{tikzpicture}[>=stealth,
    farwl/.style={thick, gray!50},
    nearwl/.style={thick, postaction={decorate, decoration={markings,
      mark=at position 0.93 with {\arrow{>}}}}},
    ring/.style={thin, gray!45, dashed},
    symfar/.style={very thick, green!42!black, opacity=0.85,
      postaction={decorate, decoration={markings,
        mark=at position 0.52 with {\arrow{>}}}}},
    symnear/.style={very thick, green!42!black,
      postaction={decorate, decoration={markings,
        mark=at position 0.52 with {\arrow{>}}}}},
    rung/.style={thick, postaction={decorate, decoration={markings,
      mark=at position 0.60 with {\arrow{>}}}}},
    vtx/.style={circle, fill, inner sep=1.15pt},
    lab/.style={font=\small}
  ]

    \newcommand{\panel}[2]{%
      \begin{scope}[xshift=#1]
        \def\R{2.15}\def\Ry{0.52}\def\Bot{-1.30}\def\Top{2.60}
        \def\Rm{2.62}\def\Rym{0.66}
        \draw[ring] (0,\Bot) ellipse [x radius=\R, y radius=\Ry];
        \foreach \t in {30,60,90,120,150}{
          \draw[farwl] ({\R*cos(\t)},{\Ry*sin(\t)+\Bot})
                    -- ({\R*cos(\t)},{\Ry*sin(\t)+\Top});
        }
        \draw[symfar] ({\Rm},{#2}) arc[start angle=0, end angle=180,
                                       x radius=\Rm, y radius=\Rym];
        \foreach \t in {0,180,210,240,270,300,330}{
          \draw[white, line width=2.9pt] ({\R*cos(\t)},{\Ry*sin(\t)+\Bot})
                                      -- ({\R*cos(\t)},{\Ry*sin(\t)+\Top});
          \draw[nearwl] ({\R*cos(\t)},{\Ry*sin(\t)+\Bot})
                     -- ({\R*cos(\t)},{\Ry*sin(\t)+\Top});
        }
        \draw[rung] ({\R*cos(240)},{\Ry*sin(240)+1.15})
                 -- ({\R*cos(270)},{\Ry*sin(270)+1.15});
        \node[vtx] at ({\R*cos(240)},{\Ry*sin(240)+1.15}) {};
        \node[vtx] at ({\R*cos(270)},{\Ry*sin(270)+1.15}) {};
        \draw[rung] ({\R*cos(300)},{\Ry*sin(300)+0.55})
                 -- ({\R*cos(330)},{\Ry*sin(330)+0.55});
        \node[vtx] at ({\R*cos(300)},{\Ry*sin(300)+0.55}) {};
        \node[vtx] at ({\R*cos(330)},{\Ry*sin(330)+0.55}) {};
        \draw[symnear] ({-\Rm},{#2}) arc[start angle=180, end angle=360,
                                         x radius=\Rm, y radius=\Rym];
        \node[lab, anchor=east, green!42!black] at ({-\Rm-0.12},{#2}) {$\mu$};
      \end{scope}
    }

    \panel{0cm}{-0.40}
    \node[lab] at (0,-2.70) {(a)};

    \node[lab] at (3.45,0.6) {$=$};

    \panel{6.90cm}{1.95}
    \node[lab] at (6.90,-2.70) {(b)};

  \end{tikzpicture}
\caption{The topological symmetry line $Y_\mu$ slides freely in the time direction, past the local Hamiltonian terms drawn as rungs between neighboring worldlines. Equivalence of (a) and (b) is precisely the statement $[H_{\text{per}},Y_\mu]=0$.}
  \label{fig:symmetry-line-commutes}
\end{figure}
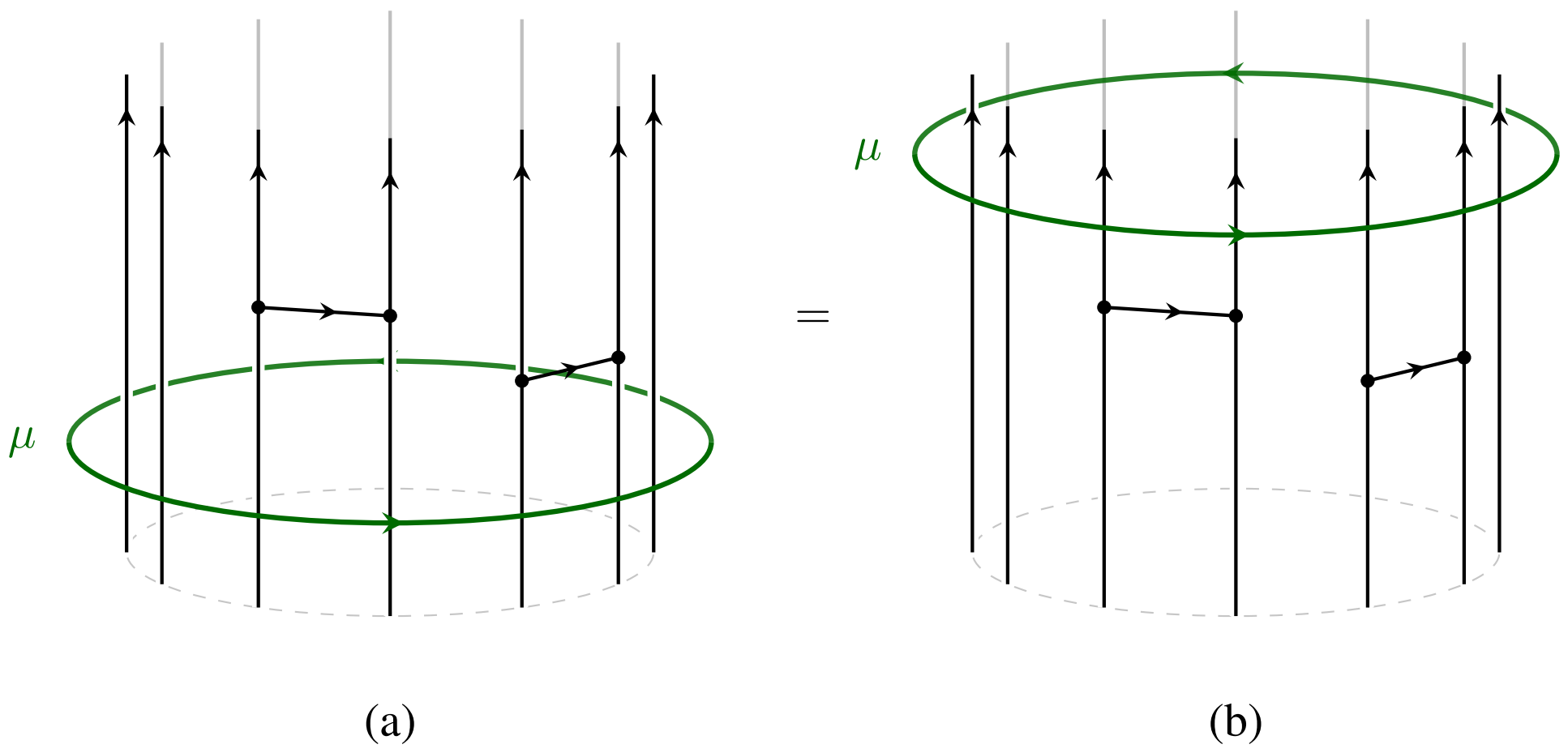

Besides translation, the periodic chain carries the non-invertible symmetries $Y_\mu$, $\mu\in P_k^+$,
forming the Verlinde algebra,
\begin{equation}
Y_\mu Y_\nu = \sum_{\lambda\in P_k^+} N_{\mu\nu}^{\lambda} Y_\lambda,
\label{eq:periodic-verlinde-algebra}
\end{equation}
These are the closed Wilson lines of Section~\ref{sec:periodic-chain} running parallel to the chain along the $a$-cycle \cite{Feiguin2007:InteractingAnyons,Gils2013:AnyonicQuantumSpin}. They commute with $U^{\mathrm{L}}_q(L\fsl_m)$ and hence with $H_{\mathrm{per}}$, which is the statement that such a line slides freely in the time direction past the local Hamiltonian terms, as in Figure~\ref{fig:symmetry-line-commutes}. Their simultaneous eigenspaces are the topological flux sectors,
\begin{equation}
\begin{gathered}
  \mathcal{H}_{\mathrm{per}}
  =\bigoplus_{\eta\in P_k^+}\mathcal{H}_{\mathrm{per}}^{(\eta)},
  \\
  Y_\mu\big|_{\mathcal{H}_{\mathrm{per}}^{(\eta)}}
  =\frac{S_{(\mu,\eta)}}{S_{(\mathbbm{1},\eta)}}\,\mathbbm{1}.
\end{gathered}
\label{eq:periodic-flux-sectors}
\end{equation}
where $S_{(a,b)}$ is the $S$-matrix of the $\widehat{\fsl}_N$ WZW model at level $k$, and the characters $\chi_{\eta}(\mu) = \frac{S_{(\mu,\eta)}}{S_{(\mathbbm{1},\eta)}}$ form representations of the Verlinde algebra of Equation~\eqref{eq:periodic-verlinde-algebra}. The label $\eta$ is the flux detected by the Wilson lines $Y_\mu$, which run along the $a$-cycle. It can be thought of as an anyon threading the $a$-cycle, running along the transverse $b$-cycle of Figure~\ref{fig:periodic-chain-torus}. Note that it is different from the basepoint label $\alpha$ of Equation~\eqref{eq:periodic-Hilbert-space}, which labels a fusion channel in a chosen cut. Each flux sector is preserved by $U^{\mathrm{L}}_q(L\fsl_m)$, which acts irreducibly on each charged subsector $\mathcal{H}^{(\eta)}_{\mathrm{per}}(Q)$.

The $q$-fermion chain also carries a baryon number symmetry $U(1)_B = U(1)_F/\bZ_N$, generated by $Q_{B}=\frac{1}{N}\sum_{j=1}^{m}n_j$. It is this quotient $U(1)_B$, rather than $U(1)_F$ itself, that acts faithfully, because the unitary operator $U(2\pi/N) \in U(1)_F$ acts trivially on the operator algebra and the states of the $q$-fermion chain. To see this, write the $U(1)_F$ symmetry operator as $U(\alpha)=\exp(i\alpha F)$. On a fusion-tree state it acts by
\begin{equation}
U(\alpha)\,|\boldsymbol{\lambda},\boldsymbol{r}\rangle
  = \exp\Bigl(i\alpha\sum_{j=1}^{m} r_j\Bigr)\,|\boldsymbol{\lambda},\boldsymbol{r}\rangle .
  \label{eq:lsm-fermion-number-action}
\end{equation}
The fusion tree imposes the constraint on the labels,
\begin{equation}
|\lambda^{(j)}| - |\lambda^{(j-1)}| = r_j \bmod N ,
  \label{eq:lsm-local-charge-selection-rule}
\end{equation}
so for periodic boundary conditions the total charge satisfies
\begin{equation}
\sum_{j=1}^{m} r_j = |\lambda^{(m)}| - |\lambda^{(0)}| = 0 \bmod N ,
  \label{eq:lsm-periodic-charge-selection-rule}
\end{equation}
Therefore, $U(2\pi/N)$ is the identity on every state. The same holds on the operator algebra. The mesons made of generators of $U^{\mathrm{L}}_q(L\fsl_m)$ are neutral under $F$, while the baryon operators of Appendix~\ref{app:baryon-algebra} carry a charge that is a multiple of $N$. No operator distinguishes $U(2\pi/N)$ from the identity. Therefore the faithfully acting symmetry group is $U(1)_B = U(1)_F/\bZ_N$.\footnote{This can also be understood from the quotient of the symmetry group of the free fermion chain because of gauging, $U(N)/SU(N) \cong U(1)/\bZ_N$.}

\begin{figure}[!htbp]
  \centering
  \begin{tikzpicture}[>=stealth,
    trunk/.style={very thick, postaction={decorate, decoration={markings,
      mark=at position 0.20 with {\arrow{>}},
      mark=at position 0.80 with {\arrow{>}}}}},
    leg/.style={thick, postaction={decorate, decoration={markings,
      mark=at position 0.60 with {\arrow{>}}}}},
    jline/.style={very thick, ForestGreen},
    jarrow/.style={very thick, ForestGreen, postaction={decorate,
      decoration={markings, mark=at position 0.78 with {\arrow{>}}}}},
    vtx/.style={circle, fill, inner sep=1.2pt},
    lab/.style={font=\small}]
    \def\H{1.25}\def\Rx{0.50}\def\Ry{0.92}
    \draw[jline] (0,0) ellipse [x radius=\Rx, y radius=\Ry];
    \draw[white, line width=3.4pt] (-\Rx-0.10,0) -- (\Rx+0.10,0);
    \draw[trunk] (-3.30,0) -- (3.30,0);
    \foreach \x in {-2.55,-1.25,1.25,2.55}{\draw[leg] (\x,\H) -- (\x,0);}
    \foreach \x in {-2.55,-1.25,1.25,2.55}{\node[vtx] at (\x,0) {};}
    \draw[jarrow] ({\Rx*cos(220)},{\Ry*sin(220)})
        arc[start angle=220, end angle=140, x radius=\Rx, y radius=\Ry];
    \node[lab,anchor=south] at (-1.25,\H+0.06) {$j$};
    \node[lab,anchor=south] at ( 1.25,\H+0.06) {$j+1$};
    \node[lab,ForestGreen,anchor=north] at (0,-\Ry-0.06) {$J$};
    \node[lab,anchor=north] at (-0.88,-0.12) {$\lambda^{(j)}$};
  \end{tikzpicture}
\caption{The operator $\mathbf{J}_{j+\frac{1}{2}}$ is the simple-current Wilson line $J$, in the representation $k\omega_1$, wrapped once around the bond between sites $j$ and $j+1$.}
  \label{fig:simple-current-operator}
\end{figure}
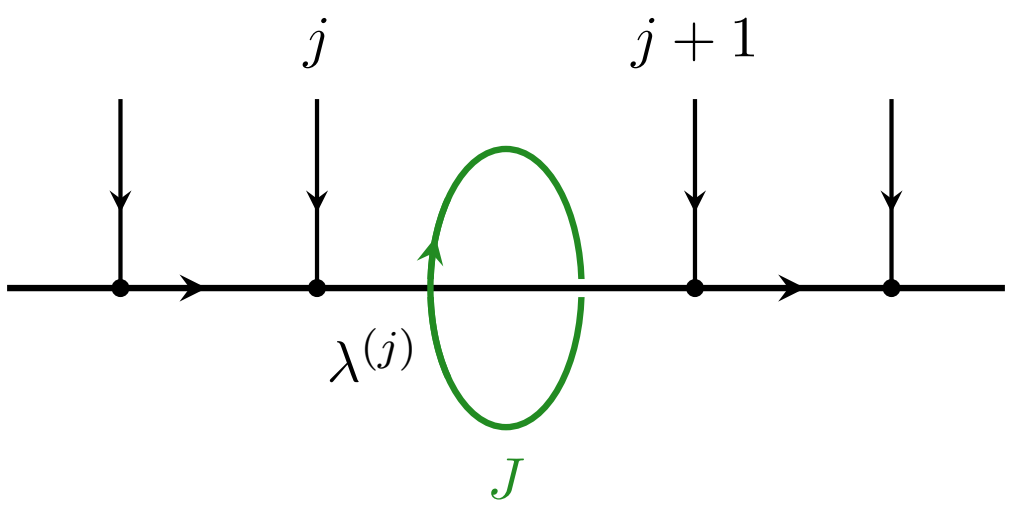

The translation and $U(1)_B$ symmetries of $H_{\mathrm{per}}$ allow an LSM argument analogous to Equation~\eqref{eq:lsm-anomaly-u1}. Unlike the free fermion case, we do not have the full color symmetry $U(N)$ to twist one flavor at a time. Instead we make use of the local operator $\mathbf{J}_{j+\frac{1}{2}}$, defined on the bond between sites $j$ and $j+1$, whose action on a state is given by
\begin{equation}
  \mathbf{J}_{j+\frac{1}{2}} |\boldsymbol{\lambda} , \boldsymbol{r} \rangle
  =
  \frac{S_{(k\omega_1,\lambda^{(j)})}}{S_{(\mathbbm{1},\lambda^{(j)})}}
  \,|\boldsymbol{\lambda} , \boldsymbol{r} \rangle
  =
  \omega^{|\lambda^{(j)}|} |\boldsymbol{\lambda} , \boldsymbol{r} \rangle \qquad \omega:=e^{2\pi i/N},
  \label{eq:lsm-simple-current-action}
\end{equation}
This operator is the simple-current Wilson line $J$ of representation $k\omega_1$, wrapped once around the bond between sites $j$ and $j+1$; see Figure~\ref{fig:simple-current-operator}. Let $|\text{GS}\rangle$ be a normalized ground state of charge $Q$, filling $\nu = Q/m$, energy $E_0$ and translation eigenvalue $e^{ip_0}$.\footnote{For comparison with the free fermion case, we let the ground state carry fermion charge $Q$ which will be a multiple of $N$} Define the baryon twist operator
\begin{equation}
  U^B_{2\pi}
  :=\mathbf{J}_{\frac{1}{2}}^{\dagger}\exp\!\left(\frac{2\pi i}{mN}\sum_{j=1}^{m}j n_j\right),
  \qquad
  U_\pm:=(U^B_{2\pi})^{\pm1},
  \qquad
  \theta:=\frac{2\pi}{mN}.
  \label{eq:lsm-baryon-twist-operator}
\end{equation}
Under translation, the exponential alone acquires a boundary factor $\omega^{n_1}$, which is canceled by the transformation of $\mathbf{J}_{\frac{1}{2}}^{\dagger}$. Using
\begin{equation}
Tn_jT^{-1}=n_{j+1}, \qquad
  T\mathbf{J}_{j-\frac{1}{2}}^{\dagger}T^{-1}=\mathbf{J}_{j+\frac{1}{2}}^{\dagger}=\omega^{-n_j}\mathbf{J}_{j-\frac{1}{2}}^{\dagger},
  \label{eq:lsm-translation-local-operators}
\end{equation}
we obtain the projective commutation relation
\begin{equation}
  TU_\pm T^{-1}=e^{\mp2\pi i\nu/N}U_\pm.
  \label{eq:lsm-twist-translation-algebra}
\end{equation}
Therefore the states $U_\pm|\text{GS}\rangle$ have definite momentum $p_0\mp2\pi\nu/N$ and are orthogonal to $|\text{GS}\rangle$ for $0<\nu<N$. We also have
\begin{equation}
  U_\pm^\dagger E_iU_\pm=e^{\pm i\theta}E_i,
  \qquad
  U_\pm^\dagger F_iU_\pm=e^{\mp i\theta}F_i,
  \label{eq:lsm-twisted-hopping-action}
\end{equation}
Defining
\begin{equation}
  \Delta E_\pm:=
  \langle\text{GS}|U_\pm^\dagger H_{\mathrm{per}}U_\pm|\text{GS}\rangle-E_0,
  \label{eq:lsm-twist-excitation-energies}
\end{equation}
and averaging over the two energy shifts gives
\begin{equation}
\begin{aligned}
  0
  \leq \min(\Delta E_+,\Delta E_-)
  &\leq \frac{\Delta E_++\Delta E_-}{2}\\
  &=(\cos\theta-1)
    \langle\text{GS}|H_{\mathrm{hop}}|\text{GS}\rangle\\
  &\leq mc_{N,k}(1-\cos\theta)
  \leq \frac{2\pi^2c_{N,k}}{mN^2}.
\end{aligned}
  \label{eq:lsm-twist-energy-bound}
\end{equation}
Here $H_{\mathrm{hop}}=-t\sum_{i}(E_i+F_i)$ is the hopping part of Equation~\eqref{eq:periodic-hamiltonian}, and $\|t(E_i+F_i)\|\leq c_{N,k}$ for every $i$ in the physical Hilbert space, with $c_{N,k}$ independent of the system size $m$. This is because each $E_i$ and $F_i$ acts on a single bond, which at fixed $N$ and $k$ carries only finitely many fusion paths. For $0<\nu<N$, the variational principle implies that there exists an energy eigenstate orthogonal to $|\text{GS}\rangle$, with momentum $p_0-2\pi\nu/N$ or $p_0+2\pi\nu/N$ and excitation energy at most $\cO(1/m)$. Hence, the periodic $q$-fermion chain at any filling $0<\nu<N$ cannot have a unique gapped ground state respecting translation and $U(1)_B$.

\begin{figure}[!htbp]
  \centering
  \begin{tikzpicture}[>=stealth,
    wl/.style={thick, postaction={decorate, decoration={markings,
      mark=at position 0.93 with {\arrow{>}}}}},
    bdyL/.style={thick, red!80!black},
    bdyR/.style={thick, blue!75!black},
    sym/.style={very thick, green!42!black,
      postaction={decorate, decoration={markings,
        mark=at position 0.52 with {\arrow{>}}}}},
    rung/.style={thick, postaction={decorate, decoration={markings,
      mark=at position 0.60 with {\arrow{>}}}}},
    vtx/.style={circle, fill, inner sep=1.15pt},
    junc/.style={circle, fill=green!42!black, inner sep=1.7pt},
    lab/.style={font=\small}
  ]

    \newcommand{\openpanel}[3]{%
      \begin{scope}[xshift=#1]
        \def\Bw{4.60}\def\Bot{-1.40}\def\Top{1.40}
        \draw[bdyL] (0,\Bot) -- (0,\Top);
        \draw[bdyR] (\Bw,\Bot) -- (\Bw,\Top);
        \node[lab, red!80!black, anchor=base] at (0,{\Top+0.16}) {$\alpha$};
        \node[lab, blue!75!black, anchor=base] at (\Bw,{\Top+0.16}) {$\beta^{\vee}$};
        \foreach \x in {0.95,1.75,2.85,3.65}{
          \draw[wl] (\x,\Bot) -- (\x,\Top);
        }
        \draw[rung] (0.95,#2) -- (1.75,#2);
        \node[vtx] at (0.95,#2) {};
        \node[vtx] at (1.75,#2) {};
        \draw[rung] (2.85,#3) -- (3.65,#3);
        \node[vtx] at (2.85,#3) {};
        \node[vtx] at (3.65,#3) {};
        \begin{scope}
          \clip (0.20,-2) rectangle ({\Bw-0.20},2);
          \draw[white, line width=3.2pt]
            (0,0) .. controls (1.30,-0.46) and ({\Bw-1.30},-0.46) .. (\Bw,0);
        \end{scope}
        \draw[sym]
          (0,0) .. controls (1.30,-0.46) and ({\Bw-1.30},-0.46) .. (\Bw,0);
        \node[junc] at (0,0) {};
        \node[junc] at (\Bw,0) {};
        \node[lab, green!42!black, anchor=east] at (-0.10,0) {$u_i$};
        \node[lab, green!42!black, anchor=west] at ({\Bw+0.10},0) {$v_j$};
        \node[lab, green!42!black, below] at ({\Bw/2},-0.38) {$\mu$};
      \end{scope}
    }

    \openpanel{0cm}{0.80}{0.45}
    \node[lab] at (2.30,-2.00) {(a)};

    \node[lab] at (5.75,0) {$=$};

    \openpanel{6.90cm}{-0.80}{-1.10}
    \node[lab] at (9.20,-2.00) {(b)};

  \end{tikzpicture}
\caption{The open-chain symmetry operator $Y_{\mu;i,j}$ is the topological line $\mu$ ending on the boundary lines $\alpha$ and $\beta^{\vee}$ at the junctions $u_i$ and $v_j$. The line and its junctions slide past the local Hamiltonian terms, drawn as rungs between neighboring worldlines. Equivalence of (a) and (b) is the commutation relation in Equation~\eqref{eq:open-symmetry-commutator}.}
  \label{fig:open-symmetry-line-action}
\end{figure}
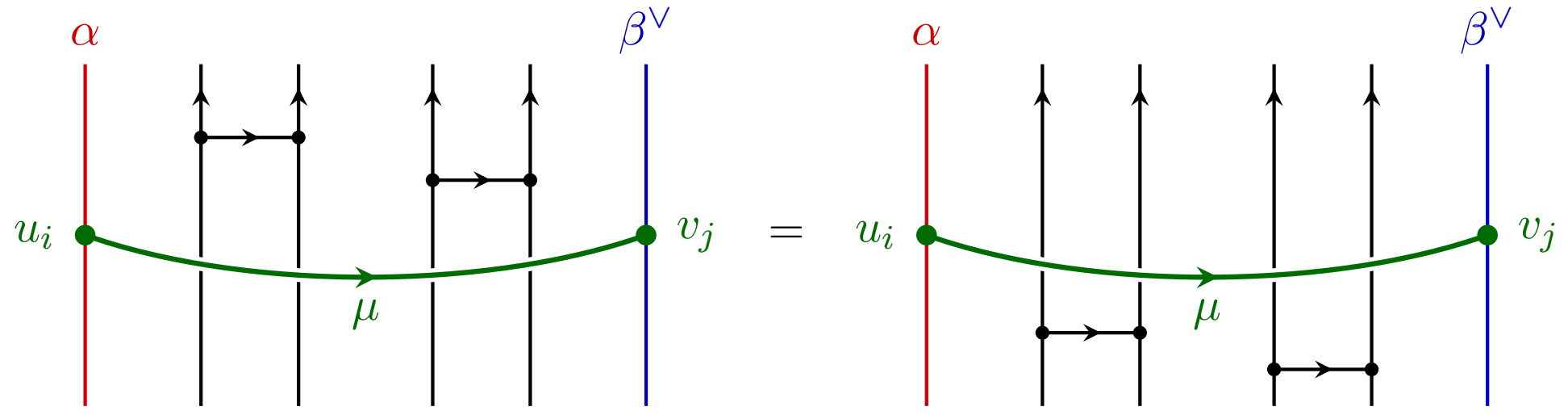

We now turn to the symmetries of the open chain. With boundary conditions $\alpha,\beta\in P_k^+$, the topological symmetry operators inherited from the bulk $SU(N)_k$ lines can no longer usefully wrap a cycle: a loop around the chain encloses a fixed boundary charge and acts on all of $\mathcal{H}_{(\alpha,\beta)}$ as a number. Such loops therefore represent the Verlinde algebra of Equation~\eqref{eq:periodic-verlinde-algebra} by a single character, a one-dimensional representation that is not faithful, in contrast to the periodic chain, where every character appears. Instead, we need to consider Wilson lines that end on the two boundary lines $\alpha$ and $\beta^{\vee}$, see Figure~\ref{fig:open-chain-sphere}. A line $\mu$ defines such an operator on $\mathcal{H}_{(\alpha,\beta)}$ only if it can end topologically on both boundary lines without changing them:
\begin{equation}
  N_{\mu,\alpha}^{\alpha}>0,
  \qquad
  N_{\mu,\beta^{\vee}}^{\beta^{\vee}}>0.
  \label{eq:open-weak-symmetry-condition}
\end{equation}
These are the conditions for the two boundaries to be \textit{weakly symmetric} under $\mu$ in the sense of \cite{Choi2023:BoundariesNoninvertibleSymmetries}. For each allowed line, choose bases of endpoint junctions
\begin{equation}
\begin{aligned}
  u_i&\in\text{Hom}_{\mathcal{C}_{N,k}}
    (\mathcal{L}_\alpha,\mathcal{L}_\mu\otimes\mathcal{L}_\alpha),
  &i&=1,\ldots,N_{\mu,\alpha}^{\alpha},\\
  v_j&\in\text{Hom}_{\mathcal{C}_{N,k}}
    (\mathcal{L}_\mu\otimes\mathcal{L}_{\beta^{\vee}},\mathcal{L}_{\beta^{\vee}}),
  &j&=1,\ldots,N_{\mu,\beta^{\vee}}^{\beta^{\vee}}.
\end{aligned}
\label{eq:open-symmetry-junctions}
\end{equation}
The resulting topological operator is denoted $Y_{\mu;i,j}$. For the open-chain hopping Hamiltonian, the line and its endpoint junctions can be moved in the time direction past the local Hamiltonian terms, as shown in Figure~\ref{fig:open-symmetry-line-action}. Consequently,
\begin{equation}
  [H_0^q,Y_{\mu;i,j}]=0.
  \label{eq:open-symmetry-commutator}
\end{equation}

Composing two such operators requires both fusing the bulk lines and composing their endpoint topological junctions. Their multiplication therefore takes the form
\begin{equation}
  Y_{\mu;i,j}\,Y_{\nu;i',j'}
  =\sum_{\rho,p,q}
    C^{(\alpha,\beta)\,\rho;p,q}_{\mu;i,j\,;\,\nu;i',j'}
    Y_{\rho;p,q}.
  \label{eq:open-symmetry-algebra}
\end{equation}
We refer the reader to \cite{Choi2024:GeneralizedTubeAlgebras,Cordova2024:RepresentationTheorySolitons,Konechny2020:OpenTopologicalDefects,Choi2023:BoundariesNoninvertibleSymmetries} for an explicit discussion of the symmetry algebra of boundary conformal field theories. Along with the non-invertible symmetries, like the periodic chain, the open chain also carries a faithfully acting baryon number symmetry $U(1)_B=U(1)_F/\bZ_N$, since the generators of $U^{\mathrm{L}}_q(\fgl_m)$ are neutral under $F$, while the baryon operators carry a charge that is a multiple of $N$. The difference lies in how $U(2\pi/N)$ acts on states. Since the states of $\mathcal{H}_{ \alpha, \beta}$ carry a nontrivial charge modulo $N$,
\begin{equation}
  Q=\sum_{j=1}^{m}r_j
  \equiv |\beta|-|\alpha|\pmod N,
  \label{eq:open-charge-selection-rule}
\end{equation}
and $U(2\pi/N)$ acts on the whole Hilbert space as a single phase,
\begin{equation}
  \left.U(2\pi/N)\right|_{\mathcal{H}_{(\alpha,\beta)}}
  =e^{2\pi i(|\beta|-|\alpha|)/N}
    \text{id}_{\mathcal{H}_{(\alpha,\beta)}}.
  \label{eq:open-charge-central-phase}
\end{equation}
This phase records the boundary-induced offset $Q_B\in\mathbb{Z}+(|\beta|-|\alpha|)/N$.
\section[Continuum limit of the \texorpdfstring{$q$}{q}-fermion chain]{Continuum limit of the \texorpdfstring{$q$}{q}-fermion chain}
\label{sec:continuum-limit}
We now turn to the main claim of this paper, the continuum limit of these $q$-fermion chains, which we conjecture to be a compact boson glued to the $W_N$ minimal model. We give three main arguments for this claim. The first is field-theoretic. As set up in Section~\ref{sec:q-fermion-chain}, the $q$-fermions form a lattice of fermion worldlines inside $SU(N)_k$ Chern--Simons theory, so their continuum limit is likely to be a non-chiral interface of $SU(N)_k$ Chern--Simons theory. Such interfaces were constructed recently in \cite{GaiottoZeng2026:InterfaceMinimalModel}, and we follow that construction closely. We expect the non-chirality to hold by analogy with the Nielsen--Ninomiya theorem~\cite{Nielsen1981:AbsenceNeutrinosI,Nielsen1981:AbsenceNeutrinosII} in the free-fermion case, and with the more recent result of~\cite{KapustinSpodyneiko2019:AbsenceEnergyCurrents} that a one-dimensional lattice Hamiltonian with bounded finite-range interactions cannot flow to a CFT with $c_L\neq c_R$ under its thermal clustering assumptions. For our fusion-constrained chain, this remains motivation for the conjecture. The second is numerical. We use exact diagonalization for momentum- and line-resolved fusion-tree spectra, and fusion-path DMRG for larger open chains. Appendix~\ref{app:continuum-numerics} compares their periodic spectra and Casimir energies with the continuum proposal.
The third connects to the literature. In Section~\ref{sec:projector-chain} we modify the $q$-fermion Hamiltonian and relate the result to existing anyon chains that flow to the $W_N$ minimal model.

\subsection{The non-chiral interface}
\label{sec:non-chiral-interface}
In Section~\ref{sec:free-fermion-dynamics} we recalled how the free Dirac fermion is realized as a particular sum of correlated modules of the chiral algebra $U(1)^{\text{lat}}_N\times SU(N)_1$, which are mutually local. We now ask the question: what theory is obtained when this system of two-dimensional Dirac fermions is coupled to Chern--Simons theory along a two-dimensional interface? We are motivated by the assumption that \textit{coupling to Chern--Simons theory} might commute with \textit{taking the continuum limit} at least for the simple Hamiltonian~\eqref{eq:intro-hamiltonian}. The $q$-fermion chain couples the lattice fermions first and takes the limit, whereas the interface construction takes the limit first. The numerical evidence in Appendix~\ref{app:continuum-numerics} supports this assumption, but only up to \textit{exactly marginal} deformations of the continuum theory, which are not fixed from the lattice. This should be understood as the natural ambiguity in the choice of a finite renormalization scheme when coupling two-dimensional Dirac fermions to Chern--Simons theory, unless some symmetry fixes the scheme. We describe this phenomenon explicitly in the rest of this section.

We first briefly review the two boundary conditions from which the interface is built. A \textit{Dirichlet} boundary condition for $SU(N)_k$ Chern--Simons theory fixes a component of the gauge field to be zero at the boundary,
\begin{equation}
    A_{\bar z}\big|_{\partial M}=0,
  \label{eq:cs-holomorphic-boundary-condition}
\end{equation}
and restricts the space of gauge transformations to be those that are identity at the boundary, $g|_{\partial M} = \mathbbm{1}$. Gauge transformations that are nontrivial at the boundary then act as global symmetries, and the boundary supports the chiral $SU(N)_k$ WZW algebra, generated by the current $J_{(k)} = k A_{z}$ \cite{ElitzurEtAl1989:CanonicalQuantization}.

$SU(N)_k$ Chern--Simons theory also admits an \textit{enriched Neumann} boundary condition, for which the space of allowed gauge transformations includes those that act nontrivially at the boundary. Such a boundary condition requires an auxiliary two-dimensional theory $\mathcal{T}$ whose local operator algebra $\mathcal{A}_{\mathcal{T}}$ contains a chiral $SU(N)_k$ WZW symmetry algebra. The $SU(N)$ anomaly of $\mathcal{T}$ then cancels the inflow from the Chern--Simons bulk \cite{GaiottoRapcak2019:VertexAlgebrasCorner,GaiottoZeng2026:InterfaceMinimalModel}. The Neumann boundary condition then imposes
\begin{equation}
  kA_{z} = J_{\mathcal{T}} ,
  \label{eq:neumann-bc}
\end{equation}
where $J_{\mathcal{T}}$ is the current generating the chiral $SU(N)_k$ WZW subalgebra in $\mathcal{A}_{\mathcal{T}}$. The algebra of gauge-invariant local chiral boundary operators is therefore the coset
\begin{equation}
    \mathcal{A}_{\mathrm N[\mathcal{T}]}
    =
    \text{Com}\!\left(SU(N)_k,\mathcal{A}_{\mathcal{T}}\right),
    \label{eq:neumann-coset}
\end{equation}
consisting of the operators in $\mathcal{A}_{\mathcal{T}}$ whose OPEs with the gauged $SU(N)_k$ currents are nonsingular. We denote the resulting boundary theory by $\mathcal{T}/SU(N)_k$.

To construct the chiral interface of interest, we stack $N$ chiral complex fermions on the Dirichlet boundary of $SU(N)_{k-1}$ Chern--Simons theory. This gives the theory $\mathcal{T} = SU(N)_{k-1} \times \mathrm{Ff}^{N}_{\bC}$.\footnote{One can also take $\mathcal{T} = SU(N)_{k-1}$ alone, but this stacks a Dirichlet and a Neumann boundary condition and gives the trivial interface.} Through their gauge anomaly, the chiral fermions raise the level by one, so $\mathcal{T}$ carries a chiral $SU(N)_{k}$ WZW symmetry. We then use $\mathcal{T}$ to define a Neumann boundary condition for $SU(N)_{k}$ Chern--Simons theory. The gauge fields on the two sides in Equation~\eqref{eq:neumann-bc} satisfy the boundary conditions,
\begin{equation}
  A^{(k-1)}_{\bar z}=A^{(k)}_{\bar z} ,
  \qquad
  k\,A^{(k)}_{z}=(k-1)\,A^{(k-1)}_{z}+\psi^{\dagger}\psi .
  \label{eq:interface-gauge-bc}
\end{equation}

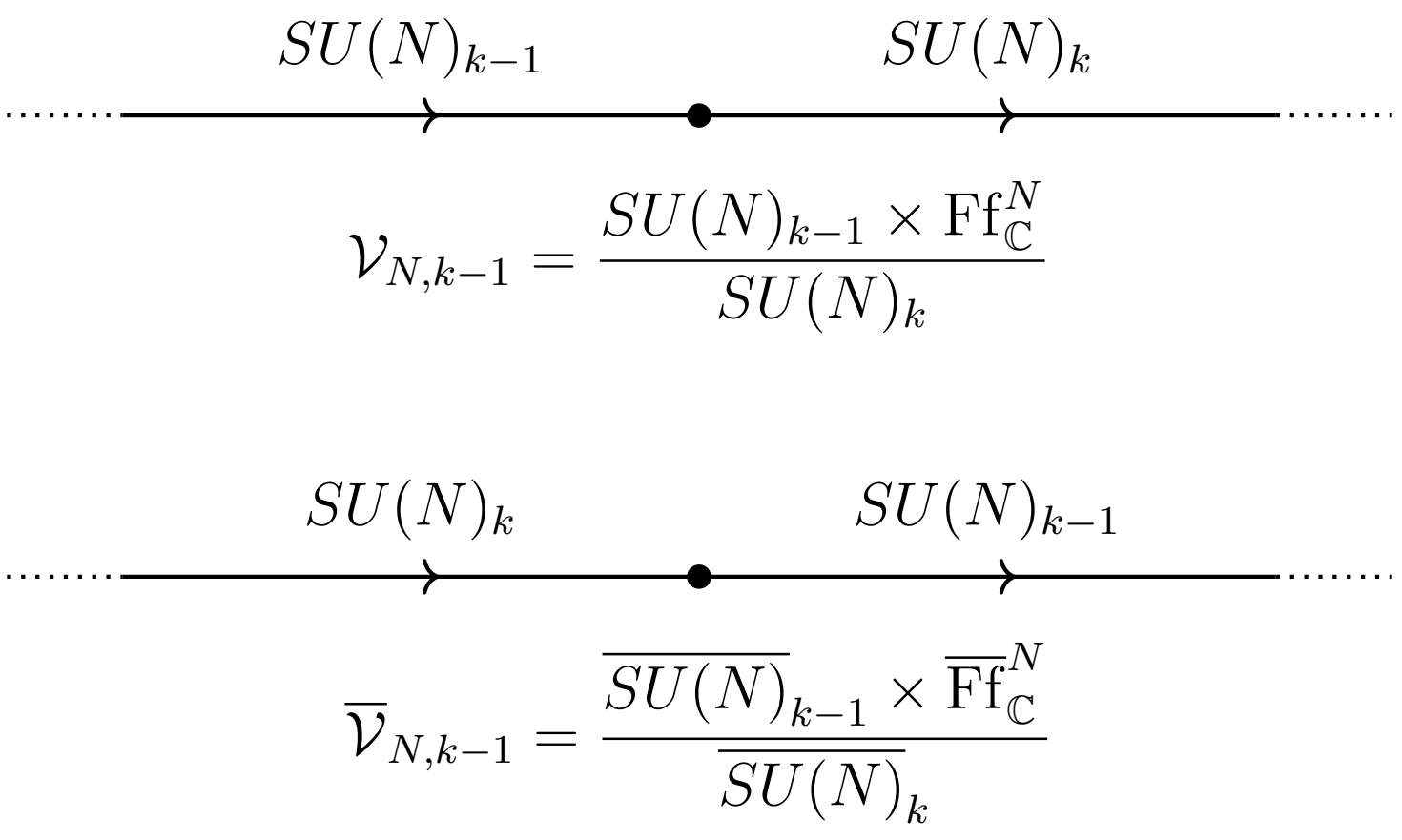
\begin{figure}[!htbp]
\centering
\begin{tikzpicture}[
  thick,
  midarr/.style={postaction={decorate,
    decoration={markings, mark=at position 0.55 with {\arrow{>}}}}}
]
  \begin{scope}
    \draw[dotted] (-0.8,0) -- (0,0);
    \draw[midarr] (0,0) -- (4,0);
    \draw[midarr] (4,0) -- (8,0);
    \draw[dotted] (8,0) -- (8.8,0);
    \fill (4,0) circle (2.4pt);
    \node[above] at (2,0.1) {$SU(N)_{k-1}$};
    \node[above] at (6,0.1) {$SU(N)_{k}$};
    \node[below] at (4,-0.3)
      {$\displaystyle\mathcal{V}_{N,k-1}
        =\frac{SU(N)_{k-1}\times \mathrm{Ff}^{N}_{\bC}}{SU(N)_{k}}$};
  \end{scope}
  \begin{scope}[yshift=-3.2cm]
    \draw[dotted] (-0.8,0) -- (0,0);
    \draw[midarr] (0,0) -- (4,0);
    \draw[midarr] (4,0) -- (8,0);
    \draw[dotted] (8,0) -- (8.8,0);
    \fill (4,0) circle (2.4pt);
    \node[above] at (2,0.1) {$SU(N)_{k}$};
    \node[above] at (6,0.1) {$SU(N)_{k-1}$};
    \node[below] at (4,-0.3)
      {$\displaystyle\overline{\mathcal{V}}_{N,k-1}
        =\frac{\overline{SU(N)}_{k-1}\times \overline{\mathrm{Ff}}^{N}_{\bC}}
              {\overline{SU(N)}_{k}}$};
  \end{scope}
\end{tikzpicture}
\caption{The chiral interface $\mathcal{V}_{N,k-1}$ (top) and the anti-chiral interface $\overline{\mathcal{V}}_{N,k-1}$ (bottom) in Chern--Simons theory. The level jumps from $k-1$ to $k$ along the orientation of the line.}
\label{fig:continuum-chiral-wall}
\end{figure}

From Equation~\eqref{eq:neumann-coset} at level $k$, the local operators on the interface are those of $SU(N)_{k-1}\times\mathrm{Ff}^{N}_{\bC}$ that commute with the gauged $SU(N)_{k}$ current,
\begin{equation}
  J^{A}_{(k)} = J^{A}_{(k-1)} +{:}\psi^{a\dagger}(T^A)_{ab}\,\psi^{b}{:}
  \label{eq:interface-total-current}
\end{equation}
where $T^A$ are the Hermitian generators of $\fsu(N)$ in the fundamental representation, normalized as in Equation~\eqref{eq:su-n-generators}. These operators form the coset chiral algebra\footnote{The coset $\mathcal{V}_{N, k-1}$ does not contain a global $SU(N)$ symmetry, since the diagonal global $SU(N)$ current of $\mathcal{T}$ is gauged, while the anti-diagonal $SU(N)$ currents are not local with $J^A_{(k)}$}
\begin{equation}
  \mathcal{V}_{N,k-1}
  =
  \frac{SU(N)_{k-1}\times \mathrm{Ff}^{N}_{\bC}}{SU(N)_{k}}
  = \text{Com}\!\left(SU(N)_{k}, SU(N)_{k-1}\times \mathrm{Ff}^{N}_{\bC}\right).
  \label{eq:continuum-chiral-interface}
\end{equation}
From Equation~\eqref{eq:free-fermion-decomposition}, the free fermion decomposes into $N$ summands. Since $U(1)^{\text{lat}}_N$ commutes with the diagonal $SU(N)_{k}$ currents, the commutant condition constrains only the $SU(N)_1$ modules. Only the vacuum summand of \eqref{eq:free-fermion-decomposition} contains an $SU(N)_{k}$ vacuum module, and therefore only the $r=0$ summand contributes to the local operator algebra on the interface.
\begin{equation}
  \mathcal{V}_{N,k-1}
  = \frac{SU(N)_{k-1} \times (\mathcal{U}_0 \otimes \mathcal{L}_{\omega_0})}{SU(N)_{k}}
  = U(1)^{\text{lat}}_N \times \mathcal{W}_{N,k-1} ,
  \label{eq:continuum-chiral-regrouped}
\end{equation}
where
\begin{equation}
  \mathcal{W}_{N,k-1}
  =
  \frac{SU(N)_{k-1}\times SU(N)_{1}}{SU(N)_{k}}
  \label{eq:continuum-wn-minimal-model}
\end{equation}
is the $W_N$ chiral algebra \cite{Zamolodchikov1985:InfiniteAdditionalSymmetries,FateevLukyanov1988:ZnSymmetryModels}, realized here as a coset \cite{Bais1988:CosetConstructionExtendedVirasoro}. It is generated by the stress tensor $T=W_2$ and the currents $W_3,\ldots,W_N$ of spins $3,\ldots,N$ \cite{BouwknegtSchoutens1993:WSymmetry}. The remaining summands describe operators attached to bulk Wilson lines, which we describe in Section~\ref{sec:operator-content}.

\begin{figure}[!htbp]
\centering
\begin{tikzpicture}[
  thick,
  midarr/.style={postaction={decorate,
    decoration={markings, mark=at position 0.55 with {\arrow{>}}}}}
]
  \begin{scope}
    \draw[dotted] (-0.8,0) -- (0,0);
    \draw[midarr] (0,0) -- (3,0);
    \draw[midarr] (3,0) -- (6.6,0);
    \draw[midarr] (6.6,0) -- (9.6,0);
    \draw[dotted] (9.6,0) -- (10.4,0);
    \fill (3,0) circle (2.4pt);
    \fill (6.6,0) circle (2.4pt);
    \node[above] at (1.5,0.1) {$SU(N)_{k}$};
    \node[above] at (4.8,0.1) {$SU(N)_{k-1}$};
    \node[above] at (8.1,0.1) {$SU(N)_{k}$};
    \node[left] at (-1.0,0) {(a)};
  \end{scope}
  \begin{scope}[yshift=-1.6cm]
    \draw[dotted] (-0.8,0) -- (0,0);
    \draw[midarr] (0,0) -- (3,0);
    \draw[midarr] (3,0) -- (6.6,0);
    \draw[midarr] (6.6,0) -- (9.6,0);
    \draw[dotted] (9.6,0) -- (10.4,0);
    \fill (3,0) circle (2.4pt);
    \fill (6.6,0) circle (2.4pt);
    \node[above] at (1.5,0.1) {$SU(N)_{k-1}$};
    \node[above] at (4.8,0.1) {$SU(N)_{k}$};
    \node[above] at (8.1,0.1) {$SU(N)_{k-1}$};
    \node[left] at (-1.0,0) {(b)};
  \end{scope}
\end{tikzpicture}
\caption{Both $M^\pm_{N,k-1}$ interfaces have the same chiral algebra, but live in different bulk Chern--Simons theories. Shrinking the intermediate region in $(a)$ and $(b)$ gives, respectively, the non-chiral interfaces $M^\pm_{N,k-1}$ inside $SU(N)_{k}$ and $SU(N)_{k-1}$ Chern--Simons theory.}
\label{fig:continuum-orientation}
\end{figure}
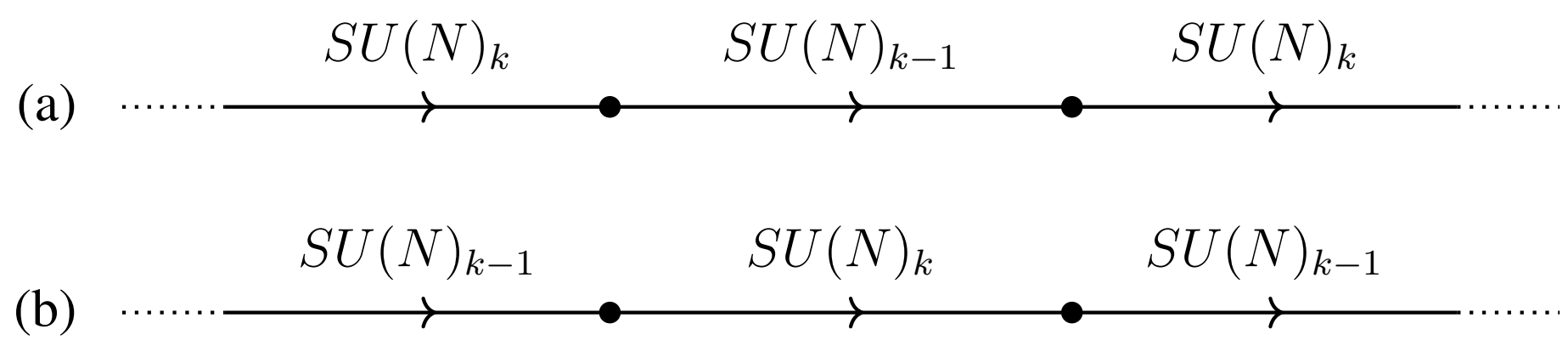

Adding the central charges of the three factors of Equation~\eqref{eq:continuum-wn-minimal-model} gives
\begin{equation}
  c_{\mathcal{W}_{N,k-1}}
  = N-1 + \frac{(N^2-1)(k-1)}{k+N-1} - \frac{(N^2-1)k}{k+N}
  = (N-1)\left(1-\frac{N(N+1)}{(N+k-1)(N+k)}\right),
  \label{eq:continuum-wn-central-charge-2}
\end{equation}
Together with the $U(1)^{\text{lat}}_N$ factor of Equation~\eqref{eq:continuum-chiral-regrouped}, the total central charge of the coset algebra $\mathcal{V}_{N,k-1}$ is therefore
\begin{equation}
  c_{N,k-1}
  = 1+(N-1)\left(1-\frac{N(N+1)}{(N+k-1)(N+k)}\right).
  \label{eq:continuum-wn-central-charge}
\end{equation}
The non-chiral interface is obtained by fusing a chiral and an anti-chiral interface, as in Figure~\ref{fig:continuum-orientation}. Depending on the orientation we obtain two possible non-chiral interfaces, which are fixed points of the RG flow inside $SU(N)_{k}$ Chern--Simons theory, denoted, in the notation of \cite{GaiottoZeng2026:InterfaceMinimalModel}, by $M^{-}_{N,k}$ and $M^{+}_{N,k-1}$. $M^-_{N,k}$ contains the chiral algebra $\mathcal{V}_{N, k} \times \overline{\mathcal{V}}_{N, k}$, while $M^+_{N,k-1}$ contains the chiral algebra $\mathcal{V}_{N,k-1} \times \overline{\mathcal{V}}_{N,k-1}$. Their exact operator content is described in Section~\ref{sec:operator-content}.

$M^{+}_{N,k-1}$ has lower central charge than $M^-_{N,k}$. We conjecture that the continuum limit of our $q$-fermion chain with the Hamiltonian given by Equation~\eqref{eq:path-qfermion-hamiltonian} flows to the interface $M^+_{N,k-1}$ with the lower central charge, up to some exactly marginal deformations. This is the sense in which the interface picture and the lattice model agree. We leave lattice realizations of $M^-_{N,k}$ for future work.

\subsection{Operator content and symmetries}
\label{sec:operator-content}
So far we have described the local operators on each chiral interface, organized by $\mathcal{V}_{N,k-1}$. The interface also carries operators living at the endpoints of bulk Wilson lines, which furnish modules of $\mathcal{V}_{N,k-1}$. Let $\Lambda_-\in P^+_{k-1}$ label a Wilson line on the $SU(N)_{k-1}$ side and $\Lambda_+\in P^+_{k}$ on the $SU(N)_{k}$ side. Endpoint operators are obtained by decomposing $\mathcal{L}^{(k-1)}_{\Lambda_-}\otimes\mathrm{Ff}^{N}_{\bC}$ under the diagonal affine $SU(N)_k$ algebra and taking the $\Lambda_+$ multiplicity space. These operators living at the endpoints of the bulk Wilson lines form the multiplicity space
\begin{equation}
  \mathcal{M}_{\Lambda_-,\Lambda_+}
  :=
  \text{Hom}_{SU(N)_k}
  \left(
  \mathcal{L}^{(k)}_{\Lambda_+},\;
  \mathcal{L}^{(k-1)}_{\Lambda_-}\otimes\mathrm{Ff}^{N}_{\bC}
  \right)
  \label{eq:interface-endpoint-space}
\end{equation}
The operators forming the chiral algebra are the case with no bulk line attached, $\Lambda_-=\Lambda_+=0$, i.e., $\mathcal{M}_{0,0}=\mathcal{V}_{N,k-1}$ as in Equation~\eqref{eq:continuum-chiral-regrouped}. To compute $\mathcal{M}_{\Lambda_-,\Lambda_+}$ we decompose the fermions as in Equation~\eqref{eq:free-fermion-decomposition} and decompose the tensor product of modules of $SU(N)_{k-1}\times SU(N)_1$ into modules of $SU(N)_k$,

\begin{equation}
  \mathcal{L}_{\Lambda_-}^{(k-1)}\otimes
  \mathcal{L}_{\omega_r}^{(1)}
  \cong
  \bigoplus_{\Lambda_+\in P_k^+}
  \mathcal{B}_{(\Lambda_-,\,\omega_r;\,\Lambda_+)}
  \otimes\mathcal{L}_{\Lambda_+}^{(k)},
  \label{eq:decomposition-modules}
\end{equation}
where $\mathcal{B}_{(\Lambda_-,\,\omega_r;\,\Lambda_+)}$ is a module of the coset $\mathcal{W}_{N,k-1}$. Demanding that the $\bZ_N$ center of $SU(N)$ acts in the same way on both sides of Equation~\eqref{eq:decomposition-modules} gives the selection rule
\begin{equation}
  |\Lambda_+|-|\Lambda_-|\equiv r\pmod N,
  \label{eq:interface-selection-rule}
\end{equation}
where $|\Lambda|$ is the number of boxes of the Young diagram \cite{GannonWalton1995:DiagonalCosetModularInvariants}. The coset modules inherit the free-fermion gluing of $\mathcal{U}_r$ to $\mathcal{L}_{\omega_r}$, so a pair of bulk labels selects one charge module,
\begin{equation}
  \mathcal{M}_{\Lambda_-,\Lambda_+}
  =
  \mathcal{U}_r\otimes\mathcal{B}_{(\Lambda_-,\,\omega_r;\,\Lambda_+)},
  \label{eq:interface-endpoint-module}
\end{equation}
and each coset module is glued to a definite $\mathcal{U}_r$. The operators in $\mathcal{M}_{\Lambda_-,\Lambda_+}$ are
\begin{equation}
  \mathcal{O}(z)
  =
  {:}e^{\,iQ\phi_L(z)/\sqrt N}{:}\;
  \Phi_{(\Lambda_-,\,\omega_r;\,\Lambda_+)}(z),
  \qquad
  Q\in r+N\bZ,
  \label{eq:interface-endpoint-operator}
\end{equation}
and their descendants, where $\Phi_{(\Lambda_-,\omega_r;\Lambda_+)}$ is the $\mathcal{W}_{N,k-1}$ primary of $\mathcal{B}_{(\Lambda_-,\omega_r;\Lambda_+)}$. Its conformal weight is
\begin{equation}
  h_{\mathcal{O}}
  =\frac{Q^2}{2N}+h_{(\Lambda_-,\,\omega_r;\,\Lambda_+)},
  \qquad
  h_{(\Lambda_-,\,\omega_r;\,\Lambda_+)}
  =\frac{C_2(\Lambda_-)}{k+N-1}+\frac{r(N-r)}{2N}-\frac{C_2(\Lambda_+)}{k+N}+n,
  \label{eq:interface-endpoint-weight}
\end{equation}
where $n\in\bZ_{\ge0}$ is the grade at which $\mathcal{L}^{(k)}_{\Lambda_+}$ appears in the branching of Equation~\eqref{eq:decomposition-modules} \cite{GoddardKentOlive1986:UnitaryVirasoro,KacPeterson1984:ThetaFunctions,DiFrancescoMathieuSenechal1997:CFT}.

\begin{figure}[!htbp]
\centering
\definecolor{chiralc}{RGB}{255,0,0}
\definecolor{antic}{RGB}{0,0,255}
\definecolor{Wilsonp}{RGB}{140,40,160}
\definecolor{Wilsong}{RGB}{0,115,0}
\definecolor{innerc}{RGB}{245,150,55}
\definecolor{outerc}{RGB}{110,190,225}
\begin{minipage}[c]{9.5cm}
\centering
\includegraphics{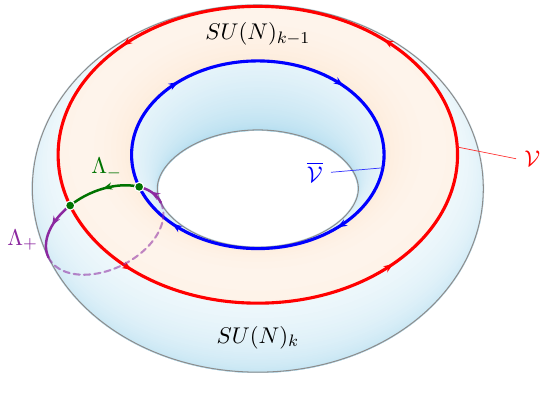}
\par\smallskip
{\footnotesize (a)}
\end{minipage}\hfill
\begin{minipage}[c]{6.0cm}
\centering
\begin{tikzpicture}[scale=0.92,
   line cap=round, line join=round, font=\small,
   surf/.style={thick},
   ileft/.style={very thick, chiralc},
   iright/.style={very thick, antic},
   gline/.style={thick, Wilsong},
   wline/.style={thick, Wilsonp},
   gvtx/.style={circle, fill=Wilsong, inner sep=1.3pt},
   lead/.style={black!65,line width=.35pt},
   tk/.style={line width=0.9pt, DarkRed},
   lab/.style={font=\small},
   reglab/.style={font=\small, inner sep=2.2pt,
                  rounded corners=1.5pt, draw=black!30, line width=0.4pt},
   dblr/.style={postaction={decorate, decoration={markings,
       mark=at position 0.42 with {\arrow{Stealth[length=4.6pt]}},
       mark=at position 0.60 with {\arrow{Stealth[length=4.6pt]}}}}},
   sglr/.style={postaction={decorate, decoration={markings,
       mark=at position 0.52 with {\arrow{Stealth[length=4.6pt]}}}}},
]
  \fill[outerc!22!white] (0,0) rectangle (1.33,5.6);
  \fill[innerc!22!white] (1.33,0) rectangle (2.42,5.6);
  \fill[outerc!22!white] (2.42,0) rectangle (3.90,5.6);

  \draw[surf] (0,0) rectangle (3.90,5.6);

  \draw[ileft, postaction={decorate, decoration={markings,
        mark=at position 0.42 with {\arrow{Stealth[length=4.6pt]}}}}]
    (1.33,0) -- (1.33,5.6);
  \draw[iright, postaction={decorate, decoration={markings,
        mark=at position 0.58 with {\arrow{Stealth[length=4.6pt]}}}}]
    (2.42,5.6) -- (2.42,0);

  \draw[wline, dblr] (0,3.53) -- (1.33,3.53);
  \draw[gline, sglr] (1.33,3.53) -- (2.42,3.53);
  \draw[wline, dblr] (2.42,3.53) -- (3.90,3.53);
  \node[gvtx] at (1.33,3.53) {};
  \node[gvtx] at (2.42,3.53) {};

  \node[lab, Wilsong, anchor=south] at (1.875,3.63) {$\Lambda_-$};
  \node[lab, Wilsonp, anchor=south west] at (2.55,3.63) {$\Lambda_+$};
  \node[lab, Wilsonp, anchor=south east] at (1.10,3.63) {$\Lambda_+$};

  \node[reglab, fill=innerc!22!white] at (1.30,6.58) {$SU(N)_{k-1}$};
  \draw[lead, ->] (1.72,6.33) .. controls (2.15,5.85) .. (1.87,5.10);
  \node[reglab, fill=outerc!22!white] at (4.10,6.58) {$SU(N)_{k}$};
  \draw[lead, ->] (4.00,6.33) .. controls (4.28,5.75) and (3.65,5.55) .. (3.28,5.10);

  \node[lab, chiralc, anchor=east] at (1.26,0.72) {$\mathcal{V}$};
  \node[lab, antic, anchor=east] at (2.35,0.72) {$\overline{\mathcal{V}}$};

  \draw[tk] (0.81,5.44) -- (1.09,5.76);
  \draw[tk] (0.81,-0.16) -- (1.09,0.16);
  \draw[tk] (-0.16,2.51) -- (0.16,2.79);
  \draw[tk] (-0.16,2.81) -- (0.16,3.09);
  \draw[tk] (3.74,2.51) -- (4.06,2.79);
  \draw[tk] (3.74,2.81) -- (4.06,3.09);

\end{tikzpicture}
\par\smallskip
{\footnotesize (b)}
\end{minipage}
\caption{The non-chiral interface $M^+_{N, k-1}$ regularized as two chiral halves $\mathcal{V}$ (red) and $\overline{\mathcal{V}}$ (blue) on the torus.}
\label{fig:interface-torus-domain}
\end{figure}
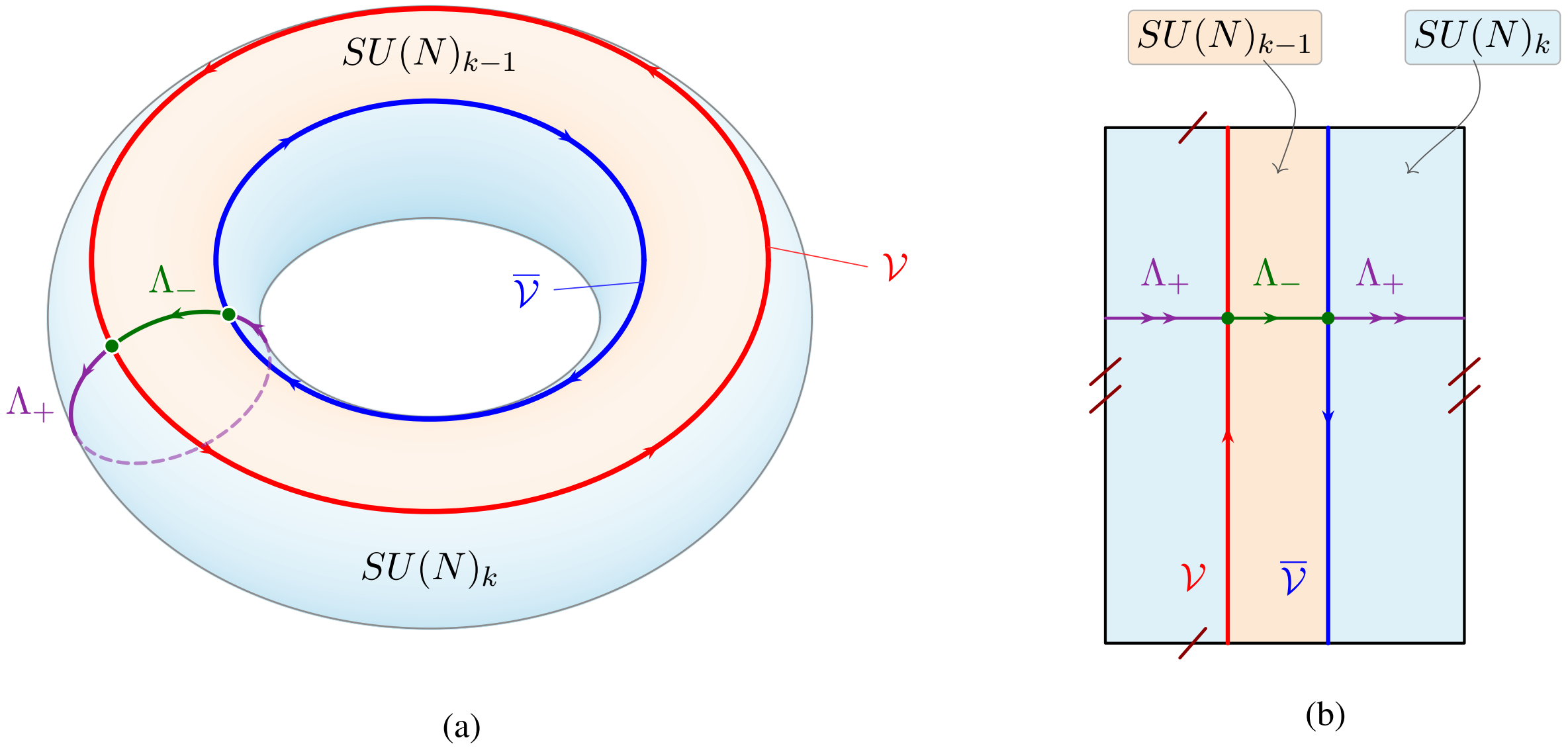

The sandwich construction on the torus, Figure~\ref{fig:interface-torus-domain}, pairs each chiral module with its dual on the anti-chiral interface. Summing over the bulk Wilson-line sectors, the Hilbert space is
\begin{equation}
\begin{aligned}
  \mathcal{H}\bigl(M^+_{N,k-1}\bigr)
  \cong{}&
  \bigoplus_{\Lambda_-\in P_{k-1}^+,\;\Lambda_+\in P_k^+}
  \mathcal{M}_{\Lambda_-,\Lambda_+}\otimes
  \overline{\mathcal{M}}_{\Lambda_-^\vee,\Lambda_+^\vee}
  \\
  ={}&
  \bigoplus_{r=0}^{N-1}
  \bigoplus_{\substack{
    \Lambda_-\in P_{k-1}^+,\;\Lambda_+\in P_k^+\\
    |\Lambda_+|-|\Lambda_-|\equiv r\; (\mathrm{mod}\;N)}}
  \Bigl(\mathcal{U}_r\otimes
  \mathcal{B}_{(\Lambda_-,\,\omega_r;\,\Lambda_+)}\Bigr)\otimes
  \Bigl(\overline{\mathcal{U}}_{N-r}\otimes
  \overline{\mathcal{B}}_{(\Lambda_-^\vee,\,\omega_{N-r};\,\Lambda_+^\vee)}\Bigr),
\end{aligned}
  \label{eq:interface-diagonal-invariant}
\end{equation}
where $a^\vee$ denotes the charge conjugate of $a$ and all charge labels are understood modulo $N$. An operator with $\Lambda_+\neq0$ ends on a Wilson line of the $SU(N)_k$ theory \textit{outside} the interface, so it is not a \textit{strictly local} operator of the non-chiral interface, that is, one attached to no Wilson line of the outer $SU(N)_k$ theory. Such an operator is nevertheless a state of $\mathcal{H}(M^+_{N,k-1})$ on the torus. The strictly local operators are therefore those with $\Lambda_+=0$, while the inner label $\Lambda_-$ remains unconstrained.\footnote{In fact, these are the only operators that can be generated under the RG flow from the lattice Hamiltonian~\eqref{eq:periodic-hamiltonian}, which is a sum of gauge-invariant strictly local terms.}

As $\mathcal{W}_{N,k-1}$-modules, the coset modules obey the field identification
\begin{equation}
  \mathcal{B}_{(\Lambda_-,\,\omega_r;\,\Lambda_+)}
  \cong
  \mathcal{B}_{(J_{k-1}\Lambda_-,\,\omega_{r+1};\,J_k\Lambda_+)},
  \label{eq:interface-field-identification}
\end{equation}
where $J_k$ is the simple current Wilson line of representation $k\omega_1$ in the UMTC $\mathcal{C}_{N,k}$. Therefore, the irreducible $\mathcal{W}_{N,k-1}$ modules are labeled by the orbits of this simultaneous action \cite{GannonWalton1995:DiagonalCosetModularInvariants}.

The field identification does not, however, identify the operators on the interface $\mathcal{M}_{\Lambda_-, \Lambda_+}$ (Figure~\ref{fig:cs-slab}). By Equation~\eqref{eq:interface-endpoint-module} the two sides of Equation~\eqref{eq:interface-field-identification} are glued to $\mathcal{U}_r$ and $\mathcal{U}_{r+1}$ respectively, and therefore correspond to different charge sectors. Therefore the interface operators are labeled by the pair $(\Lambda_-,\Lambda_+)$. If we forget the bulk Wilson lines, for instance by the folding trick \cite{FuchsRunkelSchweigert2002:TFTConstructionI,KapustinSaulina2011:SurfaceOperators3dTFT,Fuchs2013:BicategoriesBoundarySurfaceDefects}, each irreducible $\mathcal{W}_{N,k-1}$ module appears in Equation~\eqref{eq:interface-diagonal-invariant} once with each of the $N$ charge modules $\mathcal{U}_r$, and the spectrum reorganizes as
\begin{equation}
  \mathcal{H}\bigl(M^+_{N,k-1}\bigr)
  \cong
  \bigoplus_{\xi\in\text{Irr}(\mathcal{W}_{N,k-1})}
  \bigoplus_{r=0}^{N-1}
  \bigl(\mathcal{U}_r\otimes\mathcal{W}_\xi\bigr)
  \otimes
  \bigl(\overline{\mathcal{U}}_{N-r}\otimes\overline{\mathcal{W}}_{\xi^\vee}\bigr) .
  \label{eq:interface-regrouped-spectrum}
\end{equation}
Therefore there is a one-to-one map from $(\Lambda_-, \Lambda_+) \rightarrow (r, \xi)$. This is the sense in which the low-energy spectrum of the $q$-fermion chain is that of the diagonal rational conformal field theory of the chiral algebra $U(1)^{\mathrm{lat}}_N\times\mathcal{W}_{N,k-1}$.

We now turn to the symmetries of $M^+_{N,k-1}$. As in the free fermion case, the $U(1)$ factor of the chiral algebra gives a continuous family of topological lines, forming the group
\begin{equation}
  U(1)_{B} \subset G_{\text{charge}}=\frac{U(1)_L\times U(1)_R}{\bZ^{\text{diag}}_N},
  \label{eq:interface-charge-group}
\end{equation}
where $U(1)_L$ and $U(1)_R$ rotate the left and right charges. Since $M^+_{N,k-1}$ is the diagonal modular invariant of $U(1)^{\mathrm{lat}}_N\times\mathcal{W}_{N,k-1}$, every operator lives in the charge lattice,
\begin{equation}
  Q_L=r+Nn_L,\qquad Q_R=-r+Nn_R,\qquad n_L,n_R\in\bZ,
  \label{eq:interface-charge-lattice}
\end{equation}
which is generated by the vectors $(1,-1)$ and $(N,0)$. We see that $Q_L+Q_R\in N\bZ$, and therefore the elements $(\tfrac{2\pi s}{N},\tfrac{2\pi s}{N})$, $s\in\bZ_N$, act trivially on all of them, which is the quotient in Equation~\eqref{eq:interface-charge-group}. The diagonal subgroup $U(1)_{B}$ is the exact baryon-number symmetry inherited from the lattice, with charge $Q_B=(Q_L+Q_R)/N$, and the quotient is the continuum counterpart of the quotient from fermion number to baryon number in Section~\ref{sec:qchain-symmetries}. It should be contrasted with the free fermion, whose charge symmetry in Equation~\eqref{eq:ff-continuum-charge-symmetry} has no such quotient, since it contains freely moving fermions of vector charge $Q_V = 1$.

The theory $M^{+}_{N,k-1}$ also contains non-invertible symmetries which it inherits from the Wilson lines of both Chern--Simons theories: $a\in\mathcal{C}_{N,k-1}$ and $b\in\overline{\mathcal{C}_{N,k}}$. Using the folding trick to bring the external region inside while reversing its braiding and twist gives the topological lines $Y_{a,b}$ labeled by
\begin{equation}
  (a,b)\in\mathcal{C}_{N,k-1}\boxtimes\overline{\mathcal{C}_{N,k}},
  \label{eq:interface-bulk-lines}
\end{equation}
The line $Y_{a,b}$ acts on $\mathcal{M}_{\Lambda_-,\Lambda_+}$ by linking with the attached Wilson lines \cite{GannonRayhaun2026:HypergroupRelativeQFT},
\begin{equation}
  Y_{a,b}\big|_{\mathcal{M}_{\Lambda_-,\Lambda_+}}
  =
  \frac{S^{(k-1)}_{(a,\Lambda_-)}}{S^{(k-1)}_{(\mathbbm{1},\Lambda_-)}}\;
  \overline{\left(\frac{S^{(k)}_{(b,\Lambda_+)}}{S^{(k)}_{(\mathbbm{1},\Lambda_+)}}\right)}\,
  \mathrm{id},
  \label{eq:interface-line-action}
\end{equation}
The lines $Y_{0,\mu^\vee}$ are the continuum versions of the topological symmetries $Y_\mu$ of the lattice model of Section~\ref{sec:qchain-symmetries}, while the remaining symmetry lines are emergent. These symmetry lines can also be recast as Verlinde lines of the chiral algebra $U(1)^{\text{lat}}_N\times\mathcal{W}_{N,k-1}$. We define

\begin{equation}
  r_{a,b}\equiv|b|-|a|\pmod N,
  \qquad
  \xi_{a,b}=\bigl[(a,\,\omega_{r_{a,b}};\,b)\bigr]\in\text{Irr}(\mathcal{W}_{N,k-1}),
  \label{eq:interface-line-relabel}
\end{equation}

\begin{equation}
  Y_{a,b}
  =
  \mathcal{D}^{U(1)}_{r_{a,b}}\,
  \mathcal{D}^{\mathcal{W}}_{\xi_{a,b}} ,
  \label{eq:interface-line-factorization}
\end{equation}
which can be phrased as an equivalence of braided fusion categories,
\begin{equation}
  \mathcal{C}_{N,k-1}\boxtimes\overline{\mathcal{C}_{N,k}}
  \cong
  \text{Vec}^{\alpha}_{\bZ_N}\boxtimes\text{Ver}(\mathcal{W}_{N,k-1}),
  \label{eq:interface-line-equivalence}
\end{equation}
where $\text{Vec}^{\alpha}_{\bZ_N}$ is the pointed modular category that corresponds to an anomalous $\bZ_N$ symmetry, the discrete subgroup of $G_{\text{charge}}$ formed by the elements $[(\tfrac{2\pi s}{N},0)]=[(0,-\tfrac{2\pi s}{N})]$, $s\in\bZ_N$.

The interfaces $M^+_{N,k-1}$ and $M^-_{N,k}$ share the same exact symmetry, the lines $\overline{\mathcal{C}_{N,k}}$ of the outer $SU(N)_k$ Chern--Simons theory, but differ in their emergent lines, which come from the interior, $SU(N)_{k-1}$ for $M^+_{N,k-1}$ and $SU(N)_{k+1}$ for $M^-_{N,k}$,
\begin{equation}
  M^+_{N,k-1}:\;\underbrace{\mathcal{C}_{N,k-1}}_{\text{emergent}}\boxtimes\underbrace{\overline{\mathcal{C}_{N,k}}}_{\text{exact}},
  \qquad
  M^-_{N,k}:\;\underbrace{\mathcal{C}_{N,k+1}}_{\text{emergent}}\boxtimes\underbrace{\overline{\mathcal{C}_{N,k}}}_{\text{exact}}.
  \label{eq:interface-emergent-lines}
\end{equation}
Since $\mathcal{C}_{N,k+1}$ has more simple objects than $\mathcal{C}_{N,k-1}$, the interface $M^-_{N,k}$ has more emergent lines, and correspondingly $\text{Ver}(\mathcal{W}_{N,k})$ is larger than $\text{Ver}(\mathcal{W}_{N,k-1})$. This is another reason why $M^+_{N,k-1}$ is more stable than $M^-_{N,k}$, and we will see in Section~\ref{sec:continuum-symmetries} that $M^-_{N,k}$ admits an exact symmetry-allowed relevant perturbation that makes it flow to $M^+_{N,k-1}$.

\begin{figure}[!htbp]
  \centering
  \includegraphics{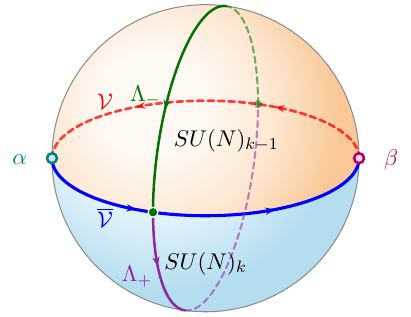}
\caption{The open case, the genus-zero analog of Figure~\ref{fig:interface-torus-domain}. The interface is the equator, cut at the two boundary-condition changing points $\alpha$ and $\beta$ into a chiral arc $\mathcal{V}$ (far half, dashed) and an anti-chiral arc $\overline{\mathcal{V}}$ (near half); the hemispheres carry $SU(N)_{k-1}$ and $SU(N)_{k}$. A non-chiral operator is one meridian loop crossing the interface twice, carrying $\Lambda_-$ above and $\Lambda_+$ below.}
  \label{fig:open-sphere-bcc}
\end{figure}

Having discussed the periodic chain on the torus, we now turn to the continuum limit of the open chain. Living on the sphere, this chain can be regularized by splitting the strip into chiral and anti-chiral pieces joined at their endpoints, with two different Chern--Simons theories on either side, as in Figure~\ref{fig:open-sphere-bcc}. As in the periodic case, there are two choices for the interior theory, $SU(N)_{k-1}$ and $SU(N)_{k+1}$, and we again focus on $SU(N)_{k-1}$ with the lower central charge. In the open case we must also specify how the two chiral halves are joined. This amounts to choosing conformal boundary conditions. We recall that, for an extended chiral algebra, the Cardy boundary conditions preserving the full chiral algebra are in one-to-one correspondence with its distinct primaries \cite{Cardy1989:BoundaryConditions}. In our case, since we can view the low-energy theory as a diagonal $U(1)_N \times \mathcal{W}_{N,k-1}$ rational conformal field theory, the boundary conditions are labeled by the pair $(r, \xi)$, which is equivalent to $(\Lambda_- , \omega_r ; \Lambda_+)$ with the gluing condition of Equations~\eqref{eq:interface-selection-rule} and \eqref{eq:interface-endpoint-module}.

On the lattice, we had the open Hilbert spaces $\mathcal{H}_{(\alpha,\beta)}$, with endpoint labels $\alpha$ and $\beta$ in $\text{Rep}(SU(N)_k)$. The natural question to ask is how these boundary conditions are related to the ones in the continuum. The open-chain calculations fix $\alpha=\mathbbm{1}$. Motivated by the interface construction and the boundary-entropy comparison below, we conjecture that the open hopping Hamiltonian~\eqref{eq:intro-hamiltonian} on $\mathcal{H}_{(\alpha,\beta)}$ flows to the boundary conditions $(\mathbbm{1},\omega_{|\alpha|}; \alpha)$ and $(\mathbbm{1},\omega_{|\beta|}; \beta)$. To reach boundary conditions outside this family one would require a more general choice of boundary interactions or additional boundary degrees of freedom \cite{BehrendPearce2001:BoundaryConditions}. This family can be motivated by comparing the $g$-functions of the boundary conditions $(\Lambda_-,\omega_r;\Lambda_+)$, since the $g$-function decreases monotonically along a boundary RG flow \cite{FriedanKonechny2004:BoundaryEntropy}. The $g$-function is computed from the modular $S$-matrix of the chiral algebra \cite{AffleckLudwig1991:GroundStateDegeneracy,Cardy1989:BoundaryConditions}. For a Cardy boundary condition labeled by a primary $a$, it is
\begin{equation}
  g_a
  = \frac{S_{(a,\mathbbm{1})}}
         {\sqrt{S_{(\mathbbm{1},\mathbbm{1})}}},
  \label{eq:boundary-cardy-entropy}
\end{equation}
where $\mathbbm{1}$ denotes the vacuum module. The modular $S$-matrix of the coset factorizes into those of its three affine factors \cite{GannonWalton1995:DiagonalCosetModularInvariants}, so for $a=(\Lambda_-,\omega_r;\Lambda_+)$ we obtain
\begin{equation}
  \frac{g_a}{g_{\mathrm{vac}}}
  = d_{\Lambda_-}^{(k-1)}d_{\Lambda_+}^{(k)},
  \qquad
  d_\lambda^{(k)}
  = \frac{S^{(k)}_{(\lambda,\mathbbm{1})}}{S^{(k)}_{(\mathbbm{1},\mathbbm{1})}}.
  \label{eq:boundary-relative-entropy}
\end{equation}
Here $d_\lambda^{(k)}$ is the quantum dimension of $\lambda\in\mathrm{Rep}(SU(N)_k)$. The $SU(N)_1$ and $U(1)^{\text{lat}}_N$ factors do not contribute, since all their primaries have quantum dimension one.

The boundary conditions $(\mathbbm{1},\omega_{|\Lambda_+|};\Lambda_+)$ proposed for the lattice chain carry the trivial label on the $SU(N)_{k-1}$ side. Comparing them with any other boundary condition that carries the same $\Lambda_+$ gives
\begin{equation}
  g_{(\mathbbm{1},\,\omega_{|\Lambda_+|};\,\Lambda_+)}
  = g_{\mathrm{vac}}\,d_{\Lambda_+}^{(k)},
  \qquad
  \frac{g_{(\Lambda_-,\,\omega_r;\,\Lambda_+)}}
       {g_{(\mathbbm{1},\,\omega_{|\Lambda_+|};\,\Lambda_+)}}
  = d_{\Lambda_-}^{(k-1)} \geq 1.
  \label{eq:boundary-endpoint-entropy-comparison}
\end{equation}
Thus, at fixed external $SU(N)_k$ label $\Lambda_+$, the vacuum on the $SU(N)_{k-1}$ side gives the smallest $g$-function. This comparison motivates the proposed boundary assignment but does not establish the boundary RG flow of the microscopic chain.

\subsection{Anomalies}
\label{sec:continuum-symmetries}

We now match the anomalies of the lattice model to those in the continuum, which the gapless theory $M^+_{N,k-1}$ saturates. As discussed in Section~\ref{sec:operator-content}, the lattice charge symmetry is the subgroup $U(1)_B=\{[(\alpha,\alpha)]\}\subset G_{\text{charge}}$, isomorphic to $U(1)_V/\bZ^{\text{diag}}_N$, and the lattice lines $Y_\mu$ become $Y_{0,\mu^\vee} \in \cC_{N, k-1} \otimes \overline{\cC_{N, k}}$. There is also lattice translation $\bZ^{\mathrm{trans}}$, which just as in the free fermion case, emanates to a discrete axial rotation, \cite{ChengSeiberg2023:LSMLuttingerTHooft}. In the continuum,
\begin{equation}
  T = e^{i a_0 P}\tau,
  \qquad
  \tau=e^{-ik_F(Q_L-Q_R)},
  \qquad
  k_F=\frac{\pi\nu}{N},
  \label{eq:emanant-translation}
\end{equation}
where $P$ generates continuum spatial translations and $k_F$ is the Fermi momentum, as in Section~\ref{sec:free-fermion-anomalies}. For an operator with chiral charges $(Q_L,Q_R)$, $\tau$ acts as
\begin{equation}
  \tau \mathcal{O} \tau^{-1}
  =
  e^{-ik_F(Q_L-Q_R)}
  \mathcal{O}
  \label{eq:continuum-translation-charge-action}
\end{equation}
It forms a discrete cyclic subgroup of $G_{\text{charge}}$,
\begin{equation}
  \tau^j=[(-j k_F,j k_F)],
  \qquad
  \langle\tau\rangle
  =
  \bigl\{[(-j k_F,j k_F)]\,:\,j\in\bZ\bigr\}
  \subset
  \frac{U(1)_L\times U(1)_R}{\bZ_N^{\mathrm{diag}}}.
  \label{eq:continuum-translation-subgroup}
\end{equation}
As discussed earlier in Equation~\eqref{eq:interface-charge-lattice}, the charge lattice of the operators is generated by $(N,0)$ and $(1,-1)$. Consequently, $\tau^j$ acts trivially on the spectrum of $M^+_{N,k-1}$ precisely when
\begin{equation}
  j\nu\in 2\bZ,
  \qquad \text{and} \qquad
  j\nu\in N\bZ.
  \label{eq:continuum-translation-order-conditions}
\end{equation}
For rational filling $\nu=a/b$,\footnote{For irrational filling in the continuum limit $\tau$ generates an infinite cyclic subgroup whose closure is $U(1)_A$} with $\gcd(a,b)=1$, we obtain
\begin{equation}
  \langle\tau\rangle \cong \bZ_{n_\tau},
  \qquad
  n_\tau=
  \begin{cases}
    \displaystyle\frac{Nb}{\gcd(N,a)},
      & N\ \text{even},\\[7pt]
    \displaystyle\frac{2Nb}{\gcd(2N,a)},
      & N\ \text{odd}.
  \end{cases}
  \label{eq:continuum-translation-order}
\end{equation}
The two cases differ in how $\langle\tau\rangle$ overlaps with $U(1)_B$. Since
\begin{equation}
  \tau^{Nb/\gcd(N,a)}=(-1)^{\mathrm{lcm}(N,a)\,Q_B},
  \label{eq:continuum-translation-fermion-parity-power}
\end{equation}
the intersection $\langle\tau\rangle\cap U(1)_B$ is $\{1,(-1)^{Q_B}\}$ when $N$ and $a$ are both odd and trivial otherwise. At unit filling, for instance, $\tau$ has order $N$ for even $N$ and $2N$ for odd $N$, and in the latter case
\begin{equation}
  \tau^N=e^{i\pi Q_B}=(-1)^F,
  \label{eq:continuum-unit-filling-translation-parity}
\end{equation}
which acts nontrivially on the fermionic baryon operators, see Appendix~\ref{app:baryon-algebra}.

As in the free fermion discussion, it is useful to combine $\tau$ with a baryon-number rotation for anomaly arguments. Since $Q_B=(Q_L+Q_R)/N$, we have
\begin{equation}
  \rho:=e^{-i\pi\nu Q_B}\tau
       =e^{-2ik_FQ_L},
  \qquad
  \rho=[(-2k_F,0)].
  \label{eq:continuum-chiral-translation-generator}
\end{equation}
At rational filling, this generator has order
\begin{equation}
  n_\rho=\frac{Nb}{\gcd(N,a)}.
  \label{eq:continuum-chiral-translation-order}
\end{equation}
\begin{equation}
\langle U(1)_B, \tau \rangle = \langle U(1)_B, \rho \rangle = U(1)_B\times\bZ_{n_\rho}
  \label{eq:continuum-charge-translation-group}
\end{equation}
Invariance under $\rho$ and $U(1)_B$ is equivalent to invariance under $\tau$ and $U(1)_B$. Then the mixed anomaly between $U(1)_B$ and $T$ on the lattice is matched by the mixed anomaly between $U(1)_B$ and $\bZ_{n_\rho}$ in the continuum.

As in the free fermion case, it is instructive to ask whether any relevant perturbation preserving the lattice symmetries could destabilize $M^+_{N,k-1}$, driving the chain to a different gapless theory or to a degenerate gapped phase. We check this case by case for the operators allowed by the exact and emanant symmetries. Neutrality under the exact lines $Y_{0,b}$ restricts the perturbations to operators with $\Lambda_+=0$, therefore the strictly local operators of Section~\ref{sec:operator-content}, and conservation of $U(1)_B$ requires opposite chiral charges $p$ and $-p$. The charge--coset gluing of Equation~\eqref{eq:interface-endpoint-module} then fixes the remaining labels,
\begin{equation}
p\in\bZ,
\qquad
r = p\!\!\mod N,
\qquad
\Lambda_-\in P^+_{k-1},
\qquad
|\Lambda_-|+r\equiv0\pmod N,
\label{eq:qf-neutral-gluing}
\end{equation}
so the candidate scalar primaries are
\begin{equation}
\mathcal{O}_{p,\Lambda_-}
:={:}\exp\!\left[
\frac{ip}{\sqrt N}(\phi_L-\phi_R)
\right]{:}
\Phi_{(\Lambda_-,\omega_r;0)}
\overline{\Phi}_{(\Lambda_-^\vee,\omega_{N-r};0)},
\label{eq:qf-neutral-scalar}
\end{equation}
together with their descendants and Hermitian conjugates. These operators transform under emanant lattice translation or $\rho \in \bZ_{n_\rho}$, Equation~\eqref{eq:emanant-translation} as
\begin{equation}
\rho\mathcal{O}_{p,\Lambda_-}(x)\rho^{-1}
=
e^{-2ipk_F}\mathcal{O}_{p,\Lambda_-}(x),
  \label{eq:continuum-neutral-operator-translation}
\end{equation}
which is invariant only if
\begin{equation}
\frac{p\nu}{N}\in\bZ.
\label{eq:qf-umklapp-selection}
\end{equation}
For every $0<\nu<N$ this excludes $p=\pm1$, so either $p=0$ or $|p|\ge2$ are the only allowed values. $\mathcal{O}_{p,\Lambda_-}$ has the scaling dimension,
\begin{equation}
\Delta_{p,\Lambda_-}(\tilde R_0)
=
\frac{p^2+r(N-r)}{N}
+\frac{2C_2(\Lambda_-)}{k+N-1}
+2n_{\Lambda_-,r},
\qquad n_{\Lambda_-,r}\in\bZ_{\ge0},
\label{eq:qf-reference-dimension}
\end{equation}
where $n_{\Lambda_-,r}$ is the grade at which the $SU(N)_k$ vacuum module first appears in the decomposition of $\mathcal{L}^{(k-1)}_{\Lambda_-}\otimes\mathcal{L}^{(1)}_{\omega_r}$ \cite{Gaberdiel2011:AdS3Dual}.

We first consider the case $p=r=0$. The allowed operators are the identity $\mathbbm{1}_{U(1)}\otimes\mathbbm{1}_{\mathcal{W}}$ with its descendants, and the primaries $\Phi_{(\Lambda_-,\omega_0;0)}\overline{\Phi}_{(\Lambda_-^\vee,\omega_0;0)}$ with $\Lambda_-\neq0$ and their descendants. The primaries are simple to rule out. Since the tensor product $\Lambda_-\otimes\mathbbm{1}$ contains no singlet, the vacuum of $SU(N)_k$ appears at grade $n_{\Lambda_-,0}\ge1$ and their scaling dimension is
\begin{equation}
\Delta_{0,\Lambda_-}
=
\frac{2C_2(\Lambda_-)}{k+N-1}
+2n_{\Lambda_-,0}>2
\qquad (\Lambda_-\ne0),
\label{eq:qf-zero-charge-bound}
\end{equation}
so they are \textit{irrelevant}.

As in the free fermion case, the descendants of the identity are more interesting. $J_L+J_R$ is \textit{relevant} and shifts the chemical potential, but since we tune $\mu$ externally in Equation~\eqref{eq:path-qfermion-hamiltonian}, we ignore it. The combinations $J_L-J_R$, ${:}J_L^2{:}-{:}J_R^2{:}$ and $T_{\mathcal W}-\bar T_{\mathcal W}$, with $T_{\mathcal W}$ the stress tensor of $\mathcal{W}_{N,k-1}$, are odd under parity or time reversal and are not generated. What remains at dimension two are ${:}J_L^2{:}+{:}J_R^2{:}$ and $J_LJ_R$, which form $H_c$ of Equation~\eqref{eq:ff-marginal-hamiltonian}, and
\begin{equation}
  H_{\mathcal W}=\frac{v_{\mathcal W}}{2\pi}\int dx\,\bigl(T_{\mathcal W}+\bar T_{\mathcal W}\bigr) .
  \label{eq:qf-coset-hamiltonian}
\end{equation}
Unlike the free fermion case, there is no $W_{1+\infty}$ symmetry that forbids $g_c J_L J_R$ from being generated, and indeed we observe numerically that the continuum limit of the $q$-fermion chain, Equation~\eqref{eq:periodic-hamiltonian}, flows to a CFT with the chiral algebra $U(1)_{\tilde{R}^2}\times\mathcal{W}_{N,k-1}$, with a radius $R(\nu)$ that is a function of the filling $\nu$. As in Section~\ref{sec:ff-mott}, the theory becomes a non-rational CFT with a $\bZ^2$ charge lattice. Writing the charges of Equation~\eqref{eq:interface-charge-lattice} as $(Q_L,Q_R)=(w+NQ_B,-w)$, with baryon number $Q_B\in\bZ$ and winding number $w\in\bZ$, the deformation keeps $Q_B$, $w$ and the coset labels fixed and changes only the vertex operator,
\begin{equation}
  {:}e^{\frac{i}{\sqrt N}(Q_L\phi_L+Q_R\phi_R)}{:}\,\Phi_{\xi}\bar{\Phi}_{\xi^\vee}
  \;\longrightarrow\;
  {:}\exp\Bigl[\tfrac{i}{R}\,Q_B\,\phi+\tfrac{iR}{4}\,(NQ_B+2w)\,\tilde\phi\Bigr]{:}\,\Phi_{\xi}\bar{\Phi}_{\xi^\vee},
  \label{eq:interface-deformed-vertex}
\end{equation}
whose scaling dimension is
\begin{equation}
  \Delta_{w,Q_B}(R)=\frac{Q_B^2}{R^2}+\frac{R^2(NQ_B+2w)^2}{16}+h_{\xi}+\bar h_{\xi^\vee},
  \label{eq:interface-deformed-dimension}
\end{equation}
while its spin $h-\bar h=Q_B(NQ_B+2w)/2+h_\xi-\bar h_{\xi^\vee}$ is independent of $R$. The local operators therefore organize into
\begin{equation}
\begin{gathered}
  \mathcal{H}\bigl(M^+_{N,k-1}\bigr)
  \cong
  \bigoplus_{\xi\in\text{Irr}(\mathcal{W}_{N,k-1})}
  \bigoplus_{w,Q_B\in\bZ}
  \bigl(\mathcal F_{\lambda_L}\otimes\mathcal{W}_\xi\bigr)
  \otimes
  \bigl(\overline{\mathcal F}_{\lambda_R}\otimes\overline{\mathcal{W}}_{\xi^\vee}\bigr),
  \\
  \lambda_{L,R}=\frac{2Q_B}{R^2}\pm\Bigl(\frac{NQ_B}{2}+w\Bigr),
\end{gathered}
  \label{eq:interface-deformed-spectrum}
\end{equation}
where $\mathcal F_\lambda$ is the $U(1)_{\tilde R^2}$ module of weight $\lambda^2/2\tilde R^2$, with $\tilde R=2/R$ as in Section~\ref{sec:free-fermion-dynamics}. At $\tilde R^2=N$ one has $\lambda_L=Q_L$ and $\lambda_R=Q_R$, and grouping the modules with $w\equiv r\bmod N$ into the $U(1)^{\text{lat}}_{N}$ modules $\mathcal{U}_r$ gives back Equation~\eqref{eq:interface-regrouped-spectrum}.

\begin{figure}[!htbp]
 \centering
 \includegraphics[width=0.72\textwidth]{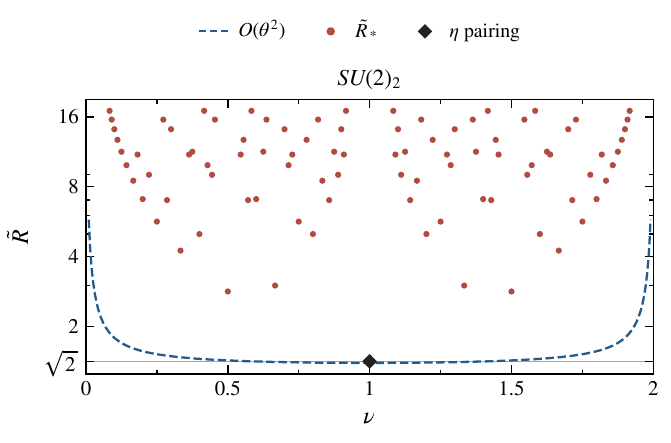}
 \caption{Radius and stability threshold for $SU(2)_2$. The dashed curve is the perturbative radius through order $\theta^2$ from Equation~\eqref{eq:numerical-radius-reference}; the expansion is nonuniform near $\nu=0,2$. Red circles are the thresholds $\tilde R_*$ of Equation~\eqref{eq:qf-critical-radius} at rational fillings. The black diamond is the exact radius $\sqrt2$ at $\nu=1$, fixed by $\eta$ pairing.}
 \label{fig:perturbative-radius}
\end{figure}

Since this is an interacting theory, the exact function $\tilde{R}(\nu)$ is hard to compute, but we can still make approximate statements using perturbation theory around the free fermion in the parameter $\theta = \pi/(k+N)$, where $q = e^{i\theta}$. In Appendix~\ref{app:charge-radius-perturbation} we perform the full derivation, which gives
\begin{equation}
\begin{aligned}
 \tilde R(\nu)&=\sqrt N\,\biggl[1+\frac{(N-1)\theta^2}{4\pi^2}\biggl(\frac{\pi(N-2\nu)}{N}\cot\frac{\pi\nu}{N}-1\biggr)\biggr]+O(\theta^4)\\
 &=\sqrt N\,\biggl[1-\frac{(N-1)\theta^2}{4\pi^2}+\frac{(N-1)\theta^2}{2N^2}\,\delta^2\biggr]+O(\theta^4,\theta^2\delta^4),
\end{aligned}
 \label{eq:numerical-radius-reference}
\end{equation}
where the second line is the expansion near half filling in $\delta=\nu-N/2$. Figure~\ref{fig:perturbative-radius} shows this curve for $SU(2)_2$.

The change of radius does not affect the conclusions above for the $p=0$ operators, since the radius enters the scaling dimensions only through the charged operators. We now redo the analysis of Equation~\eqref{eq:qf-reference-dimension}. In the deformed theory a local operator carries a winding $w$ and a baryon number $Q_B$, and $U(1)_B$ invariance forces $Q_B=0$. Its chiral charges are then $(w,-w)$, so $w$ is the charge $p$ of Equation~\eqref{eq:qf-neutral-scalar}, and the candidate scalars become
\begin{equation}
\mathcal{O}_{p,\Lambda_-}={:}e^{ip\tilde\phi/\tilde R}{:}\,\Phi_{(\Lambda_-,\omega_r;0)}\overline{\Phi}_{(\Lambda_-^\vee,\omega_{N-r};0)},
\label{eq:qf-charged-scalar}
\end{equation}
with scaling dimensions,
\begin{equation}
\Delta_{p,\Lambda_-}(\tilde R)=\frac{p^2}{\tilde R^2}+\frac{r(N-r)}{N}+\frac{2C_2(\Lambda_-)}{k+N-1}+2n_{\Lambda_-,r},
\label{eq:qf-charged-dimension}
\end{equation}
which reduces to Equation~\eqref{eq:qf-reference-dimension} at $\tilde R_0=\sqrt N$. It is enough to check the primaries $\mathcal{O}_{p,\Lambda_-}$ with $p\neq0$ as their descendants are heavier by positive integers. At the free-fermion radius $\tilde R_0=\sqrt N$,
\begin{equation}
\frac{p^2+r(N-r)}{N}\ge|p|,
  \label{eq:stability-reference-dimension-bound}
\end{equation}
so every allowed charged operator has $\Delta\ge|p|\ge2$. Equality requires $|p|\le N$, $\Lambda_-=0$ and $n_{\Lambda_-,r}=0$, which by Equation~\eqref{eq:qf-neutral-gluing} forces $r=0$ and hence $p=\pm N$. This is the \textit{umklapp} operator
\begin{equation}
\cos\!\left(\frac{N}{\tilde R}\tilde\phi\right)\otimes\mathbbm{1}_{\mathcal{W}_{N,k-1}},
\label{eq:qf-umklapp-operator}
\end{equation}
which from Equation~\eqref{eq:qf-umklapp-selection} is allowed only at integer filling, $\nu\in\bZ$. Its dimension is $\Delta=N$, so for $N>2$ every allowed charged operator is irrelevant at the free-fermion radius. For $N=2$ the umklapp is marginal, and we will treat this case separately.

The $q$-fermion chain, however, does not in general flow to the free-fermion radius. By Equation~\eqref{eq:numerical-radius-reference}, $\tilde R$ lies slightly below $\sqrt N$ at half filling and grows away from it, as confirmed numerically in Figure~\ref{fig:numerical-radius}. A larger radius lowers $p^2/\tilde R^2$ so it is possible that at some rational filling $\nu=a/b$,\footnote{At irrational filling, translation forbids every charged operator, and the $p=r=0$ analysis above suffices.} the lightest charged scalar operator allowed by the lattice symmetries, of the form
\begin{equation}
\cos\left(\frac{p\tilde\phi}{\tilde R(\nu)}\right)\,\Phi_{(\Lambda_-,\omega_r;0)}\overline{\Phi}_{(\Lambda_-^\vee,\omega_{N-r};0)},
  \label{eq:stability-generic-radius-perturbation}
\end{equation}
could then become relevant and destabilize the $U(1)_{\tilde R^2}\times\mathcal{W}_{N,k-1}$ theory. To check this, we need the smallest charge that translation allows. Equation~\eqref{eq:qf-umklapp-selection} requires
\begin{equation}
p\in p_{\min}\bZ \qquad
p_{\min}=\frac{Nb}{\gcd(N,a)}.
  \label{eq:stability-allowed-umklapp-charges}
\end{equation}
By Equation~\eqref{eq:qf-neutral-gluing}, such a charge must be glued to a coset operator, the lightest of which has scaling dimension
\begin{equation}
\Delta_r=\frac{r(N-r)}{N}+\min_{\substack{\Lambda_-\in P^+_{k-1}\\|\Lambda_-|+r\equiv0\ (\mathrm{mod}\ N)}}\frac{2C_2(\Lambda_-)}{k+N-1},
  \label{eq:stability-minimal-neutral-dressing}
\end{equation}
where $r=p\bmod N$. It can be checked that the coset operator with the lowest scaling dimension is attained at $\Lambda_-=\omega_{N-r}$, with
\begin{equation}
\Delta_r=\frac{r(N-r)(k+2N)}{N(k+N-1)}
  \label{eq:stability-neutral-dressing-dimension}
\end{equation}
for $k\ge2$, while for $k=1$ only $r=0$ is possible. Therefore, at each filling, $\tilde R(\nu)$ must stay below
\begin{equation}
\tilde R_*(\nu)=\min_{\substack{p\in p_{\min}\bZ_{>0}\\\Delta_r<2}}\frac{p}{\sqrt{2-\Delta_r}}
\label{eq:qf-critical-radius}
\end{equation}
for every allowed charged scalar to be irrelevant. Figure~\ref{fig:perturbative-radius} shows the critical value of the radius $\tilde R_*$ for some rational fillings of $SU(2)_2$.

Since $\Delta_r\ge0$, every term in Equation~\eqref{eq:qf-critical-radius} is at least $p/\sqrt2$, and $p_{\min}=Nb/\gcd(N,a)\ge Nb/a=N/\nu$, so
\begin{equation}
\tilde R_*(\nu)\ge\frac{N}{\sqrt2\,\nu}.
  \label{eq:stability-dilute-critical-radius-bound}
\end{equation}
For small $\nu$, the perturbative result of Equation~\eqref{eq:numerical-radius-reference} gives
\begin{equation}
\tilde R\simeq\sqrt N + \frac{N^{3/2}(N-1)\theta^2}{4\pi^2\nu},
  \label{eq:stability-dilute-radius-expansion}
\end{equation}
which supports stability in the dilute regime $\theta^2\ll\nu\ll1$, where the weak-deformation expansion is controlled. The numerical comparisons in Appendix~\ref{app:numerical-methods} test the stability bound beyond this perturbative regime.

\paragraph{The case of $SU(2)_k$.}

The $N=2$ $q$-fermion chain at half filling is special. It has an analog of the $SU(2)$ $\eta$-pairing symmetry of free fermions \cite{Yang1989:EtaPairing,Zhang1990:PseudospinSymmetry}, whose generators are built from the on-site baryon operators of Appendix~\ref{app:baryon-algebra},
\begin{equation}
\eta^+=\sum_j(-1)^j\,\mathsf B_j^\dagger,\qquad
\eta^-=(\eta^+)^\dagger,\qquad
\eta^z=\frac{F-m}{2},
\label{eq:qf-eta-generators}
\end{equation}
where $F$ is the fermion number and $\mathsf B_j^\dagger$ is the on-site baryon creation operator of Equation~\eqref{eq:local-determinant-junction}, which raises the occupation of site $j$ from zero to $N$. Since $\mathsf B_j^\dagger$ acts within the physical Hilbert space, its Hermitian conjugate, and hence $\eta^-$, is well defined. By the relations of Appendix~\ref{app:baryon-algebra}, each bond obeys $[E_i+F_i,\mathsf B_i^\dagger-\mathsf B_{i+1}^\dagger]=0$, so on a ring with an even number of sites,\footnote{The same holds on the open chain of any length.}
\begin{equation}
[H^q_{\mathrm{per}},\eta^\pm]=\mp2\mu\,\eta^\pm .
  \label{eq:stability-eta-commutator}
\end{equation}
The generators obey the $\mathfrak{su}(2)$ algebra $[\eta^+,\eta^-]=2\eta^z$ and $[\eta^z,\eta^\pm]=\pm\eta^\pm$, so at $\mu=0$, which is half filling, the lattice $U(1)_B$ symmetry is enlarged to an $SO(3)$. This is because, for an even number of sites, the center $e^{2\pi i\eta^z}$ of $SU(2)_\eta$ acts trivially on local operators, which carry fermion number $0$ (mesons) or in $2\bZ$ (baryons), so the group that acts faithfully is $SO(3)=SU(2)/\bZ_2$.

In the continuum, $\eta^\pm$ become the zero modes of the chiral baryon currents $J^\pm_{L,R}={:}e^{\pm i\sqrt2\phi_{L,R}}{:}$, which extend the charge currents $J^z_{L,R}=J_{L,R}/2$ to the currents $\boldsymbol J_{L,R}$ of $SU(2)_1$ WZW algebra. This is exactly the symmetry of the compact boson at the self-dual radius, where $U(1)^{\text{lat}}_2=SU(2)_1$ \cite{DiFrancescoMathieuSenechal1997:CFT}. The radius-changing deformation $J_LJ_R$ and the umklapp of Equation~\eqref{eq:qf-umklapp-operator} each individually break this lattice $SO(3)$ symmetry, but an $SO(3)$-invariant combination can be generated during the RG flow,
\begin{equation}
g_\eta\,\boldsymbol J_L\cdot\boldsymbol J_R
=g_\eta\Bigl[\tfrac14J_LJ_R+\cos(\sqrt2\,\tilde\phi)\Bigr].
  \label{eq:stability-eta-current-perturbation}
\end{equation}
It is marginally irrelevant for $g_\eta<0$, where by Equation~\eqref{eq:ff-charge-radius} its $J_LJ_R$ part lowers the radius, and marginally relevant for $g_\eta>0$. At $N=2$, $\nu=1$, perturbation theory in Equation~\eqref{eq:numerical-radius-reference} gives a radius below $\sqrt2$, so $g_\eta<0$ and the flow returns the radius to $\tilde{R}_0 = \sqrt2$ in the IR. The symmetry-fixed marginal point is shown in Figure~\ref{fig:numerical-radius-stability}.

For $N>2$, as for free fermions, no combination of the on-site baryons commutes with the Hamiltonian of Equation~\eqref{eq:periodic-hamiltonian}. So there is no symmetry that pins the radius at $\tilde R_0=\sqrt N$. For $N=3$, the numerics in Figure~\ref{fig:numerical-radius} show that near half filling the radius is slightly below $\sqrt3$, consistent with Equation~\eqref{eq:numerical-radius-reference}.

Another interesting conclusion one can make from the search for relevant operators is that the interface with higher central charge $M^{-}$ is unstable. It admits a relevant perturbation that respects all the exact and emanant symmetries and makes it flow to the stable interface with lower central charge $M^{+}$. To see this, consider the interface $M^-_{N,k-1}$, which has the chiral algebra $\mathcal{W}_{N,k-1}$ and is therefore the same theory as $M^+_{N,k-1}$, except that the roles of the exact and emergent bulks are exchanged. The $q$-fermion lattice now lives in $SU(N)_{k-1}$ Chern--Simons theory, while $SU(N)_k$ is the emergent internal bulk. The labels in Equation~\eqref{eq:qf-neutral-gluing} are unchanged, $\Lambda_-\in P^+_{k-1}$ and $\Lambda_+\in P^+_k$, but strict locality, that is, neutrality under the exact lines, now demands $\Lambda_-=0$ and leaves $\Lambda_+$ free. The neutral sector $p=0$, $r=0$ then contains the primary
\begin{equation}
\Phi_{(0,\omega_0;\text{Adj})}\overline{\Phi}_{(0,\omega_0;\text{Adj})},
  \label{eq:stability-adjoint-scalar}
\end{equation}
where $\text{Adj}=\omega_1+\omega_{N-1}$ is the adjoint representation of $SU(N)_k$. It appears at grade $n_{0,\Lambda_+}=1$ in the branching \cite{Gaberdiel2011:AdS3Dual} and has dimension
\begin{equation}
  \Delta_{0,\text{Adj}}
  =2\Bigl(1-\frac{C_2(\text{Adj})}{k+N}\Bigr)
  =2-\frac{2N}{k+N}<2 .
  \label{eq:stability-adjoint-scalar-dimension}
\end{equation}
This relevant operator therefore drives the theory away from $M^-_{N,k-1}$. In fact, for one sign of its coupling, it is exactly the operator that drives the RG flow between adjacent minimal models, from $\mathcal{W}_{N,k-1}$ to $\mathcal{W}_{N,k-2}$, which in our case is the flow from $M^-_{N,k-1}$ to $M^+_{N,k-2}$ \cite{Dunning2002:MasslessWFlows}. We stress again that this operator also has $\Delta<2$ in $M^+_{N,k-1}$, but there the exact symmetries forbid it.

\section{The Mott transition and the projector chain}
\label{sec:mott-projection}
We saw above that the continuum limit of our lattice model is a compact charge boson together with the $W_{N}$ minimal model. In this section, just as for the free fermion chain in Section~\ref{sec:free-fermion-anomalies}, we wish to gap out this compact boson mode while retaining the charge-neutral sector of the $W_N$ minimal model, by modifying the Hamiltonian of Equation~\eqref{eq:path-qfermion-hamiltonian}. One motivation is to make contact with the existing literature on the subject, which deals with anyon chains with a single anyon $a$, specifically the fundamental of $SU(N)_k$, instead of a direct sum of anyons as for the fermions. We add a Hubbard interaction or charging energy term of the form
\begin{equation}
H^q_{\text{int}} = U \sum_{j=1}^{m} (n_j - \nu)^2 ,
  \label{eq:stability-charging-perturbation}
\end{equation}
analogous to the interaction in Equation~\eqref{eq:ff-mott-hamiltonian}, which forces the local occupation number to be $\nu$ and is neutral under the $U(1)_B$, translation and the non-invertible symmetry $\text{Rep}(SU(N)_k)$. In the subsequent subsection we examine the effective Hamiltonian in the $U \rightarrow \infty$ limit.

\subsection{The projector Hamiltonian at $\nu=1$}
\label{sec:projector-chain}

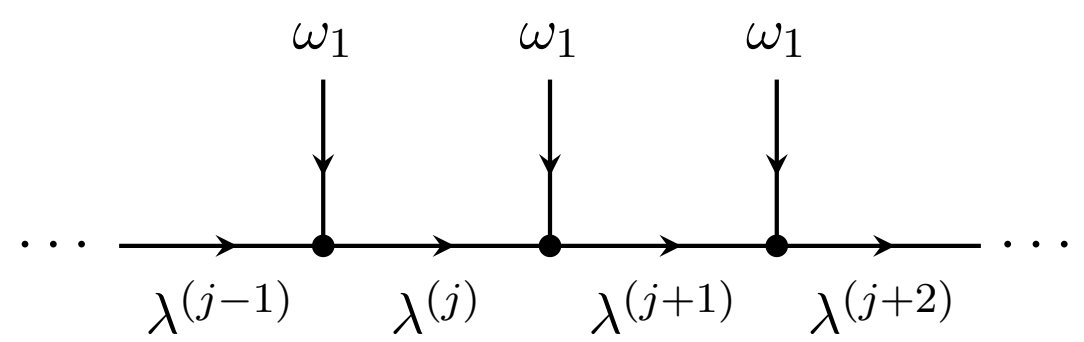
\begin{figure}[!htbp]
  \centering
  \begin{tikzpicture}[>=stealth,
    trunk/.style={thick},
    leg/.style={thick},
    mid/.style={postaction={decorate, decoration={markings,
      mark=at position 0.58 with {\arrow{>}}}}},
    vtx/.style={circle, fill, inner sep=1.5pt},
    lab/.style={font=\small}
  ]
    \node at (-0.55,0) {$\cdots$};
    \draw[trunk,mid] (-0.15,0) -- (1.20,0);
    \draw[trunk,mid] ( 1.20,0) -- (2.70,0);
    \draw[trunk,mid] ( 2.70,0) -- (4.20,0);
    \draw[trunk,mid] ( 4.20,0) -- (5.55,0);
    \node at (5.95,0) {$\cdots$};

    \foreach \x in {1.20,2.70,4.20}{
      \node[vtx] at (\x,0) {};
      \draw[leg,mid] (\x,1.10) -- (\x,0);
      \node[lab, above] at (\x,1.10) {$\omega_1$};
    }

    \node[lab, below] at (0.52,-0.08) {$\lambda^{(j-1)}$};
    \node[lab, below] at (1.95,-0.08) {$\lambda^{(j)}$};
    \node[lab, below] at (3.45,-0.08) {$\lambda^{(j+1)}$};
    \node[lab, below] at (4.90,-0.08) {$\lambda^{(j+2)}$};
  \end{tikzpicture}
\caption{The low-energy fusion tree at unit filling for $U\rightarrow\infty$. Every leg carries $\omega_1$, so only the running labels $\lambda^{(j)}\in P^+_k$ remain.}
  \label{fig:mott-fundamental-tree}
\end{figure}

At unit filling, the interaction is
\begin{equation}
H^q_{\text{int}} = U \sum_{j=1}^{m} (n_j - 1)^2 ,
\label{eq:qf-hubbard-unit-filling}
\end{equation}
exactly as in Equation~\eqref{eq:ff-mott-hamiltonian}. In the $U\rightarrow\infty$ limit the low-energy states are those in which every site carries the fundamental representation. These are the fusion trees of Figure~\ref{fig:mott-fundamental-tree}, where the only remaining degrees of freedom are the labels $\lambda^{(j)}\in P^+_k$. We call the span of these states the degenerate manifold $\mathcal{D}$. The low-energy Hamiltonian can be computed using degenerate perturbation theory,\footnote{We have dropped the chemical potential term, as it contributes only an overall constant.}
\begin{equation}
H=H^q_{\text{int}}+t V,
\qquad
V=-\sum_{i}(E_i+F_i).
\label{eq:mott-hamiltonian}
\end{equation}
In the fusion-tree basis $H^q_{\text{int}}$ is diagonal,
\begin{equation}
H^q_{\text{int}}|\boldsymbol \lambda;\boldsymbol r\rangle
=
U\left[\sum_{j}(r_j-1)^2\right]
|\boldsymbol \lambda;\boldsymbol r\rangle,
\label{eq:qf-hubbard-diagonal}
\end{equation}
so all states with $r_j=1$ are degenerate. Let $\bP^{(\omega_1)}$ be the projector onto $\mathcal{D}$ built from the site projectors,\footnote{$\bP_j^{(\omega_r)}=\prod_{s\neq r}(n_j-s)/(r-s)$, with $s=0,\ldots,N$, projects site $j$ onto $n_j=r$.}
\begin{equation}
\bP^{(\omega_1)}=\prod_{j}\bP_j^{(\omega_1)},
\label{eq:qf-hubbard-projector}
\end{equation}
and let $\overline{\bP}^{(\omega_1)}=\mathbbm{1}-\bP^{(\omega_1)}$. The quantum group generators act as
\begin{equation}
\begin{aligned}
E_i: (r_i,r_{i+1})&\longmapsto(r_i+1,r_{i+1}-1),
&
F_i: (r_i,r_{i+1})&\longmapsto(r_i-1,r_{i+1}+1),
\end{aligned}
\label{eq:mott-weight-shifts}
\end{equation}
so a single $E_i$ or $F_i$ leaves the degenerate manifold. Therefore the first-order term in perturbation theory vanishes,
\begin{equation}
\bP^{(\omega_1)} V \bP^{(\omega_1)} = 0 .
  \label{eq:mott-projected-hopping-vanishes}
\end{equation}
At second order, consider two states $|a\rangle$ and $|b\rangle$ in $\mathcal{D}$. Second-order perturbation theory gives
\begin{equation}
\left(H_{\mathrm{eff}}^{(2)}\right)_{ab}
=
t^2\sum_{n\notin\mathcal{D}}
\frac{\langle a|V|n\rangle\langle n|V|b\rangle}{E_0-E_n},
\label{eq:mott-schrieffer-wolff-components}
\end{equation}
or, in terms of the projectors of Equation~\eqref{eq:qf-hubbard-projector},
\begin{equation}
H_{\mathrm{eff}}^{(2)}
=
-t^2\,
\bP^{(\omega_1)}V\overline{\bP}^{(\omega_1)}
\frac{1}{\overline{\bP}^{(\omega_1)}H^q_{\text{int}}\overline{\bP}^{(\omega_1)}-E_0}
\overline{\bP}^{(\omega_1)}V\bP^{(\omega_1)}.
\label{eq:mott-schrieffer-wolff}
\end{equation}
which, when expanded, gives
\begin{equation}
H_{\mathrm{eff}}^{(2)}
=
-\frac{t^2}{2U}
\sum_{i}
\bP^{(\omega_1)}(E_iF_i+F_iE_i)\bP^{(\omega_1)} + \mathcal{O}(t^3/U^2) .
\label{eq:mott-second-order}
\end{equation}

\begin{figure}[!htbp]
\centering
\begin{tikzpicture}[>=stealth,
  line/.style={very thick},
  dn/.style={postaction={decorate, decoration={markings,
    mark=at position 0.58 with {\arrow{>}}}}},
  hop/.style={thick, black!70,
    postaction={decorate, decoration={markings,
      mark=at position 0.55 with {\arrow{>}}}}},
  vtx/.style={circle,fill,inner sep=1.35pt},
  lab/.style={font=\small}
]

  \begin{scope}
    \draw[line,dn] (0,2.5) -- (0,0);
    \draw[line,dn] (1.15,2.5) -- (1.15,0);
    \node[lab, left] at (-0.10,0.30) {$1$};
    \node[lab, right] at ( 1.25,0.30) {$1$};
    \node[lab, left] at (-0.10,1.25) {$2$};
    \node[lab, right] at ( 1.25,1.25) {$0$};
    \node[lab, left] at (-0.10,2.20) {$1$};
    \node[lab, right] at ( 1.25,2.20) {$1$};
    \draw[hop] (0,0.78) -- (1.15,0.78);
    \node[lab, above] at (0.575,0.82) {$E_i$};
    \draw[hop] (1.15,1.72) -- (0,1.72);
    \node[lab, above] at (0.575,1.76) {$F_i$};
    \node[lab, below] at (0,-0.06) {$i$};
    \node[lab, below] at (1.15,-0.06) {$i+1$};
    \node[anchor=north,font=\small] at (0.575,-0.60) {(a)\quad $F_iE_i$};
  \end{scope}

  \begin{scope}[xshift=3.3cm]
    \draw[line,dn] (0,2.5) -- (0,0);
    \draw[line,dn] (1.15,2.5) -- (1.15,0);
    \node[lab, left] at (-0.10,0.30) {$1$};
    \node[lab, right] at ( 1.25,0.30) {$1$};
    \node[lab, left] at (-0.10,1.25) {$0$};
    \node[lab, right] at ( 1.25,1.25) {$2$};
    \node[lab, left] at (-0.10,2.20) {$1$};
    \node[lab, right] at ( 1.25,2.20) {$1$};
    \draw[hop] (1.15,0.78) -- (0,0.78);
    \node[lab, above] at (0.575,0.82) {$F_i$};
    \draw[hop] (0,1.72) -- (1.15,1.72);
    \node[lab, above] at (0.575,1.76) {$E_i$};
    \node[lab, below] at (0,-0.06) {$i$};
    \node[lab, below] at (1.15,-0.06) {$i+1$};
    \node[anchor=north,font=\small] at (0.575,-0.60) {(b)\quad $E_iF_i$};
  \end{scope}

  \begin{scope}[xshift=7.1cm, line/.style={thick}]
    \draw[line,dn] (0,0.85) -- (-0.5,0);
    \draw[line,dn] (0,0.85) -- ( 0.5,0);
    \draw[line,dn] (0,1.65) -- (0,0.85);
    \draw[line,dn] (-0.5,2.5) -- (0,1.65);
    \draw[line,dn] ( 0.5,2.5) -- (0,1.65);
    \node[vtx] at (0,0.85) {};
    \node[vtx] at (0,1.65) {};
    \node[lab, left] at (-0.44,0.16) {$1$};
    \node[lab, right] at ( 0.44,0.16) {$1$};
    \node[lab, right] at ( 0.06,1.25) {$2$};
    \node[lab, left] at (-0.44,2.34) {$1$};
    \node[lab, right] at ( 0.44,2.34) {$1$};
    \node[anchor=north,font=\small] at (0,-0.60)
      {(c)\quad $[2]_q\,\Pi_i^{(\omega_2)}$};
  \end{scope}

\end{tikzpicture}
\caption{The operators $E_iF_i$ and $F_iE_i$ on the bond $(i,i+1)$ at unit filling, recast as the projector onto $\Lambda_q^{2}\bC_q^N$.}
\label{fig:mott-hecke-channel}
\end{figure}
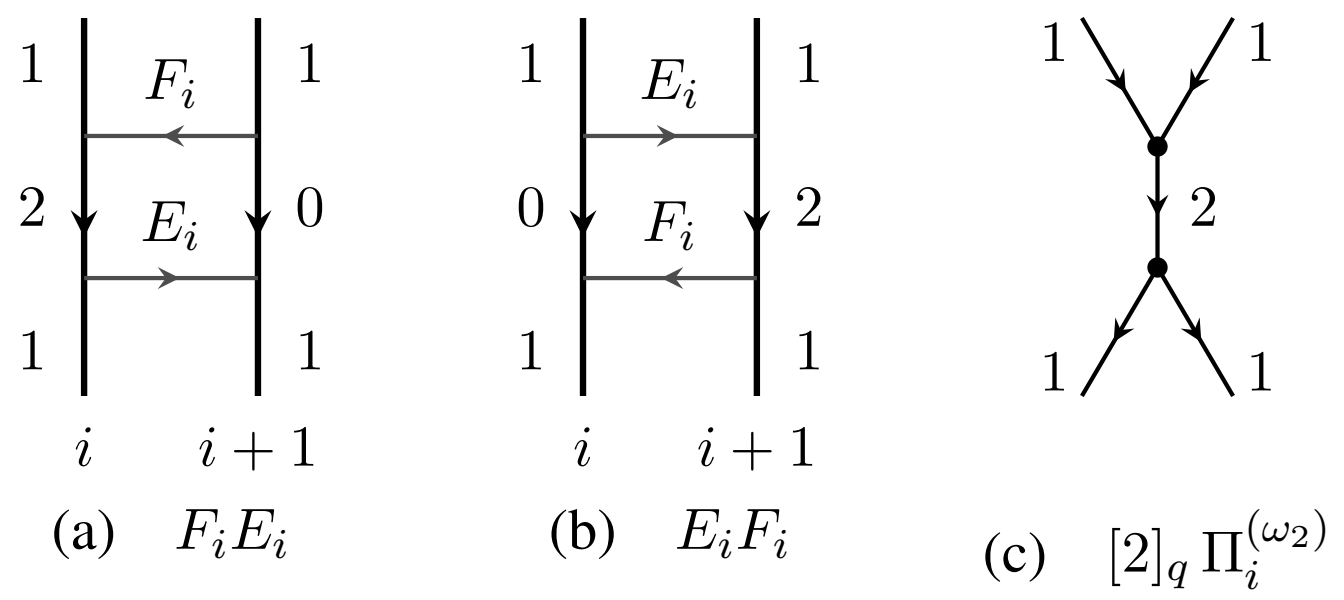

The products $E_iF_i$ and $F_iE_i$ can be understood as projectors that fuse the two adjacent fundamentals on the bond $(i,i+1)$ into the second antisymmetric power $\Lambda_q^{2}\bC_q^N$ and back, as shown in Figure~\ref{fig:mott-hecke-channel}.\footnote{On $\mathcal{D}$, $E_iF_i=F_iE_i$, since $[E_i,F_i]=[r_i-r_{i+1}]_q$ vanishes on the local occupation $(1,1)$.} We define the generator $e_i$ by
\begin{equation}
\bP^{(\omega_1)}E_iF_i\bP^{(\omega_1)}
=\bP^{(\omega_1)}F_iE_i\bP^{(\omega_1)}=e_i,
\qquad
e_i^2=[2]_q e_i,
\label{eq:mott-hecke-projector}
\end{equation}
and write $e_i=[2]_q\,\Pi_i^{(\omega_2)}$, where $\Pi_i^{(\omega_2)}$ is the orthogonal projector onto the $q$-antisymmetric channel.\footnote{Not to be confused with the site projectors $\bP_j^{(\lambda)}$ of Equation~\eqref{eq:qf-hubbard-projector}; $\Pi_i^{(\omega_2)}$ acts on the bond $(i,i+1)$.} Therefore
\begin{equation}
H_{\mathrm{eff}}
=
-\frac{t^2}{U}\sum_{i}e_i
=
-\frac{[2]_q t^2}{U}
\sum_{i}\Pi_i^{(\omega_2)} .
\label{eq:mott-effective-hamiltonian}
\end{equation}
For $N=2$ the open-chain generators $e_1,\ldots,e_{m-1}$ generate an image of the Temperley--Lieb algebra $\mathrm{TL}_m(d)$ with loop weight $d=[2]_q=2\cos\frac{\pi}{k+2}$ \cite{TemperleyLieb1971:PercolationColouring,Feiguin2007:InteractingAnyons}, whose relations are
\begin{equation}
e_i^2 = [2]_q e_i,
\qquad
e_ie_{i\pm1}e_i = e_i .
\label{eq:mott-TL-algebra}
\end{equation}
For $N>2$ the same generators generate an image of the Hecke algebra $\mathrm{H}_m(q)$ at $q=e^{i\pi/(k+N)}$, with Hecke generators $g_i=q-e_i$ \cite{Jimbo1986:qAnalogueHeckeYangBaxter}. For the periodic chain, the closing bond is obtained by translation. The case $N=2$ is worked out explicitly in Appendix~\ref{app:su2k-projected-chain}.

We point out a parallel with the free fermions. The projector Hamiltonian of Equation~\eqref{eq:mott-effective-hamiltonian} is integrable \cite{PasquierSaleur1990:CommonStructures,JimboKunibaMiwaOkado1988:AnFaceModels} and has an infinite set of conserved charges commuting with $H_{\text{eff}}$, while we believe that the $q$-fermion Hamiltonian of Equation~\eqref{eq:path-qfermion-hamiltonian} is not integrable. This is analogous to the free fermion case, where the $SU(N)$ Heisenberg chain is integrable for all $N$ \cite{Sutherland1975:ModelMulticomponentQuantum}, while the $SU(N)$ Hubbard chain is integrable only for $N=2$ \cite{LiebWu1968:Hubbard,Assaraf1999:MetalInsulatorSUN}. Coupling to Chern--Simons theory and taking the continuum limit also appear to commute exactly here, taking $SU(N)_1$ to $\mathcal{W}_{N,k-1}$ \cite{Huse1984:ExactExponents,Gils2009:CollectiveStatesInteracting}, since there is no exactly marginal coupling like the one that shifts the charge radius in Section~\ref{sec:continuum-symmetries}.

\subsection{Continuum limit, symmetries and anomalies at $\nu=1$}
\label{sec:projector-continuum}
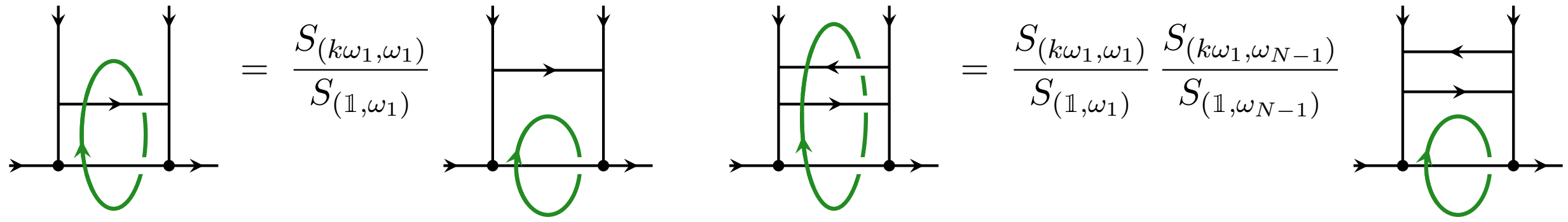
\begin{figure}[!htbp]
  \centering
  \tikzset{>=stealth,
    trunk/.style={thick, postaction={decorate, decoration={markings,
      mark=at position 0.07 with {\arrow{>}},
      mark=at position 0.93 with {\arrow{>}}}}},
    leg/.style={thick, postaction={decorate, decoration={markings,
      mark=at position 0.14 with {\arrow{>}}}}},
    rung/.style={thick, postaction={decorate, decoration={markings,
      mark=at position 0.58 with {\arrow{>}}}}},
    halo/.style={white, line width=5pt},
    jline/.style={very thick, ForestGreen},
    jarrow/.style={very thick, ForestGreen, postaction={decorate,
      decoration={markings, mark=at position #1 with {\arrow{>}}}}},
    vtx/.style={circle, fill, inner sep=1.2pt}}
  \newcommand{\jpanel}[4]{%
    \begin{tikzpicture}[scale=0.64, baseline={(0,0.90)}]
      \def\Lx{0.90}\def\Rx{0.52}
      \draw[jline] (0,{#1}) ellipse [x radius=\Rx, y radius=#2];
      \draw[halo] (-\Rx-0.12,0) -- (\Rx+0.12,0);
      \draw[trunk] (-1.70,0) -- (1.70,0);
      \foreach \x in {-\Lx,\Lx}{\draw[leg] (\x,2.60) -- (\x,0);}
      \foreach \x in {-\Lx,\Lx}{\node[vtx] at (\x,0) {};}
      #4
      \draw[jarrow=#3] ({\Rx*cos(230)},{#1+#2*sin(230)})
          arc[start angle=230, end angle=130, x radius=\Rx, y radius=#2];
    \end{tikzpicture}}
  \newcommand{\Ehop}[1]{%
    \draw[halo] (-\Rx-0.12,#1) -- (\Rx+0.12,#1);
    \draw[rung] (-\Lx,#1) -- (\Lx,#1);}
  \newcommand{\Fhop}[1]{%
    \draw[halo] (-\Rx-0.12,#1) -- (\Rx+0.12,#1);
    \draw[rung] (\Lx,#1) -- (-\Lx,#1);}
{\small
  $\jpanel{0.50}{1.20}{0.45}{\Ehop{1.00}}
   \;=\;\dfrac{S_{(k\omega_1,\omega_1)}}{S_{(\mathbbm{1},\omega_1)}}\;
   \jpanel{0}{0.80}{0.72}{\Ehop{1.55}}
   \qquad
   \jpanel{0.80}{1.50}{0.365}{\Ehop{1.00}\Fhop{1.60}}
   \;=\;\dfrac{S_{(k\omega_1,\omega_1)}}{S_{(\mathbbm{1},\omega_1)}}
        \,\dfrac{S_{(k\omega_1,\omega_{N-1})}}{S_{(\mathbbm{1},\omega_{N-1})}}\;
   \jpanel{0}{0.80}{0.72}{\Ehop{1.20}\Fhop{1.85}}$}
\caption{Moving the simple-current loop $\mathbf{J}_{j+\frac12}$ past a single hop $E_j$ costs the factor $S_{(k\omega_1,\omega_1)}/S_{(\mathbbm{1},\omega_1)}=\omega$ (left). For the product $E_jF_j$ and $F_jE_j$ of the projected Hamiltonian the factors of $\omega_1$ and $\omega_{N-1}$ cancel, so the loop passes freely (right).}
  \label{fig:mott-simple-current-symmetry}
\end{figure}

The projected chain inherits the translation $\bZ^{\text{Trans}}$ and the non-invertible lines of  $\mathrm{Rep}(SU(N)_k)$ from Section~\ref{sec:qchain-symmetries}. The $U(1)_B$ symmetry acts trivially on the operators $e_i$ and as an overall phase on every site, so the symmetry is no longer faithful. In its place there is a new symmetry. Consider the operator $\mathbf{J}_{j+\frac12}$ of Figure~\ref{fig:simple-current-operator}, the simple-current line wrapped around the bond $(j,j+1)$. It was not a symmetry of the Hamiltonian~\eqref{eq:periodic-hamiltonian}, as moving the loop past $E_j$ costs the factor $\omega = e^{2 \pi i / N}$ and conjugate factor $\omega^{-1}$, as shown in Figure~\ref{fig:mott-simple-current-symmetry}. Since the projected Hamiltonian of Equation~\eqref{eq:mott-effective-hamiltonian} is built from the products $E_jF_j$ and $F_jE_j$, the two factors cancel,
\begin{equation}
  [H_{\mathrm{eff}},\mathbf{J}_{j+\frac12}]=0 .
  \label{eq:mott-simple-current-symmetry}
\end{equation}

There are two ways to view this operator. Seen from the three-dimensional system, it is the simple current of $\mathrm{Rep}(SU(N)_k)$ linked around the chain. Seen from the two-dimensional chain, it is a \textit{topological point operator}. Like the simple current $J$ it generates a $\bZ_N$ one-form symmetry $\langle \mathbf{J} \rangle = \bZ_N$, whose elements are the lines $k\omega_r$ wrapped around the same bond. It is not entirely topological along the spatial direction, however, since it has a nontrivial commutation relation with translation,
\begin{equation}
  T\,\mathbf{J}_{j+\frac12}\,T^{-1}=\mathbf{J}_{j+\frac32}=\frac{S_{(k\omega_1,\omega_1)}}{S_{(\mathbbm{1},\omega_1)}}\,\mathbf{J}_{j+\frac12}=\omega\,\mathbf{J}_{j+\frac12}, \qquad \omega=e^{2\pi i/N} .
  \label{eq:mott-simple-current-translation}
\end{equation}
This is the descent of the relation $TU_\pm T^{-1}=e^{\mp2\pi i\nu/N}U_\pm$ at $\nu =1$. It also satisfies a projective commutation relation with the non-invertible lines $Y_\mu$,
\begin{equation}
  \mathbf{J}_{j+\frac12}\,Y_{\mu}
  =\frac{S_{(k\omega_1,\mu)}}{S_{(\mathbbm{1},\mu)}}\,Y_{\mu}\,\mathbf{J}_{j+\frac12}
  =\omega^{|\mu|}\,Y_{\mu}\,\mathbf{J}_{j+\frac12},
  \label{eq:mott-simple-current-braiding}
\end{equation}
which is the usual monodromy of the simple current with the line $\mu$.

We now follow the effects of the Mott term of Equation~\eqref{eq:qf-hubbard-unit-filling} in the continuum. It generates the symmetry-allowed interactions of Section~\ref{sec:continuum-symmetries}. For $N>2$ these include the coupling $g_cJ_LJ_R$ of Equation~\eqref{eq:ff-marginal-hamiltonian}, with $g_c\propto U$, so for $U>0$ the radius $\tilde R$ grows, as for free fermions in Section~\ref{sec:ff-mott}. At unit filling, where translation allows only $p\in N\bZ$, the first operator to become relevant is the umklapp of Equation~\eqref{eq:qf-umklapp-operator}, with dimension $N^2/\tilde R^2$. It becomes marginal at some $U_c>0$, where the radius reaches the threshold $\tilde R_*=N/\sqrt2$ of Equation~\eqref{eq:qf-critical-radius}. This is a standard BKT transition, described by the sine-Gordon action of the dual field,
\begin{equation}
  S_{\mathrm{SG}}
  =\int d^{2}x\left[
    \frac{1}{8\pi}\,\partial_\mu \tilde\phi\,\partial^{\mu}\tilde\phi
    +\mu\cos\!\left(\frac{N\,\tilde\phi}{\tilde R}\right)
  \right].
  \label{eq:sine-gordon-action}
\end{equation}
The cosine has $N$ minima, so the charge sector flows to the topological field theory
\begin{equation}
\mathcal{T}^{\text{SSB}}_{\bZ_N},
  \label{eq:mott-symmetry-breaking-topological-theory}
\end{equation}
where the emanant lattice translation $\bZ_N^\tau$ is spontaneously broken. At low energies we therefore find $N$ copies of the $\mathcal{W}_{N,k-1}$ minimal model exchanged by $\bZ_N^{\tau}$, a gapless-SSB. Each copy is protected by the non-invertible symmetry $\text{Rep}(SU(N)_k)$ \cite{Feiguin2007:InteractingAnyons,Pfeifer2012:TranslationTopologyCriticality,Gils2013:AnyonicQuantumSpin}, which is enlarged to $\text{Ver}(\mathcal{W}_{N,k-1})$.

For $N=2$, the $\eta$-pairing symmetry of Equation~\eqref{eq:qf-eta-generators} implies that the symmetry-allowed perturbation is the non-Abelian current bilinear $g_\eta\boldsymbol J_L\cdot\boldsymbol J_R$ of the $SU(2)_1$ WZW model. This operator is marginally relevant for $g_\eta>0$, which happens for $U>U_c$. Since the faithful symmetry is $SO(3)$ and not $SU(2)$, together with translation it forbids a unique gapped ground state, and the theory flows to a doubly degenerate one. As for $N>2$, the final gapped theory is therefore $\mathcal{T}^{\text{SSB}}_{\bZ_2}$.

The topological point operator $\mathbf{J}$ of Equation~\eqref{eq:mott-simple-current-symmetry} descends to a one-form symmetry $\bZ_N^{(1)}$ of the infrared theory. In the continuum $e^{ia_0P}$ acts trivially on a topological point operator, so Equation~\eqref{eq:mott-simple-current-translation} implies that $\mathbf{J}$ is charged under the emanant symmetry $\tau$ of Section~\ref{sec:continuum-symmetries},
\begin{equation}
\tau\,\mathbf{J}\,\tau^{-1}=\omega\,\mathbf{J} .
  \label{eq:mott-translation-wilson-algebra}
\end{equation}
The eigenvalue of $\mathbf{J}$ labels the $N$ vacua, and $\tau$ moves between them. The mixed anomaly of \eqref{eq:mott-simple-current-translation} is then realized in the infrared as a mixed anomaly between the emanant translation zero-form symmetry $\bZ_N^{(0)}$ and the one-form symmetry $\bZ_N^{(1)}$, and this clock and shift pair protects the SSB part of the phase.\footnote{The K\"unneth formula gives $H^3(B\bZ_N\times B^2\bZ_N;U(1))\cong H^3(B\bZ_N;U(1))\oplus H^1(B\bZ_N;U(1))\otimes H^2(B^2\bZ_N;U(1))\cong\bZ_N\oplus\bZ_N$, since $H^3(B^2\bZ_N;U(1))=0$. The second summand represents the mixed anomaly.} The lines $Y_\mu$ also shift the eigenvalue of $\mathbf{J}$, by Equation~\eqref{eq:mott-simple-current-braiding}, so the combination $\tau^{|\mu|}\,Y_\mu$ commutes with $\mathbf{J}$. It is the non-invertible symmetry operator that acts within a single copy of the $W_N$ minimal model.

On the lattice the operator acting on a single copy can then be written as $\mathsf{D}_\mu=Y_\mu T^{|\mu|}$. For $SU(2)_k$, since $Y_{\omega_1}$ commutes with $T$, by isotopy of the parallel closed Wilson line, and $Y_{\omega_1}^2=1+Y_{2\omega_1}$, we have the identity,
\begin{equation}
\mathsf{D}_{\omega_1}^2=(1+Y_{2\omega_1})\,T^2
  \label{eq:mott-fundamental-defect-fusion}
\end{equation}
For $SU(2)_2$ each copy is the Ising chain on a tensor-product Hilbert space, with $Y_{2\omega_1}$ its $\bZ_2$ spin flip and $T^2$ its translation (while $T$ exchanges the two copies) and $\mathsf{D}_{\omega_1}$ is the lattice non-invertible Kramers--Wannier symmetry \cite{SeibergShao2024:MajoranaChainIsing} acting on a single copy of the theory. The stability of the low-energy spectrum for $SU(2)_2$ is worked out in Appendix~\ref{app:su2-2-mott-example}. The case $\nu\neq1$ is discussed in Appendix~\ref{sec:other-fillings}. Figure~\ref{fig:qf-anomaly-rectangle} summarizes the anomaly matching discussed in this section.

\section*{Acknowledgments}
We are especially grateful to Davide Gaiotto for his dedication to our scientific training, generous guidance and support, and the freedom to develop this project. We thank Francisco Divi, Jaume Gomis, Tarun Advaith Kumar, Adri\'an L\'opez-Raven, Mykola Semenyakin, Valerie Wu, and Themistocles Zikopoulos for helpful discussions. We used ChatGPT and Codex (GPT-5.5 through GPT-6 Astra) and Claude Code (Opus 5 through Fable 5.1) extensively at every stage of the project. We take full responsibility for the accuracy and integrity of the manuscript. Research at Perimeter Institute is supported in part by the Government of Canada through the Department of Innovation, Science and Economic Development and by the Province of Ontario through the Ministry of Colleges and Universities.

\appendix
\addtocontents{toc}{\protect\addvspace{1em}}
\addtocontents{toc}{\protect\setcounter{tocdepth}{1}}

\newpage
\section[Quantum groups at roots of unity]{Quantum groups and their representations at roots of unity}
\label{app:uqglm-relations}

This appendix reviews the finite quantum groups, affine quantum groups and quantum loop algebras used in the paper, together with Lusztig's root-of-unity specialization, whose representation is the rung algebra that we considered in the main text. It also gives a brief overview of its semisimple representations.

\subsection{The finite quantum groups \texorpdfstring{$U_q(\mathfrak{g})$}{Uq(g)}}
With $q$ an indeterminate, $U_q(\mathfrak{g})$ is the Drinfeld--Jimbo Hopf algebra over the field of rational functions $\bC(q)$ deforming the universal enveloping algebra of semisimple Lie algebras $U(\mathfrak{g})$. For $m\geq2$, $U_q(\fgl_m)$ is generated by $E_i,F_i$, $i=1,\ldots,m-1$, and the invertible Cartan generators $L_j^{\pm1}$, $j=1,\ldots,m$, while $U_q(\fsl_m)$ is generated by $E_i,F_i,K_i^{\pm1}$, with $K_i:=L_iL_{i+1}^{-1}$. They satisfy the defining relations
\begin{equation}
\begin{aligned}
  &L_jL_\ell=L_\ell L_j,\\
  &L_jE_iL_j^{-1}=q^{\delta_{ji}-\delta_{j,i+1}}E_i,
  \qquad
  L_jF_iL_j^{-1}=q^{-\delta_{ji}+\delta_{j,i+1}}F_i,\\
  &K_iK_j=K_jK_i,\\
  &K_iE_jK_i^{-1}=q^{a_{ij}}E_j,
  \qquad
  K_iF_jK_i^{-1}=q^{-a_{ij}}F_j,\\
  &[E_i,F_j]=\delta_{ij}\frac{K_i-K_i^{-1}}{q-q^{-1}},
\end{aligned}
\label{eq:finite-defining-relations}
\end{equation}
with $a_{ij}=2\delta_{ij}-\delta_{j,i+1}-\delta_{j,i-1}$ the Cartan matrix of $\fsl_m$. They also satisfy the $q$-Serre relations
\begin{equation}
  \sum_{s=0}^{1-a_{ij}}
  (-1)^{s}
  \genfrac{[}{]}{0pt}{}{1-a_{ij}}{s}_{q}
  X_i^{\,1-a_{ij}-s}\,X_j\,X_i^{\,s}
  =0,
  \qquad
  i\neq j,
  \qquad
  X=E\ \text{or}\ F,
  \label{eq:finite-serre-relations}
\end{equation}
where
\begin{equation}
[n]_q=\frac{q^n-q^{-n}}{q-q^{-1}},
  \qquad
  [n]_q!=[1]_q\cdots[n]_q,
  \qquad
  \genfrac{[}{]}{0pt}{}{n}{s}_{q}=\frac{[n]_q!}{[s]_q!\,[n-s]_q!}.
  \label{eq:finite-quantum-number-definitions}
\end{equation}
Being a Hopf algebra, the quantum group also has the coproduct $\Delta$, counit $\varepsilon$ and antipode $S$ maps,
\begin{equation}
\begin{aligned}
  \Delta(L_j)&=L_j\otimes L_j,
  &\qquad S(L_j)&=L_j^{-1},
  &\qquad \varepsilon(L_j)&=1,\\[4pt]
  \Delta(E_i)&=E_i\otimes K_i+1\otimes E_i,
  &\qquad S(E_i)&=-E_iK_i^{-1},
  &\qquad \varepsilon(E_i)&=0,\\[4pt]
  \Delta(F_i)&=F_i\otimes1+K_i^{-1}\otimes F_i,
  &\qquad S(F_i)&=-K_iF_i,
  &\qquad \varepsilon(F_i)&=0.
\end{aligned}
  \label{eq:finite-hopf-coproduct}
\end{equation}

For the unitary Chern--Simons theory the quantum group of interest is Lusztig's divided-power form, specialized at $q$ equal to a root of unity. For the case of $\fgl_m$, it is defined by first considering the divided powers and the Cartan quantum binomial coefficients,
\begin{equation}
  E_i^{(n)}:=\frac{E_i^{\,n}}{[n]_q!},
  \qquad
  F_i^{(n)}:=\frac{F_i^{\,n}}{[n]_q!},
  \label{eq:divided-powers}
\end{equation}
\begin{equation}
  \binom{L_j;c}{t}_{q}
  :=\prod_{s=1}^{t}\frac{q^{c+1-s}L_j-1}{q^s-1},
  \qquad c\in\bZ,\quad t\in\bZ_{\geq0},
  \label{eq:cartan-quantum-binomial}
\end{equation}
with $[0]_q!=1$. Since $q$ is an indeterminate, $[n]_q!$ and $q^s-1$ are invertible in $\bC(q)$, so these are already elements of $U_q(\fgl_m)$. Set $\mathcal{A}:=\bC[q,q^{-1}]$. Lusztig's integral form of $U_q(\fgl_m)$ is the $\mathcal{A}$-subalgebra of finite $\mathcal{A}$-linear combinations of products \cite{Lusztig1993:IntroductionQuantumGroups,GavariniRakic2007:QuantizedHyperalgebras}
\begin{equation}
U^{\mathrm{L}}_{\mathcal{A}}(\fgl_m)
  :=\Bigl\langle L_j^{\pm1},\ \binom{L_j;c}{t}_{q},\ E_i^{(n)},\ F_i^{(n)}\Bigr\rangle_{\mathcal{A}}
  \subset U_q(\fgl_m).
  \label{eq:lusztig-integral-form}
\end{equation}
Extending this subalgebra with the field of rational functions gives back the original quantum group,
\begin{equation}
U^{\mathrm{L}}_{\mathcal{A}}(\fgl_m)
  \otimes_{\mathcal{A}}\bC(q)\cong U_q(\fgl_m).
  \label{eq:lusztig-generic-base-extension}
\end{equation}
Unlike $U_q(\fgl_m)$, the integral form can be specialized to a root of unity $q=\zeta$,
\begin{equation}
  U^{\mathrm{L}}_{\zeta}(\fgl_m)
  :=U^{\mathrm{L}}_{\mathcal{A}}(\fgl_m)
    \otimes_{\mathcal{A},\,q\mapsto\zeta}\bC.
  \label{eq:lusztig-specialization}
\end{equation}
The divided powers survive this specialization. At $\zeta=e^{\pi i/(k+N)}$, $[k+N]_\zeta=0$ gives $E_i^{\,k+N}=F_i^{\,k+N}=0$, while $E_i^{(k+N)}$ and $F_i^{(k+N)}$ stay nonzero \cite{Lusztig1990:QuantumGroupsRootsOf1}.

The root vectors of $U^{\mathrm{L}}_q(\fgl_m)$ for arbitrary roots are defined using the $q$-commutator
\begin{equation}
[X,Y]_q:=XY-qYX.
  \label{eq:affine-q-commutator-definition}
\end{equation}
With $E_{i,i+1}=E_i$ and $E_{i+1,i}=F_i$, the non-simple root vectors are defined recursively by
\begin{equation}
  E_{ij}:=[E_{i,j-1},E_{j-1,j}]_q,
  \qquad
  E_{ji}:=[E_{j,j-1},E_{j-1,i}]_{q^{-1}},
  \qquad
  j>i+1,
  \label{eq:root-vector-convention}
\end{equation}
and the general Cartan elements are $K_{ij}:=L_iL_j^{-1}$ for $i<j$. The coproducts following from Equation~\eqref{eq:q-fermion-coproduct} are
\begin{equation}
\begin{aligned}
  \Delta(E_{ij})
  &=
  E_{ij}\otimes K_{ij}
  +\mathbbm{1}\otimes E_{ij}
  -(q-q^{-1})
  \sum_{i<k<j}
  E_{ik}\otimes E_{kj}K_{ik},
  \\
  \Delta(E_{ji})
  &=
  E_{ji}\otimes\mathbbm{1}
  +K_{ij}^{-1}\otimes E_{ji}
  +(q-q^{-1})
  \sum_{i<k<j}
  K_{ik}^{-1}E_{jk}\otimes E_{ki},
  \qquad i<j .
\end{aligned}
\label{eq:general-root-coproduct}
\end{equation}
On single-particle states, $E_{rs}$ transfers a $q$-fermion from site $s$ to site $r$,
\begin{equation}
  E_{rs}(\bqpsi_\ell^a)
  =
  \delta_{s\ell}\,\bqpsi_r^a,
  \qquad
  L_s(\bqpsi_\ell^a)
  =
  q^{\delta_{s\ell}}\bqpsi_\ell^a ,
\label{eq:general-root-one-particle-action}
\end{equation}
The action on general $n$-particle states is given by the coproduct in Equation~\eqref{eq:general-root-coproduct}.

The $*$-structure  of Equation~\eqref{eq:path-qfermion-hermiticity-condition} is extended conjugate-linearly with $\bar q=q^{-1}$. It is compatible with the coproduct in the sense $\bigl((\cdot)^\star\otimes(\cdot)^\star\bigr)\circ\Delta^{\mathrm{op}}=\Delta\circ(\cdot)^\star$, as one checks on the generators.

\subsection{Affine and quantum loop algebras}
The affine algebra is presented in the same Chevalley form by adjoining one further pair $E_0$, $F_0$, with Cartan generator $K_0^{\pm1}$, on the same footing as the others, so that $i=0,1,\ldots,m-1$ runs over the nodes of the cyclic Dynkin diagram $A^{(1)}_{m-1}$.
With node indices understood modulo $m$, the generators satisfy the $K_i$, $E_i$, $F_i$ relations of Equation~\eqref{eq:finite-defining-relations} and the Serre relations~\eqref{eq:finite-serre-relations}, with the Cartan matrix $a_{ij}$ taken modulo $m$.\footnote{For $m\geq3$, deleting any one node of the cycle leaves the linear diagram $A_{m-1}$, giving the subalgebra $U_q(\fsl_m)\subset U_q(\widehat{\fsl}_m)$} One can check that the element
\begin{equation}
  C:=K_0K_1\cdots K_{m-1}
  \label{eq:affine-central-element}
\end{equation}
commutes with every generator. One can define $C = q^{c}$, with $c$ the canonical central element of $\widehat{\fsl}_m$. The full Kac--Moody quantum group $U_q(\widehat{\fsl}_m)$ also adjoins an invertible degree generator $D$, with
\begin{equation}
  DK_iD^{-1}=K_i,
  \qquad
  DE_iD^{-1}=q^{\delta_{i0}}E_i,
  \qquad
  DF_iD^{-1}=q^{-\delta_{i0}}F_i,
  \label{eq:affine-derivation}
\end{equation}
which keeps track of the grading of the generators analogous to $\widehat{\fsl}_m$. $U_q'(\widehat{\fsl}_m)$ denotes the algebra without the degree generator $D$. The restricted forms $U^{\mathrm{L}}_q(\widehat{\fsl}_m)$ and $U^{\mathrm{L}\prime}_q(\widehat{\fsl}_m)$ are defined exactly as in the finite case, from the divided powers of all $m$ Chevalley pairs \cite{Lusztig1993:IntroductionQuantumGroups}.\footnote{
The same algebra admits a second, \textit{current} presentation, Drinfeld's new realization: at generic $q$, $U_q'(\widehat{\fsl}_m)$ is generated by mode operators $x^{\pm}_{i,r}$ and $h_{i,s}$, with $i=1,\ldots,m-1$, $r\in\bZ$, $s\in\bZ\setminus\{0\}$, together with $C^{\pm1/2}$ and the $K_i^{\pm1}$ \cite{Drinfeld1988:NewRealization}}

On the periodic chain of Section~\ref{sec:periodic-chain}, the affine node is the closing bond,
\begin{equation}
  E_0=E_m,
  \qquad
  F_0=F_m,
  \qquad
  K_0=K_m=q^{n_m-n_1},
  \label{eq:loop-affine-node-realization}
\end{equation}
so the central element~\eqref{eq:affine-central-element} is
\begin{equation}
C=\prod_{i=1}^{m}K_i=q^{\sum_{i=1}^{m}(n_i-n_{i+1})}=1,
  \qquad n_{m+1}\equiv n_1,
  \label{eq:affine-trivial-central-element}
\end{equation}
as in Equation~\eqref{eq:periodic-level-zero}, and the chain realizes the quantum loop algebra,
 \begin{equation}
U^{\mathrm{L}}_q(L\fsl_m) = U^{\mathrm{L}\prime}_q(\widehat{\fsl}_m) / \langle C-1 \rangle
  \label{eq:affine-quantum-loop-quotient}
\end{equation}

\subsection{Representations of quantum groups at roots of unity}
\label{app:quantum-groups-at-root-of-unity}
In this subsection we give a brief overview of the representation theory of $U^{\mathrm{L}}_q(\fsl_N)$ at a root of unity, and refer the reader to \cite{Andersen1992:TensorProductsTilting,Sawin2006:QuantumGroupsRootsUnity,BakalovKirillov2001:LecturesTensorCategories} for more details. Chern--Simons theory at level $k$ replaces the representation theory of $SU(N)$ by that of Lusztig's quantum group $U^{\mathrm{L}}_q(\fsl_N)$ \cite{Witten1989:QuantumFieldTheory,Lusztig1990:QuantumGroupsRootsOf1} at the root of unity
\begin{equation}
  q=\exp\!\left(\frac{\pi i}{k+N}\right).
  \label{eq:tilting-root-of-unity}
\end{equation}
At this $q$ the finite-dimensional type-$1$ modules no longer form a semisimple category, and the category of Wilson lines is extracted from them in three steps, reviewed in \cite{BakalovKirillov2001:LecturesTensorCategories}.
\newline
\newline
\noindent
\textit{Weyl modules.} For each dominant weight $\lambda$, the irreducible module of highest weight $\lambda$ at generic $q$ contains the $\mathcal{A}$-lattice $U_{\mathcal{A}}^{-}v_\lambda$ generated from the highest-weight vector $v_\lambda$ by $U_{\mathcal{A}}^{-}$, the part of the integral form generated by the divided powers $F_i^{(n)}$ of Equation~\eqref{eq:divided-powers}. Specializing this lattice to the root of unity, $q\mapsto\zeta=\exp(\pi i/(k+N))$, as in Equation~\eqref{eq:lusztig-specialization}, gives the Weyl module $\Delta_q(\lambda)=U_{\mathcal{A}}^{-}v_\lambda\otimes_{\mathcal{A},\,q\mapsto\zeta}\bC$. It has the same character, and hence the same dimension, as the irreducible module at generic $q$, but at the root of unity it is in general reducible. It is indecomposable, with a unique simple quotient $L_q(\lambda)$, where a \textit{simple module} is one with no proper submodules. In the simplest reducible case it is the non-split extension
\begin{equation}
0\rightarrow L_q(\mu)\rightarrow\Delta_q(\lambda)\rightarrow L_q(\lambda)\rightarrow0,
  \label{eq:tilting-weyl-module-extension}
\end{equation}
where $\mu<\lambda$ lies in the orbit of $\lambda$ under the dot action $w\cdot\lambda=w(\lambda+\rho)-\rho$ of the affine Weyl group $\widehat W$. Dually, the dual Weyl module $\nabla_q(\lambda)=\Delta_q(-w_0\lambda)^{*}$, with $w_0$ the longest element of the finite Weyl group, has $L_q(\lambda)$ as its unique simple submodule. In general, a tensor product of Weyl modules is not a direct sum of Weyl modules.
\newline
\newline
\noindent
\textit{Tilting modules.} A module $T$ is tilting if it has both a filtration by Weyl modules and a filtration by dual Weyl modules, that is, a chain of submodules
\begin{equation}
0=T_0\subset T_1\subset\cdots\subset T_r=T,
\qquad
T_i/T_{i-1}\cong\Delta_q(\nu_i),
  \label{eq:tilting-weyl-filtration}
\end{equation}
and a second chain whose quotients are dual Weyl modules $\nabla_q(\nu'_j)$. The indecomposable tilting modules $T_q(\lambda)$ are labeled by dominant weights, and tilting modules have the important property of being closed under tensor products, duals and direct summands \cite{Andersen1992:TensorProductsTilting,Sawin2006:QuantumGroupsRootsUnity}. The fundamental modules $\Lambda^r_q\bC^N_q$ at each site are simple and tilting, so the site Hilbert spaces of Equation~\eqref{eq:local-qfock} and all their tensor products lie in the ribbon category $\operatorname{Tilt}(U^{\mathrm{L}}_q(\fsl_N))$.
\newline
\newline
\noindent
\textit{Semisimplification.} A morphism $f\colon X\to Y$ between tilting modules is negligible if $\operatorname{tr}_q(fg)=0$ for every $g\colon Y\to X$, with $\operatorname{tr}_q$ the quantum trace, and negligible morphisms form a tensor ideal $\operatorname{Negl}$. A module $X$ is negligible if its identity morphism $\mathrm{id}_X$ is negligible, and an indecomposable tilting module $X$ is negligible exactly when its quantum dimension $\dim_qX=\operatorname{tr}_q(\mathrm{id}_X)$ vanishes. The quantum dimension of $T_q(\lambda)$ is nonzero exactly when $\lambda$ lies in the fundamental alcove
\begin{equation}
  P^+_k=\{\lambda\ \text{dominant} : \langle\lambda+\rho,\theta^\vee\rangle<k+N\}=\{\lambda : \lambda_i\in\bZ_{\geq0},\ \lambda_1+\cdots+\lambda_{N-1}\leq k\},
  \label{eq:tilting-integrable-alcove}
\end{equation}
with $\theta$ the highest root, and there $T_q(\lambda)=\Delta_q(\lambda)=L_q(\lambda)$ \cite{Andersen1992:TensorProductsTilting,Sawin2006:QuantumGroupsRootsUnity}. The quotient $\mathcal{C}_{N,k}=\operatorname{Tilt}/\operatorname{Negl}$ of Section~\ref{sec:q-fermion-chain} is therefore semisimple, with simple objects $T_q(\lambda)=\mathcal{L}_\lambda$, $\lambda\in P^+_k$, and
\begin{equation}
  T_q(\lambda)\otimes T_q(\mu)\cong\bigoplus_{\nu\in P^+_k}\mathcal{N}_{\lambda\mu}^{\ \nu}\,T_q(\nu)\ \oplus\ (\text{negligible}),
  \label{eq:tilting-tensor-product-fusion}
\end{equation}
with $\mathcal{N}_{\lambda\mu}^{\ \nu}$ the level-$k$ fusion coefficients \cite{AndersenParadowski1995:FusionCategories,GoodmanWenzl1990:LRHeckeRootsUnity}. They are given by the Kac--Walton formula \cite{Walton1990:FusionRulesWZW},
\begin{equation}
  \mathcal{N}_{\lambda\mu}^{\ \nu}=\sum_{w\in\widehat W}\det(w)\,c_{\lambda\mu}^{\,w\cdot\nu},
  \label{eq:tilting-kac-walton-formula}
\end{equation}
for $\lambda,\mu,\nu\in P^+_k$, where $c_{\lambda\mu}^{\,\kappa}$ are the Littlewood--Richardson coefficients, that is, the tensor-product multiplicities of $\fsl_N$ representations. $\widehat W$ is the affine Weyl group and $c_{\lambda\mu}^{\,\kappa}$ is set to zero for non-dominant $\kappa$.

\paragraph{Skew Howe duality at a root of unity.}
The Fock space $\Lambda_q^\bullet(\bC_q^N\otimes\bC_q^m)$ is the tensor product over the sites of the local spaces $\Lambda_q^\bullet\bC_q^N$, each a direct sum of simple tilting modules $\mathcal{L}_{\omega_r}$ of the color quantum group $U^{\mathrm{L}}_q(\fsl_N)$. Since tilting modules are closed under tensor products~\cite{Andersen1992:TensorProductsTilting},
\begin{equation}
 \Lambda_q^\bullet(\bC_q^N\otimes\bC_q^m)
 \cong\bigoplus_{\gamma\in P_k^+}\mathcal{L}_\gamma\otimes M_\gamma
 \;\oplus\;T_{\mathrm{neg}},
 \label{eq:skew-howe-color-tilting-summands}
\end{equation}
where $T_{\mathrm{neg}}$ is a sum of indecomposable negligible tilting modules. At generic $q$ there is no $T_{\mathrm{neg}}$, and skew Howe duality, Equation~\eqref{eq:q-skew-howe-sl}, identifies the multiplicity spaces $M_\gamma$ with direct sums of simple $U^{\mathrm{L}}_q(\fgl_m)$-modules. At the root of unity the two actions still commute, but the Fock space is no longer a direct sum of simple modules, and some of the summands $\mathcal{L}_{\overline\lambda}\otimes\mathcal{L}_{\lambda^\top}$ no longer split off. This is because a simple color module $\mathcal{L}_\gamma$ with $\gamma\in P_k^+$ can also appear as a proper submodule of a negligible tilting module, $\mathcal{L}_\gamma\subset T_{\mathrm{neg}}$, from which it does not split off~\cite{Andersen1992:TensorProductsTilting,AndersenParadowski1995:FusionCategories}. The site action commutes with the color action, but it can still map a state in $\mathcal{L}_\gamma\otimes M_\gamma$ to one with a component in a copy of $\mathcal{L}_\gamma$ inside $T_{\mathrm{neg}}$. Passing to the semisimplified category $\mathcal{C}_{N,k}$ removes $T_{\mathrm{neg}}$, and with it those copies of $\mathcal{L}_\gamma$. This is the origin of the null states of Section~\ref{sec:fusion-tree-basis} whose final charge lies in the alcove. Seen from the site algebra, the full multiplicity space of $\mathcal{L}_\gamma$, counting also the copies inside $T_{\mathrm{neg}}$, is the direct sum of Weyl modules
\begin{equation}
 \widetilde M_\gamma=\operatorname{Hom}_{U^{\mathrm{L}}_q(\fsl_N)}\bigl(\mathcal{L}_\gamma,\Lambda_q^\bullet(\bC_q^N\otimes\bC_q^m)\bigr)
 \cong\bigoplus_{\overline\lambda=\gamma}\Delta_q(\lambda^\top),
 \label{eq:skew-howe-site-weyl-module}
\end{equation}
on which $U^{\mathrm{L}}_q(\fgl_m)$ acts by postcomposition with its Fock-space action in Equation~\eqref{eq:q-fermion-site-cartan-action}. Its null vectors are the maps that factor through $T_{\mathrm{neg}}$, that is, the negligible morphisms. The site generators preserve this null subspace of $\widetilde M_{\gamma}$. At each fixed fermion number $|\lambda|$, with $\overline\lambda=\gamma$, the null vectors are removed by the quotient
\begin{equation}
 0\longrightarrow\mathcal{R}_\lambda\longrightarrow\Delta_q(\lambda^\top)\longrightarrow L_q(\lambda^\top)\longrightarrow0,
 \qquad
 M_\gamma\cong\bigoplus_{\overline\lambda=\gamma}L_q(\lambda^\top),
 \label{eq:skew-howe-site-weyl-quotient}
\end{equation}
where $\mathcal{R}_\lambda$ is the unique maximal submodule of the Weyl module. In the fusion-tree basis, the simple quotient $L_q(\lambda^\top)$ has one state for each path from the vacuum to $\gamma$ with total fermion number $|\lambda|$ whose intermediate charges all lie in $P_k^+$. This is the statement of Equations~\eqref{eq:q-skew-howe-physical} and~\eqref{eq:q-fermion-vacuum-endpoint-skew-howe}, and the example below illustrates it.

For a nontrivial incoming charge $\alpha$, the multiplicity space of $\mathcal{L}_\beta$ is obtained by fusing with $\mathcal{L}_\alpha$ in $\mathcal{C}_{N,k}$,
\begin{equation}
 \mathcal{H}_{(\alpha,\beta)}\cong
 \bigoplus_{\substack{\lambda\subseteq(m^N)\\ \overline\lambda\in P_k^+}}
 \operatorname{Hom}_{\mathcal{C}_{N,k}}(\mathcal{L}_\alpha\otimes\mathcal{L}_{\overline\lambda},\mathcal{L}_\beta)
 \otimes L_q(\lambda^\top),
 \label{eq:skew-howe-general-fusion-multiplicity}
\end{equation}
a semisimple $U^{\mathrm{L}}_q(\fgl_m)$-module in which the first factor has dimension $\mathcal{N}_{\alpha\overline\lambda}^{\ \beta}$, as in Equation~\eqref{eq:q-fermion-general-endpoint-skew-howe}.

\paragraph{Example of quotient.} We illustrate how a null state arises, as a submodule of a Weyl module of the site quantum group and of a negligible tilting module of the color quantum group, in the adjoint representation of $U^{\mathrm{L}}_q(\fsl_3)$ for $SU(2)_1$.

\begin{figure}[!htbp]
  \centering
  \begin{tikzpicture}[
    x=1cm,y=1cm,font=\small,
    >={Stealth[length=1.5mm,width=1.2mm]},
    hop/.style={->,
      line width=0.65pt,shorten <=2.8pt,shorten >=2.8pt},
    pair/.style={hop,<->},
    state/.style={circle,draw,fill=white,line width=0.65pt,
      inner sep=0pt,minimum size=5.6pt},
    lab/.style={inner sep=1pt},
    gen/.style={font=\scriptsize,inner sep=1pt}
  ]
    \newcommand{\adjointpair}[2]{%
      \draw[pair] (#1) -- (#2);
    }
    \newcommand{\adjointpanel}[2]{%
      \begin{scope}[xshift=#1]
        \coordinate (w201) at ( 2.00, 0);
        \coordinate (w210) at ( 2.00, 1.60);
        \coordinate (w120) at ( 0, 1.60);
        \coordinate (w021) at (-2.00, 0);
        \coordinate (w012) at (-2.00,-1.60);
        \coordinate (w102) at ( 0,-1.60);
        \coordinate (h0) at (0,0);
        \coordinate (hf) at (1.00,0.60);

        \adjointpair{w120}{w210}
        \adjointpair{w021}{h0}
        \adjointpair{h0}{w201}
        \adjointpair{w012}{w102}
        \adjointpair{w201}{w210}
        \adjointpair{w012}{w021}
        \adjointpair{w102}{h0}
        \adjointpair{h0}{w120}

        \ifnum#2=0
          \draw[white,line width=3pt,out=35,in=-90,
            shorten <=3pt,shorten >=3pt] (w102) to (hf);
          \draw[pair] (w120) -- (hf);
          \draw[pair,out=35,in=-90] (w102) to (hf);
          \node[gen] at (1.00, 1.96) {$E_1$};
          \node[gen] at (-1.00,-1.96) {$F_1$};
          \node[gen] at (-2.34,-0.80) {$E_2$};
          \node[gen] at (-1.66,-0.80) {$F_2$};
        \else
          \draw[white,line width=3pt,out=45,in=-110,
            shorten <=3pt,shorten >=3pt] (w102) to (hf);
          \draw[white,line width=3pt,out=-70,in=25,
            shorten <=3pt,shorten >=3pt] (hf) to (w102);
          \draw[hop,out=45,in=-110] (w102) to (hf);
          \draw[hop,bend left=12] (w120) to (hf);
          \draw[hop,bend left=12,densely dashed] (hf) to (w120);
          \draw[hop,out=-70,in=25,densely dashed] (hf) to (w102);
        \fi
        \foreach \name in {w201,w210,w120,w021,w012,w102,h0}
          \node[state] at (\name) {};
        \ifnum#2=0
          \node[state] at (hf) {};
        \else
          \node[state,densely dashed] at (hf) {};
        \fi
        \node[lab,anchor=east] at (-0.17,0.40) {$|h_0\rangle$};
        \node[lab] at (1.16,0.98) {$|h_{\rm f}\rangle$};
        \node[lab,anchor=west] at ( 2.17,0) {$|201\rangle$};
        \node[lab,anchor=east] at (-2.17,0) {$|021\rangle$};
        \node[lab] at ( 2.00, 2.00) {$|210\rangle$};
        \node[lab] at ( 0, 2.00) {$|120\rangle$};
        \node[lab] at (-2.00,-2.00) {$|012\rangle$};
        \node[lab] at ( 0,-2.00) {$|102\rangle$};
      \end{scope}%
    }
    \adjointpanel{0cm}{0}
    \adjointpanel{6.40cm}{1}
    \node[lab] at (0,2.65) {Generic $q$};
    \node[lab] at (6.40,2.65) {$q=e^{i\pi/3}$, $[3]_q=0$};
  \end{tikzpicture}
  \caption{Double-headed links combine the two directions, with coefficients
  suppressed. Rightward and upward actions are $E_1,E_2$; the reverse
  actions are $F_1,F_2$. The null-state links at the root are split into
  surviving incoming arrows (solid) and vanishing outgoing arrows (dashed),
  before quotienting.}
  \label{fig:adjoint-root-of-unity-quotient}
\end{figure}
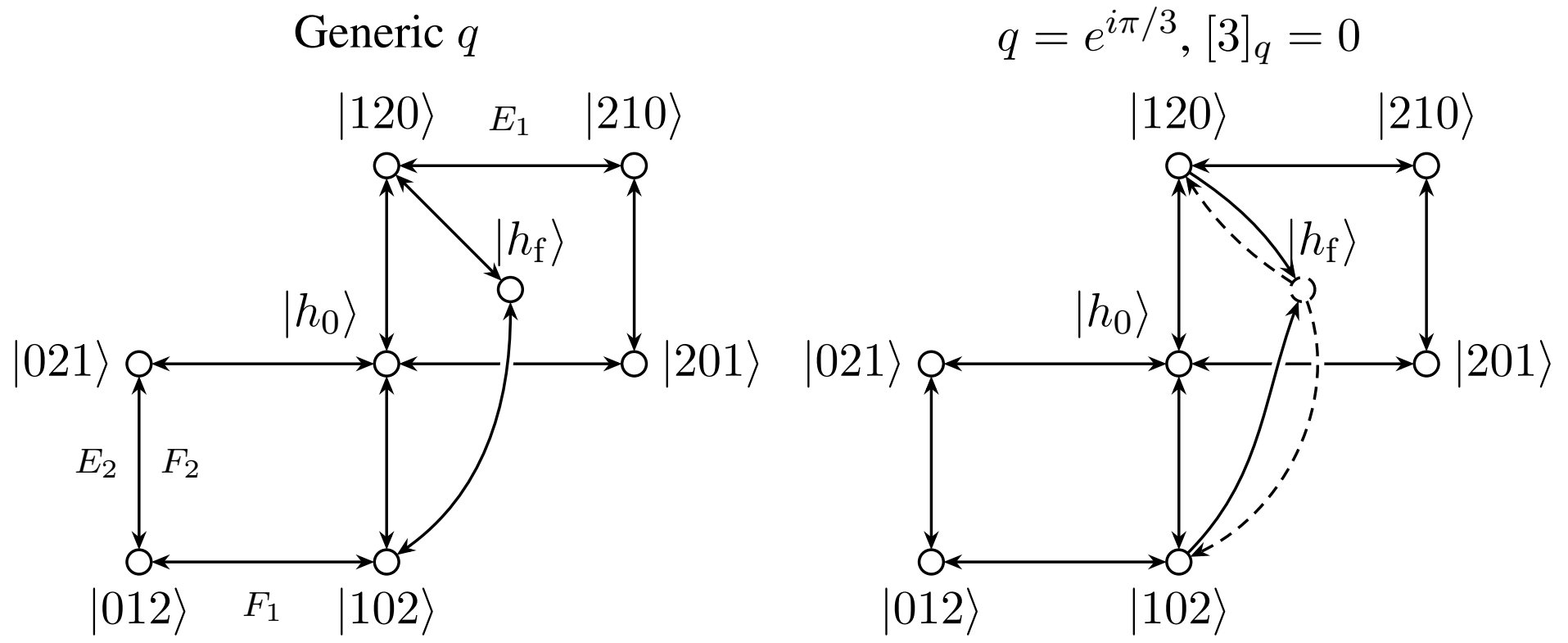

Take $N=2$ colors on $m=3$ sites. As a module of the color quantum group $U^{\mathrm{L}}_q(\fsl_2)$, the three-fermion sector decomposes as
\begin{equation}
  \Lambda_q^3(\bC_q^2\otimes\bC_q^3)\cong
  \begin{cases}
    8\,V_{\frac12}\oplus V_{\frac32}, & \text{generic } q,\\[2pt]
    7\,V_{\frac12}\oplus T_{\frac32}, & q=e^{i\pi/3},
  \end{cases}
  \label{eq:adjoint-example-color-decomposition}
\end{equation}
where $V_j=\mathcal{L}_{2j\omega_1}$ is the spin-$j$ module. By Equation~\eqref{eq:q-skew-howe-sl}, at generic $q$ the multiplicity space of $V_{\frac12}$ is the eight-dimensional adjoint of the site algebra $U^{\mathrm{L}}_q(\fsl_3)$. Six of its states have occupations that are permutations of $(2,1,0)$, and each has a single fusion path. The other two have occupation $(1,1,1)$ and differ in the intermediate color spin, $|h_0\rangle$ with path $0\to\tfrac12\to0\to\tfrac12$ and $|h_{\rm f}\rangle$ with path $0\to\tfrac12\to1\to\tfrac12$. Figure~\ref{fig:adjoint-root-of-unity-quotient} shows the simple-root transitions among the eight states.

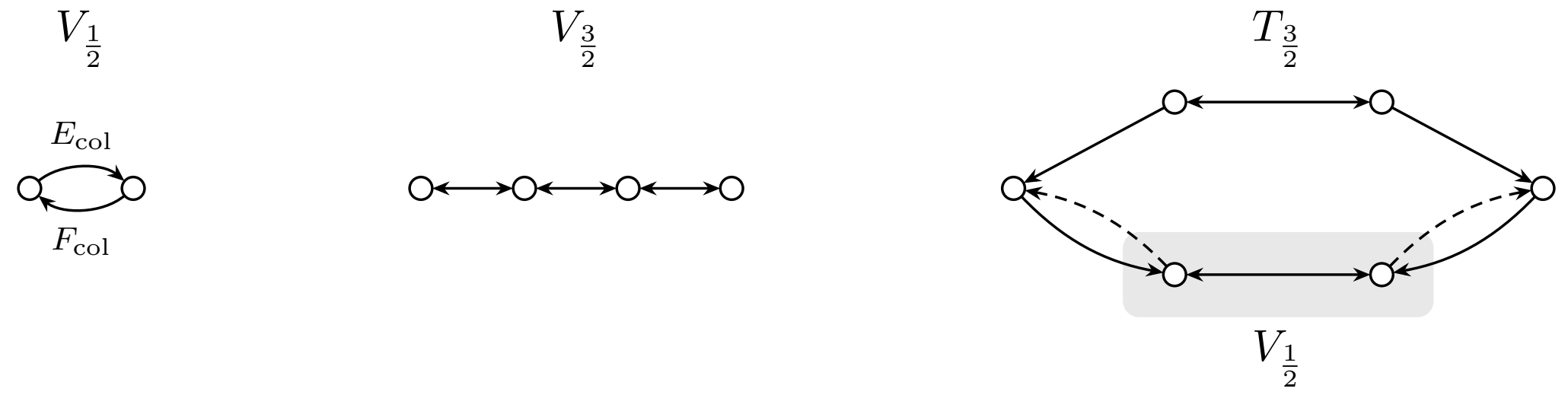
\begin{figure}[!htbp]
  \centering
  \begin{tikzpicture}[
    x=1cm,y=1cm,font=\small,
    >={Stealth[length=1.5mm,width=1.2mm]},
    hop/.style={->,
      line width=0.65pt,shorten <=2.8pt,shorten >=2.8pt},
    pair/.style={hop,<->},
    state/.style={circle,draw,fill=white,line width=0.65pt,
      inner sep=0pt,minimum size=5.6pt},
    lab/.style={inner sep=1pt},
    gen/.style={font=\scriptsize,inner sep=1pt}
  ]
    \begin{scope}[xshift=0cm]
      \coordinate (a1) at (-0.45,0);
      \coordinate (a2) at ( 0.45,0);
      \draw[hop,bend left=40] (a1) to (a2);
      \draw[hop,bend left=40] (a2) to (a1);
      \node[gen] at (0, 0.46) {$E_{\mathrm{col}}$};
      \node[gen] at (0,-0.46) {$F_{\mathrm{col}}$};
      \foreach \p in {a1,a2}
        \node[state] at (\p) {};
      \node[lab] at (0,1.30) {$V_{\frac12}$};
    \end{scope}
    \begin{scope}[xshift=4.3cm]
      \coordinate (b1) at (-1.35,0);
      \coordinate (b2) at (-0.45,0);
      \coordinate (b3) at ( 0.45,0);
      \coordinate (b4) at ( 1.35,0);
      \draw[pair] (b1) -- (b2);
      \draw[pair] (b2) -- (b3);
      \draw[pair] (b3) -- (b4);
      \foreach \p in {b1,b2,b3,b4}
        \node[state] at (\p) {};
      \node[lab] at (0,1.30) {$V_{\frac32}$};
    \end{scope}
    \begin{scope}[xshift=10.4cm]
      \coordinate (wm) at (-0.90, 0.75);
      \coordinate (wp) at ( 0.90, 0.75);
      \coordinate (um) at (-2.30, 0);
      \coordinate (up) at ( 2.30, 0);
      \coordinate (vm) at (-0.90,-0.75);
      \coordinate (vp) at ( 0.90,-0.75);
      \fill[gray!18,rounded corners=4pt] (-1.35,-1.12) rectangle (1.35,-0.38);
      \draw[pair] (wm) -- (wp);
      \draw[pair] (vm) -- (vp);
      \draw[hop] (wp) -- (up);
      \draw[hop] (wm) -- (um);
      \draw[hop,bend left=18] (up) to (vp);
      \draw[hop,bend right=18] (um) to (vm);
      \draw[hop,densely dashed,bend left=18] (vp) to (up);
      \draw[hop,densely dashed,bend right=18] (vm) to (um);
      \foreach \p in {wm,wp,um,up,vm,vp}
        \node[state] at (\p) {};
      \node[lab] at (0,1.30) {$T_{\frac32}$};
      \node[lab] at (0,-1.48) {$V_{\frac12}$};
    \end{scope}
  \end{tikzpicture}
  \caption{Color modules of the three-fermion sector for $N=2$ and $m=3$,
  Equation~\eqref{eq:adjoint-example-color-decomposition}. Arrows show selected
  color-generator transitions, $E_{\mathrm{col}}$ to the right and $F_{\mathrm{col}}$
  to the left; coefficients and additional components are omitted. The shaded
  $V_{\frac12}$ inside $T_{\frac32}$ is spanned by $|h_{\rm f}\rangle$ and
  $F_{\mathrm{col}}|h_{\rm f}\rangle$, and the dashed arrows out of it
  carry $[3]_q=0$.}
  \label{fig:tilting-diamond}
\end{figure}

At level $k=1$ we have $q=e^{i\pi/3}$ and $[3]_q=0$, so the path through color spin one leaves the alcove. The spin-$\tfrac32$ module becomes the Weyl module $\Delta_q(3\omega_1)$, which is no longer simple and sits inside the six-dimensional negligible tilting module $T_{\frac32}=V_1\otimes V_{\frac12}=T_q(3\omega_1)$ of Figure~\ref{fig:tilting-diamond}, with the character of $V_{\frac32}\oplus V_{\frac12}$, so only seven copies of $V_{\frac12}$ split off. The spin-$\tfrac12$ submodule $L_q(\omega_1)\subset\Delta_q(3\omega_1)$ of Equation~\eqref{eq:tilting-weyl-module-extension} is the multiplet of $|h_{\rm f}\rangle$, which has zero norm. In the normalization of Table~\ref{tab:su2-path-local-rules}, the transitions out of $|h_{\rm f}\rangle$ vanish, $E_2|h_{\rm f}\rangle=q^{-1}[3]_q|120\rangle=0$ and $F_2|h_{\rm f}\rangle=q^{-1}[3]_q|102\rangle=0$, and $E_1$ and $F_1$ annihilate it already. The transitions into it survive, $F_2|120\rangle=E_2|102\rangle=q(|h_{\rm f}\rangle-|h_0\rangle)/[2]_q$. Hence $|h_{\rm f}\rangle$ spans a null submodule of the adjoint Weyl module of $U^{\mathrm{L}}_q(\fsl_3)$. Passing to $\mathcal{C}_{2,1}$ removes $T_{\frac32}$, and with it $|h_{\rm f}\rangle$, so the physical adjoint sector has seven states, the six outer states and $|h_0\rangle$.

\section{The \texorpdfstring{$U(1)_k$}{U(1)} chain}
\label{app:u1-chain}
The simplest example of $q$-fermions is obtained by coupling a chain of single-flavor one-dimensional fermions to $U(1)_k$ Chern--Simons theory, as in Section~\ref{sec:q-fermion-chain}. Throughout this appendix,
\begin{equation}
 q=e^{i\pi/k} .
 \label{eq:u1-q-convention}
\end{equation}
Let $k$ be even, and $a$ denote the charge-one Wilson line of $U(1)_k$ Chern--Simons theory. The Wilson lines are $a^r$, with charge labels $r\in\bZ_k$,\footnote{For odd $k$, $U(1)_k$ is a spin theory, the labels lie in $\bZ_{2k}$, and the charge-$k$ line $a^k$ is a transparent fermion \cite{BelovMoore2005:AbelianSpinCS,SeibergWitten2016:GappedBoundaryPhases,Okuda2021:U1SpinChernSimons}.} and satisfy the abelian fusion rules
\begin{equation}
 a^r\otimes a^s\simeq a^{r+s},
 \qquad
 d_{a^r}=1 .
 \label{eq:u1-fusion-rules}
\end{equation}
The local object at each site is $\mathcal{H}_i=\mathcal{L}_{\omega_0}\oplus\mathcal{L}_{\omega_1}=\mathbbm{1}\oplus a$, as in Equation~\eqref{eq:local-hilbert-space} with $N=1$, where $a$ corresponds to a single $q$-fermion. The physical Hilbert space is
\begin{equation}
 \widehat{\mathcal{H}}^{\mathrm{phys}}
 \simeq\bigl(\Lambda_q^\bullet\bC_q\bigr)^{\otimes m}
 \simeq\Lambda_q^\bullet\bC_q^m,
 \qquad
 \dim\widehat{\mathcal{H}}^{\mathrm{phys}}=2^m .
 \label{eq:u1-open-hilbert-space}
\end{equation}
Unlike the $SU(N)_k$ case, there are no null states, $\mathcal{R}_k=\{0\}$, since the fusion rules close on $\bZ_k$ and there is no alcove boundary to truncate the fusion paths. The space is spanned by products of the $q$-fermion creation operators $\bqpsi_i$ acting on the vacuum $|0\rangle$, and for a single color the $q$-commutation relations of Equation~\eqref{eq:q-fermion-braided-antisymmetry} reduce to
\begin{equation}
 \bqpsi_i\otimes\bqpsi_j+q^{-s_{ij}}\,\bqpsi_j\otimes\bqpsi_i=0,
 \qquad
 s_{ij}=\text{sgn}(i-j),
 \label{eq:u1-exterior-relations}
\end{equation}
including $\bqpsi_i\otimes\bqpsi_i=0$. We choose the PBW basis,
\begin{equation}
 \bqpsi_{i_1}\otimes\cdots\otimes\bqpsi_{i_f}|0\rangle,
 \qquad
 1\leq i_1<\cdots<i_f\leq m,
 \qquad
 0\leq f\leq m .
 \label{eq:u1-pbw-basis}
\end{equation}
to label states of the Hilbert space. The states can be equivalently described via fusion trees as in Section~\ref{sec:fusion-tree-basis}. Since the fusion is abelian, the intermediate labels are completely fixed by the occupations $n_j\in\{0,1\}$. Writing $\lambda^{(j)}=a^{h_j}$, we have the constraint
\begin{equation}
 h_j=h_{j-1}+n_j \pmod k .
 \label{eq:u1-path-rule}
\end{equation}

\paragraph{The periodic chain.}
Closing the chain as in Section~\ref{sec:periodic-chain} joins the two ends of the fusion path across the bond $(m,1)$, so that they share a common basepoint label $\lambda^{(0)}=\lambda^{(m)}=a^p$. The periodic Hilbert space is then
\begin{equation}
 \mathcal{H}_{\mathrm{per}}=\bigoplus_{p\in\bZ_k}\mathcal{H}_{(a^p,a^p)} .
 \label{eq:u1-periodic-space}
\end{equation}
We can represent its states by the occupations and the basepoint, $|\boldsymbol{n};p\rangle=|n_1,\ldots,n_m;p\rangle$. Since
\begin{equation}
h_m-h_0\equiv F=\sum_jn_j\pmod k
  \label{eq:u1-height-charge-selection-rule}
\end{equation}
and periodic boundary conditions require $h_m=h_0$, the total fermion number for every state in $\mathcal{H}_{\mathrm{per}}$ is constrained to $F\in k\bZ$.

\paragraph{The action of \texorpdfstring{$U^{\mathrm{L}}_q(\fgl_m)$}{Uq(glm)}.}
From the single-color case of quantum skew Howe duality \cite{Cautis2014:WebsQuantumSkew}, the action of $U_q^{\mathrm{L}}(\fgl_m)$ on the open Hilbert spaces $\mathcal{H}_{(a^p, a^p)}$ is given by Equation~\eqref{eq:q-fermion-site-cartan-action} with the coproduct of Equation~\eqref{eq:q-fermion-coproduct},
\begin{equation}
\begin{aligned}
 E_i|\ldots,0,1,\ldots\rangle&=|\ldots,1,0,\ldots\rangle,
 &\qquad
 F_i|\ldots,1,0,\ldots\rangle&=|\ldots,0,1,\ldots\rangle,
 \\
 L_j|\boldsymbol{n}\rangle&=q^{n_j}|\boldsymbol{n}\rangle,
 &
 K_i|\boldsymbol{n}\rangle&=q^{n_i-n_{i+1}}|\boldsymbol{n}\rangle,
\end{aligned}
\label{eq:u1-site-action}
\end{equation}
where $E_i$ and $F_i$ act on the sites $i$ and $i+1$ and annihilate all other configurations of these two sites. This action can be realized by the ordinary fermions $c_i$ of Section~\ref{sec:free-fermion-review}. With $F_{<i}=\sum_{j<i}n_j$, the $q$-fermion can be expressed in terms of the ordinary fermions as a Jordan--Wigner string,
\begin{equation}
 \bqpsi_i=q^{-F_{<i}}\,c_i^\dagger ,
 \label{eq:u1-free-fermion-string}
\end{equation}
and because of the left ordering, the PBW state of Equation~\eqref{eq:u1-pbw-basis} is just $c_{i_1}^\dagger\cdots c_{i_f}^\dagger|0\rangle$. Since there are no null states, $\bqpsi_i$ has the adjoint $\bqpsi_i^\dagger=c_i\,q^{F_{<i}}$, as noted in Appendix~\ref{sec:non-existence-of-annihilation-operator}, and the generators become the free-fermion bilinears of Equation~\eqref{eq:ff-EF},
\begin{equation}
\begin{aligned}
 E_i&=\bqpsi_i\bqpsi_{i+1}^\dagger=c_i^\dagger c_{i+1},
 &\qquad
 F_i&=\bqpsi_{i+1}\bqpsi_i^\dagger=c_{i+1}^\dagger c_i,
 \\
 L_j&=q^{n_j},
 &
 K_i&=q^{n_i-n_{i+1}},
\end{aligned}
 \qquad
 n_j=c_j^\dagger c_j .
 \label{eq:u1-free-fermion-generators}
\end{equation}
One can check that they satisfy the relations of the quantum group given in Equations~\eqref{eq:finite-defining-relations} and~\eqref{eq:finite-serre-relations} \cite{Hayashi1990:qAnaloguesCliffordWeyl}.

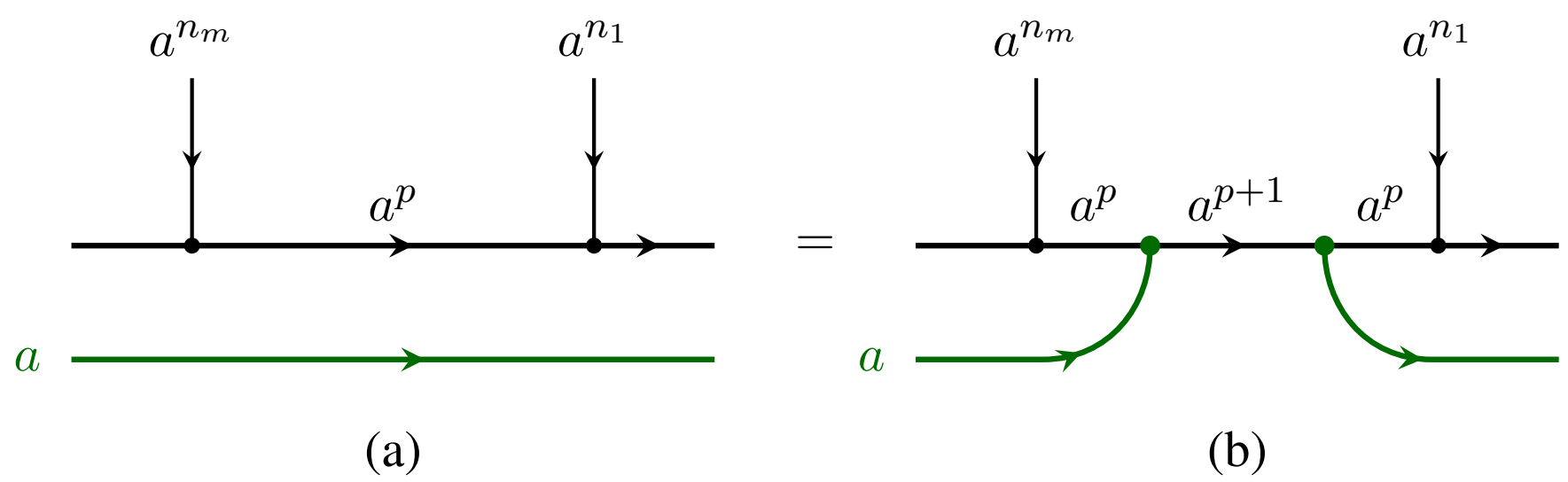
\begin{figure}[!htbp]
  \centering
  \begin{tikzpicture}[>=stealth,
    spine/.style={very thick},
    leg/.style={thick},
    mid/.style={postaction={decorate, decoration={markings,
      mark=at position 0.55 with {\arrow{>}}}}},
    sym/.style={very thick, green!42!black},
    vtx/.style={circle, fill, inner sep=1.2pt},
    junc/.style={circle, fill=green!42!black, inner sep=1.5pt},
    lab/.style={font=\small}
  ]
    \newcommand{\uoneseampanel}[1]{%
      \begin{scope}[xshift=#1]
        \draw[spine] (0,0) -- (0.9,0);
        \draw[spine,mid] (3.9,0) -- (4.8,0);
        \foreach \x/\n in {0.9/m, 3.9/1}{
          \draw[leg,mid] (\x,1.25) -- (\x,0);
          \node[vtx] at (\x,0) {};
          \node[lab, anchor=south] at (\x,1.27) {$a^{n_{\n}}$};}
        \node[lab, anchor=east, green!42!black] at (-0.08,-0.85) {$a$};
      \end{scope}}

    \uoneseampanel{0cm}
    \draw[spine,mid] (0.9,0) -- (3.9,0);
    \node[lab, anchor=south] at (2.4,0.04) {$a^{p}$};
    \draw[sym,mid] (0,-0.85) -- (4.8,-0.85);
    \node[lab] at (2.4,-1.55) {(a)};

    \node[lab] at (5.55,0) {$=$};

    \uoneseampanel{6.3cm}
    \begin{scope}[xshift=6.3cm]
      \draw[spine] (0.9,0) -- (1.75,0);
      \draw[spine,mid] (1.75,0) -- (3.05,0);
      \draw[spine] (3.05,0) -- (3.9,0);
      \draw[sym,mid] (0,-0.85) -- (0.95,-0.85)
        .. controls (1.45,-0.85) and (1.75,-0.45) .. (1.75,0);
      \draw[sym,mid] (3.05,0) .. controls (3.05,-0.45) and (3.35,-0.85) .. (3.85,-0.85)
        -- (4.8,-0.85);
      \node[junc] at (1.75,0) {};
      \node[junc] at (3.05,0) {};
      \node[lab, anchor=south] at (1.33,0.04) {$a^{p}$};
      \node[lab, anchor=south] at (2.4,0.04) {$a^{p+1}$};
      \node[lab, anchor=south] at (3.47,0.04) {$a^{p}$};
      \node[lab] at (2.4,-1.55) {(b)};
    \end{scope}
  \end{tikzpicture}
\caption{Fusing the Verlinde line $Y_a$ into the fusion path. (a) The line runs parallel to the chain, below the seam bond $(m,1)$. (b) The two lines fuse in the single channel $a^{p+1}$ with coefficient one, and sliding the two junctions around the ring until they meet raises every label $a^{h_j}$ to $a^{h_j+1}$ at fixed occupations, so that $|\boldsymbol{n};p\rangle\mapsto|\boldsymbol{n};p+1\rangle$ up to the sign of Equation~\eqref{eq:u1-cut-definition}.}
  \label{fig:u1-verlinde-fusion}
\end{figure}

\paragraph{The seam and the Wilson line \texorpdfstring{$Y_a$}{Ya}.}
The generators with $i<m$ form the quantum group $U^{\mathrm{L}}_q(\fgl_m)$. On the periodic chain we also need the affine generators $E_m$ and $F_m$ of Equation~\eqref{eq:periodic-seam-action}, which act on the bond $(m,1)$ as
\begin{equation}
\begin{aligned}
 E_m|1,n_2,\ldots,n_{m-1},0;p\rangle&=|0,n_2,\ldots,n_{m-1},1;p+1\rangle,\\
 F_m|0,n_2,\ldots,n_{m-1},1;p\rangle&=|1,n_2,\ldots,n_{m-1},0;p-1\rangle .
\end{aligned}
\label{eq:u1-seam-action}
\end{equation}
Since they act locally at the seam, they leave the intermediate labels $h_1,\ldots,h_{m-1}$ unchanged. Since $h_1=p+n_1$, this forces the shift of basepoint $p\mapsto p\pm1$. We can try to compare the action of the affine roots $E_m$ and $F_m$ of $U^{\mathrm{L}}_{q}(L\mathfrak{gl}_m)$ with the action of the ordinary fermion bilinears $c_m^\dagger c_1$ and $c_1^\dagger c_m$. Under the isomorphism $\iota$ induced by Equation~\eqref{eq:u1-free-fermion-string} we have the following identification  between $q$-fermion states and canonical fermion states,
\begin{equation}
 \iota|\boldsymbol{n};p\rangle=|p\rangle\otimes(c_1^\dagger)^{n_1}\cdots(c_m^\dagger)^{n_m}|0\rangle ,
 \label{eq:u1-fock-identification}
\end{equation}
then the bilinears act as
\begin{equation}
\begin{aligned}
 c_m^\dagger c_1|1,n_2,\ldots,n_{m-1},0;p\rangle&=(-1)^{F-1}|0,n_2,\ldots,n_{m-1},1;p\rangle,\\
 c_1^\dagger c_m|0,n_2,\ldots,n_{m-1},1;p\rangle&=(-1)^{F-1}|1,n_2,\ldots,n_{m-1},0;p\rangle ,
\end{aligned}
\label{eq:u1-naive-seam-action}
\end{equation}
where $(-1)^{F-1}$ is the sign from reordering the fermions. Notice that the basepoint $p$ is fixed, which means that all the intermediate labels change,
\begin{equation}
 h_j=p+\sum_{i=1}^{j}n_i\;\longmapsto\;h_j\mp1,
 \qquad
 1\leq j<m,
 \label{eq:u1-naive-label-shift}
\end{equation}
which implies that the naive generators act nonlocally. To relate the two actions, we use the bulk Verlinde line $Y_a$ of Equation~\eqref{eq:periodic-verlinde-algebra}, which runs parallel to the chain. It shifts the basepoint, up to a sign, and commutes with all the generators,
\begin{equation}
 Y_a|\boldsymbol{n};p\rangle=q^{F}|\boldsymbol{n};p+1\rangle ,
 \qquad
 [Y_a,E_i]=[Y_a,F_i]=[Y_a,L_i]=0 ,
 \qquad
 i=1,\ldots,m .
 \label{eq:u1-cut-definition}
\end{equation}
To see this, fuse $Y_a$ into the fusion path as in Figure~\ref{fig:u1-verlinde-fusion}. Since $U(1)_k$ is abelian with unit quantum dimensions, the lines fuse with coefficient one and every label rises by one at fixed occupations. In the standard gauge the $F$-moves at the junctions are trivial except $F^{a,a^{k-1},a}=-1$, the nontrivial associator of $U(1)_k$ \cite{Kitaev2006:AnyonsExactlySolved}, which gives a factor of $q^F=(-1)^{F/k}$. Defining the operator $\mathsf{S}:=q^{-F}Y_a$, with $\mathsf{S}|\boldsymbol{n};p\rangle=|\boldsymbol{n};p+1\rangle$, the affine generators become
\begin{equation}
 E_m=\mathsf{S}\,(-1)^{F-1}c_m^\dagger c_1,
 \qquad
 F_m=\mathsf{S}^{-1}(-1)^{F-1}c_1^\dagger c_m .
 \label{eq:u1-cut-operator}
\end{equation}
And the periodic Hamiltonian of Equation~\eqref{eq:periodic-hamiltonian} becomes
\begin{equation}
 H^{q}_{\mathrm{per}}
 =-t\Bigl[\sum_{i=1}^{m-1}\bigl(c_i^\dagger c_{i+1}+c_{i+1}^\dagger c_i\bigr)
 +(-1)^{F-1}\bigl(\mathsf{S}\,c_m^\dagger c_1+\mathsf{S}^{-1}c_1^\dagger c_m\bigr)\Bigr],
 \label{eq:u1-periodic-hamiltonian}
\end{equation}
where we have set $\mu=0$.\footnote{For $\mu\neq0$, the term $-\mu F$ commutes with all the generators and with $\mathsf{S}$, so it only fixes the filling. Since the free-fermion radius does not depend on the filling, as noted below Equation~\eqref{eq:ff-charge-radius}, the result of Equation~\eqref{eq:u1-gauged-boson} holds at every filling with $0<F<m$ and $F\in k\bZ$.}

\paragraph{Gauging of \texorpdfstring{$\bZ_k$}{Zk} and twisted sectors.}
Now consider Fourier transforming the basepoint label. The states
\begin{equation}
 |\boldsymbol{n};\ell\rangle=\frac{1}{\sqrt k}\sum_{p\in\bZ_k}e^{-2\pi i\ell p/k}\,|\boldsymbol{n};p\rangle,
 \qquad
 \ell\in\bZ_k,
 \label{eq:u1-cut-fourier}
\end{equation}
diagonalize the shift, $\mathsf{S}|\boldsymbol{n};\ell\rangle=e^{2\pi i\ell/k}|\boldsymbol{n};\ell\rangle$, and correspondingly the Wilson line $Y_a=q^{F}\mathsf{S}$ acts with eigenvalue $q^{F}e^{2\pi i\ell/k}$. This eigenvalue is nothing but $S_{(a,\eta)}/S_{(\mathbbm{1},\eta)}$ with $\eta=a^{\ell+F/2}$, where $S_{(a^r,a^s)}=e^{2\pi irs/k}/\sqrt k$, so the sectors $\ell$ are precisely the flux sectors of Equation~\eqref{eq:periodic-flux-sectors}, on which the Verlinde lines act diagonally. The generators with $i<m$ do not change $p$, and therefore do not change $\ell$. In the sector $\ell$ the seam term of Equation~\eqref{eq:u1-periodic-hamiltonian} therefore becomes
\begin{equation}
 -t\,(-1)^{F-1}\bigl(e^{2\pi i\ell/k}c_m^\dagger c_1+e^{-2\pi i\ell/k}c_1^\dagger c_m\bigr),
 \label{eq:u1-seam-twist}
\end{equation}
so we have a free-fermion Hamiltonian with the twisted boundary condition\footnote{Since $F\in k\bZ$ is even, the $\ell=0$ sector is antiperiodic, $c_{m+1}=-c_1$. The continuum fields of Equation~\eqref{eq:lattice-continuum-fields} pick up the further phase $e^{\pm ik_Fm}$ across the seam, which at half filling, $k_F=\pi/2$, is $(-1)^{m/2}$, so this sector is Neveu--Schwarz for $m\equiv0\pmod4$ and Ramond for $m\equiv2\pmod4$.}
\begin{equation}
c_{m+1}=(-1)^{F-1}e^{2\pi i\ell/k}c_1 .
  \label{eq:u1-twisted-fermion-boundary-condition}
\end{equation}

Summing over these twisted sectors while keeping only $F\equiv0\pmod k$ is precisely the act of gauging the $\bZ_k$ subgroup of $U(1)$. The periodic $q$-fermion chain is therefore the free fermion with a gauged $\bZ_k$, and in the continuum the compact boson $U(1)_1$ of the free fermion, with $\tilde R_0=1$, becomes \cite{Ginsparg1988:CuriositiesC1,Ginsparg1988:AppliedCFT}
\begin{equation}
 U(1)_1/\bZ_k=U(1)_{k^2},
 \label{eq:u1-gauged-boson}
\end{equation}
the compact boson with $\tilde R=k$.\footnote{For odd $k$, the labels, including the basepoint, lie in $\bZ_{2k}$, and $a^k$ is a transparent fermion. Closing the path then requires $F\in2k\bZ$, and the same steps, with $p,\ell\in\bZ_{2k}$, gauge the $\bZ_{2k}$ subgroup of $U(1)_F$. Since $\bZ_{2k}$ contains the fermion parity $(-1)^F$, this gives the bosonic theory $U(1)_1/\bZ_{2k}=U(1)_{4k^2}$ \cite{Okuda2021:U1SpinChernSimons}.}

\paragraph{The interface expectation.}
As for $SU(N)_k$, we can compare this result with the non-chiral interfaces obtained by applying the construction of \cite{GaiottoZeng2026:InterfaceMinimalModel} to $U(1)_k$ Chern--Simons theory, that is, by fusing the chiral interfaces of Equation~\eqref{eq:continuum-chiral-interface} for the two orientations of Figure~\ref{fig:continuum-orientation}. Their chiral algebras are the cosets
\begin{equation}
 \frac{U(1)_{k-1}\times\mathrm{Ff}^{1}_{\bC}}{U(1)_{k}}\simeq U(1)^{\text{lat}}_{k(k-1)},
 \qquad
 \frac{U(1)_{k}\times\mathrm{Ff}^{1}_{\bC}}{U(1)_{k+1}}\simeq U(1)^{\text{lat}}_{k(k+1)}.
 \label{eq:u1-interface-cosets}
\end{equation}
Each commutant is generated by the combination of the $U(1)_{k-1}$ (respectively $U(1)_k$) and $\mathrm{Ff}^{1}_{\bC}$ currents that is orthogonal to the diagonal current, and this combination is itself an abelian current of level $k(k-1)$ (respectively $k(k+1)$). The two interfaces therefore have $\tilde R^2=k(k\mp1)$, which does not agree with $\tilde R^2=k^2$ in Equation~\eqref{eq:u1-gauged-boson}. This is expected, since, as discussed in Section~\ref{sec:non-chiral-interface}, the lattice regularization of the non-chiral interface and the regularization by chiral interfaces need not agree. As seen in Appendix~\ref{app:continuum-numerics}, they only agree up to exactly marginal deformations if there is no protection from symmetry. One can then ask whether the marginal coupling can be tuned on the lattice to reproduce the values from the chiral interface picture. Since the $U(1)_k$ chain is locally a free fermion, this is indeed possible by mapping it to the integrable XXZ chain.

\paragraph{The Cartan deformation and the XXZ chain.}
We add to the periodic Hamiltonian a term built from the Cartan generators,
\begin{equation}
\begin{aligned}
 H(g)&=-t\sum_{i=1}^{m}\Bigl[E_i+F_i+ig\,\frac{K_i+K_i^{-1}}{q^{1/2}-q^{-1/2}}\Bigr]\\
 &=-t\sum_{i=1}^{m}\Bigl[E_i+F_i-2g\sin\frac{\pi}{2k}\,(n_i-n_{i+1})^2\Bigr]-\frac{mgt}{\sin\frac{\pi}{2k}},
\end{aligned}
 \qquad
 g\in\bR,
 \label{eq:u1-deformed-hamiltonian}
\end{equation}
with $n_{m+1}\equiv n_1$. The density interaction makes it the XXZ chain with $\Delta=-2g\sin\frac{\pi}{2k}$ which is critical for $|\Delta|<1$.

We stress that the gauging does not obstruct the exact solution of the XXZ chain. The Cartan term depends only on the occupations, so it commutes with $F$ and with $\mathsf{S}$. It is therefore neutral under $\bZ_k$. Gauging $\bZ_k$ thus commutes with the deformation and only rescales $\tilde R$ by $k$, which gives the factor $k^2$ in Equation~\eqref{eq:u1-exact-radius} below.

In the continuum, the density interaction generates the current terms of Equation~\eqref{eq:ff-marginal-hamiltonian} with $N=1$.\footnote{There is also a translation-neutral $4k_F$ umklapp term, which is generated but is irrelevant for $|\Delta|<1$.} Both the stress-tensor term $T+\bar T\propto{:}J_L^2{:}+{:}J_R^2{:}$ and the exactly marginal term $J_LJ_R$ are generated, and their combined effect changes the radius as in Equation~\eqref{eq:ff-charge-radius}. At half filling, $F=m/2$ with $m\in2k\bZ$, the XXZ chain allows for an exact solution from the Bethe ansatz \cite{Pereira2007:XXZStructureFactor}, which gives the radius exactly,
\begin{equation}
 \tilde R^2(g)=k^2\left[1-\frac{2}{\pi}\arcsin\!\left(2g\sin\frac{\pi}{2k}\right)\right],
 \qquad
  |\Delta| = \left|2g\sin\frac{\pi}{2k}\right|<1 .
 \label{eq:u1-exact-radius}
\end{equation}
The radius thus moves continuously away from the point $\tilde R^2=k^2$ of Equation~\eqref{eq:u1-gauged-boson} at $g=0$, and at $g=\pm\frac12$, Equation~\eqref{eq:u1-exact-radius} gives exactly $\tilde R^2=k(k\mp1)$, the values of the radius from the interface constructions in Equation~\eqref{eq:u1-interface-cosets}.

\section{The \texorpdfstring{$SU(2)_k$}{SU(2)k} examples}
\label{app:su2k}

This appendix presents, as an example, the explicit analysis of $SU(2)_k$ anyon models, following the general methodology of Sections~\ref{sec:q-fermion-chain}, \ref{sec:continuum-limit} and~\ref{sec:mott-projection}.

\subsection{The fusion trees of $SU(2)_k$}
\label{app:su2k-fusion-tree-example}
The simple objects of $SU(2)_k$ are $\mathcal{L}_a:=\mathcal{L}_{a\omega_1}$, $a=0,1,\ldots,k$, with $\omega_1$ the fundamental weight and $\mathcal{L}_0=\mathbbm{1}$. There are three possible local labels on the fusion tree,
\begin{equation}
\mathcal{L}_{\omega_0}\simeq \mathbbm{1},
\qquad
\mathcal{L}_{\omega_1},
\qquad
\mathcal{L}_{\omega_2}
=
\Lambda_q^2\bC_q^2
\simeq \mathbbm{1},
  \label{eq:su2-site-exterior-modules}
\end{equation}
and carry fermion numbers $0$, $1$, and $2$, respectively. Their fusion rules are
\begin{equation}
\begin{aligned}
\mathcal{L}_a\otimes \mathcal{L}_{\omega_0}
&\simeq \mathcal{L}_a,
\\
\mathcal{L}_a\otimes \mathcal{L}_{\omega_1}
&\simeq \mathcal{L}_{a-1}\oplus \mathcal{L}_{a+1},
\\
\mathcal{L}_a\otimes \mathcal{L}_{\omega_2}
&\simeq \mathcal{L}_a,
\end{aligned}
\label{eq:su2k-local-fusion-quiver-rules}
\end{equation}
where the summand $\mathcal{L}_{a-1}$ is omitted when $a=0$, and $\mathcal{L}_{a+1}$ is omitted when $a=k$. The fusion rules of Equation~\eqref{eq:su2k-local-fusion-quiver-rules} give the generating matrix
\begin{equation}
M(x)=(1+x^2)I+xA_{k+1}, \qquad M(x) =
\begin{pmatrix}
1+x^2&x& & & \\
x&1+x^2&x& & \\
 &x&1+x^2&\ddots& \\
 & &\ddots&\ddots&x\\
 & & &x&1+x^2
\end{pmatrix},
  \label{eq:su2-occupation-transfer-matrix}
\end{equation}
where $A_{k+1}$ is the adjacency matrix of the $\fsl_{k+2}$ Dynkin diagram. The dimension of the physical Hilbert space in Equation~\eqref{eq:physical-Hilbert-space-qfermion} with initial charge $\lambda^{(0)}=0$ is given by
\begin{equation}
S_{k,m}
:=
\sum_{a=0}^{k}\dim_{\bC}\widehat{\mathcal{H}}_{(0,a)}
=
\mathbf u^{\,T}M(1)^m\mathbf e_0 ,
  \label{eq:su2-total-path-dimension}
\end{equation}
where
\begin{equation}
\mathbf u:=(1,2,\ldots,k+1)^T,
\qquad
(\mathbf e_a)_b:=\delta_{ab},\qquad a,b=0,\ldots,k ,
  \label{eq:su2-dimension-boundary-vectors}
\end{equation}
and $\mathbf u$ collects the dimensions $\dim_{\bC}\mathcal{L}_a=a+1$ of the spin-$a/2$ representations. Diagonalizing $M(1)$ gives the exact finite-$k$ expression
\begin{equation}
S_{k,m}
=
\frac12\sum_{j=1}^{k+1}(-1)^{j+1}
\left(
2+2\cos\frac{\pi j}{k+2}
\right)^{m+1} .
\label{eq:Skm-closed-form}
\end{equation}
The truncation of the dimension can be seen from the identity $\mathbf u^{\,T}M(1)=4\mathbf u^{\,T}-(k+2)\mathbf e_k^{\,T}$, which gives the following recursion relation for the dimension,
\begin{equation}
S_{k,m+1}=4S_{k,m}-(k+2)\bigl(M(1)^m\bigr)_{k0},\qquad S_{k,0}=1 .
  \label{eq:su2-path-dimension-recursion}
\end{equation}
Since $\bigl(M(1)^n\bigr)_{k0}=0$ for $n<k$, the correction is absent for $m\leq k$ and $S_{k,m}=4^m=2^{2m}$ agrees with the dimension $2^{Nm}$ of the $q$-fermion Fock space, while at $m=k+1$ it gives
\begin{equation}
S_{k,k+1}=2^{2(k+1)}-(k+2),
  \label{eq:su2-first-truncation-dimension}
\end{equation}
so $k+2$ states are null, the color multiplet of the single fusion tree $0\to1\to\cdots\to k+1$.

\subsection{The action of $U^{\mathrm{L}}_q(\fgl_m)$ on $SU(2)_k$}
\label{sec:su2k-action-example}
We now describe the action of the site quantum group $U^{\mathrm{L}}_q(\fgl_m)$ on the $SU(2)_k$ fusion-tree basis of Appendix~\ref{app:su2k-fusion-tree-example}, with $q=e^{i\pi/(k+2)}$.

\paragraph{The map $\mu$ for $SU(2)_k$.}
We construct the map $\mu$ of Section~\ref{sec:path-to-qfermion} for the case of $N=2$, writing $X_{\mathrm{col}}:=X^1_{\mathrm{col}}$ for $X=E,F,K$. At each site $j$ the representations $\mathcal{L}_{\omega_r}$ are realized as
\begin{equation}
\mathcal{L}_{\omega_0}=\text{span}\{\mathbbm{1}_j\},
\qquad
\mathcal{L}_{\omega_1}=\text{span}\{\bqpsi^{1}_j,\bqpsi^{2}_j\},
\qquad
\mathcal{L}_{\omega_2}=\text{span}\{\bqpsi^{1}_j\bqpsi^{2}_j\} .
  \label{eq:su2-local-pbw-basis}
\end{equation}
Let $\lambda^{(j)}=a$. By Equation~\eqref{eq:path-qfermion-highest-weight-equations} the highest-weight state $\Omega^{\text{hw}}_{j}$ of $\mathcal{L}_a$ obeys
\begin{equation}
E_{\mathrm{col}}\bigl(\Omega^{\text{hw}}_{j}\bigr)=0,
\qquad
K_{\mathrm{col}}\bigl(\Omega^{\text{hw}}_{j}\bigr)=q^{a}\,\Omega^{\text{hw}}_{j},
  \label{eq:su2-path-highest-weight-action}
\end{equation}
and $\mathcal{L}_a$ is spanned by $F_{\mathrm{col}}^{\,n}\Omega^{\text{hw}}_{j}$, $n=0,\ldots,a$. There are four kinds of step when fusing with the local object $\mathcal{L}_{\omega_{r_{j+1}}}$ at site $j+1$. The overall coefficients of these states are fixed by Equation~\eqref{eq:path-qfermion-step-normalization}, which gives $1$ in the trivial, determinant and raising steps.
\begin{itemize}
\item \textbf{Trivial step.}
For $r_{j+1}=0$ nothing is added at site $j+1$,
\begin{equation}
\mathcal{L}_{a}\xrightarrow{\ \mathcal{L}_{\omega_0}\ }\mathcal{L}_{a},
\qquad
\Omega^{\text{hw}}_{j+1}=\Omega^{\text{hw}}_{j} .
\label{eq:path-qfermion-su2-identity}
\end{equation}
\item \textbf{Determinant step.}
For $r_{j+1}=2$ the site carries the color singlet $\bqpsi^{1}_{j+1}\bqpsi^{2}_{j+1}$,
\begin{equation}
\mathcal{L}_{a}\xrightarrow{\ \mathcal{L}_{\omega_2}\ }\mathcal{L}_{a},
\qquad
\Omega^{\text{hw}}_{j+1}=\Omega^{\text{hw}}_{j}\otimes\bqpsi^{1}_{j+1}\bqpsi^{2}_{j+1} .
\label{eq:path-qfermion-su2-determinant}
\end{equation}
\item \textbf{Raising step.}
For $r_{j+1}=1$ and $a\to a+1$ the two highest weights add,
\begin{equation}
\mathcal{L}_{a}\xrightarrow{\ \mathcal{L}_{\omega_1}\ }\mathcal{L}_{a+1},
\qquad
\Omega^{\text{hw}}_{j+1}=\Omega^{\text{hw}}_{j}\otimes\bqpsi^{1}_{j+1} .
\label{eq:path-qfermion-su2-raising}
\end{equation}
\item \textbf{Lowering step.}
For $r_{j+1}=1$ and $a\to a-1$ two terms have the weight $a-1$,
\begin{equation}
\mathcal{L}_{a}\xrightarrow{\ \mathcal{L}_{\omega_1}\ }\mathcal{L}_{a-1},
\qquad
\Omega^{\text{hw}}_{j+1}
=
C^{1,\{1\}}_{j}\,\bigl(F_{\mathrm{col}}\Omega^{\text{hw}}_{j}\bigr)\otimes\bqpsi^{1}_{j+1}
+C^{0,\{2\}}_{j}\,\Omega^{\text{hw}}_{j}\otimes\bqpsi^{2}_{j+1} ,
  \label{eq:su2-down-step-highest-weight-ansatz}
\end{equation}
where the first superscript counts the lowering operators $F_{\mathrm{col}}$. With the coproduct of Equation~\eqref{eq:q-fermion-coproduct} and the relations of Appendix~\ref{app:uqglm-relations},
\begin{equation}
\begin{aligned}
E_{\mathrm{col}}\bigl[\bigl(F_{\mathrm{col}}\Omega^{\text{hw}}_{j}\bigr)\otimes\bqpsi^{1}_{j+1}\bigr]
  &=q[a]_q\,\Omega^{\text{hw}}_{j}\otimes\bqpsi^{1}_{j+1},
  \\
E_{\mathrm{col}}\bigl[\Omega^{\text{hw}}_{j}\otimes\bqpsi^{2}_{j+1}\bigr]
  &=\Omega^{\text{hw}}_{j}\otimes\bqpsi^{1}_{j+1},
\end{aligned}
  \label{eq:su2-down-step-raising-action}
\end{equation}
so the highest-weight condition gives
\begin{equation}
q[a]_q\,C^{1,\{1\}}_{j}+C^{0,\{2\}}_{j}=0 .
  \label{eq:su2-down-step-coefficient-constraint}
\end{equation}
For $S=\{2\}$, with $\delta_{12}=-1$ and $a_{12}=a+1$, Equation~\eqref{eq:path-qfermion-step-normalization} sets the coefficient of the unlowered term to $C^{0,\{2\}}_{j}=[a_{12}-1]_q=[a]_q$, so that $C^{1,\{1\}}_{j}=-q^{-1}$ and
\begin{equation}
\Omega^{\text{hw}}_{j+1}
=
-q^{-1}\bigl(F_{\mathrm{col}}\Omega^{\text{hw}}_{j}\bigr)\otimes\bqpsi^{1}_{j+1}
+[a]_q\,\Omega^{\text{hw}}_{j}\otimes\bqpsi^{2}_{j+1} .
\label{eq:path-qfermion-su2-lowering}
\end{equation}
\end{itemize}

These steps completely determine how to iteratively construct $\mu$ for the case of $SU(2)_k$.

\paragraph{The action of the generators.}
Let the local region of the fusion tree at the bond $(i,i+1)$ be written as
\begin{equation}
|a,r;c,s;b\rangle
:=
\left|\,a\xrightarrow{\ \omega_r\ }c\xrightarrow{\ \omega_s\ }b\,\right\rangle .
\label{eq:su2-path-local-ket}
\end{equation}
By Equation~\eqref{eq:path-site-action-K-schematic}, $K_i$ acts diagonally,
\begin{equation}
K_i\,|a,r;c,s;b\rangle=q^{\,r-s}\,|a,r;c,s;b\rangle .
\label{eq:su2-path-local-Cartan-rule}
\end{equation}
The actions of $E_i$ and $F_i$ realize the coefficients $e_{\tilde\lambda^{(i)}}$ and $f_{\tilde\lambda^{(i)}}$ of Equations~\eqref{eq:path-site-action-E-schematic}--\eqref{eq:path-site-action-F-schematic} which we now compute.

Let $\Omega:=\Omega^{\text{hw}}_{i-1}$, with $\lambda^{(i-1)}=a$. In the chosen PBW basis $\Omega$ carries no fermions at sites $i$ and $i+1$, so that $E_i(\Omega X)=\Omega\,E_i(X)$, where we suppress the coproduct and the tensor product. From Equations~\eqref{eq:path-qfermion-su2-identity}--\eqref{eq:path-qfermion-su2-lowering}, the $q$-fermion representatives of the local states of Equation~\eqref{eq:su2-path-local-ket} are
\begin{equation}
\begin{aligned}
|a,0;a,1;a+1\rangle
&=
\Omega\,\bqpsi^{1}_{i+1},
\\
|a,0;a,1;a-1\rangle
&=
-q^{-1}(F_{\mathrm{col}}\Omega)\,\bqpsi^{1}_{i+1}
+[a]_q\,\Omega\,\bqpsi^{2}_{i+1},
\\
|a,1;a+1,2;a+1\rangle
&=
\Omega\,\bqpsi^{1}_{i}\,\bqpsi^{1}_{i+1}\bqpsi^{2}_{i+1},
\\
|a,1;a-1,2;a-1\rangle
&=
-q^{-1}(F_{\mathrm{col}}\Omega)\,\bqpsi^{1}_{i}\,\bqpsi^{1}_{i+1}\bqpsi^{2}_{i+1}
+[a]_q\,\Omega\,\bqpsi^{2}_{i}\,\bqpsi^{1}_{i+1}\bqpsi^{2}_{i+1},
\\
|a,0;a,2;a\rangle
&=
\Omega\,\bqpsi^{1}_{i+1}\bqpsi^{2}_{i+1},
\\
|a,1;a+1,1;a\rangle
&=
-q^{-1}(F_{\mathrm{col}}\Omega)\,\bqpsi^{1}_{i}\bqpsi^{1}_{i+1}
-q^{-a-1}\,\Omega\,\bqpsi^{2}_{i}\bqpsi^{1}_{i+1}
+[a+1]_q\,\Omega\,\bqpsi^{1}_{i}\bqpsi^{2}_{i+1},
\\
|a,1;a-1,1;a\rangle
&=
-q^{-1}(F_{\mathrm{col}}\Omega)\,\bqpsi^{1}_{i}\bqpsi^{1}_{i+1}
+[a]_q\,\Omega\,\bqpsi^{2}_{i}\bqpsi^{1}_{i+1}.
\end{aligned}
  \label{eq:su2-local-hopping-basis}
\end{equation}
From Equation~\eqref{eq:q-fermion-site-cartan-action}, the coproduct in Equation~\eqref{eq:q-fermion-coproduct}, and PBW reordering by Equation~\eqref{eq:q-fermion-braided-antisymmetry}, the action of $E_i$ is, for $\gamma=1,2$,
\begin{equation}
\begin{aligned}
E_i\bigl(\bqpsi^{\gamma}_{i+1}\bigr)
&=
\bqpsi^{\gamma}_{i},
\\
E_i\bigl(\bqpsi^{\gamma}_{i}\,\bqpsi^{1}_{i+1}\bqpsi^{2}_{i+1}\bigr)
&=
-\,\bqpsi^{1}_{i}\bqpsi^{2}_{i}\,\bqpsi^{\gamma}_{i+1},
\\
E_i\bigl(\bqpsi^{1}_{i+1}\bqpsi^{2}_{i+1}\bigr)
&=
-\bqpsi^{2}_{i}\bqpsi^{1}_{i+1}
+q\,\bqpsi^{1}_{i}\bqpsi^{2}_{i+1},
\\
E_i\bigl(\bqpsi^{2}_{i}\bqpsi^{1}_{i+1}\bigr)
&=
-q^{-1}\,\bqpsi^{1}_{i}\bqpsi^{2}_{i},
\\
E_i\bigl(\bqpsi^{1}_{i}\bqpsi^{2}_{i+1}\bigr)
&=
\bqpsi^{1}_{i}\bqpsi^{2}_{i}.
\end{aligned}
\label{eq:su2-path-local-E-singlet-components}
\end{equation}
Using Equation~\eqref{eq:su2-path-local-E-singlet-components} and $q^{-a-1}+[a]_q=q^{-1}[a+1]_q$ and rewriting the image in the fusion-tree basis and doing a similar computation for $F_i$ gives the entries of Table~\ref{tab:su2-path-local-rules}.
Any local action not listed in Table~\ref{tab:su2-path-local-rules} vanishes; for example, $E_i$ annihilates $\lvert a,2;a,0;a\rangle$.

\begin{table}[t]
\centering
\small
\renewcommand{\arraystretch}{1.5}
\begin{tabular}{c @{\quad$\longmapsto$\quad} c}
\toprule
Initial State & Final State \\
\midrule
\multicolumn{2}{c}{Action of $E_i$} \\
\addlinespace[1mm]
$\lvert a,0;a,1;b\rangle$
& $\lvert a,1;b,0;b\rangle$ \\
$\lvert a,1;b,2;b\rangle$
& $-\lvert a,2;a,1;b\rangle$ \\
$\lvert a,0;a,2;a\rangle$
& $\displaystyle q\,\frac{\lvert a,1;a+1,1;a\rangle-\lvert a,1;a-1,1;a\rangle}{[a+1]_q}$ \\
$\lvert a,1;a+1,1;a\rangle$
& $q^{-1}[a+2]_q\lvert a,2;a,0;a\rangle$ \\
$\lvert a,1;a-1,1;a\rangle$
& $-q^{-1}[a]_q\lvert a,2;a,0;a\rangle$ \\
\midrule
\multicolumn{2}{c}{Action of $F_i$} \\
\addlinespace[1mm]
$\lvert a,1;b,0;b\rangle$
& $\lvert a,0;a,1;b\rangle$ \\
$\lvert a,2;a,1;b\rangle$
& $-\lvert a,1;b,2;b\rangle$ \\
$\lvert a,2;a,0;a\rangle$
& $\displaystyle q\,\frac{\lvert a,1;a+1,1;a\rangle-\lvert a,1;a-1,1;a\rangle}{[a+1]_q}$ \\
$\lvert a,1;a+1,1;a\rangle$
& $q^{-1}[a+2]_q\lvert a,0;a,2;a\rangle$ \\
$\lvert a,1;a-1,1;a\rangle$
& $-q^{-1}[a]_q\lvert a,0;a,2;a\rangle$ \\
\bottomrule
\end{tabular}
\captionsetup{position=bottom}
\caption{Action of $E_i$ and $F_i$ on the fusion-tree basis for $SU(2)_k$, with $b=a\pm1$ and the normalization of Equation~\eqref{eq:path-qfermion-step-normalization}.}
\label{tab:su2-path-local-rules}
\end{table}

Also the rules of Table~\ref{tab:su2-path-local-rules} should be understood modulo null states. For example at $a=k$ the projection in Equation~\eqref{eq:path-qfermion-action-on-physical-Hilbert-space} removes the null channel $c=k+1$. Similarly at $a=0$ the lowering step should give zero, since $F_{\mathrm{col}}\Omega=0$. With $[1]_q=[k+1]_q=1$ this gives
\begin{equation}
\begin{aligned}
E_i|0,0;0,2;0\rangle
&=q\,|0,1;1,1;0\rangle,
\\
E_i|k,0;k,2;k\rangle
&=-q\,|k,1;k-1,1;k\rangle.
\end{aligned}
\label{eq:su2-path-local-boundary-rules}
\end{equation}

\paragraph{Hermiticity condition and the induced norm.}
The actions of $E_i$ and $F_i$ close on the blocks labeled by $(a,b,n = r+s)$. For $SU(2)_k$ every such block is at most two-dimensional, except for $b=a$ and $n=2$. There the intermediate label can take the values $c=a-1$, $a$ or $a+1$, and for $1\leq a\leq k-1$ the block is four-dimensional. Labeling the states as $v_1,\ldots,v_4$, we have
\begin{equation}
  v_1=\lvert a,0;a,2;a\rangle,\quad
  v_2=\lvert a,1;a-1,1;a\rangle,\quad
  v_3=\lvert a,1;a+1,1;a\rangle,\quad
  v_4=\lvert a,2;a,0;a\rangle .
  \label{eq:su2-path-block-basis}
\end{equation}
Then Table~\ref{tab:su2-path-local-rules} gives the matrices
\begin{equation}
  E_i=
  \begin{pmatrix}
    0&0&0&0\\[2pt]
    -\tfrac{q}{[a+1]_q}&0&0&0\\[2pt]
    \tfrac{q}{[a+1]_q}&0&0&0\\[2pt]
    0&-q^{-1}[a]_q&q^{-1}[a+2]_q&0
  \end{pmatrix},
  \qquad
  F_i=
  \begin{pmatrix}
    0&-q^{-1}[a]_q&q^{-1}[a+2]_q&0\\[2pt]
    0&0&0&-\tfrac{q}{[a+1]_q}\\[2pt]
    0&0&0&\tfrac{q}{[a+1]_q}\\[2pt]
    0&0&0&0
  \end{pmatrix},
  \label{eq:su2-path-block-EF}
\end{equation}
while Equation~\eqref{eq:su2-path-local-Cartan-rule} gives $K_i=\text{diag}\bigl(q^{-2},1,1,q^{2}\bigr)$. For $q$ a root of unity one has $K_i^{\dagger}=K_i^{-1}$, while $E_i$ and $F_i$ are not Hermitian conjugates of one another. Instead they are related through the diagonal Gram matrix $G=\text{diag}\bigl(\langle v_1,v_1\rangle,\ldots,\langle v_4,v_4\rangle\bigr)$ of the basis in Equation~\eqref{eq:su2-path-block-basis} by
\begin{equation}
\begin{aligned}
G\,F_i&=E_i^{\dagger}\,G, \\
  G&\propto\text{diag}\bigl(1,\;[a]_q[a+1]_q,\;[a+1]_q[a+2]_q,\;1\bigr).
\end{aligned}
\label{eq:su2-path-block-transpose}
\end{equation}

Rescaling the middle states to $\hat v_2:=v_2/\sqrt{[a]_q[a+1]_q}$ and $\hat v_3:=v_3/\sqrt{[a+1]_q[a+2]_q}$, and omitting null states, gives
\begin{equation}
\begin{aligned}
E_i\,v_1
&=
-q\sqrt{\frac{[a]_q}{[a+1]_q}}\,\hat v_2
+q\sqrt{\frac{[a+2]_q}{[a+1]_q}}\,\hat v_3,
\\
E_i\,\hat v_2
&=
-q^{-1}\sqrt{\frac{[a]_q}{[a+1]_q}}\,v_4,
\qquad
E_i\,\hat v_3
=
q^{-1}\sqrt{\frac{[a+2]_q}{[a+1]_q}}\,v_4,
\end{aligned}
\label{eq:su2-path-local-symmetric-block}
\end{equation}
and we obtain $F_i=E_i^{\dagger}$ in this basis.

The remaining blocks on which $E_i$ and $F_i$ act nontrivially have $b=a\pm1$ and $n=1,3$, and are two-dimensional. In the bases $\bigl(\lvert a,0;a,1;b\rangle,\lvert a,1;b,0;b\rangle\bigr)$ for $n=1$ and $\bigl(\lvert a,1;b,2;b\rangle,\lvert a,2;a,1;b\rangle\bigr)$ for $n=3$, Table~\ref{tab:su2-path-local-rules} gives
\begin{equation}
  E_i=\begin{pmatrix}0&0\\ \epsilon&0\end{pmatrix},
  \qquad
  F_i=E_i^{\dagger},
  \qquad
  K_i=\text{diag}\bigl(q^{-1},q\bigr),
  \label{eq:su2-path-2d-blocks}
\end{equation}
with $\epsilon=+1$ for $n=1$ and $\epsilon=-1$ for $n=3$.

\paragraph{Junction normalization.}
We now verify that the norm given by the Gram matrix in Equation~\eqref{eq:su2-path-block-transpose} agrees exactly with the junction normalization proposed in Section~\ref{sec:orthonormalization-qfermion-basis}. By Equation~\eqref{eq:suN-quantum-Weyl-dimension}, the quantum dimensions of the simple objects of $SU(2)_k$ are
\begin{equation}
 d_a=\dim_q(\mathcal{L}_a)=[a+1]_q,
 \qquad
 a=0,1,\ldots,k.
  \label{eq:su2-quantum-dimensions}
\end{equation}
For $N=2$ the only positive root is $\alpha_{12}$, with $a_{12}=a+1$, and it enters Equation~\eqref{eq:suN-root-product-junction-factor} for both steps $a\to a\pm1$, so that
\begin{equation}
 \mathcal{K}_{a,\omega_1}^{a+1}=\mathcal{K}_{a,\omega_1}^{a-1}=[a+1]_q=d_a .
 \label{eq:su2-qfermion-junction-factors}
\end{equation}
The norm of Equation~\eqref{eq:norm-fusion-tree} is then
\begin{equation}
\langle\mathrm{P}|\mathrm{P}\rangle=
  d_{\lambda^{(m)}}\prod_{\substack{1\leq j\leq m\\ r_j=1}} d_{\lambda^{(j-1)}} ,
 \label{eq:su2-open-path-norm}
\end{equation}
where the product runs over the sites with a single fermion. On the states of Equation~\eqref{eq:su2-path-block-basis} it gives $\langle v_4,v_4\rangle=\langle v_1,v_1\rangle$, $\langle v_2,v_2\rangle=d_ad_{a-1}\langle v_1,v_1\rangle$ and $\langle v_3,v_3\rangle=d_ad_{a+1}\langle v_1,v_1\rangle$, exactly as the Gram matrix of Equation~\eqref{eq:su2-path-block-transpose}.\footnote{Also notice that any fusion tree that reaches the alcove $a=k+1$ has zero norm, $d_{k+1}=[k+2]_q=0$.} For a tree with $\lambda^{(m)}=a$, the descendants of the highest-weight vector $\Omega^{\text{hw}}_m$ obtained by lowering with $F_{\mathrm{col}}$ have the norms
\begin{equation}
 \bigl\|F_{\mathrm{col}}^{\,n}\Omega^{\text{hw}}_m\bigr\|^2
 =
 [n]_q!\,
 \frac{[a]_q!}{[a-n]_q!} \bigl\lVert \Omega^{\text{hw}}_m \bigr\rVert^2 .
 \label{eq:qfermion-normalized-root-string}
\end{equation}

\paragraph{The projected chain at unit filling.}
\label{app:su2k-projected-chain}
We now describe the projected chain of Section~\ref{sec:projector-chain} for $SU(2)_k$. Since the local objects are all $\mathcal{L}_{\omega_1}$, the Hilbert spaces are
\begin{equation}
  \mathcal{H}_{(a_0,a_m)}
  =\operatorname{span}\bigl\{
    |a_0,a_1,\ldots,a_m\rangle
    \;\big|\;
    a_j\in\{0,\ldots,k\},\quad
    |a_j-a_{j-1}|=1
  \bigr\},
  \label{eq:su2-fundamental-path-hilbert-space}
\end{equation}
and the effective Hamiltonian of Equation~\eqref{eq:mott-effective-hamiltonian} becomes
\begin{equation}
\begin{gathered}
H_{\mathrm{eff}}^{SU(2)_k}
=-J\sum_i e_i
=-[2]_q J\sum_i\Pi_i^{(\mathbbm{1})},
\qquad J=\frac{t^2}{U},\\
e_i=[2]_q\Pi_i^{(\mathbbm{1})}
=\frac12\,\bP^{(\omega_1)}(E_iF_i+F_iE_i)\bP^{(\omega_1)} .
\end{gathered}
\label{eq:mott-su2-effective}
\end{equation}
In the orthonormal basis rephased as below, the Temperley--Lieb generators act on the paths as
\begin{equation}
e_i|\ldots,a_{i-1},a_i,a_{i+1},\ldots\rangle
=
\delta_{a_{i-1},a_{i+1}}
\sum_{\substack{a_i'=a_{i-1}\pm1\\0\leq a_i'\leq k}}
\frac{\sqrt{[a_i+1]_q[a_i'+1]_q}}
{[a_{i-1}+1]_q}
|\ldots,a_{i-1},a_i',a_{i+1},\ldots\rangle,
\label{eq:mott-TL-path-action}
\end{equation}
which follows from the action of $F_i$ and $E_i$ in Table~\ref{tab:su2-path-local-rules}. After the basis change $|a_0,\ldots,a_m\rangle\mapsto(-1)^{\sum_j\lfloor a_j/2\rfloor}|a_0,\ldots,a_m\rangle$, the action of $e_i$ matches the anyon-chain literature \cite{Feiguin2007:InteractingAnyons,Trebst2008:FibonacciAnyonModels}.

\paragraph{Continuum limit of \texorpdfstring{$SU(2)_2$}{SU(2)2}.}
\label{app:su2-2-mott-example}
For $SU(2)_2$ the continuum theory before the projection is
\begin{equation}
U(1)^{\text{lat}}_2 \otimes \mathrm{Ising},
  \label{eq:su2-level-two-continuum-theory}
\end{equation}
with the primaries glued as in Section~\ref{sec:operator-content}. After the Mott term of Section~\ref{sec:projector-continuum} gaps out the charge modes, we obtain the low-energy theory~\cite{Belletete2020:TopologicalDefectsPeriodicRSOS,AasenFendleyMong2020:TopologicalDefectsDualities,Shao2023:TASINonInvertibleSymmetries}
\begin{equation}
 \mathcal{T}^{\text{SSB}}_{\bZ_2}\otimes \text{Ising},
  \label{eq:su2-level-two-mott-theory}
\end{equation}
in which the emanant translation $\bZ_2^\tau$ is spontaneously broken: the simple current $\mathbf J$ labels the two sectors and $\tau$ exchanges them, $\tau\mathbf J\tau^{-1}=-\mathbf J$. Diagonalizing the primaries with respect to $\bZ_2^\tau$ and labeling them by their crystal momentum $k$, we find\footnote{We take $m\equiv0\pmod4$, which avoids frustration.}
\begin{equation}
  \begin{aligned}
  k &= 0 &&:\quad \mathbbm{1}_{0},\ \sigma_{1/2},\ \epsilon_{1},\\
  k &= \pi &&:\quad \mathbbm{1}_{1},\ \sigma_{1/2},\ \epsilon_{0},
  \end{aligned}
  \label{eq:su2-level-two-momentum-line-labels}
\end{equation}
where the subscript is the bulk Wilson line of $SU(2)_2$ attached to the primary. We therefore see that, apart from the identity, no operator is neutral under both $\bZ_2^\tau$ and $\text{Rep}(SU(2)_2)$, so this phase is stable.

\section{Baryons}
\label{app:baryon-algebra}
This appendix collects the defining relations of the baryon operators and their algebra with the mesons in the $q$-fermion chain.\label{app:free-baryons} We begin by reviewing the baryon operators of the free fermion theory of Section~\ref{sec:free-fermion-review}. Its baryon creation operators $B^{\dagger}_{i_1,i_2,\ldots,i_N}$ are invariant under the global $SU(N)$ but carry charge $N$ under $U(1)_F$, and their adjoints $B_{i_1,i_2,\ldots,i_N}=\nobreak\bigl(B^{\dagger}_{i_1,i_2,\ldots,i_N}\bigr)^{\dagger}$ are the baryon annihilation operators. Together with the mesons they generate the gauge-invariant subalgebra $\text{Cl}_{2Nm}(\bC)^{SU(N)}$,\footnote{For $SU(2)$ the baryons $B^{\dagger}_{ij}$ are bilinear and symmetric in $i$ and $j$, and together with $B_{ij}$ and $\widetilde{G}_{ij}-\delta_{ij}$ they span $\mathfrak{sp}_{2m}\cong\fgl_m\oplus\text{Sym}^2(\bC^m)\oplus\text{Sym}^2(\bC^m)^{*}$.}
\begin{equation}
\widetilde{G}_{ij}, \qquad B^{\dagger}_{i_1,i_2,\ldots,i_N}, \qquad B_{i_1,i_2,\ldots,i_N} .
  \label{eq:baryon-invariant-algebra-generators}
\end{equation}

For the $q$-fermion chain, the generators $E_{ij}$ of $U^{\mathrm{L}}_q(\fgl_m)$ were identified in Section~\ref{sec:q-fermion-chain} with the meson operators of Equation~\eqref{eq:meson-bilinear}. Since the three-dimensional bulk is $SU(N)_k$ Chern--Simons theory and not $U(N)_{k,k'}$ Chern--Simons theory, there are also baryon operators, which transform trivially under $U^{\mathrm{L}}_q(\fsl_N)$ and carry charge $N$ under $U(1)_F$. The simplest example of a baryon operator in this case is the one that creates a baryon localized at site $i$, and in terms of the $q$-fermion creation operators it reads
\begin{equation}
  B^{\dagger}_{i}
  :=
  \prod_{a=1}^{N}\bqpsi^{a}_{i}
  \;\equiv\;
  \bqpsi^{1}_{i}\,\bqpsi^{2}_{i}\cdots\bqpsi^{N}_{i} ,
  \label{eq:baryon-operator}
\end{equation}
in the PBW ordering of the $q$-fermions. The color generators of $U^{\mathrm{L}}_q(\fsl_N)$ annihilate $B^{\dagger}_{i}$,
\begin{equation}
  E^{ab}_{\mathrm{col}} (B^{\dagger}_{i}) = 0,
  \label{eq:baryon-color-invariance}
\end{equation}
as follows from the color action, Equation~\eqref{eq:q-fermion-site-cartan-action}, and the coproduct of Equation~\eqref{eq:q-fermion-coproduct}, so the baryon is gauge invariant. It carries the charge $Q=N$ under $U(1)_F$, $[F,B^{\dagger}_{i}]=NB^{\dagger}_{i}$, and therefore moves between charge sectors without changing the $SU(N)_k$ labels,
\begin{equation}
  B^{\dagger}_{i} :\;
  \mathcal{H}_{(\alpha, \beta)}(Q)
  \longrightarrow
  \mathcal{H}_{(\alpha, \beta)}(Q+N),
  \label{eq:baryon-open}
\end{equation}
\begin{equation}
  B^{\dagger}_{i} :\;
  \mathcal{H}_{\mathrm{per}}(Q)
  \longrightarrow
  \mathcal{H}_{\mathrm{per}}(Q+N) .
  \label{eq:baryon-periodic}
\end{equation}

The action of $B_i^\dagger$ is simplest in the fusion-tree basis. Let $r_i\in\{0,\ldots,N\}$ denote the filling at site $i$, and define
\begin{equation}
    F_{<i}
    :=
    \sum_{j<i}r_j,
    \qquad
    \chi_i
    :=
    \left((-1)^Nq^{-1}\right)^{F_{<i}} ,
  \label{eq:baryon-pbw-string}
\end{equation}
where $\chi_i$ arises from restoring the PBW order after left multiplication by $B_i^\dagger=\bqpsi_i^1\cdots\bqpsi_i^N$. On the orthonormal fusion-tree basis,
\begin{equation}
    B_i^\dagger
    \lvert\ldots,r_i,\ldots\rangle
    =
    \begin{cases}
        \chi_i\,
        \lvert\ldots,N,\ldots\rangle,
        & r_i=0,\\
        0,
        & r_i\neq0,
    \end{cases}
    \label{eq:baryon-creation-fusion-tree-action}
\end{equation}
so the baryon changes the local occupation from $r_i=0$ to $r_i=N$, up to the phase $\chi_i$, and annihilates states with any other local filling. The $SU(N)_k$ fusion labels are unchanged as $B_i^\dagger$ is a $U^{\mathrm{L}}_q(\fsl_N)$ singlet, so it leaves the integrability of the fusion tree unchanged. Therefore it preserves the radical,\footnote{Notice that although the baryon is built from $q$-fermion creation operators that do not themselves descend to the physical quotient, the baryon does descend. We note that gauge invariance, $[\mathcal{O},U^{\mathrm{L}}_q(\fsl_N)]=0$, and descent to the physical quotient are distinct conditions. For example the color generators descend as operators in the color-extended quotient $\widehat{\mathcal{H}}^{\mathrm{phys}}$ without being gauge invariant.}
\begin{equation}
    B_i^\dagger(\mathcal{R}_k)
    \subseteq
    \mathcal{R}_k .
  \label{eq:baryon-radical-invariance}
\end{equation}
It therefore defines an operator on the physical Hilbert space, with the physical Hermitian adjoint $B_i:=\bigl(B_i^\dagger\bigr)^{\dagger_{\mathrm{phys}}}$. Since $B_i^\dagger$ preserves the intermediate fusion labels its adjoint is simply,
\begin{equation}
    B_i
    \lvert\ldots,r_i,\ldots\rangle
    =
    \begin{cases}
        \chi_i^{-1}\,
        \lvert\ldots,0,\ldots\rangle,
        & r_i=N,\\
        0,
        & r_i\neq N,
    \end{cases}
    \label{eq:baryon-annihilation-fusion-tree-action}
\end{equation}
where we used $|q|=1$, so that $\chi_i^{*}=\chi_i^{-1}$. It follows immediately that
\begin{equation}
    \left(B_i^\dagger\right)^2=0,
    \qquad
    B_i^2=0.
    \label{eq:baryon-nilpotency}
\end{equation}
At distinct sites the baryons obey the braided exchange relations,
\begin{equation}
    B_i^\dagger B_j^\dagger
    =
    (-1)^N
    q^{-N\text{sgn}(i-j)}
    B_j^\dagger B_i^\dagger,
    \qquad i\neq j,
    \label{eq:baryon-creation-creation-exchange}
\end{equation}
\begin{equation}
    B_iB_j
    =
    (-1)^N
    q^{-N\text{sgn}(i-j)}
    B_jB_i,
    \qquad i\neq j,
    \label{eq:baryon-annihilation-annihilation-exchange}
\end{equation}
\begin{equation}
    B_iB_j^\dagger
    =
    (-1)^N
    q^{N\text{sgn}(i-j)}
    B_j^\dagger B_i,
    \qquad i\neq j,
    \label{eq:baryon-annihilation-creation-exchange}
\end{equation}
\begin{equation}
    B_i^\dagger B_j
    =
    (-1)^N
    q^{N\text{sgn}(i-j)}
    B_jB_i^\dagger,
    \qquad i\neq j.
    \label{eq:baryon-creation-annihilation-exchange}
\end{equation}
For $N=1$ these reduce to the relations of the $U(1)$ $q$-fermions of Appendix~\ref{app:u1-chain}, as in Equation~\eqref{eq:u1-exterior-relations}.

Since $B_i^\dagger$ and $B_i$ carry the string $\chi_i$, which depends on all fillings to the left of site $i$, they look nonlocal on the $q$-fermion Fock space. It is more useful to define the \textit{on-site baryon} by dressing $B^{\dagger}_i$ with the string $q^{F_{<i}}=\prod_{j<i}L_j$,
\begin{equation}
 \mathsf B_i^\dagger:=q^{F_{<i}}B_i^\dagger,
 \qquad
 \mathsf B_i:=\bigl(\mathsf B_i^\dagger\bigr)^{\dagger_{\mathrm{phys}}}=B_i\,q^{-F_{<i}} .
 \label{eq:local-determinant-junction}
\end{equation}
By Equations~\eqref{eq:baryon-creation-fusion-tree-action} and~\eqref{eq:baryon-annihilation-fusion-tree-action}, the string cancels the powers of $q$ in $\chi_i$,
\begin{equation}
 \mathsf B_i^\dagger\lvert\ldots,0,\ldots\rangle=(-1)^{NF_{<i}}\lvert\ldots,N,\ldots\rangle,
 \qquad
 \mathsf B_i\lvert\ldots,N,\ldots\rangle=(-1)^{NF_{<i}}\lvert\ldots,0,\ldots\rangle,
 \label{eq:onsite-baryon-action}
\end{equation}
with all other local fillings annihilated and all fusion labels fixed. For even $N$ the sign is one, and for odd $N$ it is the usual fermion-ordering sign. In place of Equations~\eqref{eq:baryon-creation-creation-exchange}--\eqref{eq:baryon-creation-annihilation-exchange}, the on-site baryons therefore obey the graded relations
\begin{equation}
\begin{aligned}
 \mathsf B_i^\dagger\mathsf B_j^\dagger&=(-1)^N\mathsf B_j^\dagger\mathsf B_i^\dagger,
 &\qquad
 \mathsf B_i\mathsf B_j&=(-1)^N\mathsf B_j\mathsf B_i,
 \\
 \mathsf B_i\mathsf B_j^\dagger&=(-1)^N\mathsf B_j^\dagger\mathsf B_i,
 &\qquad
 \mathsf B_i^\dagger\mathsf B_j&=(-1)^N\mathsf B_j\mathsf B_i^\dagger,
 \qquad i\neq j .
\end{aligned}
\label{eq:onsite-baryon-exchange}
\end{equation}
With the site projector $\bP_i^{(r)}:=\bP_i^{(\omega_r)}$ onto the local filling $r_i=r$, as in Equation~\eqref{eq:qf-hubbard-projector}, we have
\begin{equation}
 \bigl(\mathsf B_i^\dagger\bigr)^2=\mathsf B_i^2=0,
 \qquad
 \mathsf B_i\mathsf B_i^\dagger=\bP_i^{(0)},
 \qquad
 \mathsf B_i^\dagger\mathsf B_i=\bP_i^{(N)},
 \label{eq:baryon-same-site-products}
\end{equation}
so that
\begin{equation}
 \bigl\{\mathsf B_i,\mathsf B_i^\dagger\bigr\}=\bP_i^{(0)}+\bP_i^{(N)},
 \qquad
 \bigl[\mathsf B_i,\mathsf B_i^\dagger\bigr]=\bP_i^{(0)}-\bP_i^{(N)} .
 \label{eq:baryon-same-site-brackets}
\end{equation}
With the Cartan generators $L_j=q^{n_j}$ and the fermion number $F$,
\begin{equation}
 L_j\mathsf B_i^\dagger=q^{N\delta_{ij}}\mathsf B_i^\dagger L_j,
 \qquad
 L_j\mathsf B_i=q^{-N\delta_{ij}}\mathsf B_iL_j,
 \qquad
 [F,\mathsf B_i^\dagger]=N\mathsf B_i^\dagger,
 \qquad
 [F,\mathsf B_i]=-N\mathsf B_i .
 \label{eq:cartan-baryon-exchange}
\end{equation}

The algebra of the baryons with the mesons does not close on the on-site baryons alone. Commuting a meson through $B_\ell^\dagger$ moves one constituent to another site and gives the \textit{polarized baryons}
\begin{equation}
  B_{r,s}^\dagger
  :=
  \frac{1}{[N]_q}\sum_{a=1}^{N}
  \begin{cases}
  (-1)^{a-1}q^{N-a}\,\bqpsi_r^a\,\bqpsi_s^1\cdots\widehat{\bqpsi_s^a}\cdots\bqpsi_s^N, & r\leq s,\\[4pt]
  (-1)^{N-a}q^{a-1}\,\bqpsi_s^1\cdots\widehat{\bqpsi_s^a}\cdots\bqpsi_s^N\,\bqpsi_r^a, & r>s,
  \end{cases}
  \label{eq:oriented-polarized-baryon}
\end{equation}
where the hat denotes the omitted factor and $B_{s,s}^\dagger=B_s^\dagger$. For the simple roots, the relations of Appendix~\ref{app:uqglm-relations} give
\begin{equation}
\begin{aligned}
  \relax [E_i,B_\ell^\dagger]
  &=
  [N]_q\,\delta_{\ell,i+1}\,
  B_{i,i+1}^\dagger K_i,
  \\
  [F_i,B_\ell^\dagger]
  &=
  [N]_q\,\delta_{\ell i}\,
  B_{i+1,i}^\dagger
  +
  \left[
  q^{-N(\delta_{i\ell}-\delta_{i+1,\ell})}-1
  \right]
  B_\ell^\dagger F_i , \\
  [L_s,B_\ell^\dagger] &= (q^{N\delta_{s\ell}}-1)B_\ell^\dagger L_s .
\end{aligned}
\label{eq:simple-root-baryon-commutators}
\end{equation}
By quantum skew Howe duality they form a single multiplet, $U^{\mathrm{L}}_q(\fgl_m)\triangleright B^{\dagger}_{j}\cong\text{Sym}^{N}_{q}(\bC_q^m)$ \cite{Cautis2014:WebsQuantumSkew}.

\section{Derivations for the \texorpdfstring{$q$}{q}-fermion chain}
\label{app:derivations}

This appendix collects the explicit derivations used in Sections~\ref{sec:q-fermion-chain}, \ref{sec:continuum-limit} and~\ref{sec:mott-projection}.

\subsection{The \texorpdfstring{$q$}{q}-fermion anticommutation relations}
\label{app:q-fermion-braided-antisymmetry}
Following \cite{Cautis2014:WebsQuantumSkew}, we derive the braided $q$-fermion anticommutation relations in Equation~\eqref{eq:q-fermion-braided-antisymmetry}. We begin by recalling the free-fermion case in Equation~\eqref{eq:canonical-anticommutation}. Let $V\simeq \bC^m$ be the fundamental representation of $U(\fsl_m)$, with basis $\{\bpsi_i\}_{i=1}^m$. $U(\fsl_m)$ has the coproduct
\begin{equation}
    \Delta(x)=x\otimes\mathbbm{1}+\mathbbm{1}\otimes x ,
  \label{eq:exterior-classical-coproduct}
\end{equation}
which is cocommutative:
\begin{equation}
\label{eq:q-fermion-coproduct-fundamental}
    \Delta^{\mathrm{op}}(x)=\Delta(x).
\end{equation}
Consequently, the permutation operator $\tau_{V,W}:V\otimes W\rightarrow W\otimes V$, defined by
\begin{equation}
\tau_{V,W}(v\otimes w)=w\otimes v
  \label{eq:exterior-permutation-map}
\end{equation}
is a well-defined $U(\fsl_m)$-module map.\footnote{With the trivial universal $R$-matrix $\mathcal{R}=\mathbbm{1}\otimes\mathbbm{1}$, the braid operator on $V\otimes V$ is simply
\begin{equation}
    \check R_{V,V}
    =\tau\circ(\rho_V\otimes\rho_V)(\mathcal{R})
    =\tau.
  \label{eq:exterior-classical-braiding}
\end{equation}}
Moreover, $(\tau + \mathbbm{1})(\tau - \mathbbm{1})=0$ so as a $U(\fsl_m)$-module, the tensor product $V\otimes V$ decomposes as
\begin{equation}
    V\otimes V=S^2V\oplus\Lambda^2V .
  \label{eq:exterior-classical-tensor-square}
\end{equation}
Schur's lemma implies that $\tau$ acts by a scalar on each irreducible summand. Hence
\begin{equation}
    \tau=\Pi_S-\Pi_\Lambda,
  \label{eq:exterior-permutation-projector-decomposition}
\end{equation}
where
\begin{equation}
    \Pi_S=\tfrac12\bigl(\mathbbm{1}+\tau\bigr),
    \qquad
    \Pi_\Lambda=\tfrac12\bigl(\mathbbm{1}-\tau\bigr)
  \label{eq:exterior-classical-projectors}
\end{equation}
are the projectors onto the symmetric and antisymmetric subspaces. These subspaces are spanned, respectively, by
\begin{equation}
\begin{aligned}
\Pi_S(\bpsi_i \otimes \bpsi_j)&=\frac{1}{2}(\bpsi_i\otimes\bpsi_j+\bpsi_j\otimes\bpsi_i) \quad (i\leq j) , \\
    \Pi_\Lambda(\bpsi_i \otimes \bpsi_j)&=\frac{1}{2}(\bpsi_i\otimes\bpsi_j-\bpsi_j\otimes\bpsi_i) \quad (i<j) .
\end{aligned}
  \label{eq:exterior-classical-projected-basis}
\end{equation}
The fermionic Fock space is constructed as the quotient of the tensor algebra by the symmetric quadratic subspace:
\begin{equation}
    \Lambda^\bullet V := T(V)\Big/\big\langle S^2V\big\rangle .
  \label{eq:exterior-classical-quotient}
\end{equation}
For $U_q(\fsl_m)$, in contrast, the coproduct is no longer cocommutative; see Appendix~\ref{app:uqglm-relations}. To obtain a braiding operator that is a module map, we need to choose a nontrivial universal $R$-matrix $\mathcal{R} \in U_q(\fsl_m)\otimes U_q(\fsl_m)$ satisfying
\begin{equation}
  \mathcal{R}\Delta(x)=\Delta^{\mathrm{op}}(x)\mathcal{R} ,
  \label{eq:exterior-universal-r-matrix-intertwiner}
\end{equation}
where the $R$-matrix of $U_q(\fsl_m)$ in the fundamental representation is \cite{Jimbo1986:qAnalogueHeckeYangBaxter}
\begin{equation}
(\rho_V \otimes \rho_V )(\mathcal{R})=q\sum_{i=1}^{m}E_{ii}\otimes E_{ii}
 +\sum_{i\neq j}E_{ii}\otimes E_{jj}
 +(q-q^{-1})\sum_{i<j}E_{ij}\otimes E_{ji},
  \label{eq:exterior-fundamental-r-matrix}
\end{equation}
where, as in Equation~\eqref{eq:q-fermion-site-cartan-action}, $E_{ij}(\bqpsi_k)=\delta_{jk}\bqpsi_i$. The braid operator on $V \otimes V$ is then given by
\begin{equation}
    \check R_{V,V} = \tau \circ (\rho_V \otimes \rho_V)(\mathcal{R}) ,
  \label{eq:exterior-quantum-braiding}
\end{equation}
which satisfies the Hecke relation\footnote{$\check R_{V,V} = \check R$.}
\begin{equation}
    (\check R-q\,\mathbbm{1})
    (\check R+q^{-1}\mathbbm{1})=0.
  \label{eq:exterior-hecke-relation}
\end{equation}
As in the free fermion case, $V \otimes V = S_q^2V \oplus \Lambda_q^2V$. The $q$-fermion Fock space is constructed as the quotient of the tensor algebra by the $q$-symmetric quadratic subspace, that is by imposing the relations
\begin{equation}
  (\check R+q^{-1}\mathbbm{1})(\bqpsi_k\otimes\bqpsi_\ell)
  =q^{\delta_{k\ell}}\,\bqpsi_\ell\otimes\bqpsi_k
  +\bigl[q^{-1}+(q-q^{-1})\,\mathbf 1_{\ell<k}\bigr]\,\bqpsi_k\otimes\bqpsi_\ell
  =0 .
  \label{eq:q-fermion-braided-antisymmetry-fundamental}
\end{equation}
Upon imposing Equation~\eqref{eq:q-fermion-braided-antisymmetry-fundamental} and writing $s_{ij}:=\text{sgn}(i-j)$, so that $\mathbf 1_{\ell<k}=\tfrac12(1-s_{\ell k}-\delta_{\ell k})$, we obtain the $q$-fermion anticommutation relations
\begin{equation}
  \bqpsi_i \otimes\bqpsi_j + q^{-s_{ij}}\,\bqpsi_j\otimes\bqpsi_i = 0 .
  \label{eq:one-color-braided-antisymmetry}
\end{equation}
The same argument can be generalized to the case of multiple colors. For example, $V = \bC_q^N \otimes \bC_q^m$ is a fundamental representation of $U_q(\fsl_N) \times U_q(\fsl_m)$ spanned by $\bqpsi^{a}_{i}$ for $a=1,2,\ldots,N$ and $i=1,2,\ldots,m$. The $q$-symmetric quadratic subspace decomposes as \cite{BerensteinZwicknagl2008:BraidedSymmetricExterior}
\begin{equation}
S_q^2(V)
\simeq
\bigl(S_q^2(\bC_q^N)\otimes S_q^2(\bC_q^m)\bigr)
\oplus
\bigl(\Lambda_q^2(\bC_q^N)\otimes\Lambda_q^2(\bC_q^m)\bigr) .
\label{eq:multicolor-quadratic-decomposition}
\end{equation}
In the basis $\bqpsi^{a}_{i}=\bqpsi^a\otimes\bqpsi_i$, the two spaces are spanned by
\begin{equation}
\begin{split}
  S_q^2(\bC_q^N)
  & =\text{span}\bigl\{
    \bqpsi^a\otimes\bqpsi^a,\;
    \bqpsi^a\otimes\bqpsi^b+q\,\bqpsi^b\otimes\bqpsi^a
  \bigr\},\\
  \Lambda_q^2(\bC_q^N)
  &=\text{span}\bigl\{
    q\,\bqpsi^a\otimes\bqpsi^b-\bqpsi^b\otimes\bqpsi^a
  \bigr\} ,
\end{split}
\label{eq:multicolor-quadratic-squares}
\end{equation}
where $a < b$, and analogously for $\bC_q^m$ with $\bqpsi_i$. Consider first the generators where either the color or the site index is repeated, that is\ $a=b$ or $i=j$. Since $\Lambda_q^2$ contains no element with a repeated index, only the first summand of Equation~\eqref{eq:multicolor-quadratic-decomposition} contributes, and one obtains the familiar relations,
\begin{equation}
\begin{split}
  \bqpsi^{a}_{i}\otimes\bqpsi^{a}_{i}
  &=0,\\
  \bqpsi^{a}_{i}\otimes\bqpsi^{a}_{j}
  &=-q^{-s_{ij}}\,\bqpsi^{a}_{j}\otimes\bqpsi^{a}_{i},
  \qquad i\neq j,\\
  \bqpsi^{a}_{i}\otimes\bqpsi^{b}_{i}
  &=-q^{-s_{ab}}\,\bqpsi^{b}_{i}\otimes\bqpsi^{a}_{i},
  \qquad a\neq b .
\end{split}
\label{eq:multicolor-repeated-index-relations}
\end{equation}
The remaining case, in which the color and site indices both differ, is not immediate: both summands of Equation~\eqref{eq:multicolor-quadratic-decomposition} now contribute. Fix $a<b$ and $i<j$, and abbreviate
\begin{equation}
  A=\bqpsi^{a}_{i}\otimes\bqpsi^{b}_{j},
  \quad
  B=\bqpsi^{a}_{j}\otimes\bqpsi^{b}_{i},
  \quad
  C=\bqpsi^{b}_{i}\otimes\bqpsi^{a}_{j},
  \quad
  D=\bqpsi^{b}_{j}\otimes\bqpsi^{a}_{i}.
  \label{eq:multicolor-distinct-index-basis}
\end{equation}
Expanding the two summands of Equation~\eqref{eq:multicolor-quadratic-decomposition} with Equation~\eqref{eq:multicolor-quadratic-squares}, and identifying $(\bqpsi^a\otimes\bqpsi^b)\otimes(\bqpsi_i\otimes\bqpsi_j)$ with $\bqpsi^{a}_{i}\otimes\bqpsi^{b}_{j}$, gives the two generators of $S_q^2(V)$,
\begin{equation}
A+q(B+C)+q^2D,
  \qquad
  q^2A-q(B+C)+D.
  \label{eq:multicolor-symmetric-quadratic-generators}
\end{equation}
Assuming $1+q^2\neq0$, their sum and difference give the relations
\begin{equation}
\begin{aligned}
\bqpsi^{a}_{i}\otimes\bqpsi^{b}_{j}
   &=- \bqpsi^{b}_{j}\otimes\bqpsi^{a}_{i} ,
   \\
  \bqpsi^{a}_{j}\otimes\bqpsi^{b}_{i}
  &=- \bqpsi^{b}_{i}\otimes\bqpsi^{a}_{j}
   + (q-q^{-1})\,\bqpsi^{a}_{i}\otimes\bqpsi^{b}_{j} ,
\end{aligned}
\label{eq:multicolor-quadratic-relations}
\end{equation}
for $a<b,\ i<j$. Putting together all the relations in Equations~\eqref{eq:multicolor-repeated-index-relations} and \eqref{eq:multicolor-quadratic-relations}, we obtain
Equation~\eqref{eq:multicolor-braided-antisymmetry}.
\begin{equation}
\begin{aligned}
\bqpsi^{a}_{i}\otimes\bqpsi^{b}_{j}
 ={}&-
 \bigl(1-\delta_{ij}\delta_{ab}\bigr)
 q^{-s_{ij}\delta_{ab}-s_{ab}\delta_{ij}}
 \bqpsi^{b}_{j}\otimes\bqpsi^{a}_{i}
 \\
 &+\frac{q-q^{-1}}{2}
 \Bigl[
 s_{ij}(1-\delta_{ab})
 -s_{ab}(1-\delta_{ij})
 \Bigr]
 \bqpsi^{a}_{j}\otimes\bqpsi^{b}_{i}.
\end{aligned}
\label{eq:multicolor-braided-antisymmetry}
\end{equation}

\subsection{Consistency checks for the junction factor}
\label{app:junction-factor-weights}
This section checks some consistency conditions for the junction factor conjectured in Equation~\eqref{eq:suN-root-product-junction-factor}.

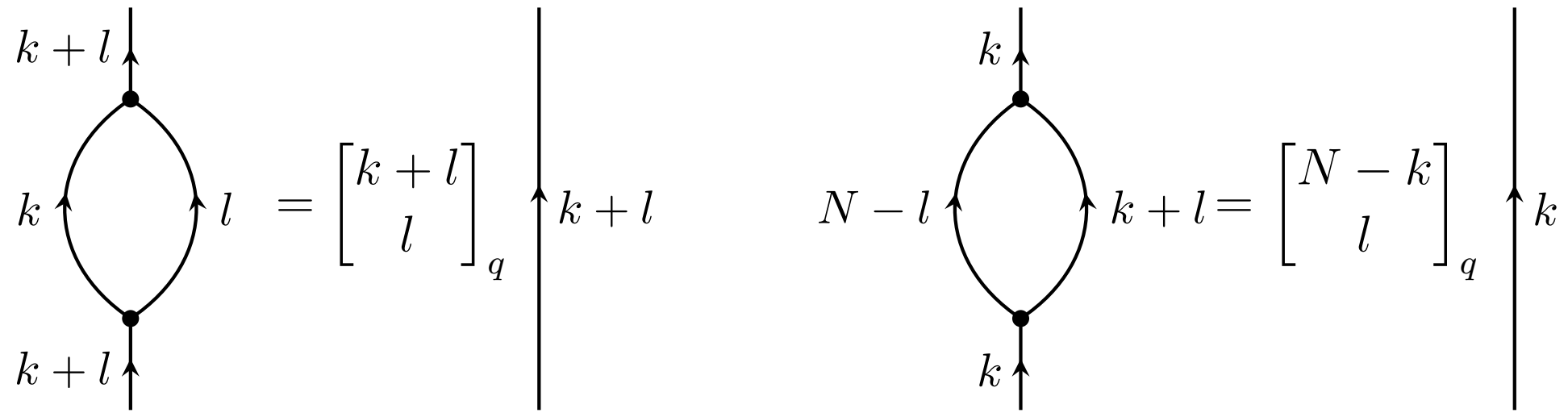
\begin{figure}[!htbp]
\centering
\begin{tikzpicture}[>=stealth,
  web/.style={thick},
  arr/.style={postaction={decorate,decoration={markings,
    mark=at position 0.56 with {\arrow{>}}}}},
  vtx/.style={circle,fill,inner sep=1.4pt},
  lab/.style={font=\small}
]
  \begin{scope}[xshift=0cm]
    \coordinate (b) at (0,0);
    \coordinate (t) at (0,1.8);
    \draw[web,arr] (0,-0.75) -- (b);
    \draw[web,arr] (b) .. controls (-0.72,0.48) and (-0.72,1.32) .. (t);
    \draw[web,arr] (b) .. controls ( 0.72,0.48) and ( 0.72,1.32) .. (t);
    \draw[web,arr] (t) -- (0,2.55);
    \node[vtx] at (b) {};
    \node[vtx] at (t) {};
    \node[lab,left] at (-0.58,0.90) {$k$};
    \node[lab,right] at ( 0.58,0.90) {$l$};
    \node[lab,left] at (0,-0.42) {$k+l$};
    \node[lab,left] at (0, 2.22) {$k+l$};
    \node at (1.35,0.90) {$=$};
    \node at (2.35,0.90)
      {$\displaystyle\begin{bmatrix}k+l\\ l\end{bmatrix}_{q}$};
    \draw[web,arr] (3.35,-0.75) -- (3.35,2.55);
    \node[lab,right] at (3.35,0.90) {$k+l$};
  \end{scope}
  \begin{scope}[xshift=7.3cm]
    \coordinate (b) at (0,0);
    \coordinate (t) at (0,1.8);
    \draw[web,arr] (0,-0.75) -- (b);
    \draw[web,arr] (b) .. controls (-0.72,0.48) and (-0.72,1.32) .. (t);
    \draw[web,arr] (b) .. controls ( 0.72,0.48) and ( 0.72,1.32) .. (t);
    \draw[web,arr] (t) -- (0,2.55);
    \node[vtx] at (b) {};
    \node[vtx] at (t) {};
    \node[lab,left] at (-0.58,0.90) {$N - l$};
    \node[lab,right] at ( 0.58,0.90) {$k+l$};
    \node[lab,left] at (0,-0.42) {$k$};
    \node[lab,left] at (0, 2.22) {$k$};
    \node at (1.75,0.90) {$=$};
    \node at (2.90,0.90)
      {$\displaystyle\begin{bmatrix}N-k\\ l\end{bmatrix}_{q}$};
    \draw[web,arr] (4.05,-0.75) -- (4.05,2.55);
    \node[lab,right] at (4.05,0.90) {$k$};
  \end{scope}
\end{tikzpicture}
\caption{Elementary bubble removal relations of \cite{Cautis2014:WebsQuantumSkew}.}
\label{fig:ckm-bigon-removal}
\end{figure}

\paragraph{Bubble removal.}
Equation~\eqref{eq:suN-root-product-junction-factor} reproduces the two bubble removal relations of \cite{Cautis2014:WebsQuantumSkew}, drawn in Figure~\ref{fig:ckm-bigon-removal}. With $\mu=\lambda+\nu_S$ as in Equation~\eqref{eq:path-qfermion-step-normalization}, we use the pairings
\begin{equation}
 \langle\rho,\alpha_{ij}^{\vee}\rangle=j-i,
 \quad
 \langle\omega_r,\alpha_{ij}^{\vee}\rangle
 =
 \begin{cases}
 1, & i\leq r<j,\\
 0, & \text{otherwise},
 \end{cases}
 \quad
 \delta_{ij}:=\langle\nu_S,\alpha_{ij}^{\vee}\rangle=\mathbf{1}_S(i)-\mathbf{1}_S(j).
 \label{eq:type-A-root-pairings}
\end{equation}
For the first bubble, $\lambda=\omega_k$ and $\mu=\omega_{k+l}$ with $k+l\leq N$, so $S=\{k+1,\ldots,k+l\}$. The pairs with $\delta_{ij}=-1$, $i\leq k<j\leq k+l$, have $a_{ij}=j-i+1$, and those with $\delta_{ij}=+1$, $k<i\leq k+l<j$, have $a_{ij}=j-i$, so their factor $[a_{ij}]_q/[j-i]_q$ is $1$. Thus
\begin{equation}
 \mathcal{K}_{\omega_k,\omega_l}^{\omega_{k+l}}
 =
 \prod_{i=1}^{k}
 \prod_{j=k+1}^{k+l}
 \frac{[j-i+1]_q}{[j-i]_q}
 =
 \prod_{s=1}^{l}\frac{[k+s]_q}{[s]_q}
 =
 \begin{bmatrix}
 k+l\\ l
 \end{bmatrix}_{q}.
 \label{eq:root-product-first-ckm-bigon}
\end{equation}
For the second bubble, $\lambda=\omega_{k+l}$ and $\mu=\omega_k$, so $S=\{1,\ldots,k\}\cup\{k+l+1,\ldots,N\}$ has $N-l$ elements. Now the pairs with $\delta_{ij}=-1$, $k<i\leq k+l<j$, have $a_{ij}=j-i+1$, and those with $\delta_{ij}=+1$, $i\leq k<j\leq k+l$, have $a_{ij}=j-i$, so their factor is again $1$. Hence
\begin{equation}
 \mathcal{K}_{\omega_{k+l},\omega_{N-l}}^{\omega_k}
 =
 \prod_{i=k+1}^{k+l}
 \prod_{j=k+l+1}^{N}
 \frac{[j-i+1]_q}{[j-i]_q}
 =
 \prod_{s=1}^{l}
 \frac{[N-k-s+1]_q}{[l-s+1]_q}
 =
 \begin{bmatrix}
 N-k\\ l
 \end{bmatrix}_{q}.
 \label{eq:root-product-second-ckm-bigon}
\end{equation}

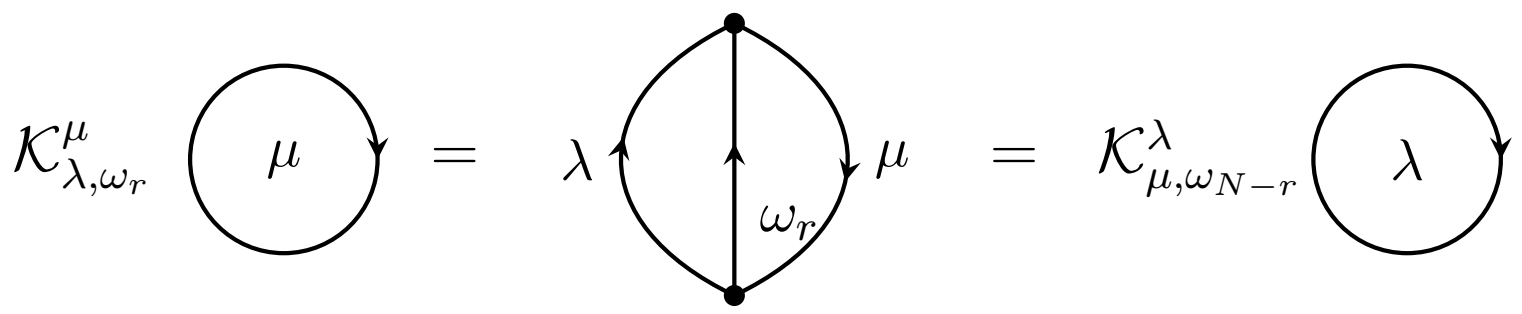
\begin{figure}[!htbp]
\centering
\begin{tikzpicture}[>=stealth,
  web/.style={thick},
  arr/.style={postaction={decorate,decoration={markings,
    mark=at position 0.56 with {\arrow{>}}}}},
  loop/.style={postaction={decorate,decoration={markings,
    mark=at position 0.25 with {\arrow{>}}}}},
  vtx/.style={circle,fill,inner sep=1.4pt},
  lab/.style={font=\small}
]
  \node[anchor=east] at (-3.70,0.90) {$\mathcal{K}_{\lambda,\omega_r}^{\mu}$};
  \draw[web,loop] (-2.98,1.52) arc[start angle=90, end angle=-270, radius=0.62];
  \node[lab] at (-2.98,0.90) {$\mu$};
  \node at (-1.85,0.90) {$=$};
  \coordinate (b) at (0,0);
  \coordinate (t) at (0,1.8);
  \draw[web,arr] (b) .. controls (-1.0,0.48) and (-1.0,1.32) .. (t);
  \draw[web,arr] (b) -- (t);
  \draw[web,arr] (t) .. controls ( 1.0,1.32) and ( 1.0,0.48) .. (b);
  \node[vtx] at (b) {};
  \node[vtx] at (t) {};
  \node[lab,left] at (-0.78,0.90) {$\lambda$};
  \node[lab,right] at (0.02,0.48) {$\omega_r$};
  \node[lab,right] at (0.78,0.90) {$\mu$};
  \node at (1.85,0.90) {$=$};
  \node[anchor=west] at (2.25,0.90) {$\mathcal{K}_{\mu,\omega_{N-r}}^{\lambda}$};
  \draw[web,loop] (4.45,1.52) arc[start angle=90, end angle=-270, radius=0.62];
  \node[lab] at (4.45,0.90) {$\lambda$};
\end{tikzpicture}
\caption{The theta graph, evaluated by removing the bubble of $\lambda$ and $\omega_r$ (left) or of $\mu$ and $\omega_r$ (right). Each loop gives a quantum dimension, Figure~\figref{fig:bubble-diagram}{b}, so the two sides are Equation~\eqref{eq:theta-graph-consistency}.}
\label{fig:theta-graph}
\end{figure}

\paragraph{The theta graph.}
The theta graph with edges $\lambda$, $\omega_r$ and $\mu$ in Figure~\ref{fig:theta-graph} can be evaluated by removing either the bubble of $\lambda$ and $\omega_r$ or that of $\mu$ and $\omega_r$, and both must agree. The first gives $\mathcal{K}_{\lambda,\omega_r}^{\mu}\,d_\mu$. The second uses the reversed step $\mu \otimes \omega_{N-r} \to\lambda$, which adds $\nu_{S^c}=-\nu_S$, a weight of $\mathcal{L}_{\omega_{N-r}}$, and gives $\mathcal{K}_{\mu,\omega_{N-r}}^{\lambda}\,d_\lambda$. Therefore we have the identity,
\begin{equation}
 \mathcal{K}_{\lambda,\omega_r}^{\mu}\,d_{\mu}
 =
 \mathcal{K}_{\mu,\omega_{N-r}}^{\lambda}\,d_{\lambda} .
 \label{eq:theta-graph-consistency}
\end{equation}
Both junction factors run over the pairs with $\delta_{ij}\neq0$ and have the same denominators $[j-i]_q$. They differ only in their numerators, which are $[a_{ij}]_q$ for the step from $\lambda$ and $[\langle\mu+\rho,\alpha_{ij}^\vee\rangle]_q=[a_{ij}+\delta_{ij}]_q$ for the step from $\mu$. Hence
\begin{equation}
 \frac{\mathcal{K}_{\mu,\omega_{N-r}}^{\lambda}}{\mathcal{K}_{\lambda,\omega_r}^{\mu}}
 =
 \prod_{\substack{1\leq i<j\leq N\\ \delta_{ij}\neq0}}\frac{[a_{ij}+\delta_{ij}]_q}{[a_{ij}]_q}
 =
 \frac{d_\mu}{d_\lambda},
 \label{eq:theta-graph-dimension-ratio}
\end{equation}
by the quantum Weyl dimension formula, Equation~\eqref{eq:suN-quantum-Weyl-dimension}.

\subsection{The site action from \texorpdfstring{$F$}{F}-moves and bubble removal}
\label{app:f-move-action}
\begin{figure}[!htbp]
\centering
\begin{tikzpicture}[
  x=0.8cm,y=0.8cm,baseline=0,
  line width=0.55pt,
  strand/.style={postaction={decorate},decoration={markings,
    mark=at position 0.53 with {\arrow{Stealth[length=1.5mm,width=1.2mm]}}}},
  every node/.style={font=\small,inner sep=2pt}
]
\node at (-0.7,3.55) {(a)};
\begin{scope}[shift={(4.2,2.9)}]
  \draw[strand] (0,0)--(0.9,0);
  \draw[strand] (0.9,0)--(2.1,0);
  \draw[strand] (2.1,0)--(3.0,0);
  \draw[strand] (0.9,1.2)--(0.9,0);
  \draw[strand] (2.1,1.2)--(2.1,0);
  \node[below] at (.45,0) {$u$};
  \node[below] at (1.5,0) {$e$};
  \node[below] at (2.55,0) {$z$};
  \node[above] at (0.9,1.2) {$v$};
  \node[above] at (2.1,1.2) {$w$};
\end{scope}
\node at (9.0,3.45) {$=\displaystyle\sum_f\,[F^{uvw}_z]_{ef}$};
\begin{scope}[shift={(10.8,2.9)}]
  \draw[strand] (0,0)--(1.5,0);
  \draw[strand] (1.5,0)--(3.0,0);
  \draw[strand] (0.9,1.2)--(1.5,.6);
  \draw[strand] (2.1,1.2)--(1.5,.6);
  \draw[strand] (1.5,.6)--(1.5,0);
  \node[below] at (.75,0) {$u$};
  \node[below] at (2.25,0) {$z$};
  \node[above] at (0.9,1.2) {$v$};
  \node[above] at (2.1,1.2) {$w$};
  \node[right] at (1.5,.3) {$f$};
\end{scope}
\node at (-0.7,.55) {(b)};
\begin{scope}
  \draw[strand] (0,0)--(0.9,0);
  \draw[strand] (0.9,0)--(2.7,0);
  \draw[strand] (2.7,0)--(3.6,0);
  \draw[strand] (0.9,1.35)--(0.9,.65);
  \draw[strand] (0.9,.65)--(0.9,0);
  \draw[strand] (2.7,1.35)--(2.7,.65);
  \draw[strand] (2.7,.65)--(2.7,0);
  \draw[strand] (0.9,.65)--(2.7,.65);
  \node[below] at (.45,0) {$a$};
  \node[below] at (1.8,0) {$c$};
  \node[below] at (3.15,0) {$b$};
  \node[left] at (0.9,.3) {$r$};
  \node[right] at (2.7,.3) {$s$};
  \node[above] at (0.9,1.35) {$r+1$};
  \node[above] at (2.7,1.35) {$s-1$};
  \node[above] at (1.8,.65) {$E_j$};
\end{scope}
\node at (4.2,.55) {$\xrightarrow{\;F\;}$};
\begin{scope}[shift={(4.8,0)}]
  \draw[strand] (0,0)--(0.9,0);
  \draw[strand] (0.9,0)--(1.35,0);
  \draw[strand] (1.35,0)--(2.7,0);
  \draw[strand] (2.7,0)--(3.6,0);
  \draw[strand] (0.9,1.35)--(0.9,0);
  \draw[strand] (1.35,0) .. controls (1.35,.5) and (1.85,.65) .. (2.7,.65);
  \draw[strand] (2.7,1.35)--(2.7,.65);
  \draw[strand] (2.7,.65)--(2.7,0);
  \node[below] at (.45,0) {$a$};
  \node[below] at (1.12,0) {$c'$};
  \node[below] at (2.02,0) {$c$};
  \node[below] at (3.15,0) {$b$};
  \node[right] at (2.7,.3) {$s$};
  \node[above] at (0.9,1.35) {$r+1$};
  \node[above] at (2.7,1.35) {$s-1$};
\end{scope}
\node at (9.0,.55) {$\xrightarrow{\;F^{-1}\;}$};
\begin{scope}[shift={(9.6,0)}]
  \draw[strand] (0,0)--(0.9,0);
  \draw[strand] (0.9,0)--(1.35,0);
  \draw[strand] (1.35,0)--(2.25,0);
  \draw[strand] (2.25,0)--(2.7,0);
  \draw[strand] (2.7,0)--(3.6,0);
  \draw[strand] (0.9,1.35)--(0.9,0);
  \draw[strand] (2.7,1.35)--(2.7,0);
  \draw[strand] (1.35,0) .. controls (1.35,.62) and (2.25,.62) .. (2.25,0);
  \node[below] at (.45,0) {$a$};
  \node[below] at (1.12,0) {$c'$};
  \node[below] at (1.8,0) {$c$};
  \node[below] at (2.48,0) {$c''$};
  \node[below] at (3.15,0) {$b$};
  \node[above] at (1.8,.47) {$1$};
  \node[above] at (0.9,1.35) {$r+1$};
  \node[above] at (2.7,1.35) {$s-1$};
\end{scope}
\node at (13.8,.55) {$\xrightarrow{\;\mathcal K\;}$};
\begin{scope}[shift={(14.4,0)}]
  \draw[strand] (0,0)--(0.9,0);
  \draw[strand] (0.9,0)--(2.7,0);
  \draw[strand] (2.7,0)--(3.6,0);
  \draw[strand] (0.9,1.35)--(0.9,0);
  \draw[strand] (2.7,1.35)--(2.7,0);
  \node[below] at (.45,0) {$a$};
  \node[below] at (1.8,0) {$c'$};
  \node[below] at (3.15,0) {$b$};
  \node[above] at (0.9,1.35) {$r+1$};
  \node[above] at (2.7,1.35) {$s-1$};
\end{scope}
\end{tikzpicture}
\caption{(a) The $F$-move. (b) The rung $E_j$ is removed by $F$-moves at sites $j$ and $j+1$, which slide its ends onto the intermediate line, and by closing the bubble there with $\mathcal{K}^{c'}_{c,\omega_1}$. The sums over $c'$ and $c''$ are understood.}
\label{fig:qf-two-F-moves}
\end{figure}
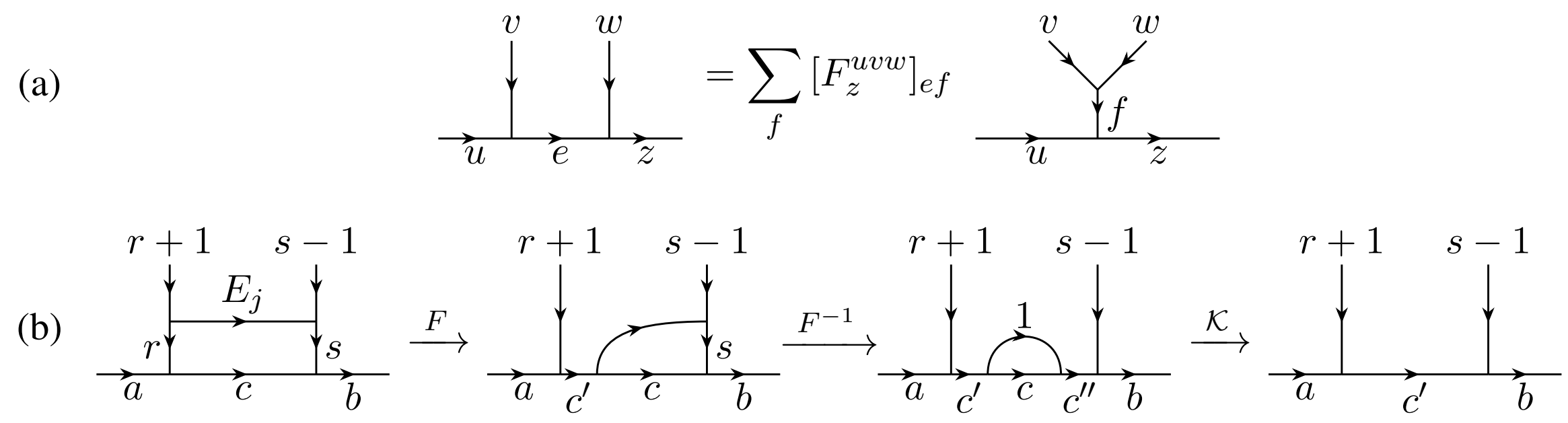

The action of the generators of $U^{\mathrm{L}}_q(\fgl_m)$ on the fusion trees can also be obtained graphically, using the $F$-matrices of $\cC_{N,k}$. An $F$-move changes the basis of fusion trees by rebracketing the tensor product, as in Figure~\figref{fig:qf-two-F-moves}{a}. The gauge of the $F$-matrices has to be chosen so that they reproduce the action of Section~\ref{sec:path-to-qfermion}. With this choice and the junction normalization of Section~\ref{sec:orthonormalization-qfermion-basis}, the steps of Figure~\figref{fig:qf-two-F-moves}{b} give the action of $E_j$,
\begin{equation}
\begin{aligned}
  E_j\,\lvert\cdots;a,r;c,s;b;\cdots\rangle
  &=\sum_{c'}e_{c'c}\,\lvert\cdots;a,r+1;c',s-1;b;\cdots\rangle ,\\
  e_{c'c}
  &=[F^{a\,\omega_r\,\omega_1}_{c'}]_{c\,\omega_{r+1}}\,\frac{\mathcal{K}^{\omega_{r+1}}_{\omega_r,\omega_1}}{\mathcal{K}^{c'}_{c,\omega_1}}\;[(F^{c\,\omega_1\,\omega_{s-1}}_{b})^{-1}]_{\omega_s c'}\;\mathcal{K}^{c'}_{c,\omega_1}\\
  &=[r+1]_q\,[F^{a\,\omega_r\,\omega_1}_{c'}]_{c\,\omega_{r+1}}\,[(F^{c\,\omega_1\,\omega_{s-1}}_{b})^{-1}]_{\omega_s c'} ,
\end{aligned}
\label{eq:f-move-E-action}
\end{equation}
where the first step of Figure~\figref{fig:qf-two-F-moves}{b} removes the bubble at site $j$, giving $\mathcal{K}^{\omega_{r+1}}_{\omega_r,\omega_1}=[r+1]_q$ of Equation~\eqref{eq:root-product-first-ckm-bigon}, and uses a completeness relation on the intermediate line, giving $1/\mathcal{K}^{c'}_{c,\omega_1}$, which the closing bubble $\mathcal{K}^{c'}_{c,\omega_1}$ of the last step cancels. The action of $F_j$ follows in the same way.

\subsection{The adjoint of the \texorpdfstring{$q$}{q}-fermion creation operator}
\label{sec:non-existence-of-annihilation-operator}
As alluded to in the main text, the $q$-fermion creation operators acting by left multiplication, for $N\geq2$ and $m\geq k+1$, do not descend to the physical Hilbert space at the root of unity $q=\exp(\pi i/(k+N))$, as shown below, and therefore have no adjoints on that quotient. Products whose lengths are multiples of $N$ \textit{can} become well-defined operators on the physical Hilbert space, for example the baryon operators of Appendix~\ref{app:baryon-algebra}.

In this subsection we give an explicit example of a null state that a $q$-fermion creation operator sends to a non-null state. Note that descending to the physical Hilbert space and being gauge invariant are two different conditions: an operator can descend to the physical Hilbert space $\widehat{\mathcal{H}}^{\mathrm{phys}}$ without being gauge invariant, as $E^a_{\mathrm{col}}$ does.

Consider $N\geq2$ and $m\geq k+1$ and the state
\begin{equation}
 \nu_{N,k}:=\bigl(\bqpsi_1^1\bqpsi_1^2\cdots\bqpsi_1^{N-1}\bigr)\bqpsi_2^1\bqpsi_3^1\cdots\bqpsi_{k+1}^1|0\rangle .
 \label{eq:general-sun-null-state}
\end{equation}
Its highest-weight fusion path, $0\to\omega_{N-1}\to\omega_{N-1}+\omega_1\to\cdots\to\omega_{N-1}+k\omega_1$, ends at a weight of level $k+1$, outside the alcove $P_k^+$, so $\nu_{N,k}\in\mathcal{R}_k$. Acting with $\bqpsi_1^N$ and using the same-site exchange relation of Equation~\eqref{eq:q-fermion-braided-antisymmetry},
\begin{equation}
 \bqpsi_1^N\nu_{N,k}=\left(-q^{-1}\right)^{N-1}\bigl(\bqpsi_1^1\bqpsi_1^2\cdots\bqpsi_1^N\bigr)\bqpsi_2^1\cdots\bqpsi_{k+1}^1|0\rangle .
 \label{eq:general-sun-creator-on-null}
\end{equation}
The first site now carries the determinant representation $\Lambda_q^N\bC_q^N\simeq\mathbbm{1}$, so the fusion path becomes $0\to0\to\omega_1\to\cdots\to k\omega_1$, which stays inside $P_k^+$, and this state has nonzero norm. Hence, $\bqpsi_1^N$ does not preserve $\mathcal{R}_k$. The same holds for every creator $\bqpsi_i^a$, acting by left multiplication. The action of Equation~\eqref{eq:q-fermion-site-cartan-action} and the coproduct of Equation~\eqref{eq:q-fermion-coproduct}, together with their color analogues, give
\begin{equation}
 \bqpsi_i^a=\bigl[E_i,\bqpsi_{i+1}^a\bigr]K_i^{-1},
 \qquad
 \bqpsi_i^{b+1}=F^b_{\mathrm{col}}\,\bqpsi_i^b-q^{-1}\,\bqpsi_i^b\,F^b_{\mathrm{col}} .
 \label{eq:creator-ladder-relations}
\end{equation}
The site generators preserve $\mathcal{R}_k$ by Equation~\eqref{eq:q-fermion-radical-invariance}, and so do the color generators, which descend to $\widehat{\mathcal{H}}^{\mathrm{phys}}$. If some $\bqpsi_i^a$ preserved $\mathcal{R}_k$, applying the second relation $N-a$ times and then the first $i-1$ times would show that $\bqpsi_1^N$ preserves $\mathcal{R}_k$, a contradiction. Hence no $\bqpsi_i^a$ descends to the physical Hilbert space. Without this adjoint, norms cannot be computed as for free fermions. There, for example, we have relations such as
\begin{equation}
 \|c^\dagger|0\rangle\|^2=\langle0|cc^\dagger|0\rangle=\langle0|(1-c^\dagger c)|0\rangle=\langle0|0\rangle ,
 \label{eq:ff-vacuum-norm-contraction}
\end{equation}
and in general the anticommutation relations move every annihilation operator onto the vacuum, so that every norm reduces to the norm of the vacuum. For $q$-fermions this is not possible. The norms must instead be computed from the fusion trees, as in Section~\ref{sec:orthonormalization-qfermion-basis}.

\subsection{Quantum-group generators in terms of canonical fermions}
\label{app:canonical-fermion-generators}
Motivated by the $U(1)_k$ chain of Appendix~\ref{app:u1-chain}, where the $q$-fermions have a simple expression in terms of canonical fermions, one can ask whether the quantum-group generators of the general $SU(N)_k$ case admit a similar expression. Such an expression was written down, in different conventions, in \cite{FrauMonteiroSciuto1994:qDeformedLieAnyonicRealization} and more recently in \cite{DalipiFelder2022:HoweDualityDynamicalWeyl}. We summarize their results here and point out a subtlety in the $*$-structure that arises when the $q$-fermion problem is expressed in terms of canonical fermions.

Let $\mathcal{H}$ be the Fock space of the canonical fermions $c_i^a$ of Equations~\eqref{eq:canonical-anticommutation} and~\eqref{eq:ff-fock-space}. We define the linear map $\iota:\Lambda_q^\bullet(\bC_q^N\otimes\bC_q^m)\to\mathcal{H}$ on the PBW basis of Equation~\eqref{eq:q-fermion-pbw-basis} by
\begin{equation}
 \iota\bigl(\bqpsi_{i_1}^{a_1}\otimes\bqpsi_{i_2}^{a_2}\otimes\cdots\otimes\bqpsi_{i_r}^{a_r}|0\rangle\bigr)
 =c_{i_1}^{a_1\dagger}c_{i_2}^{a_2\dagger}\cdots c_{i_r}^{a_r\dagger}|0\rangle ,
 \qquad
 (i_1,a_1)<(i_2,a_2)<\cdots<(i_r,a_r) .
 \label{eq:canonical-pbw-map}
\end{equation}
Since $\iota$ maps a basis to a basis, it is an isomorphism of vector spaces, consistent with the classical dimension. It is the analog of the map of Equation~\eqref{eq:u1-fock-identification}. Note that $\iota\bqpsi_i^a\iota^{-1}$ is in general not $c_i^{a\dagger}$ but, as in Equation~\eqref{eq:u1-free-fermion-string}, a dressed expression, which for $N\geq2$ can also contain color-exchanging terms.

Transporting the action of Equations~\eqref{eq:q-fermion-site-cartan-action} and~\eqref{eq:q-fermion-coproduct} with $\iota$, we find
\begin{equation}
\begin{aligned}
 \iota E_i\iota^{-1}&=\sum_{a=1}^{N}c_i^{a\dagger}c_{i+1}^{a}\,q^{\sum_{b>a}(n_{i+1}^{b}-n_i^{b})},
 &\qquad
 \iota L_j\iota^{-1}&=q^{n_j},
 \\
 \iota F_i\iota^{-1}&=\sum_{a=1}^{N}c_{i+1}^{a\dagger}c_i^{a}\,q^{\sum_{b<a}(n_i^{b}-n_{i+1}^{b})},
 &
 \iota K_i\iota^{-1}&=q^{n_i-n_{i+1}},
\end{aligned}
 \label{eq:canonical-site-generators}
\end{equation}
for $i=1,\ldots,m-1$ and $j=1,\ldots,m$, where $n_i^a:=c_i^{a\dagger}c_i^a$ and $n_i=\sum_a n_i^a$ as in Equation~\eqref{eq:ff-EF}. Notice that $c_i^{a\dagger}c_{i+1}^a$ is dressed by the colors $b>a$ in $E_i$ while $c_{i+1}^{a\dagger}c_i^a$ in $F_i$ is dressed by $b<a$.\footnote{For $N=1$ these generators reduce to those of Equation~\eqref{eq:u1-free-fermion-generators}, and at $q=1$ the simple-root generators become the free-fermion bilinears of Equation~\eqref{eq:ff-EF}.} One can check that the operators of Equation~\eqref{eq:canonical-site-generators} satisfy the relations of $U^{\mathrm{L}}_q(\fgl_m)$.

Similarly, the color generators of Section~\ref{sec:q-fock-space} can be written in terms of the canonical fermions as
\begin{equation}
\begin{aligned}
 \iota E^a_{\mathrm{col}}\iota^{-1}&=\sum_{j=1}^{m}c_j^{a\dagger}c_j^{a+1}\,q^{\sum_{l>j}(n_l^{a}-n_l^{a+1})},
 \\
 \iota F^a_{\mathrm{col}}\iota^{-1}&=\sum_{j=1}^{m}c_j^{(a+1)\dagger}c_j^{a}\,q^{\sum_{l<j}(n_l^{a+1}-n_l^{a})},
 \\
 \iota K^a_{\mathrm{col}}\iota^{-1}&=q^{\sum_{j=1}^{m}(n_j^{a}-n_j^{a+1})},
\end{aligned}
 \label{eq:canonical-color-generators}
\end{equation}
for $a=1,\ldots,N-1$. Using the canonical anticommutation relations of Equation~\eqref{eq:canonical-anticommutation}, one can show that the site and color generators of Equations~\eqref{eq:canonical-site-generators} and~\eqref{eq:canonical-color-generators} commute \cite{Aboumrad2022:SkewHoweDualityClifford}. The site generators written with canonical fermions are therefore gauge invariant under the color quantum group $U^{\mathrm{L}}_q(\fsl_N)$.

Now consider $|q|=1$, so that $\bar q=q^{-1}$. The \textit{canonical adjoint}, defined by the Hermitian structure $(c_i^a)^*=c_i^{a\dagger}$ of Section~\ref{sec:free-fermion-review}, gives
\begin{equation}
 \bigl(\iota E_i\iota^{-1}\bigr)^\dagger
 =\sum_{a=1}^{N}c_{i+1}^{a\dagger}c_i^{a}\,q^{\sum_{b>a}(n_i^{b}-n_{i+1}^{b})} ,
 \label{eq:canonical-site-adjoint}
\end{equation}
which differs from $\iota F_i\iota^{-1}$ of Equation~\eqref{eq:canonical-site-generators} in the range of the dressing, $b>a$ instead of $b<a$. This highlights the important point that, for $N\geq2$ and $q=\exp(\pi i/(k+N))$, the canonical adjoint does not realize the $*$-structure of Equation~\eqref{eq:path-qfermion-hermiticity-condition}.\footnote{For $N=1$ the canonical adjoint does give the correct $*$-structure, as Equation~\eqref{eq:u1-free-fermion-generators} shows.}

Another way to establish that the canonical adjoint is not the physical one is to note that the inner product compatible with it would be that of $\mathcal{H}$ pulled back by $\iota$, $\langle\xi,\eta\rangle_c:=\langle\iota\xi|\iota\eta\rangle$. As $\iota$ maps the PBW basis to an orthonormal basis of $\mathcal{H}$, every PBW state would then have unit norm, contradicting the existence of null states for $N\geq2$ and $m>k$, see Section~\ref{sec:fusion-tree-basis}.
We stress that, although the expressions in terms of canonical fermions do not capture the null states, they can still be used for perturbative estimates in the parameter $\theta=\pi/(k+N)$, as we do in Appendix~\ref{app:charge-radius-perturbation}.

\subsection{The charge radius in perturbation theory}
\label{app:charge-radius-perturbation}
Using the canonical-fermion expressions in Equation~\eqref{eq:canonical-site-generators} for the Hamiltonian~\eqref{eq:periodic-hamiltonian}, we estimate the change in the compact boson radius $\tilde R$ perturbatively in $\theta=\pi/(k+N)$, with $q=e^{i\theta}$. To begin, we write $H_c$ of Equation~\eqref{eq:ff-marginal-hamiltonian} in terms of the density $\delta n=J_L+J_R$ and the current $j_0=J_R-J_L$,

\begin{equation}
 H_c=\frac{\pi}{2N}\int dx\,\bigl[v_N(\delta n)^2+v_J\,j_0^2\bigr],
 \qquad
 v_N=u_c+\frac{Ng_c}{2\pi},
 \qquad
 v_J=u_c-\frac{Ng_c}{2\pi} .
 \label{eq:density-velocity-hamiltonian}
\end{equation}

Let $\varepsilon(\nu,A)$ be the ground-state energy per site of a ring at filling $\nu=F/m$, with a flux $A$ per bond, which in the free case only affects the $j_0$ term. For a gapless charge sector with Hamiltonian $H_c$ as in Equation~\eqref{eq:density-velocity-hamiltonian}, the curvatures are $B:=\partial^2\varepsilon/\partial\nu^2=\pi v_N/N$ and $D_c:=\partial^2\varepsilon/\partial A^2=Nv_J/\pi$, and Equation~\eqref{eq:ff-charge-radius} gives
\begin{equation}
 \frac{\tilde R^{4}}{N^2}
 =\frac{v_N}{v_J}
 =\frac{2\pi u_c+Ng_c}{2\pi u_c-Ng_c}
 =\frac{N^2}{\pi^2}\frac{B}{D_c}.
 \label{eq:qf-radius-stiffness}
\end{equation}
At the free point, where $u_c=v_F$ and $g_c=0$, we have $B_0=\pi v_F/N$ and $D_0=Nv_F/\pi$, so $\tilde R_0=\sqrt N$.

For the calculation, we take $t>0$ and set the lattice spacing $a_0=1$.\footnote{We also drop $-\mu F$, which only adds a constant at fixed filling.} We suppress $\iota$, define $d_i^a:=n_i^a-n_{i+1}^a$, and conjugate by the diagonal operator $S=e^{i\theta\sum_in_i(n_i-1)/4}$, so that both simple roots carry the same dressing,\footnote{Note that since $S$ is unitary, this still does not contradict the conclusion that $E_i^{\dagger} \neq F_i$ for the canonical adjoint. The perturbative corrections themselves are real.}
\begin{equation}
\begin{aligned}
 SE_iS^{-1}&=\sum_a c_i^{a\dagger}c_{i+1}^{a}\,e^{-i\theta D_i^a/2},
 \\
 SF_iS^{-1}&=\sum_a c_{i+1}^{a\dagger}c_i^{a}\,e^{-i\theta D_i^a/2},
\end{aligned}
 \label{eq:qf-symmetrically-dressed-generators}
\end{equation}
with $D_i^a:=\sum_{b>a}d_i^b-\sum_{b<a}d_i^b$. Since $S$ is unitary, conjugating the Hamiltonian by $S$ does not change the dynamics. Inserting a flux phase $A$ on every bond, we obtain the Hamiltonian
\begin{equation}
 \widehat H(\theta,A)=-t\sum_{i,a}T_i^a(A)\,e^{-i\theta D_i^a/2},
 \qquad
 T_i^a(A)=e^{iA}c_i^{a\dagger}c_{i+1}^{a}+e^{-iA}c_{i+1}^{a\dagger}c_i^{a} .
 \label{eq:qf-perturbative-canonical-hamiltonian}
\end{equation}
Expanding gives $\widehat H(\theta,A)=H_0+\theta H_1+\theta^2H_2+O(\theta^3)$, with $H_0=-t\sum_{i,a}T_i^a(A)$ and
\begin{equation}
 H_1=\frac{it}{2}\sum_{i,a}T_i^a(A)\,D_i^a,
 \qquad
 H_2=\frac{t}{8}\sum_{i,a}T_i^a(A)\,\bigl(D_i^a\bigr)^2 .
 \label{eq:qf-perturbative-hamiltonian-coefficients}
\end{equation}

At $A=0$, the ground state $|\text{G.S}\rangle$ is the free Fermi sea of Equation~\eqref{eq:ff-momentum-space-hamiltonian}, in which each color fills the momenta $|p|<k_F=\pi\nu/N$ of the band $\epsilon(p)=-2t\cos p$. With $z=\sin k_F/\pi$ and $f=\frac{\nu}{N}\bigl(1-\frac{\nu}{N}\bigr)$, its energy density is $\varepsilon_0=-2Nt\,z$, and the Fermi velocity is $v_F=2t\sin k_F$. We now compute the correction to the ground-state energy density $\varepsilon$ via perturbation theory. Since $\langle n_i^a-n_{i+1}^a\rangle_{\text{G.S}}=0$ by translation invariance, $\langle H_1\rangle_{\text{G.S}}=0$, and the leading correction is of order $\theta^2$. Second-order perturbation theory gives\footnote{Here $\dagger$ denotes the canonical adjoint, and $H_1$ is anti-Hermitian: $H_1^{\dagger}=-H_1$.}
\begin{equation}
 E^{(2)}=\langle H_2\rangle_{\text{G.S}}+\sum_{n\neq\text{G.S}}\frac{|\langle n|H_1|\text{G.S}\rangle|^2}{E_n^{(0)}-E_{\text{G.S}}^{(0)}} .
 \label{eq:qf-second-order-energy}
\end{equation}
Let $\varepsilon(\nu,0)=\varepsilon_0+\theta^2\varepsilon_2+O(\theta^4)$, where $\varepsilon_2=\varepsilon_{2,\mathrm{dir}}+\varepsilon_{2,\mathrm{virt}}$ collects the two terms of $E^{(2)}/m$ as $m\to\infty$. Using
\begin{equation}
 \langle n_i^a\rangle_{\text{G.S}}=\frac{\nu}{N},
 \qquad
 \langle c_i^{a\dagger}c_{i+1}^a\rangle_{\text{G.S}}=\frac{\sin k_F}{\pi}=z,
 \qquad
 \langle n_i^an_{i+1}^a\rangle_{\text{G.S}}=\Bigl(\frac{\nu}{N}\Bigr)^2-z^2,
 \label{eq:qf-free-sea-correlators}
\end{equation}
we have
\begin{equation}
 \langle d_i^a\rangle_{\text{G.S}}=0,
 \qquad
 \langle(d_i^a)^2\rangle_{\text{G.S}}=2(f+z^2),
 \qquad
 \langle(D_i^a)^2\rangle_{\text{G.S}}=2(N-1)(f+z^2),
 \label{eq:qf-free-sea-density-fluctuations}
\end{equation}
where the last equality follows because different colors factorize. Since $\langle T_i^a(0)(D_i^a)^2\rangle_{\text{G.S}}$ also factorizes, we have
\begin{equation}
 \varepsilon_{2,\mathrm{dir}}=\lim_{m\to\infty}\frac{\langle H_2\rangle_{\text{G.S}}}{m}=\frac{tN(N-1)}{2}\,z(f+z^2) .
 \label{eq:qf-second-order-direct-energy-density}
\end{equation}
For the virtual term, expanding $H_1$ gives the four-fermion interactions
\begin{equation}
 H_1=\frac{it}{2}\sum_{a<b}\sum_i\bigl(T_i^a(0)\,d_i^b-T_i^b(0)\,d_i^a\bigr).
 \label{eq:qf-perturbative-color-pair-decomposition}
\end{equation}
Acting on the free ground state $|\text{G.S}\rangle$, $H_1$ scatters a fermion of color $a$ from $p$ to $p+Q$ and, on the same bond, a fermion of color $b$ from $p'$ to $p'-Q$. The corresponding intermediate states can be written as
\begin{equation}
 |n\rangle
 =\bigl(c_{p+Q}^{a\dagger}c_p^a\bigr)
  \bigl(c_{p'-Q}^{b\dagger}c_{p'}^b\bigr)|\text{G.S}\rangle,
 \label{eq:qf-virtual-intermediate-state}
\end{equation}
where $p,p'$ are occupied momenta and $p+Q,p'-Q$ are unoccupied momenta. Each state has one particle-hole pair in each of the two colors and zero total momentum relative to $|\text{G.S}\rangle$. Different color pairs give orthogonal intermediate states. Defining $p=P+x$, $p'=P-x$, $p+Q=P+y$ and $p'-Q=P-y$, the matrix element is $\langle n | H_1 | \text{G.S}\rangle = v_1/m$ up to a sign, and the excitation energy is $E^{(0)}_n - E^{(0)}_{\text{G.S}} = \Delta=\epsilon(p+Q)+\epsilon(p'-Q)-\epsilon(p)-\epsilon(p')$, where
\begin{equation}
 v_1=2t\cos P\,(\sin y-\sin x),
 \qquad
 \Delta=4t\cos P\,(\cos x-\cos y) .
 \label{eq:qf-virtual-pair-kinematics}
\end{equation}
Evaluating the momentum integrals over occupied holes, $|P|+|x|\le k_F$, and empty particle states, $k_F+|P|\le|y|\le\pi$, and summing over all $\binom N2$ color pairs gives, for $0<\nu\le N/2$,\footnote{Particle-hole symmetry extends the result to $\nu>N/2$.}
\begin{equation}
 \varepsilon_{2,\mathrm{virt}}
 =\frac{tN(N-1)\sin k_F}{2\pi^3}\bigl[k_F(\pi-k_F)-\sin^2k_F\bigr]
 =\frac{tN(N-1)}{2}\,z(f-z^2).
 \label{eq:qf-second-order-virtual-energy-density}
\end{equation}
From Equations~\eqref{eq:qf-second-order-direct-energy-density} and~\eqref{eq:qf-second-order-virtual-energy-density}, we have
\begin{equation}
 \varepsilon(\nu,0)
 =-\frac{2Nt}{\pi}\sin\left(\frac{\pi\nu}{N}\right)
 \left[1-\frac{(N-1)\theta^2}{2}\,\frac{\nu}{N}\left(1-\frac{\nu}{N}\right)\right]
 +O(\theta^4).
 \label{eq:qf-energy-density}
\end{equation}

For the flux response, we compute the stiffness $D_c=\lim_{m\to\infty}m^{-1}\,\partial_A^2E(A)|_{A=0}$ from the ground-state energy $E(A)$. The phase shift gives $\langle T_i^a(A)\rangle_{\text{G.S}}=2z\cos A$ in the direct term and replaces $\cos P$ by $\cos(P+A)$ in $v_1$ and $\Delta$. In the integral of $v_1^2/\Delta$ over $P$, the odd term proportional to $\sin P$ integrates to zero. We therefore obtain an overall factor of $\cos A$,
\[
\varepsilon(\nu,A)=[\varepsilon_0+\theta^2\varepsilon_2]\cos A+O(\theta^4).
\]
Thus,
\begin{equation}
 D_c
 =\frac{2Nt}{\pi}\sin k_F\,\left[1-\frac{(N-1)\theta^2}{2}\,f\right]+O(\theta^4) .
 \label{eq:qf-perturbative-charge-stiffness}
\end{equation}
Similarly, differentiating Equation~\eqref{eq:qf-energy-density} twice with respect to $\nu$ gives
\begin{equation}
 B=\frac{2\pi t}{N}\sin k_F
 \left[1+(N-1)\theta^2\left(-\frac{f}{2}+\frac{1}{\pi}\left(1-\frac{2\nu}{N}\right)\cot k_F-\frac{1}{\pi^2}\right)\right]+O(\theta^4) .
 \label{eq:qf-perturbative-density-stiffness}
\end{equation}
Hence, Equation~\eqref{eq:qf-radius-stiffness} gives
\begin{equation}
 \frac{\tilde R(\nu)}{\sqrt N}
 =1+\frac{(N-1)\theta^2}{4\pi^2}
 \left[\pi\left(1-\frac{2\nu}{N}\right)\cot\left(\frac{\pi\nu}{N}\right)-1\right]+O(\theta^4).
 \label{eq:qf-perturbative-radius-expansion}
\end{equation}

\subsection{The projector chain at other fillings}
\label{sec:other-fillings}
\begin{figure}[!htbp]
  \centering
  \begin{tikzpicture}[>=stealth,
    trunk/.style={thick},
    leg/.style={thick},
    mid/.style={postaction={decorate, decoration={markings,
      mark=at position 0.58 with {\arrow{>}}}}},
    vtx/.style={circle, fill, inner sep=1.5pt},
    lab/.style={font=\small}
  ]
    \node at (-0.55,0) {$\cdots$};
    \draw[trunk,mid] (-0.15,0) -- (1.20,0);
    \draw[trunk,mid] ( 1.20,0) -- (2.70,0);
    \draw[trunk,mid] ( 2.70,0) -- (4.20,0);
    \draw[trunk,mid] ( 4.20,0) -- (5.55,0);
    \node at (5.95,0) {$\cdots$};

    \foreach \x in {1.20,2.70,4.20}{
      \node[vtx] at (\x,0) {};
      \draw[leg,mid] (\x,1.10) -- (\x,0);
      \node[lab, above] at (\x,1.10) {$\omega_p$};
    }

    \node[lab, below] at (0.52,-0.08) {$\lambda^{(j-1)}$};
    \node[lab, below] at (1.95,-0.08) {$\lambda^{(j)}$};
    \node[lab, below] at (3.45,-0.08) {$\lambda^{(j+1)}$};
    \node[lab, below] at (4.90,-0.08) {$\lambda^{(j+2)}$};
  \end{tikzpicture}
\caption{The low-energy fusion tree for the charging term in Equation~\eqref{eq:mott-p-charging-term}, the generalization of Figure~\ref{fig:mott-fundamental-tree} away from unit filling. Every leg carries the antisymmetric power $\omega_p$ and the horizontal pieces carry the running labels $\lambda^{(j)}\in P^+_k$.}
  \label{fig:mott-p-fusion-tree}
\end{figure}
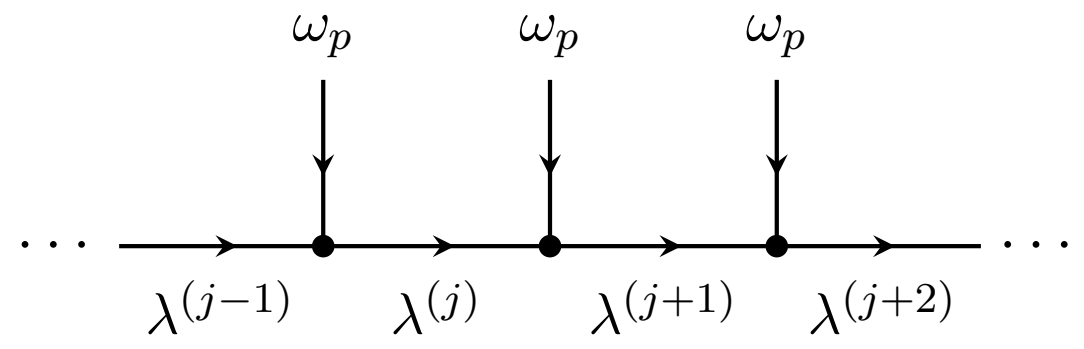

Section~\ref{sec:mott-projection} considers only unit filling, $\nu=1$. More generally, as we saw in Section~\ref{sec:ff-mott}, one can gap out the compact boson mode at any integer filling $\nu \in \bZ$ with $1 \leq \nu \leq N-1$. These choices of $\nu$ correspond to projecting onto the sets of states carrying different antisymmetric powers $\Lambda_q^{r_j}\bC_q^N$, with $r_j = \nu$, and they lead to different and interesting RG flows in the infrared, which we discuss below. Consider then the Mott term that favors the antisymmetric power $\Lambda_q^{p}\bC_q^N=\mathcal{L}_{\omega_p}$ at every site $j$,
\begin{equation}
H^q_{\text{int}}
=
U\sum_{j}(n_j-p)^2,
\label{eq:mott-p-charging-term}
\end{equation}
and define the projector $\bP^{(\omega_p)}$ onto the fusion trees with the local labels shown in Figure~\ref{fig:mott-p-fusion-tree}.

On this degenerate subspace the effective Hamiltonian at second order in perturbation theory is
\begin{equation}
H^{q}_{\mathrm{eff}}
=
-\frac{t^2}{2U}
\sum_{j}\bP^{(\omega_p)}(E_jF_j+F_jE_j)\bP^{(\omega_p)};
\label{eq:mott-p-effective}
\end{equation}
for $p=1$ the intermediate module contains the single channel $\mathcal{L}_{\omega_2}$, giving the single projector $\Pi_{(\omega_2)}$ of Equation~\eqref{eq:mott-effective-hamiltonian}. For $p>1$ several intermediate channels generally contribute, and their relative coefficients can be computed using two-site quantum skew Howe duality. Writing $X_r:=\mathcal{L}_{\omega_r}$, the action of $E_iF_i$ can be described schematically as
\begin{equation}
  X_p\otimes X_p
  \xrightarrow{\ F_i\ }
  X_{p-1}\otimes X_{p+1}
  \xrightarrow{\ E_i\ }
  X_p\otimes X_p .
  \label{eq:mott-p-virtual-hopping-path}
\end{equation}
In the same way as for unit filling, the combined action of $E_iF_i$ factors through several intermediate projector channels, corresponding to the allowed Young diagrams appearing in the fusion of $\Lambda_q^{p}\bC_q^N \otimes \Lambda_q^{p}\bC_q^N$. These are
\begin{equation}
  \lambda_s=\omega_{p-s}+\omega_{p+s},
  \qquad
  0\leq s\leq d:=\min(p,N-p),
  \label{eq:mott-p-fusion-channel-weights}
\end{equation}
Since $E_i$ and $F_i$ commute with the color quantum group, Schur's lemma implies that their composition acts by a scalar $c_s$ in each fusion channel $\lambda_s$:
\begin{equation}
  \left.E_iF_i\right|_{(p,p)}
  =\sum_{s=0}^{d}c_s\,\Pi_i^{(\lambda_s)}.
  \label{eq:mott-p-channel-projectors}
\end{equation}
We emphasize that we sum only over the admissible fusion channels $\lambda_s$, and $\Pi_i^{(\lambda_s)}$ is the orthogonal projector onto the channel $\lambda_s$.

\begin{figure}[!htbp]
  \centering
  \begin{tikzpicture}[>=stealth,
    worldline/.style={very thick},
    dn/.style={postaction={decorate, decoration={markings,
      mark=at position 0.55 with {\arrow{>}}}}},
    rung/.style={thick,
      postaction={decorate, decoration={markings,
        mark=at position 0.58 with {\arrow{>}}}}},
    leg/.style={thick,
      postaction={decorate, decoration={markings,
        mark=at position 0.60 with {\arrow{>}}}}},
    junction/.style={circle, fill, inner sep=1.4pt},
    num/.style={font=\small, inner sep=5pt}
  ]

    \begin{scope}
      \def\W{1.9}
      \foreach \x in {0,\W}{
        \draw[worldline,dn] (\x,1.60) -- (\x,0.60);
        \draw[worldline,dn] (\x,0.60) -- (\x,-0.60);
        \draw[worldline,dn] (\x,-0.60) -- (\x,-1.60);
      }
      \draw[rung] (\W,-0.60) -- (0,-0.60);
      \draw[rung] (0,0.60) -- (\W,0.60);
      \foreach \y in {-0.60,0.60}{
        \node[junction] at (0,\y) {};
        \node[junction] at (\W,\y) {};
      }
      \node[num, above] at ({0.5*\W},-0.54) {$1$};
      \node[num, above] at ({0.5*\W}, 0.66) {$1$};
      \node[num, left] at (0,-1.10) {$p$};
      \node[num, left] at (0, 0) {$p-1$};
      \node[num, left] at (0, 1.10) {$p$};
      \node[num, right] at (\W,-1.10) {$p$};
      \node[num, right] at (\W, 0) {$p+1$};
      \node[num, right] at (\W, 1.10) {$p$};
      \node[num, below] at (0,-1.60) {$i$};
      \node[num, below] at (\W,-1.60) {$i+1$};
    \end{scope}

    \node at (3.40,0) {$=$};
    \node at (4.30,0) {$\displaystyle\sum_{s}\,c_s$};

    \begin{scope}[xshift=6.25cm]
      \draw[leg] (0,-0.60) -- (-0.55,-1.60);
      \draw[leg] (0,-0.60) -- ( 0.55,-1.60);
      \draw[leg] (0,0.60) -- (0,-0.60);
      \draw[leg] (-0.55,1.60) -- (0,0.60);
      \draw[leg] ( 0.55,1.60) -- (0,0.60);
      \node[junction] at (0,-0.60) {};
      \node[junction] at (0, 0.60) {};
      \node[num, left] at (-0.48,-1.28) {$p$};
      \node[num, right] at ( 0.48,-1.28) {$p$};
      \node[num, right] at ( 0.08, 0) {$\lambda_s$};
      \node[num, left] at (-0.48, 1.28) {$p$};
      \node[num, right] at ( 0.48, 1.28) {$p$};
    \end{scope}

  \end{tikzpicture}
\caption{Equation~\eqref{eq:mott-p-channel-projectors} as webs, read upward. On the left the two rungs of label $1$ are $F_i$ and then $E_i$, so between them site $i$ carries $p-1$ and site $i+1$ carries $p+1$. On the right the same operator is resolved one channel at a time, the middle segment carrying $\lambda_s$ and each term weighted by its eigenvalue $c_s$.}
  \label{fig:mott-p-ladder-channels}
\end{figure}
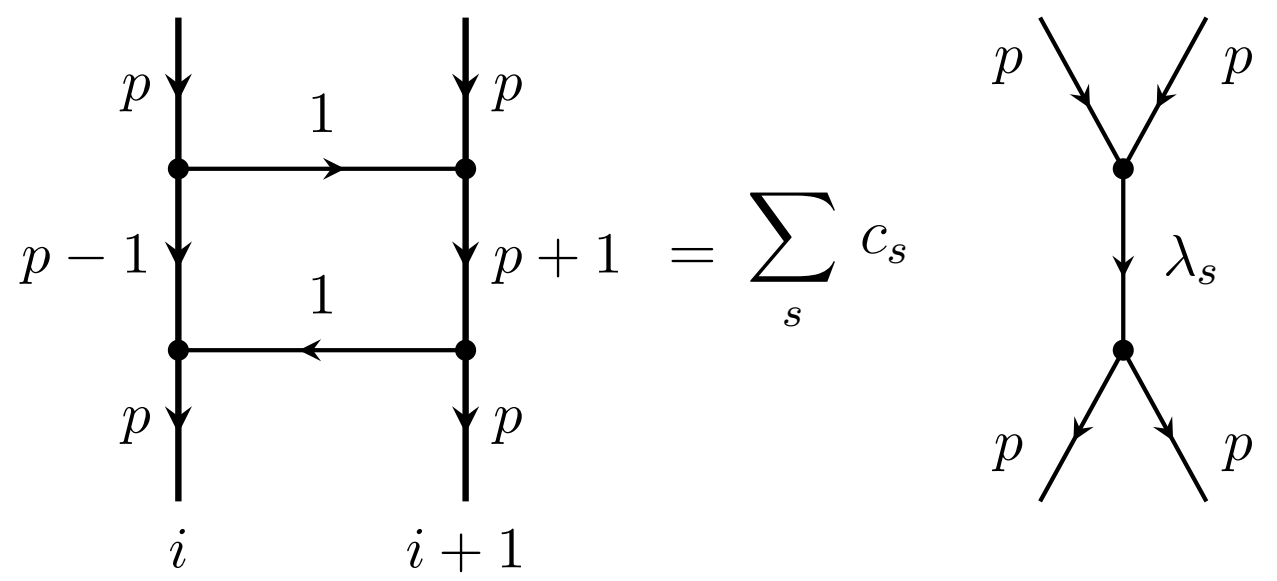

The coefficients $c_s$ can be computed by using two-site quantum skew Howe duality~\cite{Cautis2014:WebsQuantumSkew}. The color representation $\lambda_s$ is paired with the site $U^{\mathrm{L}}_q(\fgl_2)$ representation of highest weight
\begin{equation}
  \lambda_s^{T}=(p+s,p-s),
  \label{eq:mott-p-site-highest-weight}
\end{equation}
whose restriction to $U^{\mathrm{L}}_q(\fsl_2)$ is the spin-$s$ representation, with states $|s,m_z\rangle$. The exact state in this multiplet corresponding to the local $q$-fermion state occupation $(p,p)$ is fixed by the action of $K_i$,
\begin{equation}
\begin{aligned}
K_i|a,p;b,p;c\rangle &= |a,p;b,p;c\rangle, \\
  K_i|s,m_z\rangle &= q^{2m_z}|s,m_z\rangle
\end{aligned}
  \label{eq:mott-p-cartan-weight-matching}
\end{equation}
so it is the weight-zero vector $|s,0\rangle$ that is selected. On that state $E_iF_i$ acts as the Casimir.\footnote{The equality of $E_iF_i$ and $F_iE_i$ is given by the square switch relation of \cite{Cautis2014:WebsQuantumSkew}.}
\begin{equation}
    \left.E_iF_i\right|_{(p,p)}
    =\left.F_iE_i\right|_{(p,p)}
    =\sum_{s=0}^{d}[s]_q[s+1]_q\,\Pi_i^{(\lambda_s)}.
  \label{eq:mott-p-channel-eigenvalues}
\end{equation}
Therefore the effective Hamiltonian is
\begin{equation}
  H^{q}_{\mathrm{eff}}
  =-\frac{t^2}{2U}\sum_i
    \left.(E_iF_i+F_iE_i)\right|_{(p,p)}
  =-\frac{t^2}{U}\sum_i\sum_{s=0}^{\min(p,N-p)}
    [s]_q[s+1]_q\,\Pi_i^{(\lambda_s)}.
  \label{eq:mott-p-channel-expansion}
\end{equation}

At filling $\nu=p$ the infrared theory could also be identified directly in the continuum, as in Section~\ref{sec:mott-projection}, by finding the most relevant operator of the theory $M^{+}_{N,k-1}$ \cite{GaiottoZeng2026:InterfaceMinimalModel} allowed by symmetries that the charging term of Equation~\eqref{eq:mott-p-charging-term} generates and following the RG flow. Comparing the IR theory of this flow with the continuum limit of Equation~\eqref{eq:mott-p-channel-expansion} would test the projector chain away from unit filling, and we leave it for future work.

\FloatBarrier
\section{Numerical results}
\label{app:numerical-methods}

Here we compare the numerical finite-chain spectra with the continuum theories proposed in Sections~\ref{sec:continuum-limit} and~\ref{sec:mott-projection} using exact diagonalization on physical fusion spaces and fusion-path DMRG for larger sizes.

\subsection{Finite-size spectra}
\label{app:methods}

A finite critical chain has discrete excitation energies even when its infinite-volume limit is gapless. A wave traveling around a circle of length $m$ has smallest nonzero momentum $2\pi/m$, and hence an energy of order $2\pi v/m$. Conformal invariance organizes the low-energy states into towers: a primary field gives the lowest state of a tower, and its descendants give the successive long-wavelength excitations. Their energies in units of $2\pi v/m$ measure scaling dimensions.

Our proposed theory has both a charge boson and a neutral minimal model. These modes can propagate with different velocities, $v_c$ and $v_{\mathcal W}$. At fixed fermion number $F=\nu m$, their contributions to a periodic-chain gap are
\begin{equation}
 E_n-E_0=\frac{2\pi}{m}\bigl(v_c x_c+v_{\mathcal W}x_{\mathcal W}\bigr)+o(m^{-1}),
 \label{eq:numerical-energy-momentum-dictionary}
\end{equation}
where $x_c$ and $x_{\mathcal W}$ are the charge and neutral parts of the scaling dimension. Thus a single energy rescaling cannot turn the whole spectrum into scaling dimensions unless the two velocities agree \cite{Poilblanc2012:FractionalizationItinerantAnyons}. We determine the charge scale from the lowest density wave, whose gap we denote by $\Delta E_J=2\pi v_c/m$. A neutral level then fixes the second scale within a proposed minimal model. Throughout we set the lattice spacing and hopping amplitude to one.

\FloatBarrier
\subsection{Periodic spectra}
\label{app:continuum-numerics}

\begin{figure}[!htbp]
 \centering
 \includegraphics[width=\textwidth]{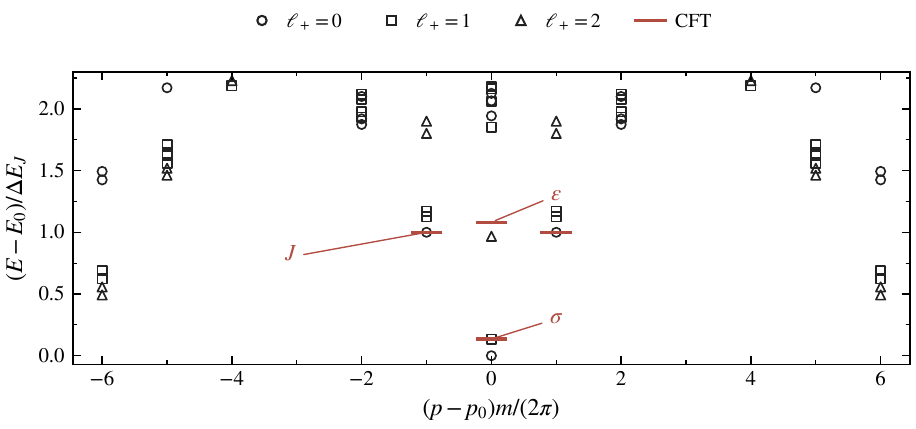}
 \caption{Fixed-charge periodic spectrum of $SU(2)_2$ at $U=0$, $\nu=1$, $m=\NumericsOperatorSize$. Markers distinguish Wilson-line sectors. The red segments mark the CFT predictions for $\sigma$, $\epsilon$ and $J$. $J$ and $\sigma$ gaps calibrate the charge and neutral velocities. Momentum is measured relative to the ground state in units of $2\pi/m$.}
 \label{fig:numerical-operator-spectrum}
\end{figure}

Take $SU(2)_2$ as an example, for which the proposed neutral theory is Ising. Its two nontrivial primary fields are $\sigma$ and $\epsilon$, of dimensions $1/8$ and $1$. Figure~\ref{fig:numerical-operator-spectrum} compares this prediction with the spectrum at unit filling. The density wave $J$ fixes $v_c$, and assigning the lowest neutral gap to $\sigma$ fixes $v_{\mathcal W}$. The position of $\epsilon$ is then a prediction. In charge-current units the neutral energies are $(v_{\mathcal W}/v_c)x_{\mathcal W}$, which explains why they need not sit at the Ising dimensions themselves.

The spectrum also carries the Wilson-line quantum number of Section~\ref{sec:operator-content}. A line surrounding the chain commutes with the Hamiltonian, so its eigenvalue separates states into sectors. The charge current $J$ leaves the neutral vacuum unchanged and has $\ell_+=0$. The neutral $\sigma$ and $\epsilon$ representatives have spin triples $(\tfrac12,0;\tfrac12)$ and $(0,0;1)$ in the coset $SU(2)_1\times SU(2)_1/SU(2)_2$. Their last entries are the exterior spins $j_+$, giving $\ell_+=2j_+=1,2$, respectively.

At $m=12$, the $\epsilon$ candidate is the zero-momentum triangle at $(E_\epsilon-E_0)/\Delta E_J\simeq0.9675$, below the predicted $v_{\mathcal W}/v_c\simeq1.0821$. Rescaling by the neutral velocity gives dimension estimates $0.809,0.860,0.894$ at $m=8,10,12$, consistent with finite-size corrections towards the continuum $x_\epsilon=1$.

The ground-state energy provides a complementary test. Its leading term is extensive, but the finite circle also changes the zero-point energy of the gapless modes. This universal Casimir correction measures their central charges:
\begin{equation}
 E_0(m)=e_\infty m-\frac{\pi (v_c+v_{\mathcal W}c_{\mathcal W})}{6m}+\cdots.
 \label{eq:numerical-casimir}
\end{equation}
The charge boson contributes one unit of central charge. Subtracting $v_c$ and dividing by $\Delta E_1$, the lowest neutral gap at the ground-state momentum, cancels $v_{\mathcal W}$,
\begin{equation}
 \mathcal D=\frac{v_{\mathcal W}c_{\mathcal W}}{m\Delta E_1/(2\pi)}=\frac{c_{\mathcal W}}{x_1}.
 \label{eq:numerical-coset-discriminator}
\end{equation}
For $SU(2)$, $x_1$ is chosen in the observed momentum and Wilson-line sector. For $SU(3)$, we use the smallest scalar dimension in each candidate model. The resulting ratios favor $M^+_{N,k-1}$ over $M^-_{N,k}$ across the tested levels and fillings. We fit $E_0(m)=e_\infty m+b/m+d/m^3$ on consecutive three-size windows and extrapolate the last two $\mathcal D$ estimates linearly in $m_{\max}^{-2}$. Table~\ref{tab:numerical-casimir-ratios} compares the numerical extrapolation of $\mathcal D$ with the two interface predictions.

\begin{table}[!htbp]
 \centering
 \begin{tabular}{@{}ccccc@{}}
\toprule
Chain & $\nu$ & $\mathcal D_{M^+}$ & $\mathcal D_{M^-}$ & $\mathcal D_{\rm num}$ (estimate) \\
\midrule
$SU(2)_{2}$ & $1$ & $4.00$ & $9.33$ & $4.00$ \\
$SU(2)_{3}$ & $1$ & $9.33$ & $16.00$ & $9.53$ \\
$SU(2)_{4}$ & $1$ & $16.00$ & $24.00$ & $16.59$ \\
$SU(3)_{2}$ & $\frac{3}{2}$ & $6.00$ & $13.50$ & $6.06$ \\
\bottomrule
\end{tabular}

 \caption{Casimir ratios $\mathcal D=c_{\mathcal W}/x_1$ for the two candidate interfaces and extrapolated numerical estimates. The latter use even chain lengths $m=8$--$14$, $6$--$12$, $6$--$12$ and $4$--$10$, respectively.}
 \label{tab:numerical-casimir-ratios}
\end{table}

\FloatBarrier
\subsection{Charge radius and stability}

The charge spectrum contains one continuous parameter, the compactification radius $\tilde R$. We determine it by comparing the energy cost of changing the total charge with that of a long-wavelength density wave. The hopping deformation changes the relative cost of density and current fluctuations, so the radius is generally dynamical.

The exception is $N=2$, $\nu=1$, where the enhanced $\eta$-pairing symmetry fixes the gapless radius to $\sqrt2$. The spectrum is unchanged by $\theta\to-\theta$, so the leading shift is quadratic in $\theta=\pi/(k+N)$, as in Equation~\eqref{eq:numerical-radius-reference}.

\begin{figure}[!htbp]
 \centering
 \includegraphics[width=\textwidth]{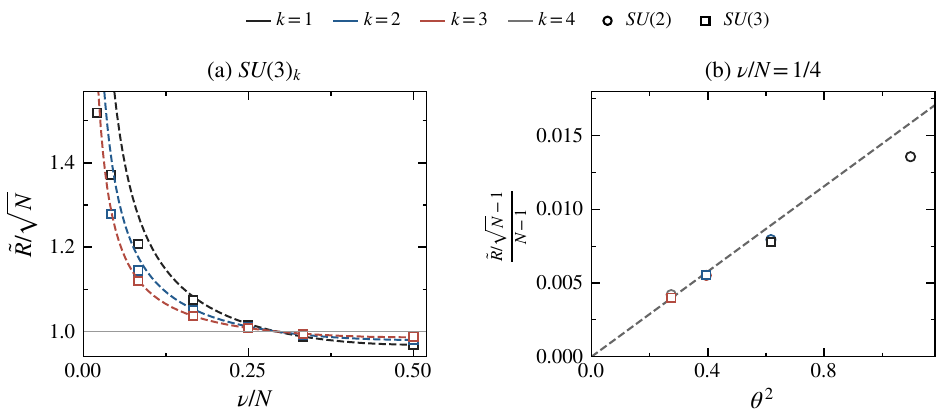}
 \caption{Charge radius at $U=0$. Markers are numerical estimates; dashed curves are the perturbative predictions through order $\theta^2$ from Equation~\eqref{eq:numerical-radius-reference}. (a) Filling dependence for $SU(3)_k$. (b) Scaled radius shifts for $SU(2)_k$ and $SU(3)_k$ at $\nu/N=1/4$.}
 \label{fig:numerical-radius}
\end{figure}

Figure~\ref{fig:numerical-radius} compares the measured radius with the $\theta$ expansion: it lies below $\sqrt N$ near half filling and grows as the chain is diluted. Dividing the shift by $N-1$, the number of other colors, brings the two ranks close to the same predicted $\theta^2$ dependence. At very low density the expansion becomes nonuniform and overestimates the shift.

\begin{figure}[!htbp]
 \centering
 \includegraphics[width=\textwidth]{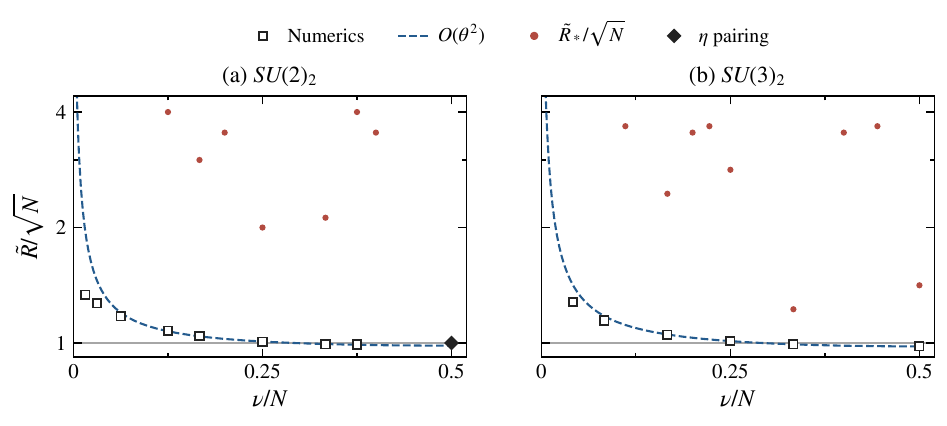}
 \caption{Charge radius and stability at $U=0$ for (a) $SU(2)_2$ and (b) $SU(3)_2$. Squares are numerical radii, dashed curves are the predictions through order $\theta^2$, and red circles are the sufficient thresholds $\tilde R_*$ at rational fillings. The black diamond is the symmetry-fixed marginal point of $SU(2)_2$.}
 \label{fig:numerical-radius-stability}
\end{figure}

A larger radius lowers the dimensions of symmetry-allowed scattering operators, which can then become relevant. Equation~\eqref{eq:qf-critical-radius} gives a sufficient stability threshold $\tilde R_*(\nu)$. The measured radii in Figure~\ref{fig:numerical-radius-stability} remain below these thresholds, supporting stability at the tested nonmarginal points. The symmetry-fixed $SU(2)$ point at $\nu=1$ saturates the bound and remains marginal.

\FloatBarrier
\subsection{Mott limit}
\label{app:mott-numerics}

\begin{figure}[!htbp]
 \centering
 \includegraphics[width=\textwidth]{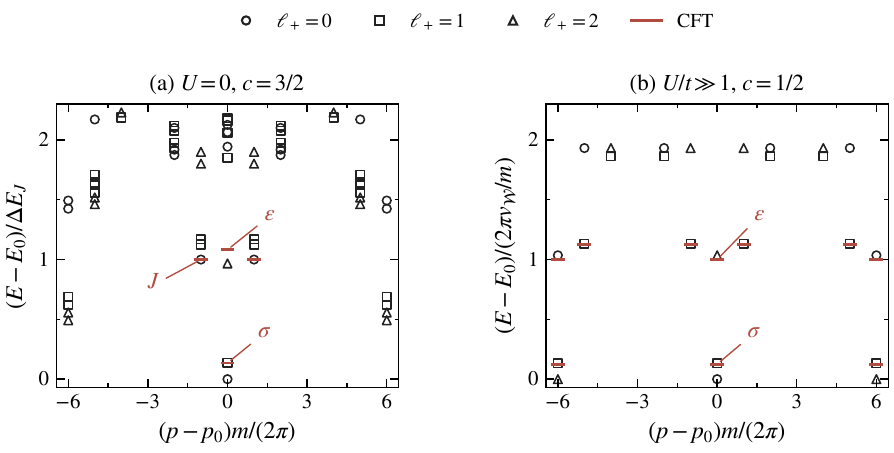}
 \caption{Periodic $SU(2)_2$ chain at unit filling. (a) Charged hopping and (b) neutral projector spectra at $m=12$. Symbols label Wilson-line sectors; red segments are CFT predictions. Energies use the charge-current gap in (a) and the neutral primary--descendant spacing in (b).}
 \label{fig:numerical-mott-spectrum}
\end{figure}

At unit filling, a sufficiently large charging energy freezes the occupation of each site. The charge boson becomes massive, while the fusion channels still fluctuate. The infrared central charge therefore decreases by one, $c=1+c_{\mathcal W}\longrightarrow c=c_{\mathcal W}$. Figure~\ref{fig:numerical-mott-spectrum} compares the charged and neutral limits for $SU(2)_2$. Virtual hopping still mixes fusion channels, with an interaction of order $t^2/U$ described by Equation~\eqref{eq:mott-effective-hamiltonian}. Thus the neutral levels move towards zero in hopping units as $U$ grows, but retain their Ising tower structure when measured in units of their own propagation velocity. The charge-current tower is absent from this low-energy spectrum. The two translated vacua, visible at momenta $0$ and $\pi$ in panel (b), support the same neutral CFT, giving the two copies predicted in Section~\ref{sec:projector-continuum}.

\FloatBarrier

\clearpage
{\hypersetup{urlcolor=black}\bibliography{Refs}}

\end{document}